\documentclass[sigconf, nonacm]{acmart}

\newcommand\vldbdoi{10.14778/3746405.3746429}
\newcommand\vldbpages{3077 - 3092}
\newcommand\vldbvolume{18}
\newcommand\vldbissue{9}
\newcommand\vldbyear{2025}
\newcommand\vldbauthors{\authors}
\newcommand\shorttitle{}
\newcommand\vldbtitle{\shorttitle} 
\newcommand\vldbavailabilityurl{https://github.com/zjuDBSystems/SVBench}
\newcommand\vldbpagestyle{empty} 

\usepackage[a-2b]{pdfx} 
\usepackage{balance}
\usepackage{fontawesome}
\usepackage[colorinlistoftodos,prependcaption,textsize=tiny]{todonotes}
\tikzset{every node/.append style={inner sep=5pt}} 

\usepackage{tcolorbox}
\usepackage{colortbl}
\definecolor{mygray}{gray}{.9}
\definecolor{mypink}{rgb}{.99,.91,.95}
\definecolor{mycyan}{cmyk}{.3,0,0,0}

\usepackage{enumerate}
\usepackage{enumitem}
\usepackage[edges]{forest}
\usepackage{tikz}
\usetikzlibrary{trees}
\tikzset{parent/.style={align=center,text width=0.5cm,rounded corners=3pt},
    child/.style={align=center,text width=1.2cm,rounded corners=3pt},
    grandchild/.style={align=center,text width=1.2cm,rounded corners=3pt}
    }
\usepackage{pifont}
\usepackage{hhline}
\usepackage{multirow}
\usepackage{tabularx}
\usepackage{makecell}
\usepackage{diagbox}
\usepackage{subfigure}
\usepackage{cleveref}
\usepackage{threeparttable}
\usepackage{multicol}
\usepackage{fancyhdr}
\usepackage{amsmath} 
\usepackage[T1]{fontenc}

\AtBeginDocument{\pagestyle{fancy}}

\setitemize[1]{itemsep=0pt,partopsep=0pt,parsep=0pt,topsep=0pt,leftmargin=10pt}

\begin{document}

\title{A Comprehensive Study of Shapley Value in Data Analytics}
\setlength{\tabcolsep}{2pt}

\author[1]{Hong Lin}
\affiliation{%
  \institution{The State Key Laboratory of Blockchain and Data Security, Zhejiang University}
}
\email{honglin@zju.edu.cn}
\authornote{Also affiliated with Hangzhou High-Tech Zone (Binjiang) Institute of Blockchain and Data Security.}

\author[1]{Shixin Wan}
\affiliation{
  \institution{The State Key Laboratory of Blockchain and Data Security, Zhejiang University}
}
\email{wansx@zju.edu.cn}
\authornotemark[1]

\author[1]{Zhongle Xie}
\affiliation{
  \institution{The State Key Laboratory of Blockchain and Data Security, Zhejiang University}
}
\email{xiezl@zju.edu.cn}
\authornotemark[1]
\authornote{Zhongle Xie and Lidan Shou are the corresponding authors.}

\author{Ke Chen}
\affiliation{%
  \institution{The State Key Laboratory of Blockchain and Data Security, Zhejiang University}
}
\email{chenk@zju.edu.cn}
\authornotemark[1]

\author{Meihui Zhang}
\affiliation{%
  \institution{School of Computer Science \& Technology, Beijing Institute of Technology}
}
\email{meihui_zhang@bit.edu.cn}

\author{Lidan Shou}
\affiliation{%
  \institution{The State Key Laboratory of Blockchain and Data Security, Zhejiang University}
}
\email{should@zju.edu.cn}
\authornotemark[1]
\authornotemark[2]

\author{Gang Chen}
\affiliation{%
  \institution{The State Key Laboratory of Blockchain and Data Security, Zhejiang University}
}
\email{cg@zju.edu.cn}
\authornotemark[1]

\begin{abstract}
Over the recent years, Shapley value (SV), a solution concept from cooperative game theory, has found numerous applications in data analytics (DA). This paper presents the first comprehensive study of SV used throughout the DA workflow, clarifying the key variables in defining DA-applicable SV and the essential functionalities that SV can provide for data scientists. We condense four primary challenges of using SV in DA, namely computation efficiency, approximation error, privacy preservation, and interpretability, disentangle the resolution techniques from existing arts in this field, then analyze and discuss the techniques w.r.t. each challenge and the potential conflicts between challenges. We also implement \textit{SVBench}, a modular and extensible open-source framework for developing SV applications in different DA tasks, and conduct extensive evaluations to validate our analyses and discussions. Based on the qualitative and quantitative results, we identify the limitations of current efforts for applying SV to DA and highlight the directions of future research and engineering. 
\end{abstract}

\maketitle

\pagestyle{\vldbpagestyle}
\begingroup\small\noindent\raggedright\textbf{PVLDB Reference Format:}\\
\vldbauthors. \vldbtitle. PVLDB, \vldbvolume(\vldbissue): \vldbpages, \vldbyear.\\
\href{https://doi.org/\vldbdoi}{doi:\vldbdoi}
\endgroup
\begingroup
\renewcommand\thefootnote{}\footnote{\noindent
This work is licensed under the Creative Commons BY-NC-ND 4.0 International License. Visit \url{https://creativecommons.org/licenses/by-nc-nd/4.0/} to view a copy of this license. For any use beyond those covered by this license, obtain permission by emailing \href{mailto:info@vldb.org}{info@vldb.org}. Copyright is held by the owner/author(s). Publication rights licensed to the VLDB Endowment. \\
\raggedright Proceedings of the VLDB Endowment, Vol. \vldbvolume, No. \vldbissue\ %
ISSN 2150-8097. \\
\href{https://doi.org/\vldbdoi}{doi:\vldbdoi} \\
}\addtocounter{footnote}{-1}\endgroup

\ifdefempty{\vldbavailabilityurl}{}{
\vspace{.3cm}
\begingroup\small\noindent\raggedright\textbf{PVLDB Artifact Availability:}\\
The source code, data, and/or other artifacts have been made available at 
\url{https://github.com/zjuDBSystems/SVBench}. 
\endgroup
}

\section{Introduction} \label{sec: intro}

Data analytics (DA), exploring data to mine insightful information for problem-solving, has garnered significant attention in industry and academia over the past few years~\cite{KadhimJawad2022BigDA, Shi2022AdvancesIB}. 
The global DA market was valued at USD 64.99 billion in 2024 and is projected to reach USD 402.70 billion by 2032, with a compound annual growth rate of 25.5\% from 2024 to 2032~\cite{DA_market_size}. 
A typical DA workflow generally follows three key stages: 
(1) \textbf{Data Fabrication} \cite{Hechler2023, data_faric}, encompassing tasks such as \textit{data collection (DC)} \cite{CHENG201631, Azcoitia2022ASO} to identify, retrieve, and transfer data from diverse data sources to the analysis platform, and \textit{data orchestration (DO)} \cite{ZHANG2022101626, data_orchestration} to cleanse and transform data to align with downstream analytical requirements. 
(2) \textbf{Data Exploration}, including \textit{data valuation (DV)} to preserve high-value, refined data, and \textit{data mining (DM)} to uncover patterns and insights using techniques such as Machine Learning (ML). 
(3) \textbf{Result Reporting}, involving \textit{result interpretation (RI)} \cite{Cuzzocrea2023AttributionMA, chen2022explaining, Bertossi2023TheSV} to render the analytical outcomes comprehensible, and \textit{result trading (RT)} to bargain and exchange DA derivatives such as trained ML models in data marketplaces \cite{xu2024model, 10.1145/3299869.3300078}.

\begin{figure}[t]
    \centering
    \includegraphics[width=\linewidth]{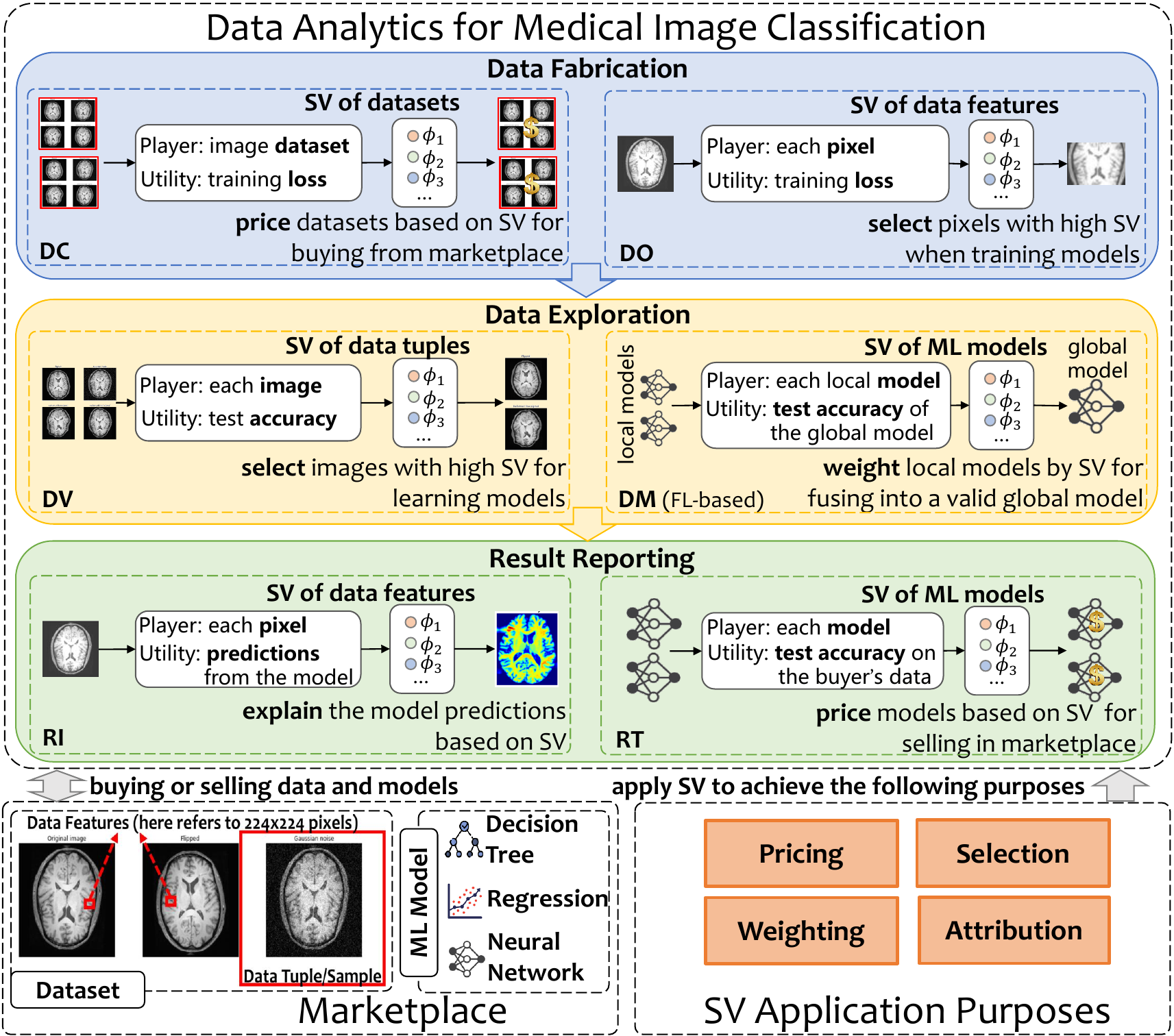}
    \caption{
    An example of using Shapley values ($\phi_1,\phi_2, \phi_3,\cdots$) throughout the data analytics workflow. 
    }
    \label{fig:SV_in_DA}
\end{figure}

Recently, data scientists have applied Shapley value (SV), a method derived from cooperative game theory for fairly distributing the total gains generated by the coalition of all players \cite{shapley1953value}, to numerous tasks throughout the DA workflow. 
\Cref{fig:SV_in_DA} depicts a series of examples applying SV to the tasks analyzing medical images, where the analytical objects include pixels, images, image sets, ML models, etc. 
The application purposes can be summarized into four categories: 
(1) \textbf{pricing}, to determine the net worth of analytical objects for trading, such as buying image datasets in the DC task and selling well-trained models in the RT task; 
(2) \textbf{selection}, to select qualified and important analytical objects for exploration, for example, selecting pixels important to reduce training losses to learn models in the DO task; 
(3) \textbf{weighting}, to assign reasonable weights to analytical objects collected from multiple sources for valid fusion of those objects, for instance, weighting local models collected in the DM task, where federated learning (FL) \cite{McMahan2016CommunicationEfficientLO,liu2022gtg} is used to protect privacy within medical images, for fusing as a valid global model whose accuracy exceeds a certain threshold; 
(4) \textbf{attribution}, to explain data exploration outputs, e.g., explaining how pixels impact the model predictions in the RI task.

To better understand the usage of SV in the DA domain, several surveys have been proposed \cite{ijcai2022p778, sv_related_survey, Li2024ShapleyVF, Bertossi2023TheSV, sim2022data, pei2020survey, tang2020abrief, 10.1007/s10618-024-01016-z}. 
However, they do not cover the entire life cycle of DA, leaving a gap in a holistic guidance for applying SV to diverse DA tasks.
As summarized in Table 1, prior works predominantly concentrated on 
ad-hoc SV implementations, mainly addressing computation efficiency instead of other challenges such as interpretability. 
Furthermore, they analyze SV applications or computing algorithms as monolithic units rather than decomposing them into reusable building blocks, which is a limitation that stifles the development of modular extensible frameworks from a DA application perspective.
For instance, existing tools like SHAP \cite{Lundberg2017AUA} (tailored to RI tasks) and DataShapley \cite{ghorbani2019data} (bound to DV tasks) are constrained by rigid task-specific configurations, sacrificing flexibility and universality. These gaps leave practitioners ill-equipped to navigate SV’s foundational design principles (such as cooperative game modeling) or resolve conflicting challenges (e.g., approximation error vs. efficiency) when adapting SV to new DA tasks.

\setcounter{table}{0}
\begin{table}[t]

    \centering 
    \scalebox{0.85}{
    \begin{tabular}{c|c|c|c|c|c|c|c|c|c|c}
    \toprule
         \multirow{2}{*}{\textbf{Survey}} & 
         \multicolumn{4}{c|}{\textbf{Application Purposes}} & 
         \multicolumn{4}{c|}{\textbf{Solutions For}}
         & \textbf{Units of} & \textbf{Frame-}\\
         
         \cline{2-9} 
         & \textbf{pric.} & \textbf{sele.} & \textbf{weig.} & \textbf{attr.} &  \textbf{eff.} &  \textbf{err.} & \textbf{priv.} & \textbf{int.} & \textbf{Analysis} &\textbf{work}\\
    \midrule
         \small{\cite{pei2020survey,tang2020abrief}}&  \faCheck  & \faTimes & \faTimes & \faTimes  & \faCheck & \faTimes & \faTimes & \faTimes  & \multirow{6}{*}{\shortstack{\small{Applications}\\\small{or}\\\small{Algorithms}}} & \faTimes  \\
         
         \small{\cite{ijcai2022p778}} & \faTimes & \faCheck & \faTimes & \faCheck &\faCheck & \faCheck & \faTimes & \faTimes  & & \faTimes   \\
         
         \small{\cite{10.1145/3626246.3654680}} & \faCheck & \faCheck & \faTimes & \faCheck & \faCheck & \faCheck & \faTimes & \faTimes  & & \faTimes   \\

         \small{\cite{sim2022data}} & \faTimes & \faCheck & \faTimes & \faTimes & \faCheck & \faTimes & \faCheck & \faTimes   &  & \faTimes   \\
         
         \small{\cite{sv_related_survey,10.1007/s10618-024-01016-z,Bertossi2023TheSV}} &\faTimes & \faTimes & \faTimes & \faCheck & \faCheck & \faTimes & \faTimes & \faTimes  & & \faTimes  \\

         \small{\cite{Li2024ShapleyVF}} &\faCheck & \faCheck & \faTimes & \faCheck & \faCheck 
         & \faTimes & \faTimes & \faTimes &  & \faTimes  \\

         \midrule
         \small{ours}&\faCheck & \faCheck & \faCheck & \faCheck & \faCheck & \faCheck & \faCheck & \faCheck  & \shortstack{\small{Techniques} 
         } & \faCheck  \\   
    \bottomrule
    \end{tabular}
    }
    \caption{
    Comparison with related works. 
    Our work focuses on the techniques (or called building blocks of SV applications and algorithms) solving four challenges of SV: computation efficiency (eff.), approximation error (err.), privacy preservation (priv.), and interpretability (int.).
    }
    \label{tab:comparison_with_previous_work}
\end{table}

In this work, we endeavor to bridge existing gaps through an in-depth survey analyzing the application of SV across the entire spectrum of DA. 
By synthesizing the insights, this paper contributes to not only a deeper understanding of the SV's potential to enhance DA but also the implementation of a modular extensible framework for SV application development, laying the groundwork for both the practical implementation of SV in real-world DA systems and the advancement of academic research in this exciting field. 
We expect this paper to be a helpful resource for both newcomers to the field and seasoned experts seeking a consolidated and systematic update on current developments. 

The contributions of this paper are outlined as follows:

\begin{itemize}
\item We present the first comprehensive survey of SV applied throughout the DA workflow, clarify the key variables in defining DA-applicable SV, and reveal the essential functionalities that SV can provide for DA. We also condense four technical challenges of using SV in DA, analyze and discuss the building blocks of solutions 
w.r.t. each challenge and potential conflicts between challenges. (\S\ref{sec: SVinDA})

\item We propose \textit{SVBench}, a modular and extensible open-source framework for developing SV applications.
\textit{SVBench} integrates abundant techniques addressing key challenges of using SV in DA and allows a flexible usage of these techniques to build up new algorithms. 
The developers, as well as the academicians, could extend the framework with flexible APIs for succeeding studies on SV. (\S\ref{sec: implementation})

\item We conduct extensive evaluations to consolidate our analyses and discussions, involving possible technical combinations for efficient computation, trade-offs in handling different challenges, and key factors affecting mainstream SV interpretations. Experiment results also validate the usability and modularity of \textit{SVBench}. 
Through experiments, we reveal limiting factors of existing efforts and highlight future research and engineering directions. 
(\S\ref{sec: exp})
\end{itemize}

\section{Preliminaries} \label{sec: preliminaries}

This section introduces a general definition of the SV used in DA. For a more accessible presentation, we start by introducing the cooperative game, an important concept used in SV definition.

A cooperative game is composed of \textit{a player set} and \textit{a utility function} that defines the utility of each coalition (i.e., a subset of the player set).
The formal definitions are stated as follows.

\textbf{Definition 1. \textit{Player set, coalition, and cooperative game.}} \textit{Let $\mathcal{N} = \{p_1, \cdots, p_n\}$ be a finite set of players. A coalition is a nonempty subset $\mathcal{S} \subseteq \mathcal{N}$ and the grand coalition is $\mathcal{N}$ itself.  
A cooperative game, denoted by $C (\mathcal{N}, U)$, consists of a player set $\mathcal{N}$ and a utility function $U(\cdot)$ that maps each coalition to a scalar value, i.e., $U: \mathcal{P}(\mathcal{N}) \rightarrow \mathbb{R}$, where $\mathcal{P}(\mathcal{N})$ is the power set of $\mathcal{N}$ and $U(\emptyset)=0$.
}

For any $\mathcal{S} \subseteq \mathcal{N}$, $U(\mathcal{S})$ represents the sum of the expected utility that the members of $\mathcal{S}$ can achieve through cooperation and is available for distribution among the members of $\mathcal{S}$.

SV is a method in cooperative game theory designed to fairly allocate the overall utility generated by the collective efforts of all players within a game. SV assigns a value to each player in the game, based on the player's marginal contribution to each possible coalition's utility, especially considering the case where the player is not part of the coalition. Intuitively, SV captures the essence of how much one coalition's utility increases (or decreases) with the inclusion of a new player, providing a fair and quantifiable measure of each player's influence on the game's overall utility. 

SV has already been widely applied in numerous DA tasks modeled as cooperative games. According to the application purpose, existing works fall into 4 high-level categories, i.e., pricing, selection, weighting, and attribution, as presented in \S\ref{sec: intro}. Within each category, SV applications can be further classified into 8 finer-grained subcategories based on the definition of  player and utility in the cooperative game, as shown in \Cref{fig:cooperative_games_in_DA}. 
Through a comprehensive review of these applications, we formalize a general definition of SV used in DA as follows. 
A detailed discussion of SV applied for different purposes will be presented in the next section. 

\textbf{Definition 2. \textit{Shapley value in data analytics.}} \textit{Given a task modeled as $C(\mathcal{N}, U)$, where each player $p_i \in \mathcal{N}$ is an analytical object 
and the utility $U(\cdot)$ refers to an analytical outcome or the outcome evaluation score, 
the Shapley value $\phi_i$ is a numerical value representing the weighted average of marginal contributions made by the player $p_i$ to $U(\mathcal{S})$ 
produced by each coalition $\mathcal{S} \subseteq \mathcal{N}\textbackslash \{p_i\}$.} 

\begin{align} 
    \phi_i &= \sum_{\mathcal{S} \subseteq \mathcal{N} \textbackslash \{p_i\}} \underbrace{\frac{|\mathcal{S}|!(n-|\mathcal{S}|-1)!}{n!}}_{weight\ factor} \underbrace{[U(\mathcal{S} \cup \{p_i\}) - U(\mathcal{S})]}_{marginal\ contribution} \label{eq:sv} \\
    &= \sum_{\mathcal{O} \in \pi(\mathcal{N}) } \frac{1}{n!} \underbrace{[U(P(\mathcal{O}, p_i) \cup \{p_i\}) - U(P(\mathcal{O}, p_i))]}_{marginal\ contribution} .\label{eq:sv2} 
\end{align}

Equations~\ref{eq:sv} and \ref{eq:sv2} show the mathematical formulation of SV from different perspectives. Equation~\ref{eq:sv} is given based on Definition 2, while Equation~\ref{eq:sv2} is given from the perspective of expectation calculation. 
The two formulas are interchangeable in SV computation. However, Equation~\ref{eq:sv2} would be preferred if a developer expects to compute SV using expectation calculation techniques in mathematical statistics (even if not originally devised for SV, e.g., Monte Carlo methods). In Equation~\ref{eq:sv2}, $\pi(\mathcal{N})$ is the set of all possible permutations of players in the grand coalition, and $P(\mathcal{O}, p_i)$ is the set of predecessors of player $p_i$ in a specific permutation $\mathcal{O}\in\pi(\mathcal{N})$.

\section{The Shapley Value in Data Analytics} \label{sec: SVinDA}

In this section, we first introduce cooperative game modeling, a critical step to apply SV to DA tasks (\S\ref{ssec: application_overview}), and then discuss the challenges of using SV in DA and the corresponding solutions (\S\ref{ssec: challenges_and_solutions}).

\begin{figure}
    \centering
\tikzset{every picture/.style={line width=0.75pt}} 
\scalebox{0.6}{
\hspace{-0.3cm}

}
    \caption{
    Cooperative game modeling for applying SV in the DA domain (\ding{182}, \ding{183}, \ding{184} refer to the three functionalities summarized in Finding 3). 
    }
    \label{fig:cooperative_games_in_DA}
\end{figure}

\subsection{Cooperative Game Modeling} \label{ssec: application_overview}

The fundamental step of applying SV for a DA task is cooperative game modeling.
The core is to properly match the player and the utility to the elements in the task.
\Cref{fig:cooperative_games_in_DA} summarizes existing combinations of the player and utility in DA tasks and their associated SV application purposes. 

From the figure, we can see that the player is taxonomized into four types: (1) the data \textit{feature}, (2) the data \textit{tuple} (or called sample) composed of several features and label(s), (3) the \textit{dataset} composed of several data tuples, or (4) the data \textit{derivative}, e.g., the ML model trained on several datasets.

\textbf{Finding 1. Players may differ from the analytical object.} 
While players in SV typically align with the analytical objects of a DA task, they can also represent sub-components of the analytical object for specialized objectives.

\textbf{Discussion.} 
\Cref{fig:SV_in_DA} demonstrates how the players align with the analytical objects of each task: in DO, the players are pixels (i.e., data features); in DV, they are images (i.e., data tuples); in DC, they correspond to image sets (namely datasets); in RT, players are ML models (data derivatives).
Such alignments ensure that SV quantifies contributions at the granularity of the task's core analytical object.

In contrast, \citet{wang2019measure} defined the players as the sub-components of the analytical object in a multi-group data valuation task.
The task is to evaluate contributions to Cervical cancer prediction across five disjoint groups, where each group's dataset contains three unique data features (e.g., age) with no feature overlap between groups.
Here, the player is the data feature, a sub-component of the core analytical object (each group's dataset).
The flexibility in defining players as either the analytical object or its sub-components underscores the adaptability to DA task-specific interpretability needs.

The utility function has two kinds of outputs: 
(1) the \textit{goodness-of-fit score} between the DA task's outputs and ground truth facts, including but not limited to test accuracy, training loss, confidence score, etc., (2) the \textit{task outputs} themselves, more specifically, predictive outputs from the ML models learned in DA tasks or the answers to queries/questions on the data used in the DA workflow.

\textbf{Finding 2. Utility definitions are task-dependent.} 
The choice of utility is dictated by the objectives of the DA task. When maximizing or minimizing goodness-of-fit scores (e.g., accuracy), the utility directly reflects those scores. Conversely, when explaining task outputs or quantifying data impacts on those outputs, the utility is defined by the outputs themselves.

\textbf{Discussion.} 
\Cref{fig:SV_in_DA} exemplifies this duality.
In the DV task, utility is tied to test accuracy to optimize a medical image classifier, while in the RI task, utility derives from predictive outputs to explain how pixel-level variations influence results.
Notably, a single player-utility pairing, such as defining players as ML models and utility as test accuracy, can serve multiple DA tasks (e.g., DM and RT in the figure).
Similarly, one task may leverage SV for diverse purposes: the DM task, for instance, can use SV both to weight local models and to assign pricing incentives, ensuring higher model quality while aligning contributions with economic rewards.
This adaptability reveals SV’s capacity to address heterogeneous analytical goals through context-aware utility design. 

\textbf{Finding 3. The real-world functionality of applying SV lies in three aspects:} (1) to construct fair marketplaces for data and related products, incentivizing data sharing; (2) to improve the quality of DA outputs while reducing the economic cost; (3) to transfer DA outputs into actions to solve real-world problems.

\textbf{Discussion.} 
\Cref{fig:SV_in_DA} exemplifies the three functionalities that different DA tasks target to achieve. 
DC and RT tasks seek to provide helpful results to construct fair marketplaces for data in the medical domain. 
DV and DM tasks strive to improve the test accuracy of a medical image classification model while reducing the learning cost. 
The RI task intends to transfer predictions from a medical image classification model into actions assisting diagnosis. 
As summarized in \Cref{fig:cooperative_games_in_DA}, 
to serve the first functionality, the purpose of applying SV is typically pricing. 
For the second functionality, the purposes of applying SV include pricing, selection, and weighting. 
To fulfill the last functionality, the purpose of applying SV is primarily attribution. 
Demanders can design SV applications that satisfy their specific requirements according to \Cref{fig:cooperative_games_in_DA} and Findings 1-3.

\subsection{Challenges and Solutions of Applying SV} \label{ssec: challenges_and_solutions}

In this section, we taxonomize existing studies on applying SV in the DA domain according to the challenges they tackle, namely, \textit{computation efficiency}, \textit{approximation error}, \textit{privacy preservation}, and \textit{interpretability}.
Subsequently, we conduct a thorough analysis and discussion of the countermeasures, 
highlighting key takeaways and remaining questions 
that we will study further through experiments in \S\ref{subsec: exp_efficiency}--\S\ref{subsec: exp_interpretability}.

\subsubsection{Computation Efficiency} \label{sssec: computation_efficiency}
Based on Definition 2 and Equation~\ref{eq:sv}, the total cost of SV computation is determined by $N_{uc}\times T_{uc}$, where $N_{uc}$ is the total number of utility computations, and $T_{uc}$ is the average time cost of every utility computation. 
To compute exact SV, the algorithm generally enumerates all possible coalitions of players in the cooperative game by \textit{iterations}, with each iteration computing the utility on one coalition.

Exact SV computing algorithm is known costly~\cite{ijcai2022p778, sv_related_survey, Bertossi2023TheSV}, with the number of iterations reaching an exponential level, i.e., $N_{uc} = 2^n$. Thus, the computation complexity of exact SV is $\textit{O}(2^n)$. 
In DA tasks, the quantity of players (i.e.,$n$) is much higher than expected, resulting in the prohibitive computation cost. Moreover, as highlighted by~\citet{jethani2021fastshap}, the total cost of SV computing in DA can be further increased when DA tasks involve ML models, since each iteration typically necessitates retraining these models, associated with non-trivial expenses. 
For example, the DO task in \Cref{fig:SV_in_DA} analyzes $224\times224$ players (pixels), thus $N_{uc}=2^{224\times224}$. Depending on the task's model size and computation resources, $T_{uc}$ can reach hours to months \cite{Cottier2024TheRC}, much longer than milliseconds in traditional game theory. 

To efficiently compute SV, the research focus has shifted from the exact computation to the approximate algorithms. 
Abundant approximate SV computing algorithms have been proposed and most existing surveys \cite{ijcai2022p778, sv_related_survey, Li2024ShapleyVF, Bertossi2023TheSV, sim2022data, pei2020survey, tang2020abrief, 10.1007/s10618-024-01016-z} classified the algorithms into two categories: task-agnostic, i.e., covering general usage for different DA tasks, and task-specific, i.e., relying on certain assumptions about DA tasks.

Different from previous surveys, our paper disentangles the vital techniques that can be used flexibly to develop new efficient SV computing algorithms from existing applications. 
Generally, these techniques fall into two categories: 
(1) iteration reduction for reducing $N_{uc}$, and (2) ML speedup for reducing $T_{uc}$.

\begin{table}[t]
    
    \centering
    \scalebox{0.85}{
    \begin{tabular}{c|c|c|c|c}
    \toprule
        \textbf{Tech.} & \textbf{Main Idea} & 
        \textbf{Complexity} & \textbf{Extra Cost} & \textbf{Use Scenarios} \\
    \midrule
         \multirow{5}{*}{MC} & \multirow{9}{*}{sampling} 
         & \multirow{5}{*}{$\textit{O}(n^2 \log(n))$} & \multirow{5}{*}{\faTimes } & \multirow{5}{*}{\shortstack{
         DC{\footnotesize\cite{zhang2023dynamic, xia2023equitable, liu2021dealer, agarwal2019marketplace, counterfactual, NEURIPS2022_c50c42f8, pmlr-v119-sim20a}}, \\
         DO{\footnotesize\cite{cohen2005feature, sebastian2024feature, marcilio2020explanations, fryer2021shapley}}, \\
         DV{\footnotesize\cite{jia2019towards, ghorbani2019data, ghorbani2020distributional, kwon2021efficient}},\\
         FL{\footnotesize\cite{wang2019measure, wang2020principled, nagalapatti2021game, liu2020fedcoin, 10.1007/978-981-99-8082-6_41, fan2022improving}},\\
         EL{\footnotesize \cite{rozemberczki2021shapley, bimonte2024mortality}}, RI{\footnotesize \cite{mitchell2022sampling}} 
         }} \\
         & & & & \\
         & & & & \\
         & & & & \\
         & & & & \\
    \cline{1-1} \cline{3-5}
         RE &  
         & /  & \faTimes  & RI{\footnotesize\cite{Lundberg2017AUA, covert2020improving, jethani2021fastshap}} \\
    \cline{1-1} \cline{3-5}
         MLE&  
         & Polynomial  & \faTimes  & EL{\footnotesize\cite{rozemberczki2021shapley}}, RI{\footnotesize\cite{Okhrati2020AMS, mitchell2022sampling}}\\
    \cline{1-1} \cline{3-5}
         GT&  
         & $\textit{O}(\sqrt{n}\log(n)^2)$  & \faTimes  & DV{\footnotesize\cite{jia2019towards}}, FL{\footnotesize\cite{wang2020principled}}\\
    \cline{1-1} \cline{3-5}
         CP&   
         & $\textit{O}(n\log\log(n))$ & \faTimes  & DV{\footnotesize\cite{jia2019towards}}\\
    \midrule
         \multirow{2}{*}{TC} & \multirow{2}{*}{truncation} 
         & \multirow{2}{*}{/}  & \multirow{2}{*}{\faTimes}  & \multirow{2}{*}{\shortstack{
         DC{\footnotesize\cite{tian2022data, azcoitia2022computing, luo2024fast}}, DV{\footnotesize\cite{pandl2021trustworthy, xia2024p, luo2024fast}},\\
         FL{\footnotesize\cite{ghorbani2022data, yang2022wtdp, liu2022gtg, Liu_Chen_Zhao_Yu_Liu_Bao_Jiang_Nie_Xu_Yang_2022, 10219590, yang2022wtdp, liu2022gtg, Liu_Chen_Zhao_Yu_Liu_Bao_Jiang_Nie_Xu_Yang_2022}}
         }}\\
         & & & & \\
    \midrule
         PL & fitting 
         & $\textit{O}(1)$  & \faCheck  & RI{\footnotesize\cite{jethani2021fastshap}}\\
    \bottomrule
    \end{tabular}
    }
    \caption{
    Summary of iteration reduction techniques disentangled from \textit{task-agnostic} SV computing algorithms. Based on learning paradigms, our work further classifies the DM tasks having applied SV into five types: semi-supervised learning (SSL), active learning (AL), continuous learning (CL), ensemble learning (EL), and federated learning (FL).  
    }
    \label{tab:iteration_reduction}
\end{table}

\Cref{tab:iteration_reduction} summarizes the iteration reduction techniques disentangled from task-agnostic algorithms, which can be outlined into three types: (1) sampling-based, to estimate SV using randomly-sampled coalitions or permutations, involving Monte Carlo, regression, multilinear extension, group testing, compressive permutation sampling, (2) truncation, to avoid computing the marginal contribution of new players who join a coalition unnecessarily, (3) fitting-based, to train a predictive ML model whose outputs are the estimations of the targeted SVs, including SV predictor learning. 

\underline{Monte Carlo}(MC) method computes SV based on Equation~\ref{eq:sv2} by sampling random permutations iteratively. In each iteration, the marginal contribution of every player $p_i (\in \mathcal{N})$ within a randomly sampled permutation, $U(P(\mathcal{O},p_i))$, is computed. 
The final estimation of SV for each player is the average of the marginal contributions calculated for that player in all iterations.

\underline{Regression}(RE) method models SV as the solution to a weighted least squares problem that minimizes $\sum_{\vec{z}} w(\vec{z}) (U(\vec{0}) + \vec{z}^\top \vec{\phi} - U(\vec{z}))^2$ subject to $\vec{1}^\top \vec{\phi} = U(\vec{1}) - U(\vec{0})$, where $\vec{\phi} = <\phi_i,\cdots, \phi_n>$ is the SV vector, $\vec{z}$ is a binary vector representing the coalition $\mathcal{S}=\{p_i \ | \ \vec{z}[i] =1 \}$, $U(\vec{z})$ is the utility of this coalition, and $w(\vec{z})$ is the weight computed by $\frac{(n-1) |\mathcal{S}|! (n-|\mathcal{S}|)! }{ n! |\mathcal{S}|(n-|\mathcal{S}|)}$. RE method samples the random vectors of $\vec{z}$ and leverages Karush–Kuhn–Tucker conditions \cite{kuhn1951nonlinear} to derive the unbiased estimation of SV. 

\underline{Multilinear extension}(MLE) technique formulates SV by a closed-form integral expression $\phi_i = \int_0^1 e_i(q) dq$, where $q$ is the probability of each player $p_i$ to be included in a random coalition $\mathcal{S}$ 
and $e_i(q) = \mathbb{E}[U(\mathcal{S}\cup \{p_i\})-U(\mathcal{S})]$. MLE computes unbiased SV estimations by sampling over $q$ at a fixed interval and estimating $e_i(q)$ for every sampled $q$ value. 

\underline{Group testing}(GT) technique models SV as the solution to a feasibility problem with $\sum_{i=1}^{n} \hat{\phi}_i = U(\mathcal{N})$ and $|(\hat{\phi}_i-\hat{\phi}_j)-\Delta U_{i, j}| \leq \frac{\epsilon}{2\sqrt{n}}, \forall p_i, p_j \in \mathcal{N}$, where  $\Delta U_{i, j}$ is an unbiased estimation of the SV difference between two players $p_i$ and $p_j$, and $\epsilon$ is an error threshold. GT computes $\Delta U_{i, j}$ by sampling binary random variables $\beta_1, \cdots, \beta_n$ (representing the coalition $\mathcal{S}=\{p_i \ | \ \beta_i =1 \}$) from a smartly designed distribution iteratively. 
The final SV approximations are derived from the $(\beta_i-\beta_j)U(\mathcal{S})$ values computed in all iterations. 

\underline{Compressive permutation sampling}(CP) method formulates SV by  $\overline{\phi}+\Delta \phi_i$, where $\overline{\phi} = \frac{U(\mathcal{N})}{n}$ is a mean value, and $\Delta \phi_i$ is the player-specific variance generated by solving a convex optimization problem using the marginal contributions computed in the same way as MC. CP is based on a sparse SV assumption that, given a cooperative game, the SVs of most players are concentrated around their mean value, and only a few players have significant SVs. 

\underline{Truncation}(TC) technique evaluates the necessity of marginal contribution computations encountered in the runtime of SV approximation and truncates those unnecessary ones. 
The key idea is to avoid computing the marginal contribution of new players who join a coalition unnecessarily, particularly when that coalition's utility,  $U(\mathcal{S})$, is close to the overall utility achieved by the grand coalition, $U(\mathcal{N})$. 

\underline{SV predictor learning}(PL) aims to train a predictive ML model whose outputs are the estimations of the targeted SVs. \citet{jethani2021fastshap} implemented this technique by using the RE technique to design a custom loss function that enables efficient gradient-based optimization without the need for massive (computationally intractable) ground-truth SVs.

\textbf{Finding 4. For task-agnostic iteration reduction, sampling-based techniques gain the leading place in practical applications.} 
Each sampling-based technique can be deployed independently as a base SV computing algorithm.

\textbf{Discussion.} 
Among the sampling-based iteration techniques, MC gains the widest application. 
Theoretically, GT and CP can further reduce the complexity of SV computation, but these two techniques do not gain much wider application than MC. 
We speculate that GT is limited by its assumption 
on the correctness of mirroring the difference between the SVs of any two players, while CP is limited by its strong reliance on the sparse SV assumption. The other sampling-based iteration techniques, RE and MLE, are also relatively limited in practical applications compared with MC. These two techniques are heuristic and the corresponding big-O notations 
are not given by existing arts. 

Compared with sampling-based techniques, fitting-based techniques can generate SV at a much lower complexity. However, these techniques work effectively only with access to abundant ground-truth SVs (or high-quality surrogates of the ground-truth) for training the SV predictor, and the training incurs extra cost.

\textbf{Finding 5. The truncation-based technique tends to be employed in conjunction with a sampling-based technique.}

\textbf{Discussion.} 
Many works \cite{zhang2023dynamic, luo2024fast, azcoitia2022computing, tian2022data, cohen2005feature,ghorbani2019data, xia2024p, pandl2021trustworthy,yang2022wtdp, liu2022gtg} have attempted the combination of \textit{MC+TC} to develop hybrid\footnote{In this paper, we call the SV computing algorithm adopting different techniques for efficient computation as the \textit{hybrid} algorithm.} SV computing algorithms. 
As a flexible technique, TC is also compatible with other sampling-based iteration reduction techniques, e.g., RE, MLE, GT, and CP. However, these combinations have not been evaluated yet, since prior works analyze SV applications or computing algorithms as monolithic units rather than decomposing them into reusable building blocks. Later in \S\ref{subsec: exp_efficiency}, we will evaluate the performance of TC integrated with different sampling-based techniques to validate its flexibility.

\begin{table}[t]

    \centering
    \scalebox{0.85}{
    \begin{tabular}{c|c|c|c}
    \toprule
        \textbf{Tech.} & \textbf{Main Idea} & \textbf{Complexity} & \textbf{Use Scenario}\\
    \midrule
         Linear-based & \multirow{5}{*}{\shortstack{model\\structure}} & $\textit{O}(n)$ & RI{\footnotesize\cite{ Lundberg2017AUA}} \\
    \cline{1-1} \cline{3-4}
         Tree-based &  & $\textit{O}(TLD^2)^{*}$  & RI{\footnotesize\cite{Yang2021FastTA, Muschalik_Fumagalli_Hammer_Hüllermeier_2024, yu2022linear, lundberg2020local}}\\
    \cline{1-1} \cline{3-4}
         \multirow{2}{*}{KNN-based} & & \multirow{2}{*}{$\textit{O}(n log(n))$} & \multirow{2}{*}{\shortstack{DV{\footnotesize\cite{wang2023threshold}}, SSL{\footnotesize\cite{courtnage2021shapley}},\\
         AL{\footnotesize\cite{ghorbani2022data}}, CL{\footnotesize\cite{shim2021online}} 
         }}
         \\
         & & &
         \\
    \cline{1-1} \cline{3-4}
         DNN-based &  & / & RI{\footnotesize\cite{wang2021shapley, ICML-2019-AnconaOG}}\\
    \midrule
         Uniform division & stable learning & $\textit{O}(1)$ & DV{\footnotesize\cite{jia2019towards}}\\
    \midrule
         Influence function & smooth utility &  $\textit{O}(n)$ & DV{\footnotesize\cite{jia2019towards}}\\
    \bottomrule
    \end{tabular}
    }
    \begin{flushleft}
        {\footnotesize* $T, L, D$ denote the number of trees, the maximum number of leaves in a tree, and the maximum depth of a tree, respectively.} 
    \end{flushleft}
    \caption{
    Summary of iteration reduction techniques disentangled from \textit{task-specific} SV computing algorithms.
    }
    \label{tab:iteration_reduction_task_specific}
\end{table}

\Cref{tab:iteration_reduction_task_specific} summarizes the iteration reduction techniques disentangled from task-specific algorithms, including linear-based, tree-based, K-nearest-neighbor(KNN)-based, deep-neural-network(DNN)-based, uniform division, and influence function. 

\underline{Linear-based} techniques leverage the additive property of linear models to efficiently compute SV of data features \cite{10.5555/3295222.3295230, Lundberg2017AUA}. These techniques take $w_i (x_i-\mathbb{E}(x_i))$ as the SV of a data feature $x_i$, where $w_i$ is the weight of this feature in the linear model, $\mathbb{E}(x_i)$ is the mean (or baseline) value of this feature in the training data.

\underline{Tree-based} methods exploit the structure of decision trees and tree ensembles (such as random forests and gradient-boosted trees) to compute SV of data features in polynomial time \cite{Yang2021FastTA, Muschalik_Fumagalli_Hammer_Hüllermeier_2024, yu2022linear, lundberg2020local}. Instead of evaluating all possible feature subsets, tree-based methods leverage the tree’s hierarchical splits to compute marginal contributions efficiently. 

\underline{KNN-based} techniques provide an efficient way to estimate SV of data samples leveraging KNN’s local structure. For a given data sample, KNN identifies its nearest neighbors in the training dataset \cite{wang2023threshold}. 
Instead of evaluating all possible coalitions of data samples, the SV of each sample is estimated using only the selected neighbors. 

\underline{DNN-based} techniques adapt SV computation for DNNs by back-propagating contributions layer-by-layer \cite{wang2021shapley, ICML-2019-AnconaOG, 10.5555/3305890.3306006}. These techniques define a set of rules, such as approximating ReLU/sigmoid activations as linear functions for gradient propagation, modified gradient rules, etc., to simplify the computation in DNN in order to generate SV efficiently. The adopted rules, on the other hand, can lead to bias in the resultant SV, depending on the specific model and dataset.

\underline{Uniform division} is designed for the DV tasks that train ML models based on stable learning algorithms, e.g., the algorithms with Tikhonov regularization \cite{jia2019towards}, and analyze data samples used in model training. 
Due to the inherent insensitivity of a stable learning algorithm to the training
data, the SV of each training data sample is similar to one another. Therefore, the complexity of computing SV for $n$ data samples is reduced to $\mathcal{O}(1)$.

\underline{Influence function} is also designed for DV tasks, adopting influence functions as an efficient approximation of model parameter changes after adding or removing one training data sample, thus eliminating the need for re-training models \cite{jia2019towards}. This technique takes $U(\mathcal{N})-U(\mathcal{N}\ \backslash \ \{p_i\})$ as the approximate SV of each training data sample $p_i$ in a heuristic manner, reducing the SV computation complexity to $\mathcal{O}(n)$.

\textbf{Finding 6. For task-specific iteration reduction, leveraging simple-structured ML models or loss-bounded learning algorithms is the key to lowering the complexity of SV computation.
} However, with the rising popularity of large-scale pre-trained models in the DA domain, the usability of this solution direction might be compromised.

\textbf{Discussion.} 
Although some task-specific techniques are designed based on DNN, the complexity of those techniques is determined by the number of hidden layers in the model and the number of parameters in each layer. Therefore, if a DA task relies on large-scale pre-trained models (such as GPT-4 \cite{openai2023chatgpt}), those techniques may not work efficiently to generate accurate SV approximation results in this case.


\Cref{tab:ml_speedup} summarizes the ML speedup techniques: gradient approximation, test sample skip, and model appraiser. All these techniques are disentangled from task-specific algorithms and are heuristic (thus not given with big-O notations). 

\underline{Gradient approximation}(GA) expedites the model training by approximating the gradients produced by the costly multi-step optimization algorithms with easy-to-obtain surrogate gradients \cite{jia2019towards, ghorbani2022data, sun2023shapleyfl, 9006327, tastan2024redefining, Xu2021, 10056725, 9797864, wei2020efficient, 9521339, 10.1145/3477314.3507050, ZHU2023817, 10475986, zheng2022secure, yang2022wtdp, liu2022gtg, Liu_Chen_Zhao_Yu_Liu_Bao_Jiang_Nie_Xu_Yang_2022}. 

\underline{Test sample skip}(TSS), aiming to accelerate the model inference process, evaluates ML models on only ambiguous test data whose predictive results vary significantly under different models, while skipping the evaluations on the other unambiguous data \cite{zheng2022secure}. 

\underline{Model appraiser}(MA), also targeted to accelerate model inference, trains a prediction model whose outputs are estimations of each given ML model's performance score \cite{tian2022private, ghorbani2020distributional, xu2024model}. 

\begin{table}[t]

    \centering
    \scalebox{0.85}{
    \begin{tabular}{c|c|c|c}
    \toprule
         \textbf{Tech.} & \textbf{Objective}  & \textbf{Main Idea} & \textbf{Use Scenarios}\\
    \midrule
         \multirow{3}{*}{GA} & \multirow{3}{*}{\shortstack{training \\ speedup}}& replace gradients from costly  & \multirow{3}{*}{\shortstack{
         DV{\footnotesize\cite{jia2019towards}}, AL{\footnotesize\cite{ghorbani2022data}}, \\ FL{\footnotesize\cite{sun2023shapleyfl, tastan2024redefining, Xu2021}}
         }} \\
         & &  multi-step computations with  & \\
         & & easy-to-obtain surrogates & \\
    \midrule
         \multirow{3}{*}{TSS} & \multirow{6}{*}{\shortstack{inference \\ speedup}} & evaluate on only ambiguous & \multirow{3}{*}{FL{\footnotesize\cite{zheng2022secure}}
         } \\
         & & test data whose predictive    & \\
         & & results vary across models  & \\
    \cline{1-1} \cline{3-4}
         \multirow{3}{*}{MA} &  & train a model whose outputs & \multirow{3}{*}{DC{\footnotesize\cite{tian2022private}}, DV {\footnotesize\cite{ghorbani2020distributional}}, RT{\footnotesize\cite{xu2024model}}
         } \\
         & &   are estimations of the given & \\
         & &   model’s performance score & \\
    \bottomrule
    \end{tabular}
    }
    \caption{Summary of ML speedup techniques.} 
    \label{tab:ml_speedup}
\end{table}

\textbf{Finding 7. ML speedup techniques are compatible with iteration reduction techniques.} 

\textbf{Discussion.} 
There are many works on hybrid SV computing algorithms integrating these two types of techniques. 
An example is \textit{MC+GA}, utilized by both DV tasks selecting high-quality data tuples from the UK Biobank dataset to train logistic regression \cite{ghorbani2019data} and FL tasks selecting high-quality local models to generate the global model for image classification \cite{yang2022wtdp, liu2022gtg, 10.1145/3477314.3507050}.
\textit{MC + GA + TSS} is another typical combination designed for tasks with the players being ML models \cite{zheng2022secure}. 
We note that integrating ML speedup with iteration reduction is well-suited for tasks, like DV or FL, which have (one of) the following characteristics. The first characteristic is that computing a utility involves the costly multi-step gradient descent for model training. Another characteristic is that the utility computation relies on numerous test data samples or complicated models, e.g., pre-trained models with billions of parameters. 

Though heaps of hybrid algorithms have been proposed, providing empirical usage of SV, a question remains open for rigorous demanders: \textbf{Can the hybrid SV computing algorithms always ensure higher efficiency than the algorithm using only one of the integrated techniques? }
Concluding this subsection with the question, we attempt to answer this question later in \S\ref{subsec: exp_efficiency}.

\subsubsection{Approximation Error}  \label{sssec:approximation_error}
SV approximate computation, though faster than exact computation, introduces the variance caused by the randomness in sampling the player coalitions and the bias caused by incomplete exploration of all the possible coalitions.
Hence, it poses a new challenge that the approximate SV may not be an accurate and unbiased estimation of the exact SV, and thus may fail to serve its expected application purposes. The key to reducing the SV approximation error is variance reduction. \Cref{tab:different_sampling} summarizes the techniques falling into three types: stratified, antithetic, and kernel-based.

\underline{Stratified} technique divides the player coalitions or permutations into several disjoint strata and then randomly selects coalitions or permutations from each stratum in a fixed proportion for SV approximation  \cite{wu2023variance,jia2019towards, zhang2023efficient, Watson2022DifferentiallyPS,10623283,kolpaczki2024approximating,10.5555/3618408.3619283, otmani:hal-04512814,mitchell2022sampling, kolpaczki2024approximating,10.1007/978-3-031-63797-1_25}. The coalitions are stratified based on the number of players included in each coalition, while the permutations are stratified according to the position of a player. 

\underline{Antithetic} technique generates negatively correlated coalitions or permutations for SV approximation \cite{otmani:hal-04512814,covert2020improving, Okhrati2020AMS, mitchell2022sampling}. Given a coalition $\mathcal{S}$, its negatively correlated counterpart is $\mathcal{N} \textbackslash \mathcal{S}$. Given any two permutations, they are negatively correlated if the order of players is completely reversed.

\underline{Kernel-based} technique defines the distance between permutations by kernels and sample permutations with good distributions relative to kernels when approximating SV \cite{mitchell2022sampling}. 

\begin{table}[t]

    \centering
    \scalebox{0.85}{
    \begin{tabular}{c|c|c|c}
    \toprule
         \textbf{Tech.} & \textbf{Main Idea} 
         & \textbf{Compatible With} & \textbf{Use Scenarios}\\
    \midrule
         \multirow{3}{*}{stratified}  & sample permutations or & \multirow{6}{*}{\shortstack{MC, RE, MLE, \\GT, CP }} & \multirow{3}{*}{\shortstack{
         DC{\footnotesize\cite{wu2023variance}}, DO{\footnotesize\cite{kolpaczki2024approximating}},\\ DV{\footnotesize\cite{jia2019towards, zhang2023efficient, Watson2022DifferentiallyPS}},\\
         FL{\footnotesize\cite{10.5555/3618408.3619283, otmani:hal-04512814}}, RI{\footnotesize\cite{mitchell2022sampling, kolpaczki2024approximating,10.1007/978-3-031-63797-1_25}}
         }}  \\
         & coalitions from disjoint &  & \\
         & strata proportionally &  & \\
    \cline{1-2} \cline{4-4}
         \multirow{3}{*}{antithetic} & sample negatively  & &  \multirow{3}{*}{FL{\footnotesize\cite{otmani:hal-04512814}},RI{\footnotesize\cite{covert2020improving, Okhrati2020AMS, mitchell2022sampling}}} \\
         &  correlated permutations & & \\    
         &  or coalitions  & & \\    
    \midrule
         \multirow{3}{*}{\shortstack{kernel-\\based}} & sample permutations &  \multirow{3}{*}{MC, CP} & \multirow{3}{*}{RI{\footnotesize\cite{mitchell2022sampling}}} \\
         &  with good distributions &  &\\
         &  relative to kernels  &  &\\
    \bottomrule
    \end{tabular}
    }
    \caption{Summary of variance reduction techniques.
    }
    \label{tab:different_sampling}
\end{table}

\textbf{Finding 8. Stratified and antithetic techniques are compatible with sampling-based iteration reduction techniques, regardless of the sampling objects being coalitions or permutations.}  The two techniques can be integrated seamlessly with MC, RE, MLE, GT, or CP. However, the kernel-based technique is designed exclusively for permutation sampling, e.g., in algorithms based on MC or CP. 

\textbf{Discussion.} The stratified technique is applicable regardless of the sampling objects, because there exist suitable criteria to divide coalitions or permutations into disjoint strata. 
The antithetic technique also works for sampling the two objects, since given any coalition or permutation, there exists a negatively correlated counterpart. 
The kernel-based technique, in contrast, only defines the distance and similarity between permutations, and thus cannot work when the sampling object is a coalition.

\textbf{Finding 9. Reducing the approximation error of SV might compromise its computation efficiency.}

\textbf{Discussion.} The mainstream algorithms for computing approximate SV rely on sampling. However, according to the law of large numbers \cite{wiki}, no matter which sampling strategy is adopted, it is inevitable to sample more coalitions and compute their utilities to reduce approximation error, which, on the other hand, increases computation complexity. 
This leaves a riddle: \textbf{Given a sampling strategy, how to strike a balance between SV approximation error and computation efficiency?} 
Similarly, we conduct certain experiments in \S\ref{subsec: exp_impactsOfError} to bridge this gap.

\subsubsection{Privacy Preservation}\label{sssec:privacy_challenge}

Applying SV in DA can raise privacy concerns when the data in analysis contain sensitive or personal information \cite{wang2023threshold}. The privacy issues come from two aspects: (1) computing SV in DA, especially in distributed tasks like FL, requires exposure of data or data derivatives such as ML models; and (2) attackers can leverage SV to infer private and sensitive information about individuals in the dataset when the SV is reported by private data owners themselves or a cloud service. The current research has explored the potential of SV for feature inference attacks \cite{Luo2022FeatureIA} and membership inference attacks \cite{wang2023threshold}.

\underline{SV-driven feature inference attack}(FIA) aims to reconstruct private data by deducing the features of those data using the feature's SV. Prior work \cite{Luo2022FeatureIA} has explored two attack models, in both of which the attackers have access to an explanation service and can steal the SV returned from this service to the other users. 
Attackers in the first case are assumed to own the auxiliary data following the same distribution as the private data owned by the other user. 
They attempt to reconstruct this user's data in three steps. 
First, the attacker sends the features of the auxiliary data to the explanation service and obtains SVs corresponding to those features. 
Then, the attacker trains a regression model based on the received SVs and the auxiliary data features. 
Finally, the attacker feeds the stolen SVs into the learned regression model. 
The output results are the deduced features, which can be used for data reconstruction. 
Attackers in the other case do not need knowledge about data distribution. 
They first generate a set of arbitrary features and send those features to the explanation service to obtain the feature's SV. 
Then, the SV of each feature is compared with the stolen SV corresponding to the same feature dimension. 
When the two SVs are far from each other, the attackers discard the feature. 
Finally, the attackers take an average of the remaining features in each dimension for data reconstruction.

\underline{SV-driven membership inference attack}(MIA) targets to detect the presence or absence of data tuples in a private dataset  \cite{wang2023threshold}. 
The attackers are assumed to have access to a server that generates the data tuple's SV based on its maintained private dataset. 
The attack is performed as follows. 
Firstly, the attackers create a shadow dataset by randomly sampling from a predefined data distribution. 
Then, the attackers repeat this sampling process for several rounds. 
In each round, the attackers calculate the SV of the target data tuple under the coalition of the shadow dataset with and without it, respectively.
After certain rounds, the attackers estimate the distributions of SVs in the two cases (denoted as IN and OUT distributions).
Finally, the attackers query the server to obtain SV of the target data tuple and observe which distribution the returned SV follows to determine whether the target data tuple is in the private dataset or not.

Studies of SV-driven privacy issues are aimed at two kinds of objectives: (1) exposure elimination, which safeguards raw data and derivatives during SV computation, and (2) inference prevention, which handles privacy inference attacks.
As cataloged in \Cref{tab:privacy_protection}, countermeasures include non-perturbation masking, homomorphic encryption, and secure multiparty computation for the first objective, quantization and dimensionality reduction for the second objective, and differential privacy which serves both.

\begin{table}[t]

    \centering
    \scalebox{0.85}{
    \begin{tabular}{c|c|c|c|c}
    \toprule
         \textbf{Tech.} & \textbf{Objective} 
         & \textbf{Lightweight} & \textbf{Rigorous} & \textbf{Use Scenarios}\\
    \midrule
         NPM & \multirow{3}{*}{\shortstack{exposure \\ elimination}} & \faCheck  & \faTimes  & RI {\footnotesize\cite{technologies10060125}} \\
    \cline{1-1} \cline{3-5}
         HE & & \faTimes  & \faCheck  & FL{\footnotesize\cite{zheng2022secure}}\\
    \cline{1-1} \cline{3-5}
         SMPC & & \faTimes  & \faCheck  & DC{\footnotesize\cite{tian2022private}} \\
         
    \midrule
         QT & \multirow{2}{*}{\shortstack{inference \\ prevention}} & \faCheck  & \faTimes  & RI{\footnotesize\cite{Luo2022FeatureIA}} \\
         
    \cline{1-1} \cline{3-5}
         DR & & \faCheck  & \faTimes  & RI{\footnotesize\cite{Luo2022FeatureIA}}\\
    \midrule
         DP & both 
         & \faCheck  &  \faCheck  & DV{\footnotesize\cite{Watson2022DifferentiallyPS,wang2023threshold}}, RI{\footnotesize\cite{Luo2022FeatureIA,technologies10060125}}, RT{\footnotesize\cite{liu2021dealer}}\\
    \bottomrule
    \end{tabular}
    }
    \caption{Summary of privacy protection techniques.}
    \label{tab:privacy_protection}
\end{table}

\underline{Non-perturbation masking}(NPM) involves techniques of clustering, singular value decomposition, principal component analysis, non-negative matrix factorization, etc., which search for the most informative components in data in order to reduce the details of data to be exposed \cite{technologies10060125}. These techniques have been adopted by \citet{technologies10060125} to modify the features of the data used to train ML models before exposing those data to a third-party service for SV computation.

\underline{Homomorphic encryption}(HE) is a cryptographic primitive that enables arithmetic operations over encrypted data \cite{zheng2022secure}. A fully homomorphic encryption system can support additions and multiplications between ciphertexts or between a ciphertext and a plaintext. The effectiveness of this technique for secure SV computation has been tested in FL tasks \cite{zheng2022secure}.

\underline{Secure Multiparty computing}(SMPC) is a method for cryptographic computing that allows several parties holding private data to evaluate a public function (such as the SV's utility function) on their aggregate data without revealing anything other than the function output \cite{tian2022private}. This technique has been utilized for secure SV computation in DC tasks \cite{tian2022private}.

\underline{Quantization}(QT) maps the continuous SVs to a set of discrete values for reducing the mutual information between the private players and their corresponding SVs \cite{Luo2022FeatureIA}. \citet{Luo2022FeatureIA} utilized this method in RI tasks for preventing SV-driven FIA.

\underline{Dimension reduction}(DR) evaluates the importance of each player based on the variance of the player's marginal contributions \cite{Luo2022FeatureIA}. 
SVs are released for only a small number of top important players to reduce the amount of information accessible to attackers. Similar to QT, DR was used by \citet{Luo2022FeatureIA} in RI tasks for preventing SV-driven FIA.

\underline{Differential privacy}(DP) is a state-of-the-art privacy-preserving mechanism that can protect private inputs via a rigorous theoretical guarantee. The key idea is that an arbitrary change in data input into an algorithm should not be reflected in the algorithm's outputs. 
For privacy exposure elimination, DP adds noise (usually Gaussian noise) to the analytical object of the DA task before exposing the object for SV computation \cite{liu2021dealer}. For example, adding noise to local models analyzed during FL in \Cref{fig:SV_in_DA}. When aiming to prevent privacy inference, DP adds noise to SV, making the SVs of any two different private players indistinguishable and thus hard to exploit for inference attacks \cite{Luo2022FeatureIA,wang2023threshold,Watson2022DifferentiallyPS}. For instance, noising the SV computed in the RI task to prevent inference attacks on private medical images. 
\citet{wang2023threshold} utilized DP to prevent SV-driven MIA. 
\citet{Watson2022DifferentiallyPS} also adopted this technique but did not specify any attack model.

\textbf{Finding 10. Lightweight privacy protection techniques, including NPM, QT, DR, and DP, balance efficiency and protection, while rigorous techniques, containing HE and SMPC, prioritize security at a higher computation cost.} 
To achieve both privacy preservation and secure computation, hybrid schemes that combine HE or SMPC with QT, DR, or DP are recommended. 

\textbf{Discussion.} Lightweight and rigorous theoretical privacy guarantees are two primary factors affecting the selection of privacy protection measures for SV. 
For applications related to cross-institutional highly-sensitive data (e.g., cross-hospital medical images in \Cref{fig:SV_in_DA}), rigorous techniques like HE or SMPC are needed. 
While if computing resources are constrained (e.g., in applications deployed across edge devices), the lightweight NPM, QT, or DR techniques are preferred. 
We note that DP, possessing the lightweight property and rigorous theoretical guarantees simultaneously, is applicable to both cases.

\textbf{Finding 11. Adopting privacy-preserving measures may compromise SV's computation efficiency and effectiveness.}

\textbf{Discussion.} 
The impacts on the efficiency can originate from three factors: (1) the noise (introduced by measures such as DP and NPM) which may slow down computation convergence, (2) the extra time cost needed for encrypting and decrypting data (introduced by measures such as HE), and (3) the extra time cost for multiparty interactions (needed by measures such as SMCP).  
Several studies \cite{zheng2022secure, tian2022private, Watson2022DifferentiallyPS, wang2023threshold, Luo2022FeatureIA, liu2021dealer} have attempted to achieve a compromise between SV's computation efficiency and privacy preservation. Their key idea is to combine privacy-preserving measures with hybrid SV computing algorithms which improve efficiency by integrating techniques summarized in \S\ref{sssec: computation_efficiency}. For example, \citet{tian2022private} adopted SMCP on top of a hybrid scheme that computes SV using both MC and MA techniques, ensuring the efficiency of SV computation on data held by different data owners while eliminating the need to expose data before buyers pay for them. 

Privacy protection can also influence the effectiveness of SV.
The mainstream arts rely on the scaling \cite{10475986, 10056725, sun2023shapleyfl, 10.1007/978-3-030-93049-3_14, 10219590,franses2024shapley,app13127010} or ranking \cite{jia2019towards,meepaganithage2024feature,kwon2021efficient,ghorbani2019data,Zhu2023ShapleyvaluebasedCE,tian2022private} of final SV results to perform pricing, selection, weighting, and attribution in DA. 
However, the scaled SV results or the ranking results can be easily altered \cite{technologies10060125} when integrated with techniques like DP, QT, and DR. 
We notice a scarcity of quantitative results for answering the question: \textbf{Can a balance be achieved between the effectiveness of privacy protection with the effectiveness of SV?} 
Therefore, we offer insights into this question with the experimental study in \S\ref{subsec: exp_privacy}.

Overall, we advocate a deeper cost-benefit study analyzing the marginal gains in privacy protection against the computation cost and effectiveness loss of SV, both qualitatively and quantitatively. 
Besides, a dynamic adjustment on the efficiency, privacy, and effectiveness of SV according to DA task settings, system constraints, user requirements, etc., merits further investigation.

\subsubsection{Interpretability} \label{sssec:interpretation_challenge}

In addition to the aforementioned challenges, applying SV to DA also faces the trouble of how to properly translate the obtained SVs to exact actions (e.g., adding or deleting a data sample, normalizing data features in some dimensions, etc.) in the DA workflow \cite{kumar2020problems}. 

For understandable SV interpretations, researchers have relied on three paradigms: utility-based, characteristic-based, and counterfactual. 

\underline{Utility-based} paradigm points out that players with high SV are those who have more impact on the overall utility of the targeted DA tasks. For example, altering the value of data features with high SV tends to incur more changes in model predictions \cite{kumar2021shapley, kumar2020problems, Kumar2021ShapleyRQ}. Similarly, high-valued data tuples result in more increase in test accuracy \cite{ghorbani2019data, jia2019towards, shobeiri2021shapley, shobeiri2022shapley11}.

\underline{Characteristic-based} paradigm seeks to reveal the relationship between the intrinsic characteristics of each player and its SV. It has been shown that the data tuples with low SV usually belong to noisy data with false labels \cite{jia2019towards}, outliers \cite{ghorbani2019data}, or corruptions \cite{ghorbani2019data}, while the ML models with high SV are those that have high test accuracy and certainty \cite{xu2024model, rozemberczki2021shapley}.

\underline{Counterfactual} paradigm aims to find out the minimum change between two players that can flip the direction of the inequality between the SVs of those players. \citet{counterfactual} utilized this paradigm to explain why a dataset has a larger SV than another. Given a dataset (defined as the player), the work searched for the minimum set of tuples in the given dataset such that transferring the found set from the given dataset to another dataset can flip the direction of the inequality between the SVs of those two datasets.

\begin{table}[t]
    \centering
    
    \scalebox{0.85}{
    \begin{tabularx}{0.55\textwidth}{
    >{\hsize=.2\hsize\linewidth=\hsize\centering\arraybackslash}X |
    >{\hsize=.8\hsize\linewidth=\hsize\raggedright\arraybackslash}X 
    }
    \toprule
         \textbf{Tech.} & \multicolumn{1}{c}{\textbf{Interpretations}} \\
    \midrule
         \multirow{4}{*}{Utility-based}& 
         Altering the value of data features with high SV tends to incur more changes in model predictions \cite{kumar2021shapley, kumar2020problems, Kumar2021ShapleyRQ}. 
         \\
         & High-valued data tuples result in more increase in test accuracy \cite{ghorbani2019data, jia2019towards, shobeiri2021shapley, shobeiri2022shapley11}.
         \\
    \midrule
         \multirow{4}{*}{\shortstack{Characteristic-\\based}}& Data tuples with low SV are inclined to be noisy data with false labels \cite{jia2019towards}, outliers \cite{ghorbani2019data}, or corruptions \cite{ghorbani2019data}. \\
         & ML models with high SV are those that have high test accuracy and certainty \cite{xu2024model, rozemberczki2021shapley}. \\
         
    \midrule
         \multirow{4}{*}{Counterfactual}& Given a dataset, there exists a minimum set of tuples in this dataset such that transferring the found set from this dataset to another dataset can flip the direction of the inequality between the SVs of those two datasets \cite{counterfactual}.  \\
    \bottomrule
    \end{tabularx}
    }
    \caption{Summary of SV interpretations.}
    \label{tab:SV_interpretation_paradigms}
\end{table}

\textbf{Finding 12. The utility-based interpretation paradigm is the most universal for interpreting SV in DA.} 
In contrast, the characteristic-based paradigm needs sufficient expert knowledge to elaborate intrinsic characteristics of data, and the counterfactual explanation needs costly computations.

\textbf{Discussion.} With requirements on expert knowledge or extra computations, the characteristic-based paradigm and the counterfactual explanation did not gain a wider application than the utility-based paradigm. 
Despite the popularity of utility-based interpretations, many works \cite{kumar2020problems, kumar2021shapley, zhao2024erroranalysisshapleyvaluebased, pmlr-v127-ma20a} have claimed that the player with a larger SV may have less influence on the overall utility of the associated DA task, which contradicts the mainstream interpretations.
However, these works studied only the cases defining data features as the player and thus cannot fully answer the following questions: \textbf{Can the SVs of the four types of players in DA be correctly interpreted by the mainstream paradigm? If cannot, why? Is there any general reason applicable to all four types of players?} 
We endeavor to answer these questions through a comprehensive evaluation in \S\ref{subsec: exp_interpretability}.

\subsubsection{Summary of Findings}\label{ssec:comparision_with_previous_survey}

Firstly, our findings, summarized from a wide range of the arts, provide \textit{general and comprehensive} guidelines for academicians and engineers to study and develop SV applications when confronted with DA tasks. 
Secondly, our findings analyze the arts in this field through \textit{a finer-grained perspective}. 
We analyze the pros and cons of the resolution techniques disentangled from complete algorithms in addressing four challenges of using SV in DA (Findings 4, 6, 10, 12), propose the possible combinations of the techniques for developing new algorithms (Findings 5, 7, 8), and discuss the potential conflicts between different challenges, e.g., computation efficiency vs. approximation error (Finding 9) or privacy preservation (Finding 11). 
Finally, as a consequence of the finer-grained understanding of SV, our findings can instruct the design of a \textit{modular and extensible} framework for developing SV applications, presented in the next section. 

\section{\textit{SVBench}} \label{sec: implementation}
In this section, we propose \textit{SVBench}, a modular and extensible framework for developing SV applications for DA tasks, aiming to bridge the gap in a unified library supporting flexible integrations of state-of-the-art countermeasures to different challenges of SV. 

\begin{figure}[t]
    \centering
    \includegraphics[width=.9\columnwidth]{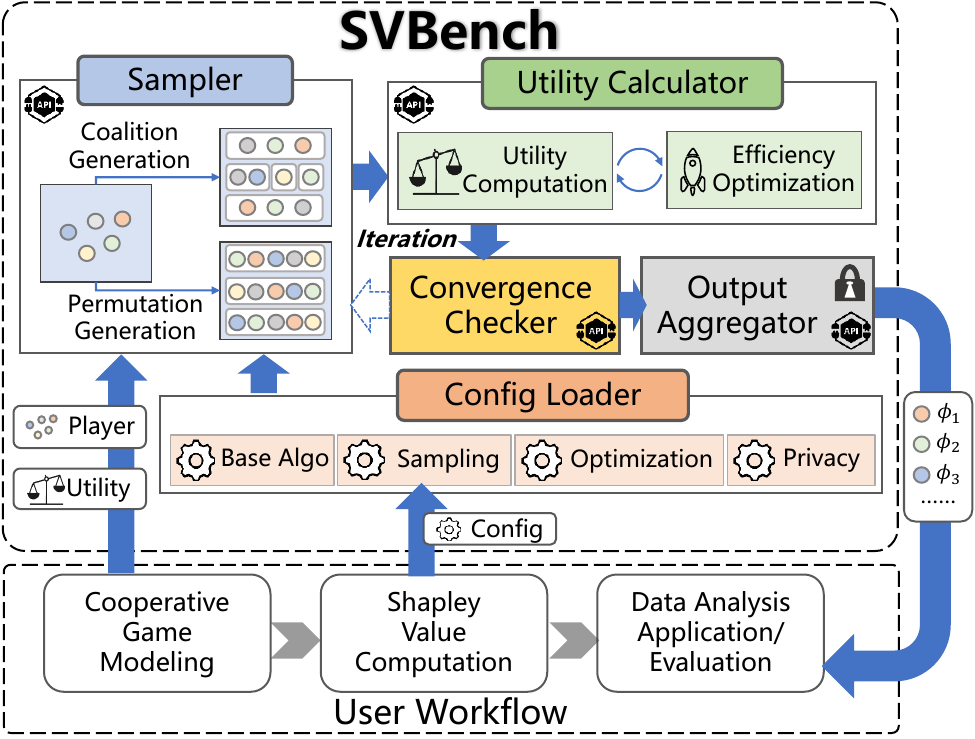}
    \caption{The \textit{SVBench} overview. %
    }
    \label{fig:benchmark_overview}
\end{figure}

\textbf{Overview.} 
As shown in \Cref{fig:benchmark_overview}, \textit{SVBench} consists of a configuration loader, a sampler, a utility calculator, a convergence checker, and an output aggregator. 
The configuration loader loads the SV computing parameters specified by the users. 
The sampler generates the coalitions or permutations of players based on the configured sampling strategy. 
The utility calculator takes the sampled coalitions or permutations as the input to a utility function and drives the computation.
When users specify an efficiency optimization strategy, the utility calculator will use that strategy to accelerate the computation. 
The convergence checker determines whether to terminate the SV computation based on the convergence criterion specified in the configuration. 
An \textit{iteration} of SV calculation is conducted starting from the sampler and ending at the convergence checker. 
Once the convergence criterion is not met, another iteration will be initiated as demonstrated in the figure (with dashed arrow). 
The output aggregator generates the final SV of each player.
If users specify privacy protection measures, the aggregator will execute those measures before reporting the final results.

\begin{table}[t]
    \centering
    \scalebox{0.9}{
    \begin{tabular}{l|l}
    \toprule
        \textbf{Config} & \textbf{Parameters (Default setting is in italics.)} \\
    \hline
        Base Algo & \textit{MC}, RE, MLE, GT, CP, user-specific\\
    \hline
        Sampling & None, \textit{random}, stratified, antithetic, user-specific\\
    \hline
        \multirow{2}{*}{Optimization} & \textit{None}, TC, GA, TC+GA, GA+TSS,\\
        &TC+GA+TSS, user-specific \\
    \hline
        Privacy & \textit{None}, DP, QT, DR, user-specific\\
    \bottomrule
    \end{tabular}
    }
    \caption{Configuration parameters used by \textit{SVBench}. 
    }
    \label{tab:configuration_parameters}
\vspace{-1em}
\end{table}

\textbf{Usage Instructions.} Users of \textit{SVBench}, including seasoned engineers and researchers seeking updates on the building blocks of SV applications and newcomers to this field, need to initialize the framework in three steps. 
The first step is \textit{cooperative game modeling}, in which the users properly define the players and the utility function of their targeted DA task. 
\textit{SVBench} suggests users perform this step according to \Cref{fig:cooperative_games_in_DA} and also allows new definitions of player and utility not included in this figure. 
The second step is \textit{SV computation}, in which users specify the configuration parameters listed in \Cref{tab:configuration_parameters}.
The default settings are marked in the table. 
The final step is \textit{data analysis application/evaluation}, in which users utilize SV results to perform pricing, selection, weighting, or attribution on data or data derivatives according to the specific DA task requirements.

\textbf{Use Case.} Take the DV task in \Cref{fig:SV_in_DA} as an example. Suppose the user plans to use \textit{SVBench} to implement a hybrid SV algorithm, which combines MC and TC techniques with DP for privacy protection, to generate SVs of 6000 images from a `MedicalImg' dataset in DV. The user defines 6000 images as the players and a `DV-MI' function, which takes images as inputs to train a classification model and outputs the test accuracy of this model, as the utility function. Next, the user configures \textit{SVBench} with the following parameters: task=`DV', dataset=`medicalImg', player=`tuple', utility\_function=`DV-MI', base\_algo=`MC', optimization\_strategy= `TC', privacy\_protection\_measure=`DP', then invokes the SV computing function in \textit{SVBench} to obtain SVs. After that, the user selects the images assigned with the top-50\% SV results to learn the classification model. More use cases can be found at GitHub \textsuperscript{\ref{fn:github_url}}.

\begin{table}[t]
    
    \centering
    \scalebox{0.9}{
    \begin{tabular}{c|c|c|c|c}
    \toprule
        \multicolumn{4}{c|}{\textbf{Configuration of \textit{SVBench}}} & \multirow{2}{*}{\shortstack{\textbf{Section} \\ \textbf{Index}}}  \\
        \cmidrule(l{1pt}r{1pt}){1-4}
        \textbf{Base Algo} & \textbf{Sampling} & \textbf{Optimization} & \textbf{Privacy} &  \\
    \hline
        \multirow{3}{*}{\shortstack{MC / RE$^\spadesuit$/\\ MLE$^\spadesuit$ / GT$^\spadesuit$/ \\ CP$^\spadesuit$ }} & \multirow{3}{*}{random} & None / TC / GA / & \multirow{3}{*}{None} & \multirow{3}{*}{\S\ref{subsec: exp_efficiency}}  \\
        & & TC+GA / GA+TSS /& & \\
        & & TC+GA+TSS & & \\
    \hline
        \multirow{5}{*}{MLE} & random/ & \multirow{3}{*}{None} & \multirow{3}{*}{None} & \multirow{3}{*}{\S\ref{subsec: exp_impactsOfError}}  \\
        & stratified/ &  & &  \\
        & antithetic & & &   \\
    \hhline{|~|-|-|-|-|}
        & \multirow{1}{*}{random} & \multirow{1}{*}{None} & DP / QT / DR & \multirow{1}{*}{\S\ref{subsec: exp_privacy}}  \\
    \hhline{|~|-|-|-|-|}
        & random & None & None & \S\ref{subsec: exp_interpretability}\\
    \bottomrule
    \end{tabular}
    }
    \caption{
    The configuration of \textit{SVBench} for implementing different SV computing algorithms in \S\ref{subsec: exp_efficiency}--\S\ref{subsec: exp_interpretability}. 
    }
    \label{tab:exp_settings_method}
\end{table}

\textbf{Extension Supports.} Besides implementing the mainstream techniques for computing SV, we provide several APIs in the modules of \textit{SVBench} (marked by an API icon in \Cref{fig:benchmark_overview}) for users to extend this framework. 
One can configure the module he expects to extend by a user-specific parameter at the SV computation step and submit the new functions corresponding to that module. 
For example, in the above use case, the user submits a user-specific utility function namely DV-MI and configures the utility calculator module by setting utility\_function=`DV-MI'. Similarly, the user can also configure a user-specific aggregator module by submitting a new privacy protection function (e.g., namely newMaskSV, which takes original SV computing results as inputs and outputs masked SV results) and setting privacy\_protection\_measure=`newMaskSV'. \textit{SVBench} will check the legitimacy and validity of the received functions and use the valid functions to execute the operations in the corresponding modules. 
Moreover, with the user permission, \textit{SVBench} will embed the valid new functions into their corresponding modules to provide more development choices for future use.

\textbf{Summary.} 
The application of SV in DA requires to be flexible in engineering implementation, yet most existing works overlook such demands. 
The application should feature \textit{a modular architecture with configurable parameters}, empowering engineers to tailor the usage of SV according to task-specific requirements.
Moreover, the application should possess the ability to be \textit{parallelized and disaggregated}. In this manner, it can leverage diverse computing architectures, including multi-core CPUs, GPUs, and distributed computing paradigms, to enhance scalability and efficiency. 
With these considerations, we propose \textit{SVBench}. 
While \textit{SVBench} has demonstrated usability, modularity, and flexibility through the success of implementing dozens of algorithms in the next section, we anticipate future extensions to further enhance its potential to develop efficient, secure, and effective SV applications in DA.

\begin{table}[t]
    \centering
    \scalebox{0.85}{
    \begin{tabular}{l|c|c|c}
    \toprule
         \textbf{Task} & \textbf{Dataset} & 
         \textbf{Player} & \textbf{Utility} \\ 
    \hline
         \multirow{5}{*}{RI} & Iris \cite{UCI_dataset} & 
         Data Feature ($n=4$) & \multirow{5}{*}{Model Output} \\ 
         &  Wine \cite{UCI_dataset} & 
         Data Feature ($n=13$) & \\ 
        & Adult \cite{UCI_dataset} & 
         Data Feature ($n=14$) & \\ 
         &  Ttt \cite{UCI_dataset}  & 
         Data Feature ($n=9$) & \\
         &  2Dplanes \cite{OpenML2013,OpenML2020}  & 
         Data Feature ($n=10$) & \\         
    \hline
    
         \multirow{5}{*}{DV} & Iris \cite{UCI_dataset} & 
         Data Tuple ($n=18$) & \multirow{5}{*}{Test Accuracy} \\ 
         & Wine \cite{UCI_dataset} &  
         Data Tuple ($n=15$) & \\ 
         & Bank \cite{UCI_dataset}  & 
         Data Tuple ($n=18$) & \\
         &  Ttt \cite{UCI_dataset} & 
         Data Tuple ($n=22$) & \\
         &  Wind \cite{OpenML2013,OpenML2020} & 
         Data Tuple ($n=18$) & \\  
    \hline
    
        \multirow{5}{*}{DSV} & Mnist \cite{lecun-mnisthandwrittendigit-2010}  &
        \multirow{5}{*}{\shortstack{Dataset \\($n=10$)}} & \multirow{5}{*}{Test Accuracy} \\ 
         & Cifar-10 \cite{Krizhevsky2009LearningML} &   & \\ 
        & Bank \cite{UCI_dataset}  & &  \\ 
         &  Dota2 \cite{UCI_dataset}  & & \\
         &  2Dplanes \cite{OpenML2013,OpenML2020} & & \\  
         
    \hline
         \multirow{5}{*}{FL} & Mnist \cite{lecun-mnisthandwrittendigit-2010}  & 
         \multirow{5}{*}{\shortstack{ML model \\($n=10$)}} & \multirow{5}{*}{Test Accuracy} \\ 
         & Cifar-10 \cite{Krizhevsky2009LearningML} &  &  \\ %
         &  Adult \cite{UCI_dataset} & &  \\  
         &  Dota2 \cite{UCI_dataset} & & \\  
         &  Wind \cite{OpenML2013,OpenML2020} & & \\  
    \bottomrule
    \end{tabular}
    }
    \caption{Task settings. $n$: total number of players. Each dataset is split into a training set with 80\% samples from the original dataset and a test set with the remaining 20\% samples. } 
    \vspace{-.5cm}
    \label{tab:exp_task_settings}
\end{table}

\section{Evaluation} \label{sec: exp}

\begin{figure*}
    \centering
    \begin{minipage}{1.2\columnwidth}
        \centering 
        \hspace{-35pt} 
        \includegraphics[width=1.1\columnwidth]{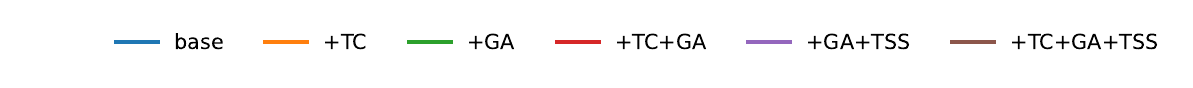}
    \end{minipage}
    \vspace{-5pt}
    \\
    \mbox{
        \includegraphics[width=0.215\columnwidth, height=0.16\columnwidth]{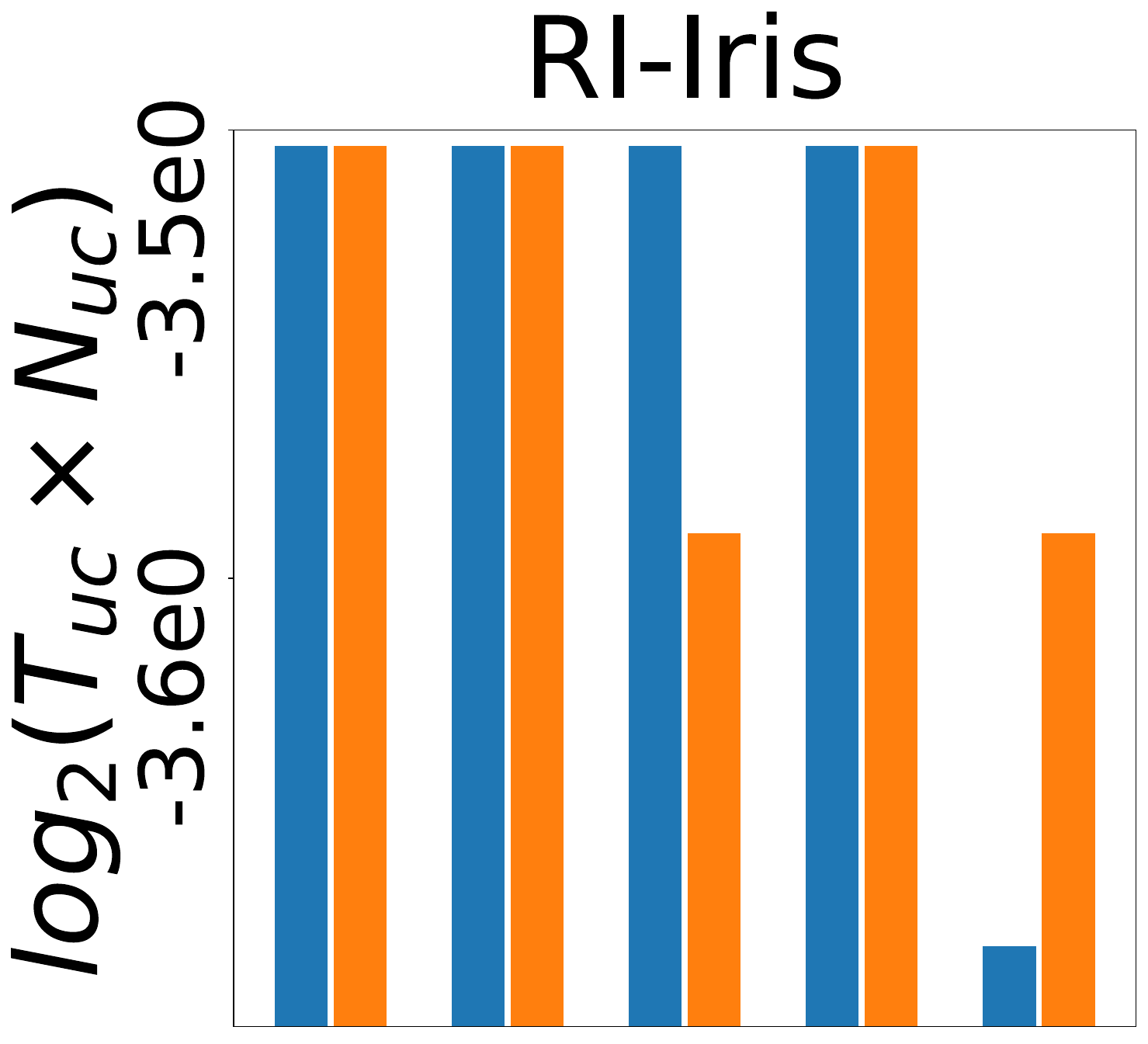}
        \includegraphics[width=0.2\columnwidth, height=0.16\columnwidth]{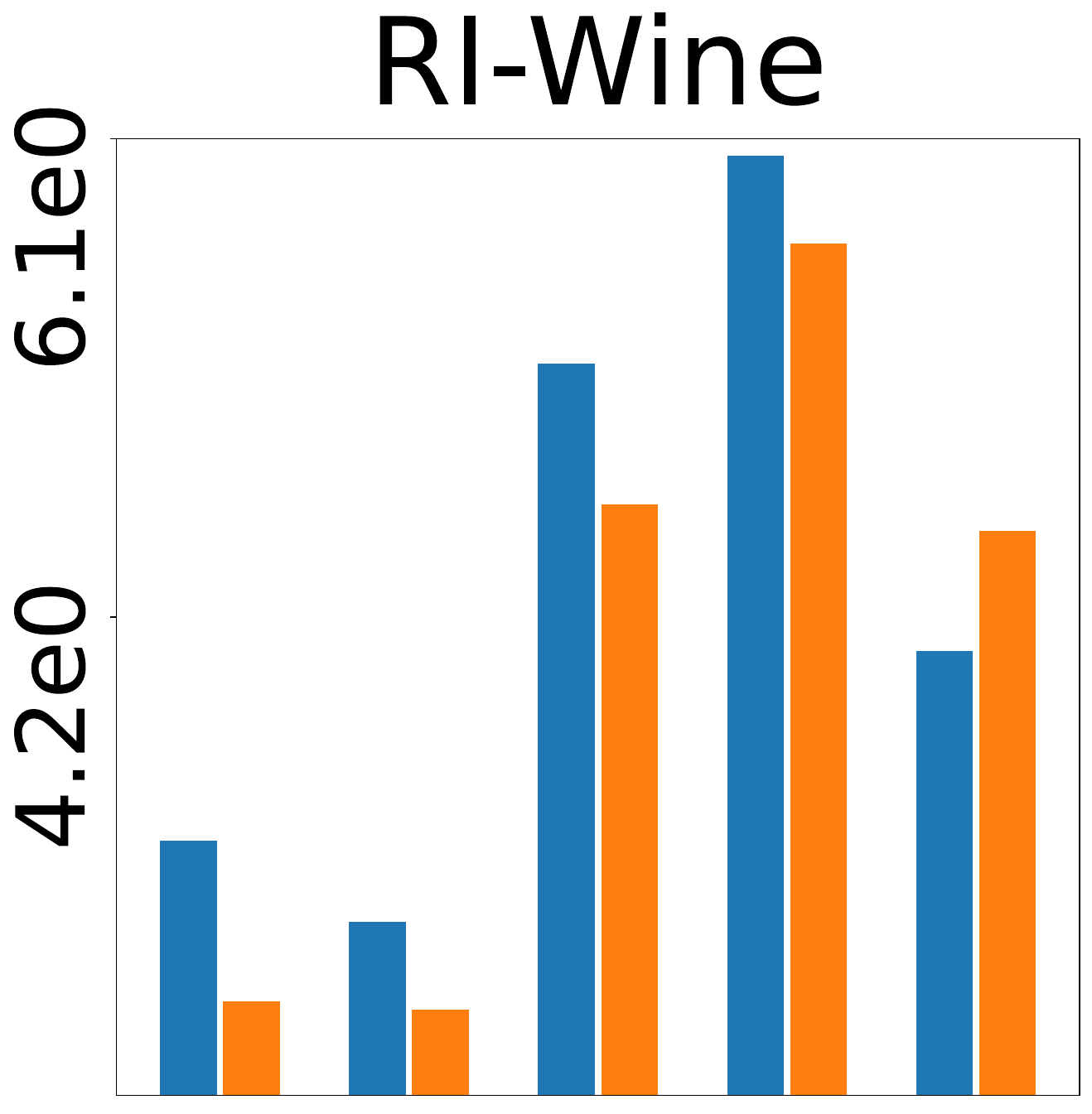}
        \includegraphics[width=0.2\columnwidth, height=0.16\columnwidth]{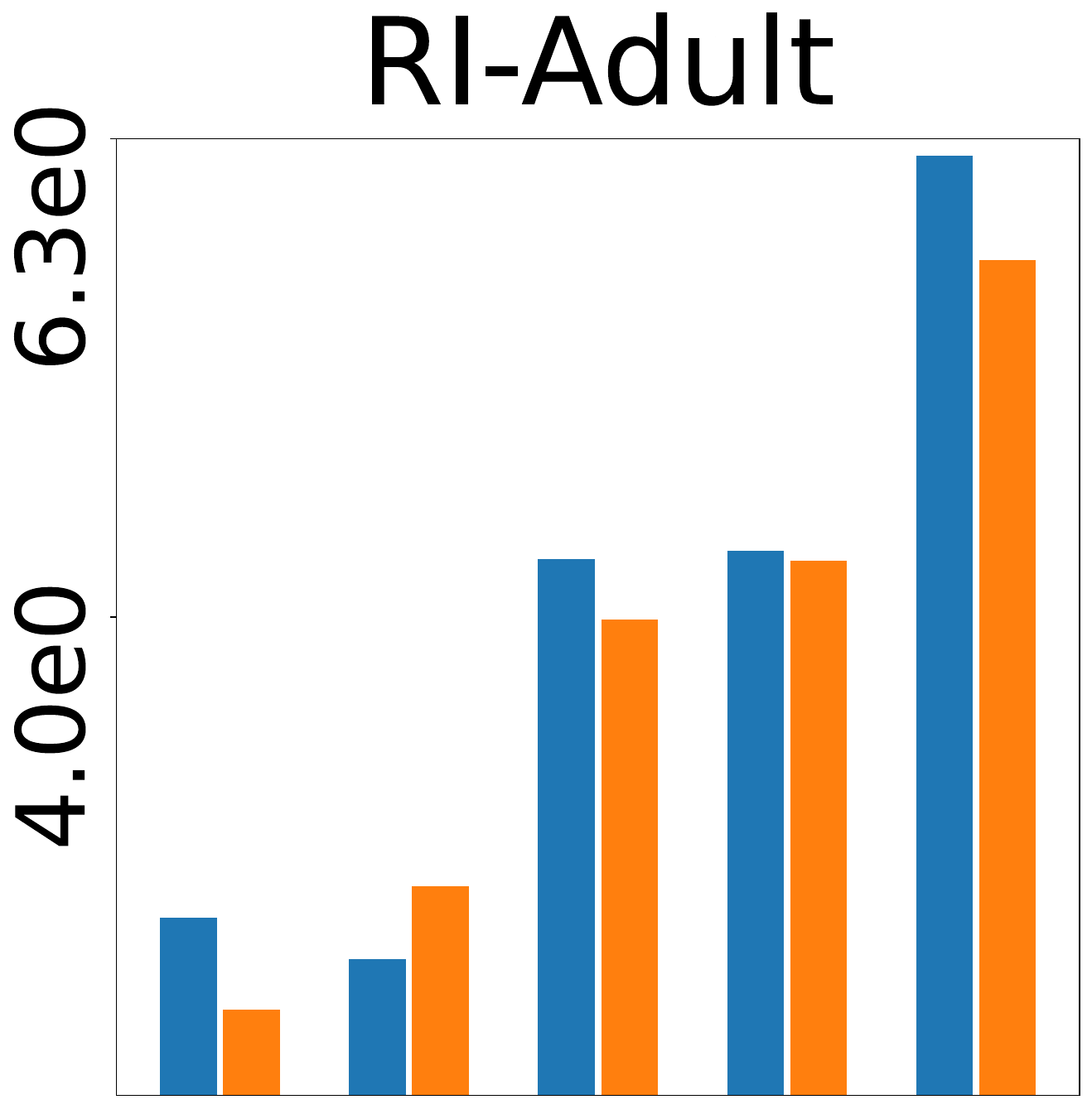}
        \includegraphics[width=0.2\columnwidth, height=0.16\columnwidth]{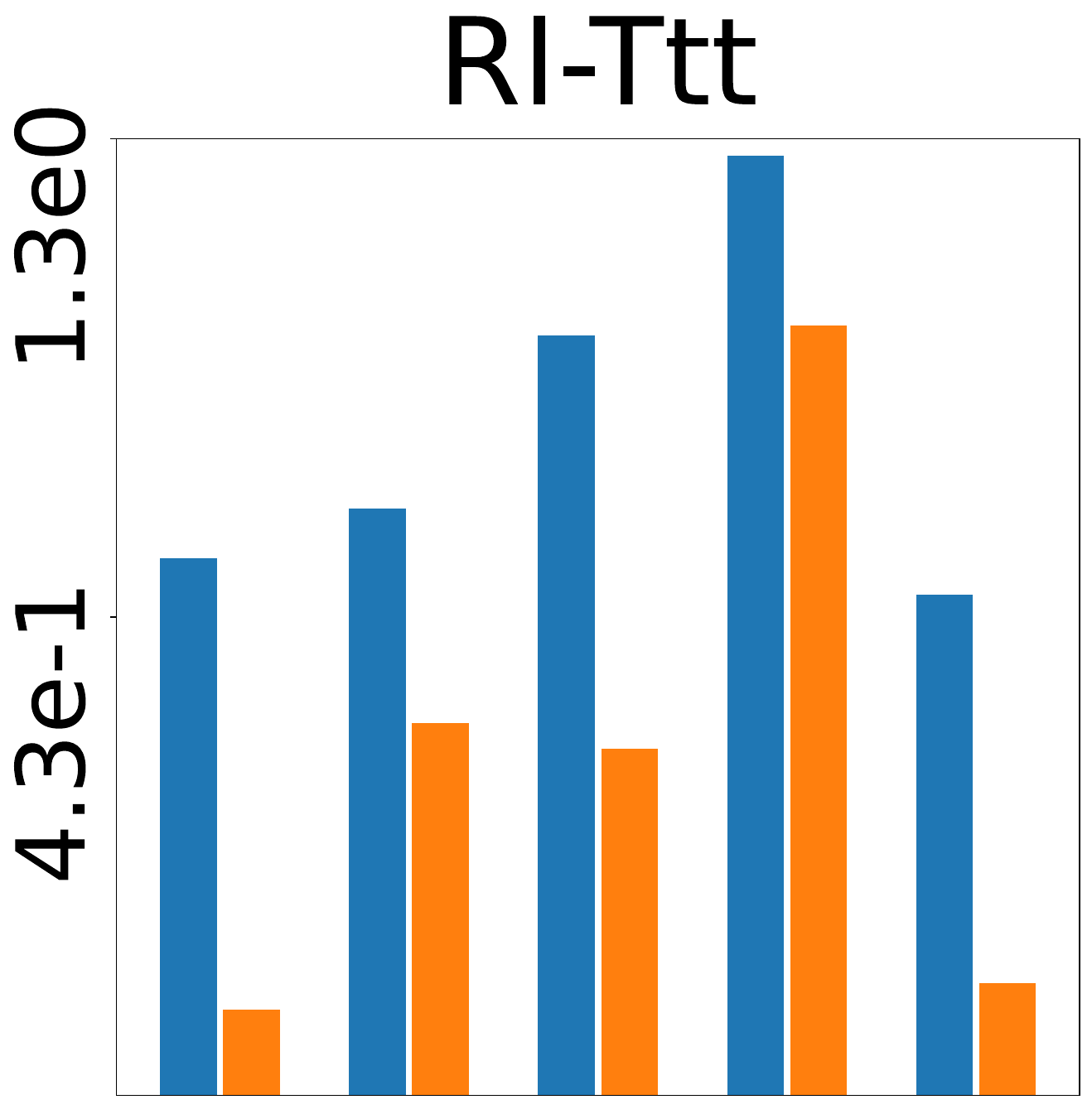}
        \includegraphics[width=0.2\columnwidth, height=0.16\columnwidth]{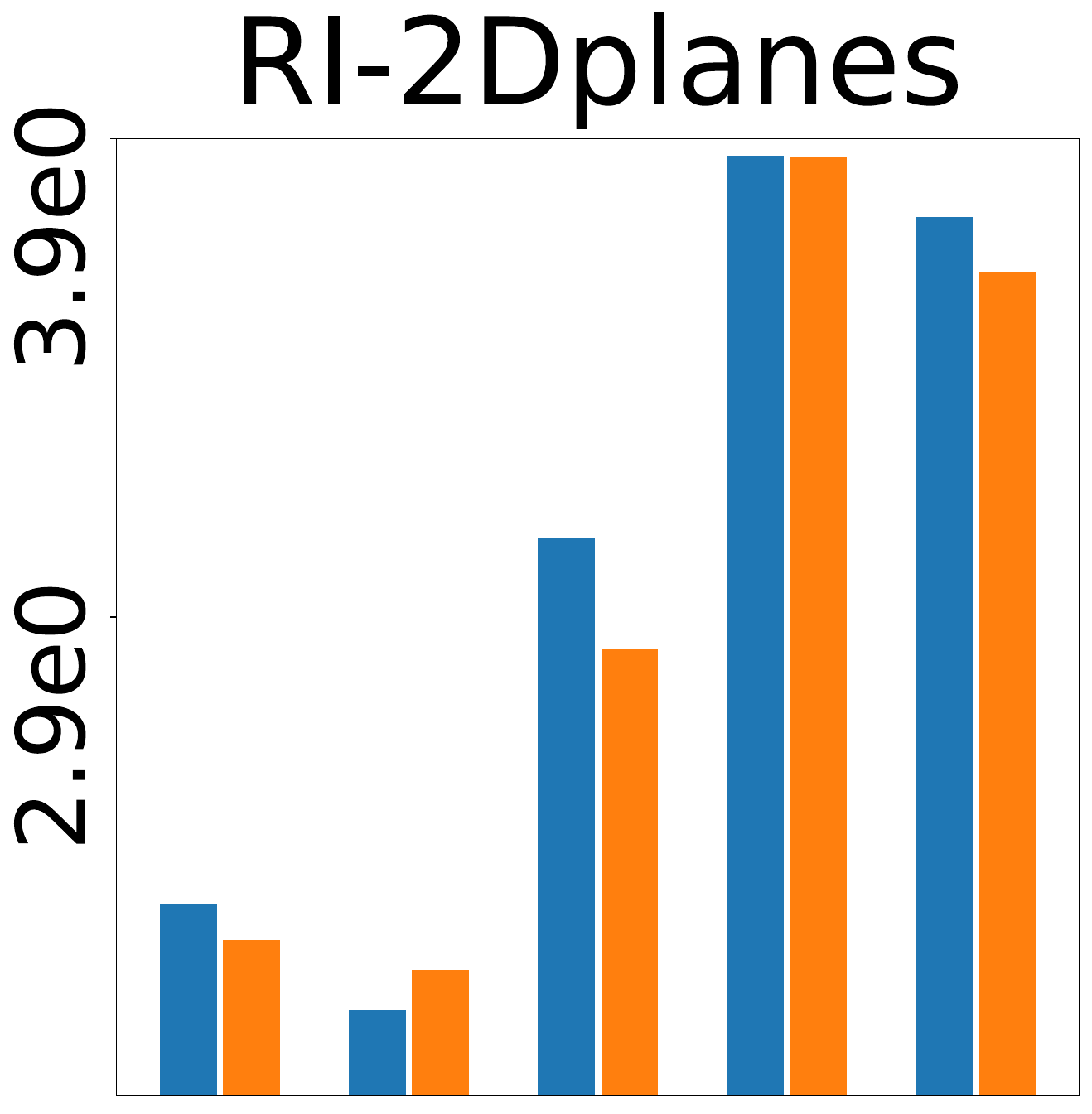}

        \includegraphics[width=0.2\columnwidth, height=0.16\columnwidth]{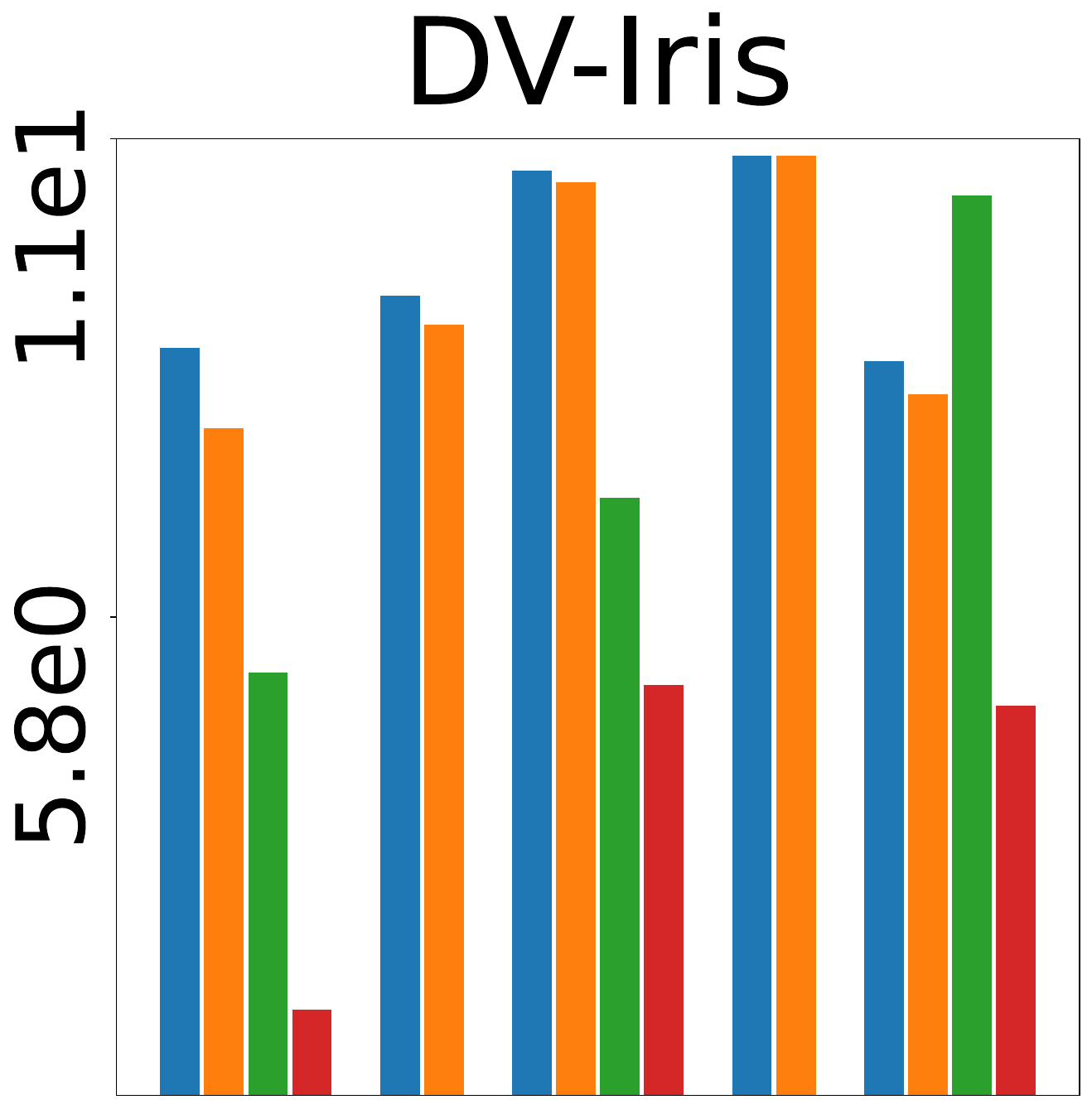}
        \includegraphics[width=0.2\columnwidth, height=0.16\columnwidth]{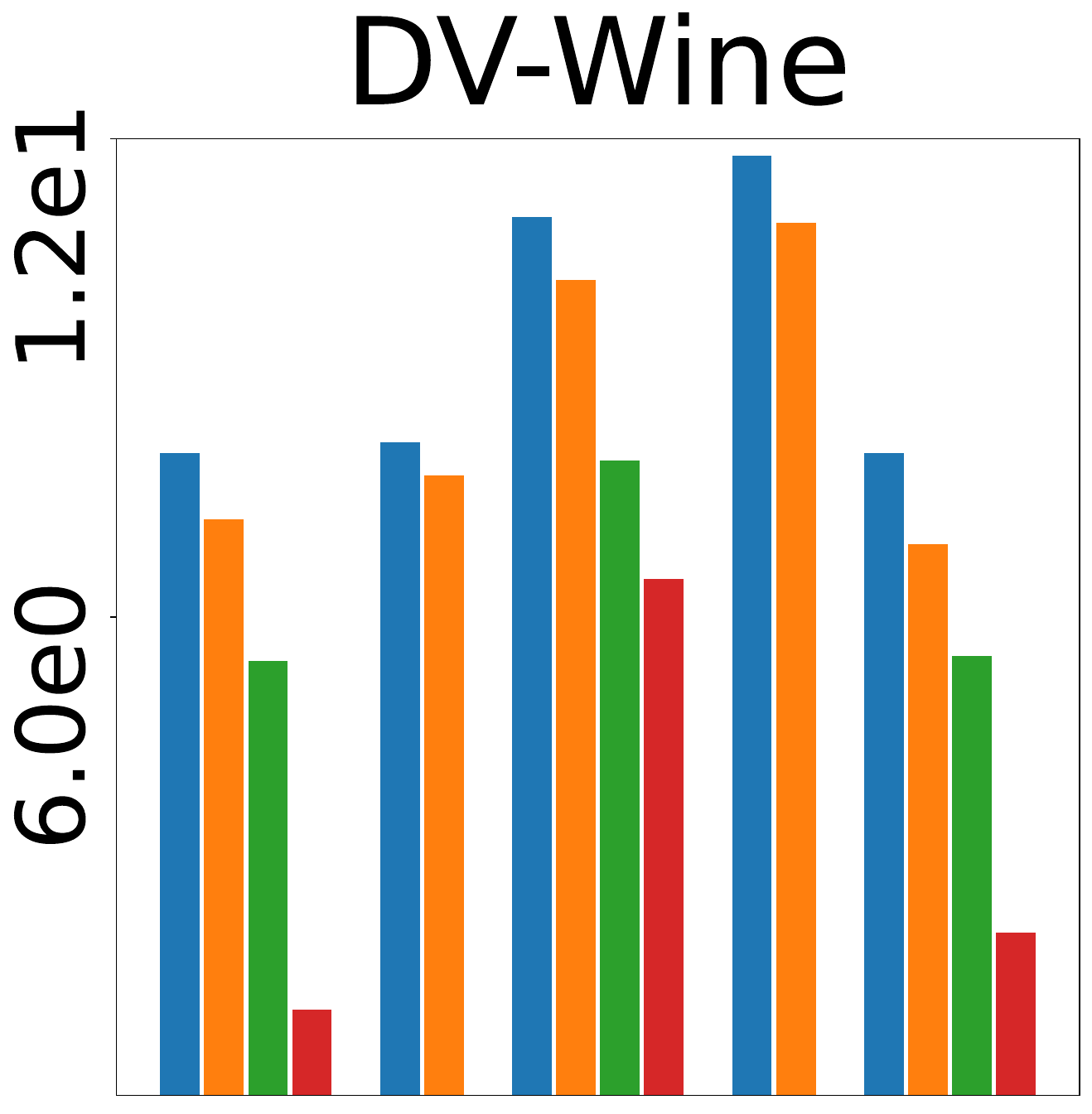}
        \includegraphics[width=0.2\columnwidth, height=0.16\columnwidth]{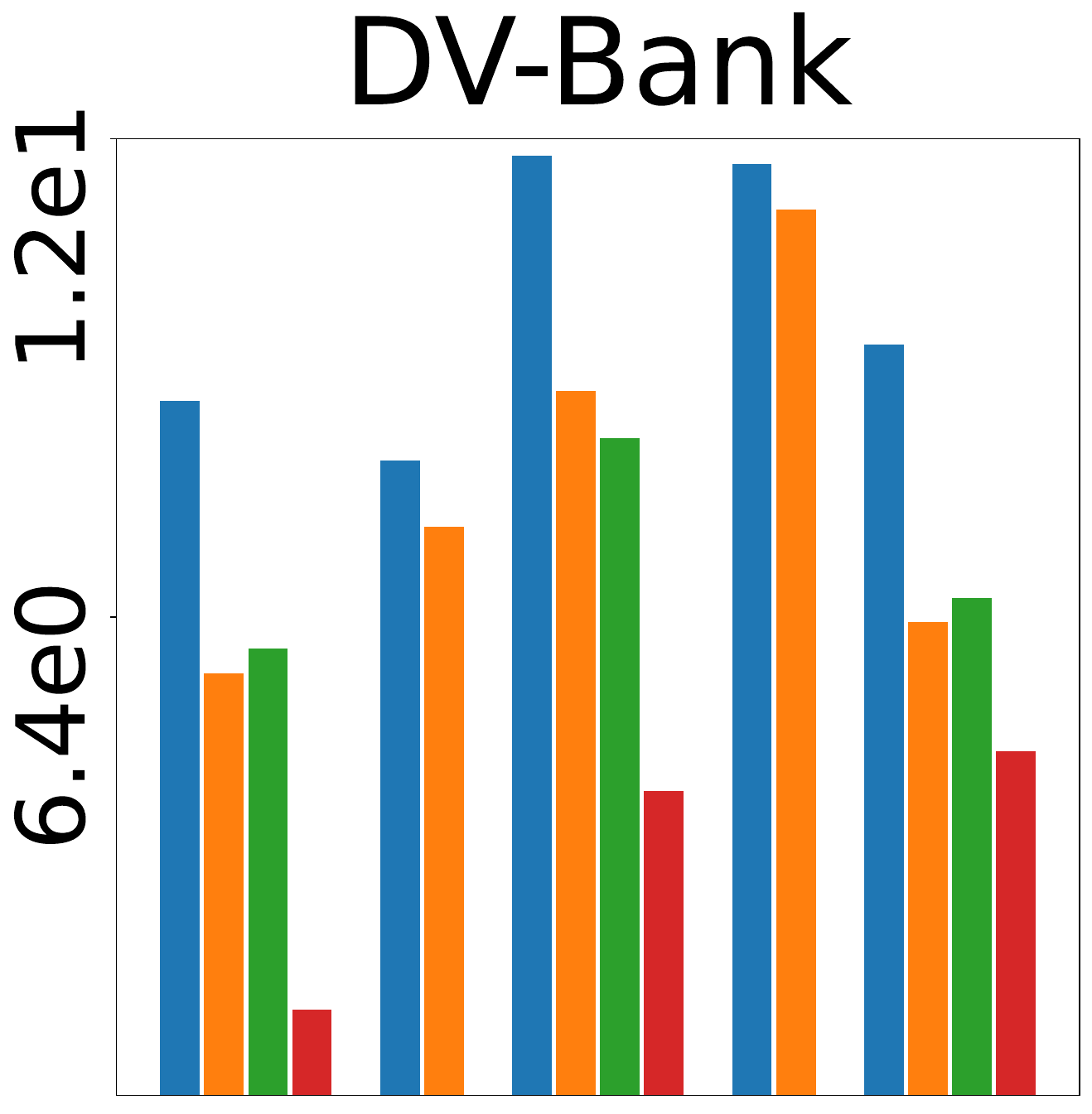}
        \includegraphics[width=0.2\columnwidth, height=0.16\columnwidth]{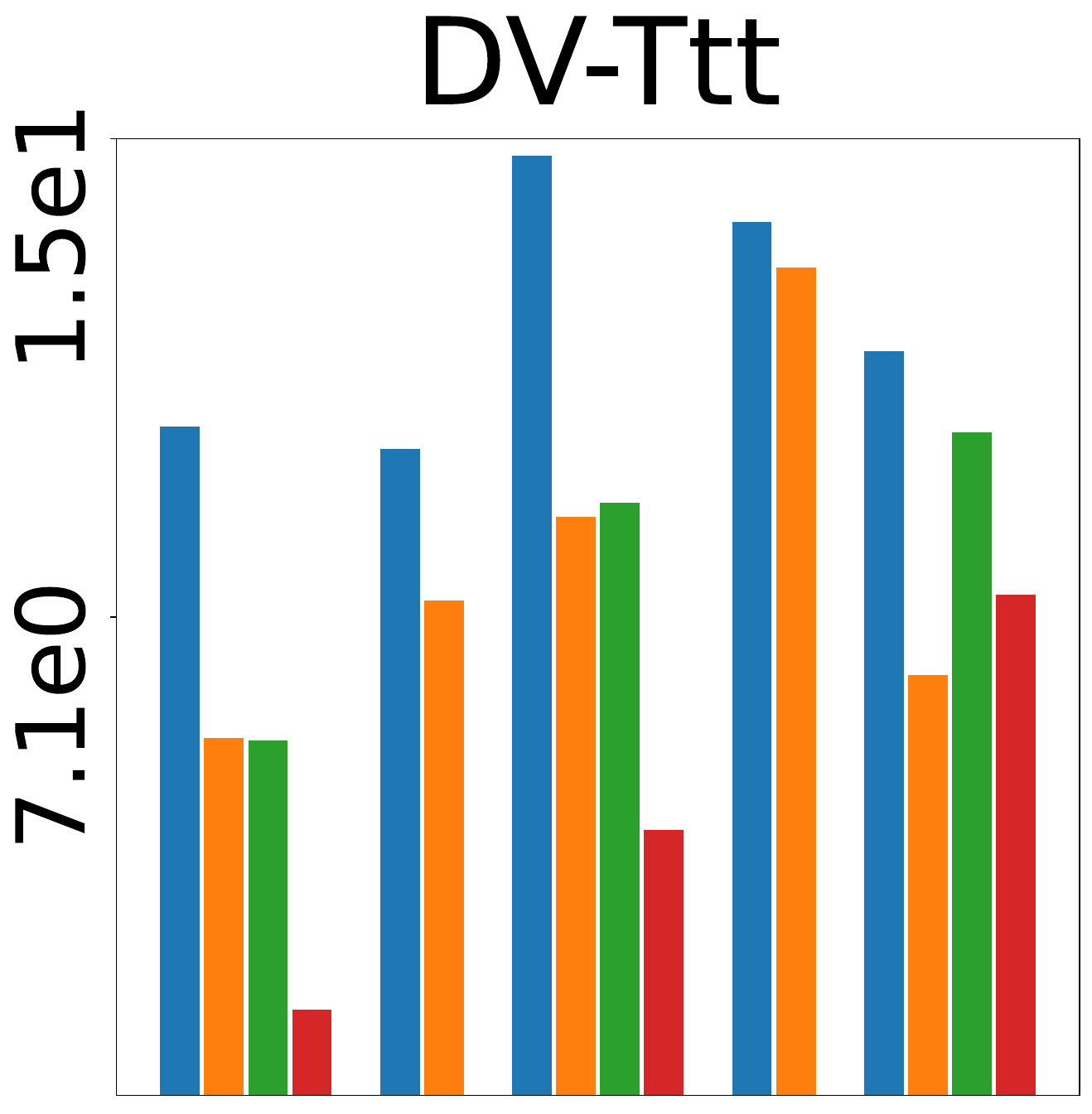}
        \includegraphics[width=0.2\columnwidth, height=0.16\columnwidth]{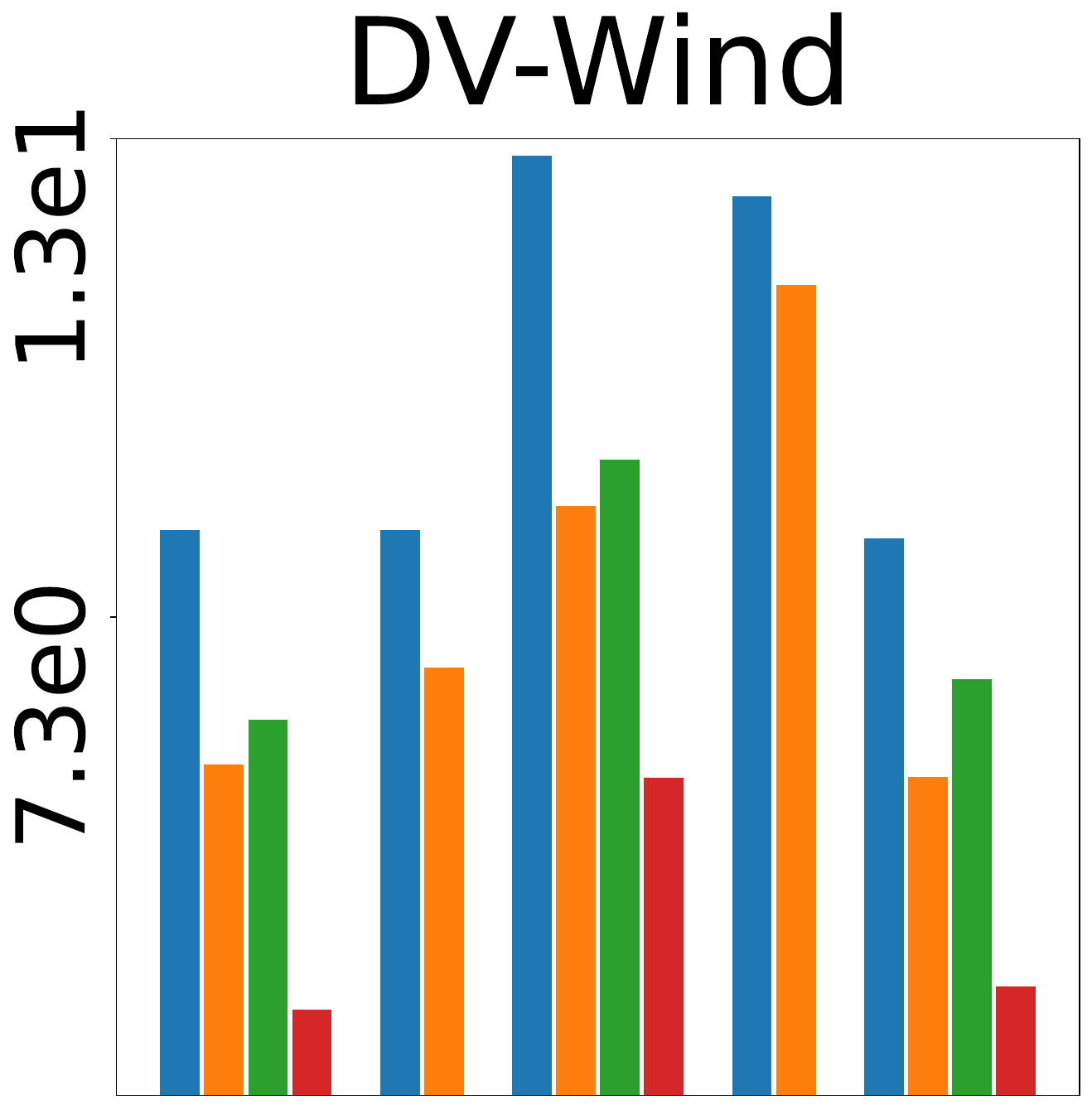}
    }
    \\
    
    \subfigure[The total time cost, $T_{uc} \times N_{uc}$.] {
        \includegraphics[width=0.215\columnwidth, height=0.2\columnwidth]{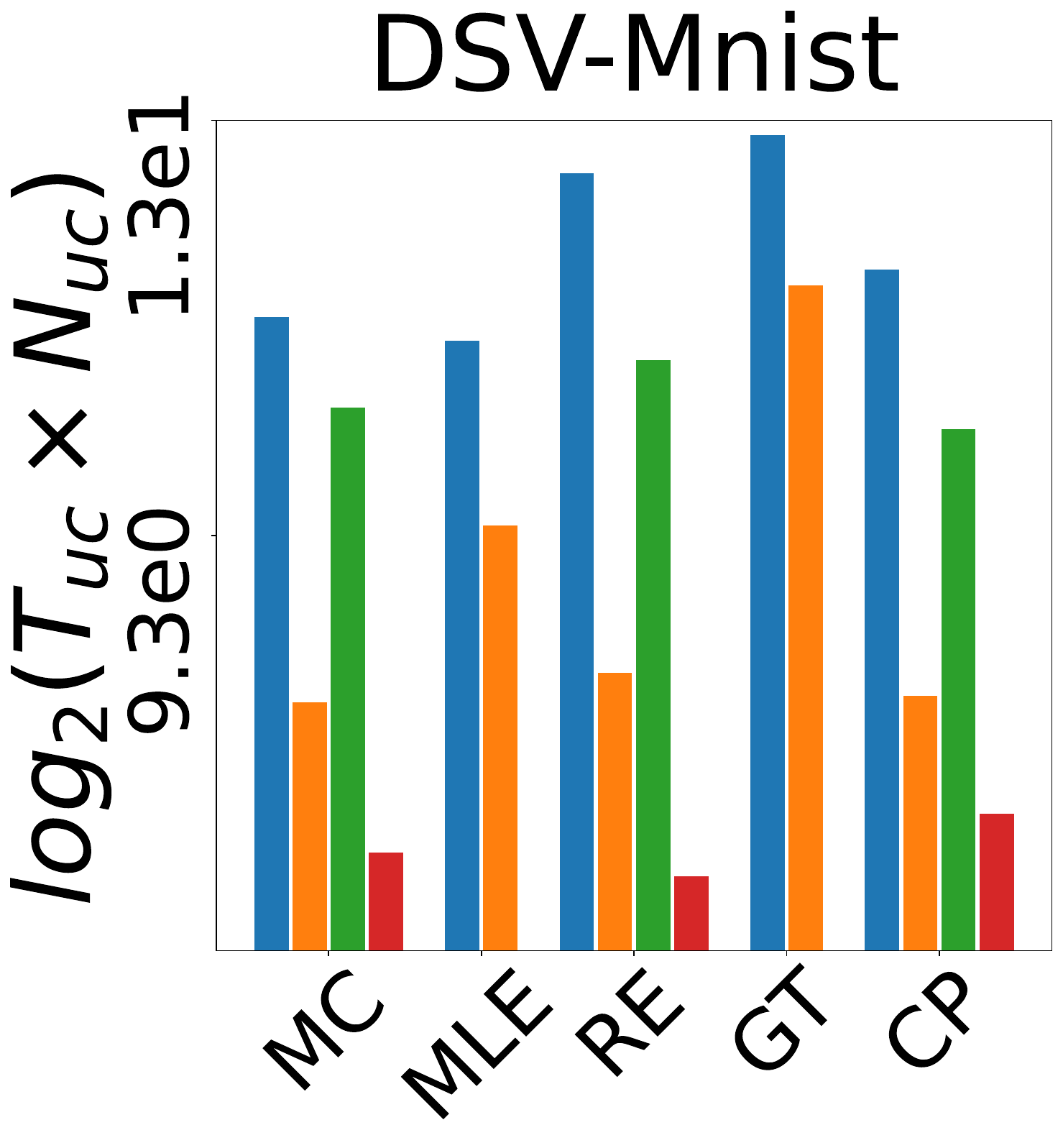}
        \includegraphics[width=0.2\columnwidth, height=0.2\columnwidth]{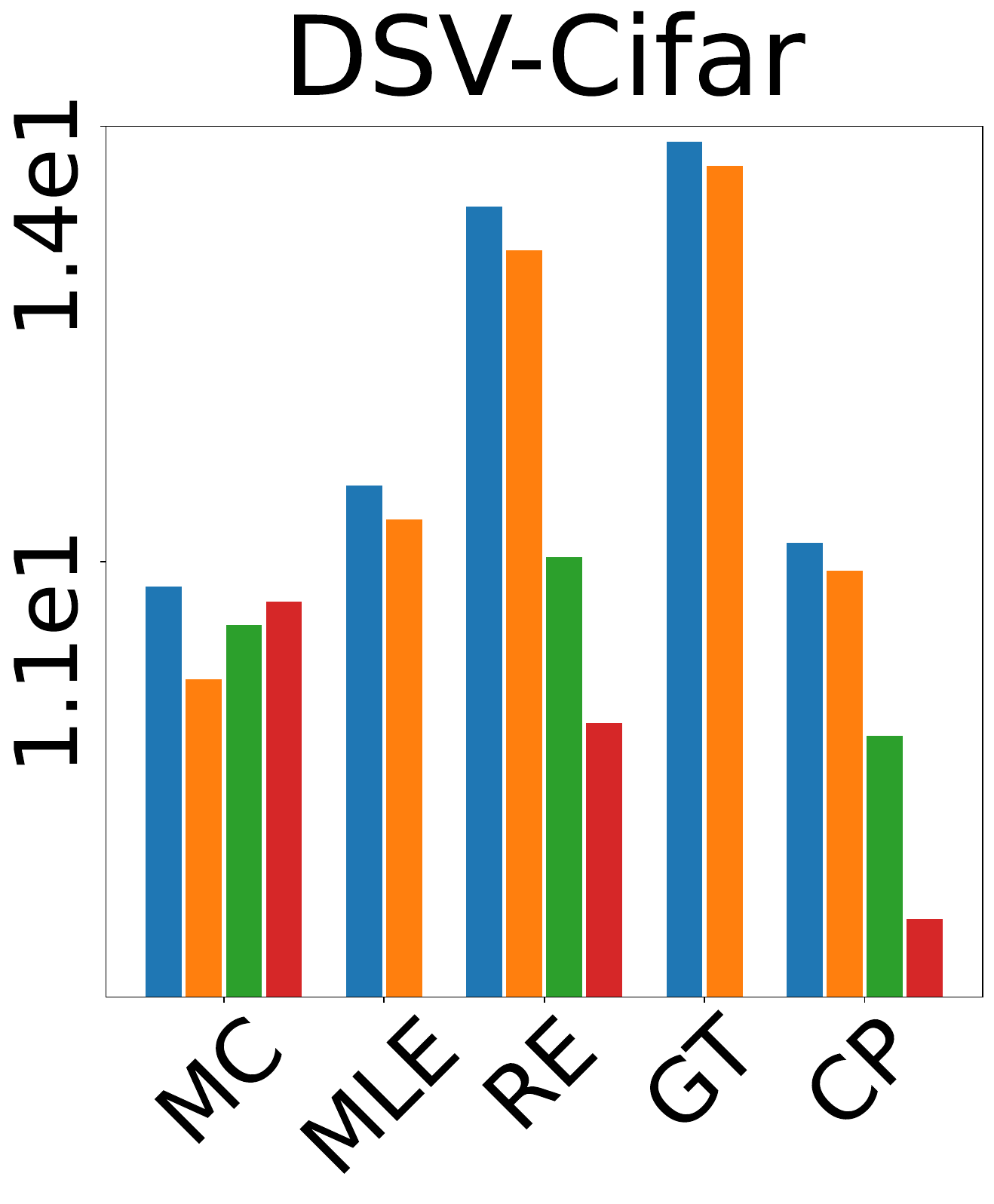}
        \includegraphics[width=0.2\columnwidth, height=0.2\columnwidth]{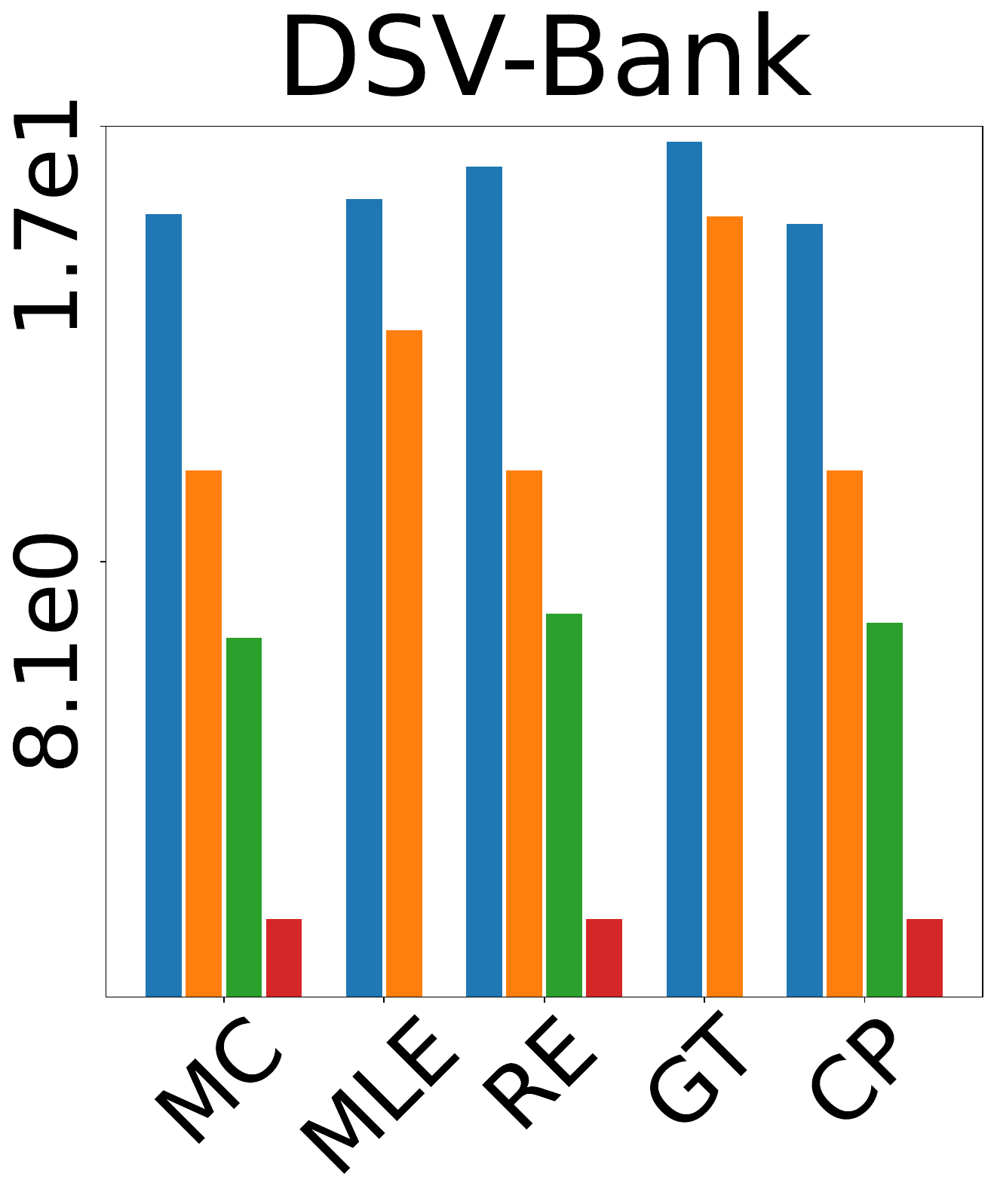}
        \includegraphics[width=0.2\columnwidth, height=0.2\columnwidth]{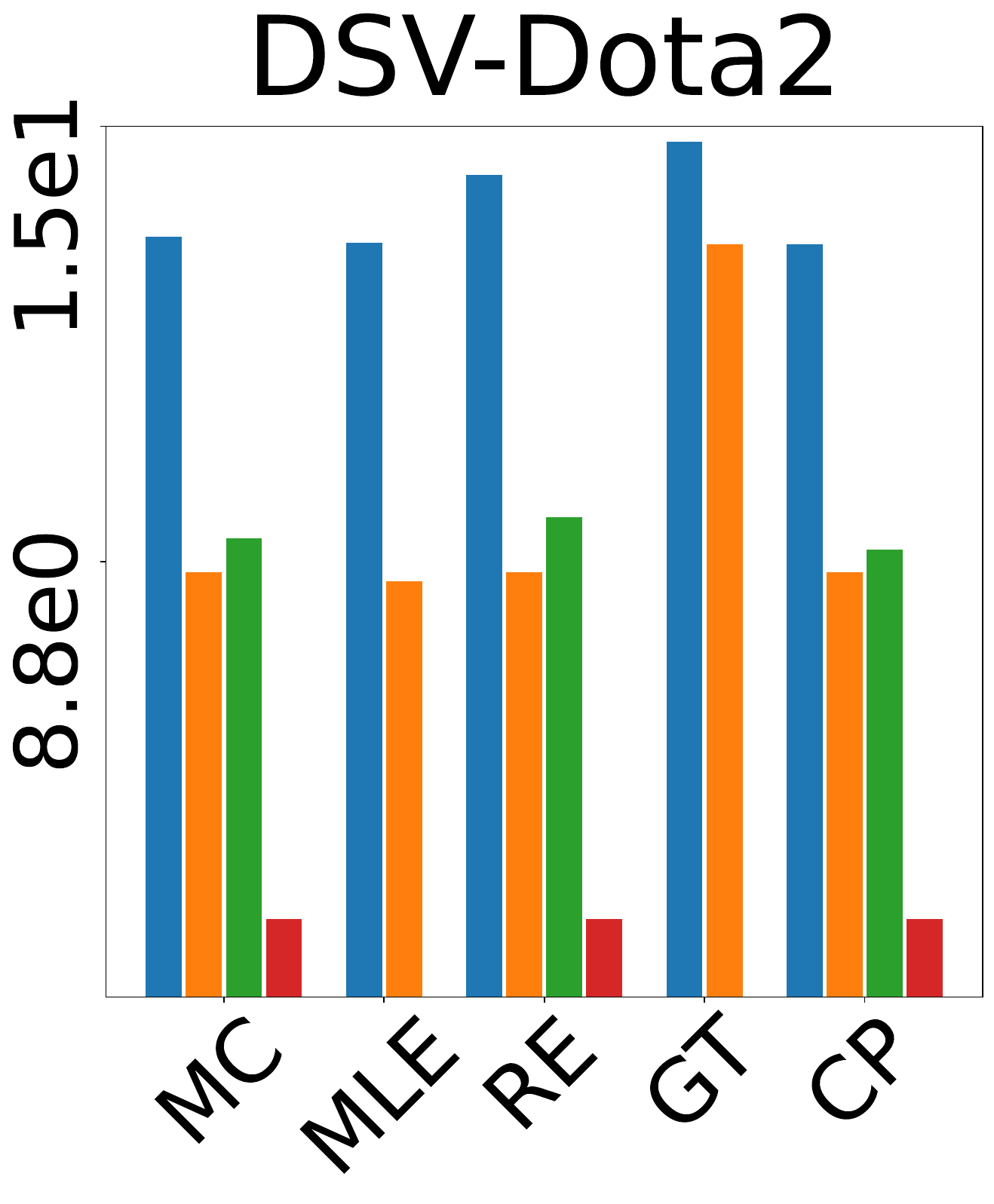}
        \includegraphics[width=0.2\columnwidth, height=0.2\columnwidth]{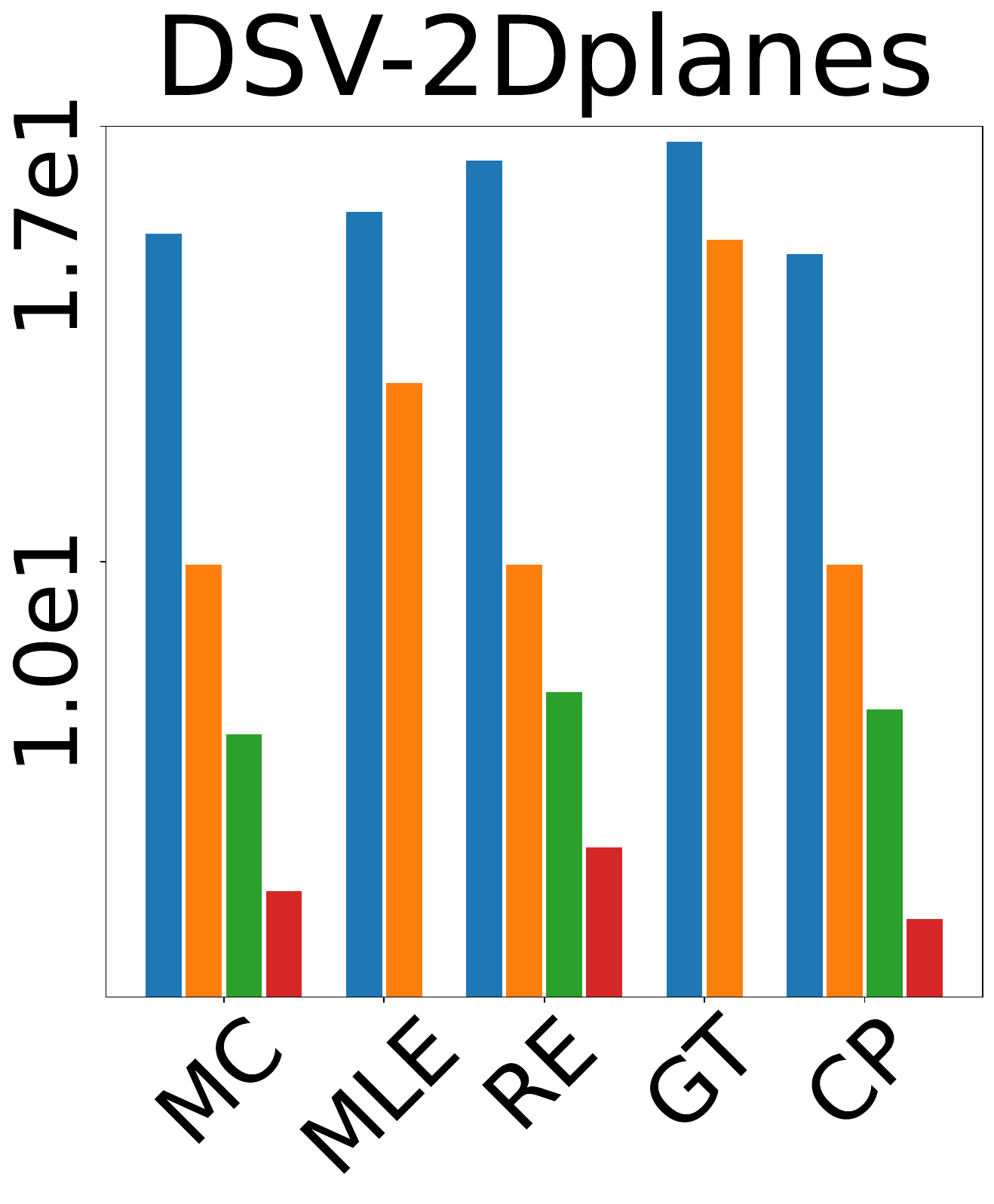}
        
        \includegraphics[width=0.2\columnwidth, height=0.2\columnwidth]{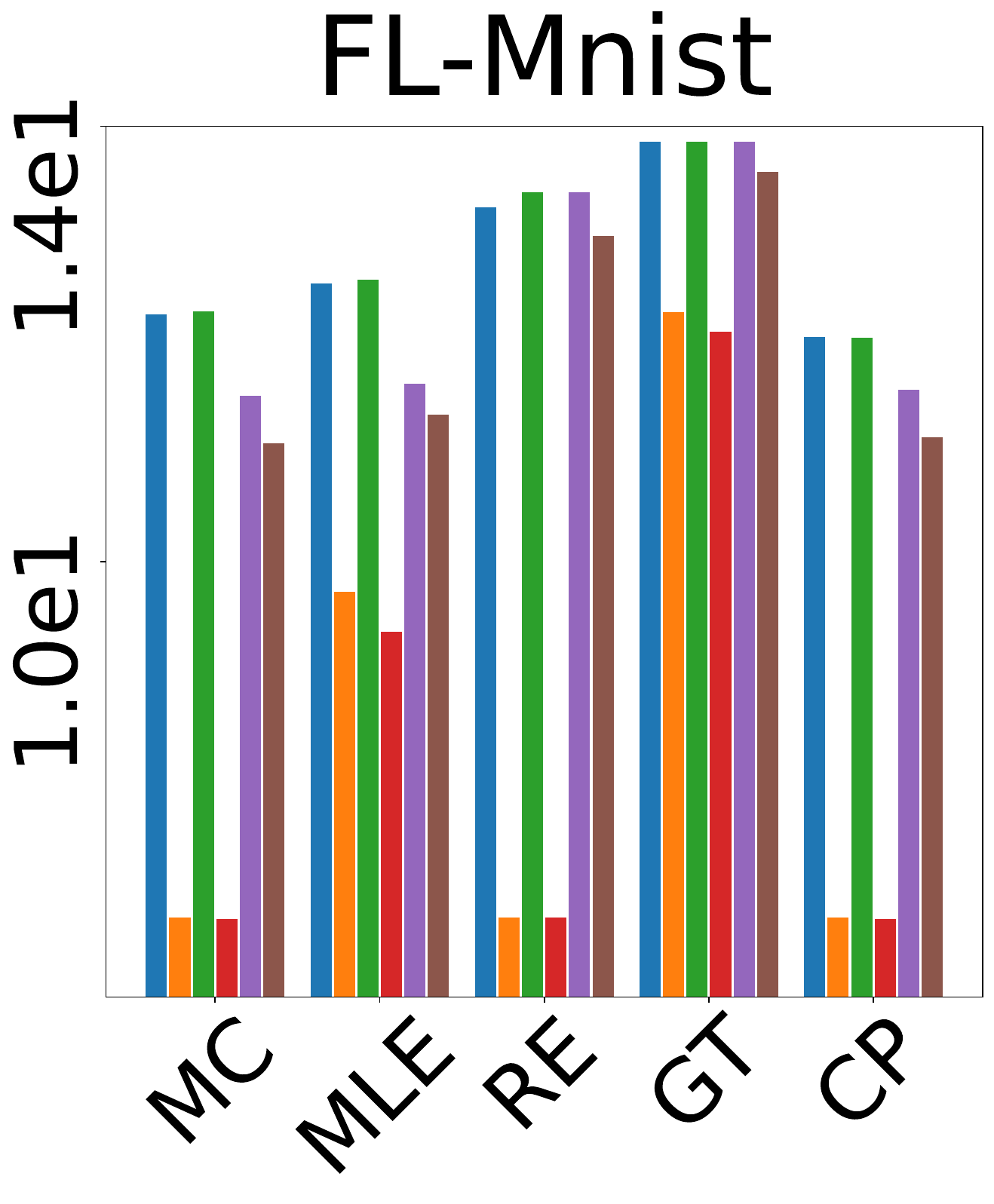}
        \includegraphics[width=0.2\columnwidth, height=0.2\columnwidth]{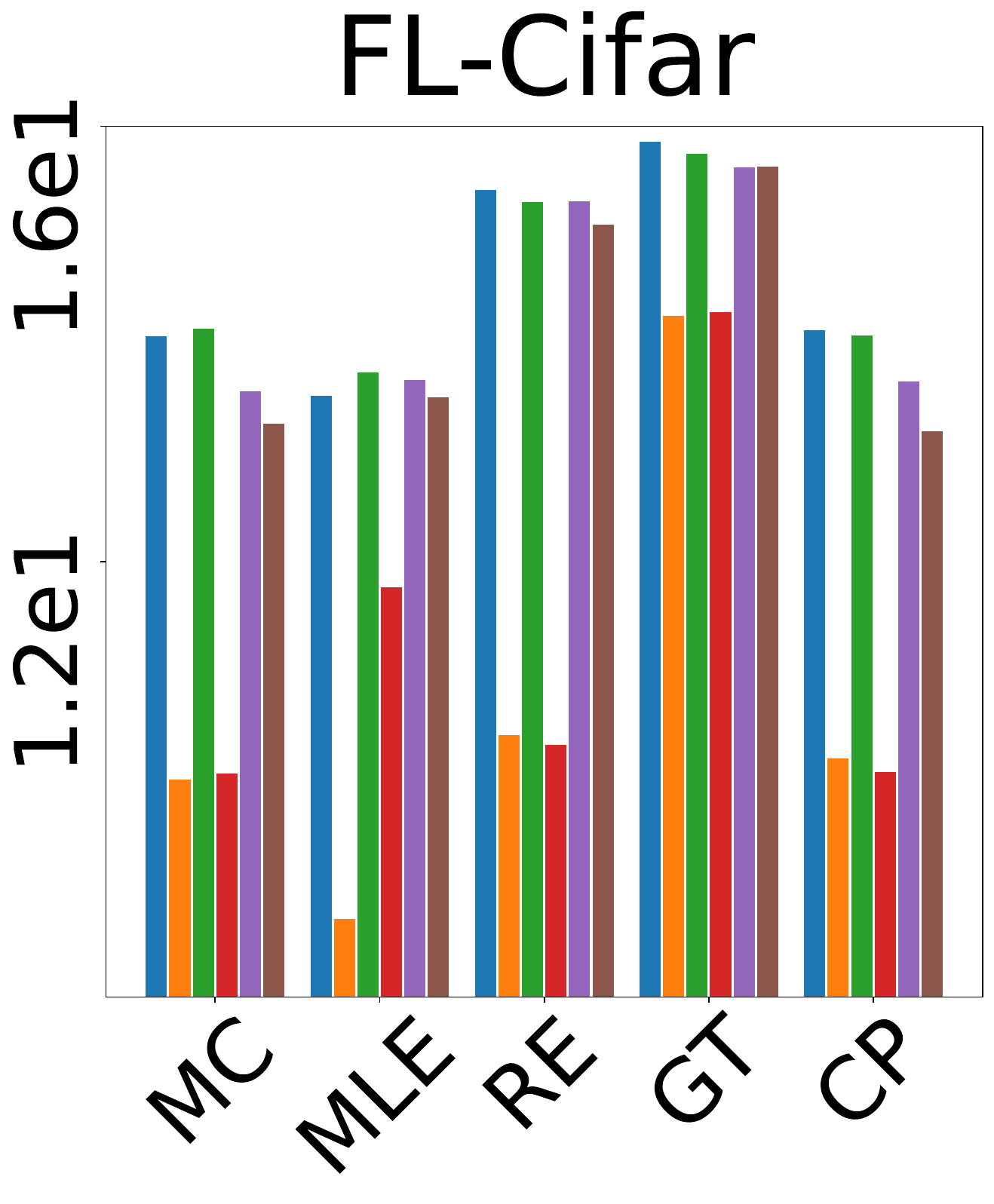}
        \includegraphics[width=0.2\columnwidth, height=0.2\columnwidth]{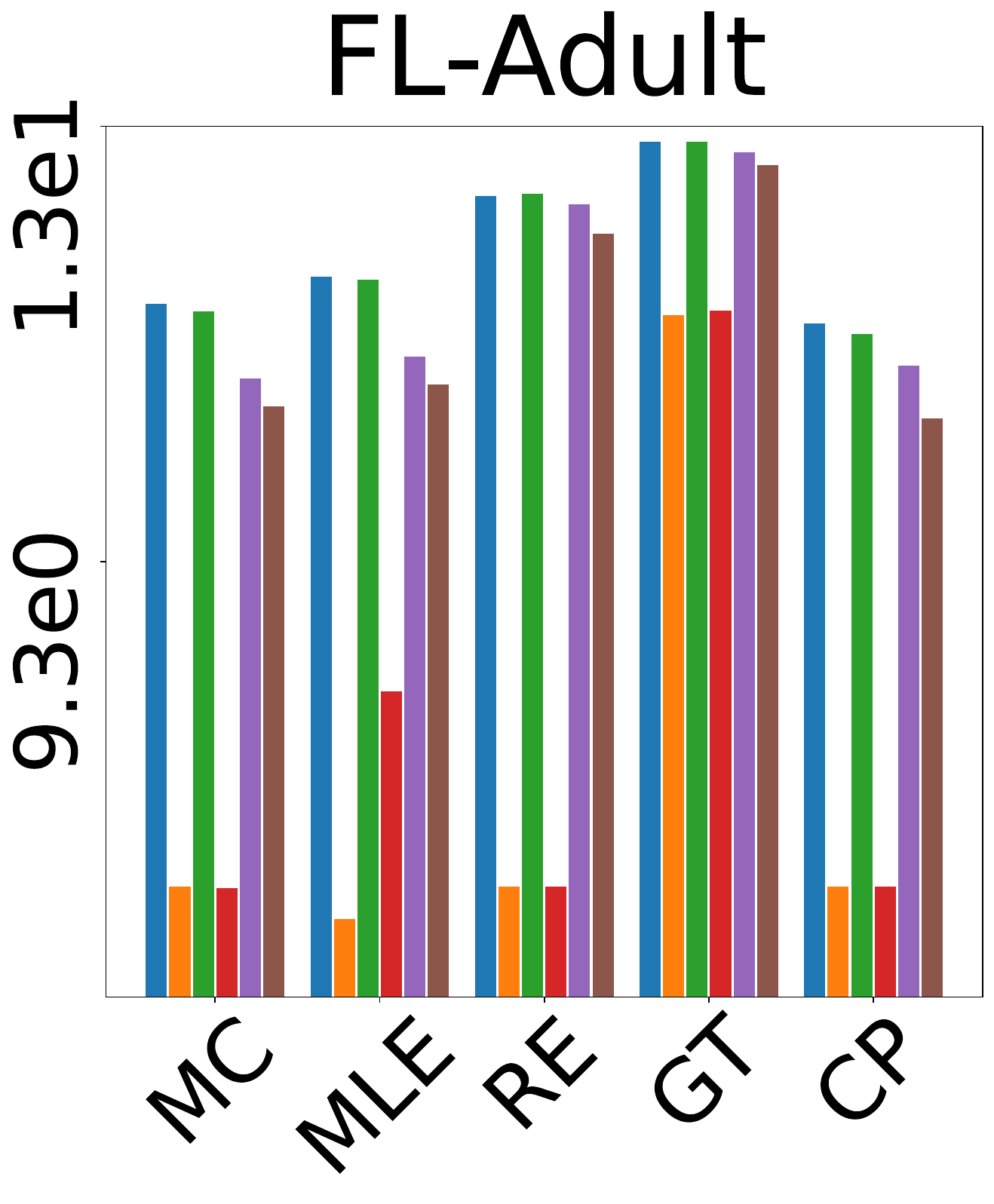}
        \includegraphics[width=0.2\columnwidth, height=0.2\columnwidth]{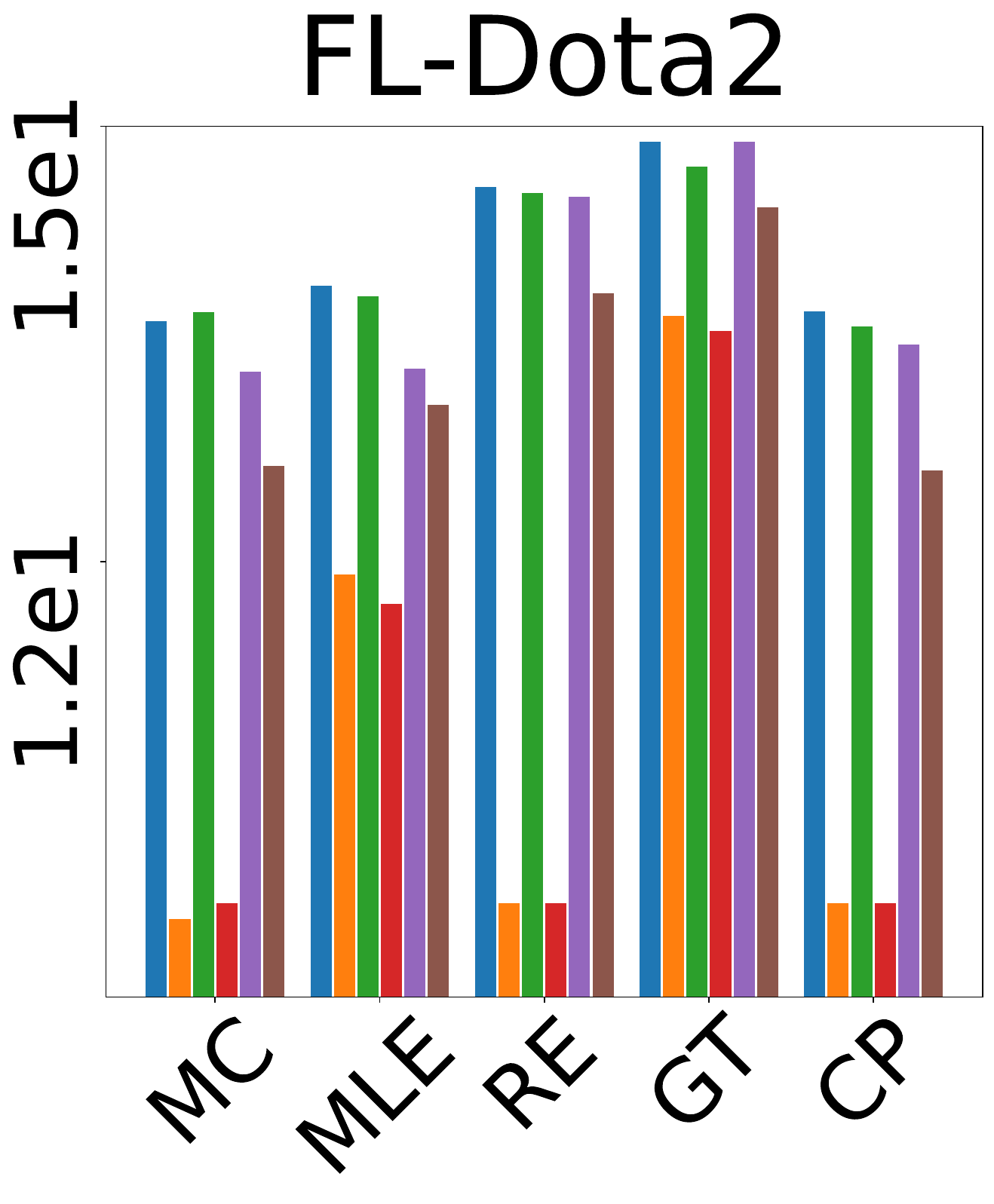}
        \includegraphics[width=0.2\columnwidth, height=0.2\columnwidth]{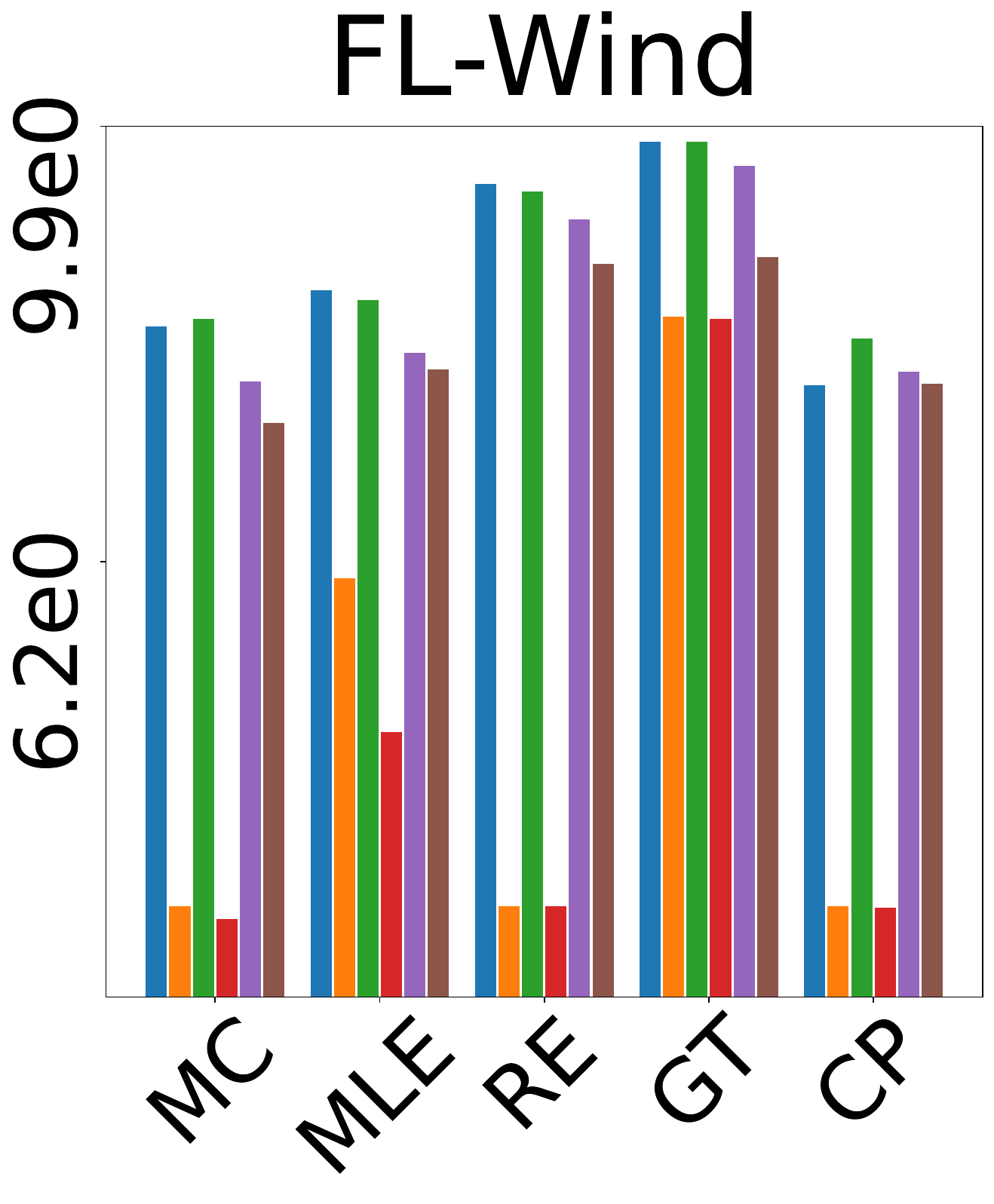}
    }  
    \\
    \mbox{
        \includegraphics[width=0.215\columnwidth, height=0.16\columnwidth]{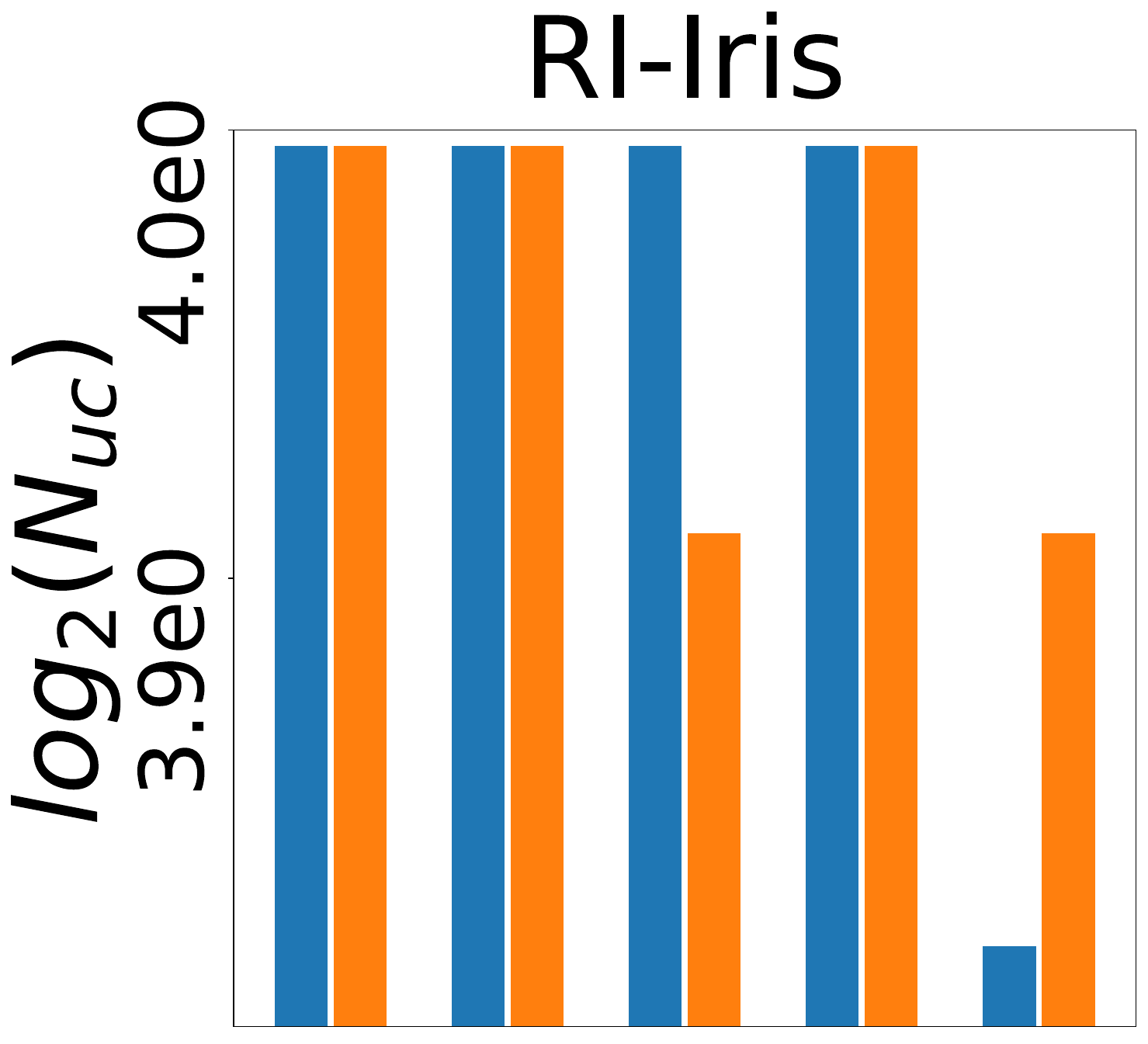}
        \includegraphics[width=0.2\columnwidth, height=0.16\columnwidth]{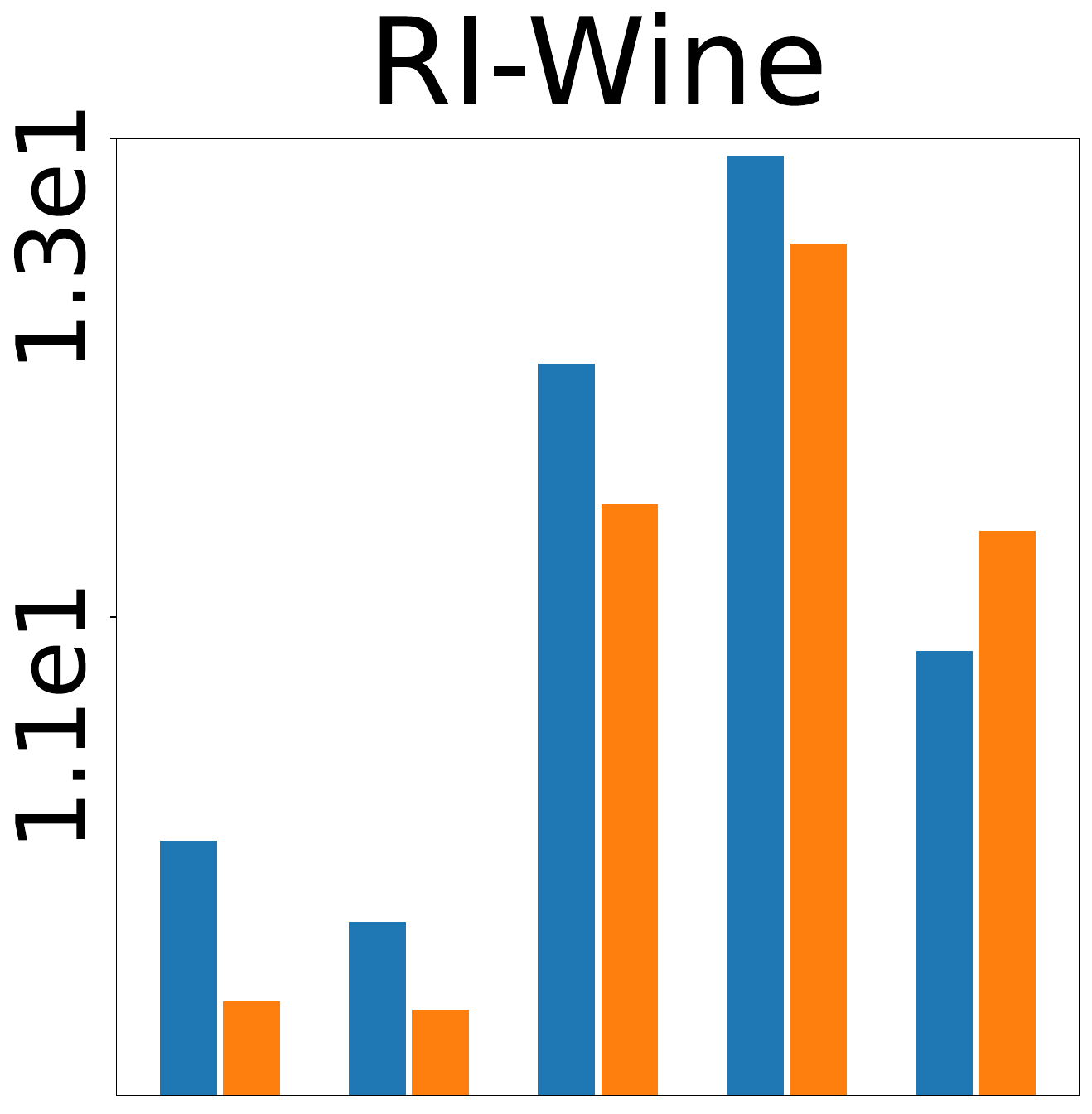}
        \includegraphics[width=0.2\columnwidth, height=0.16\columnwidth]{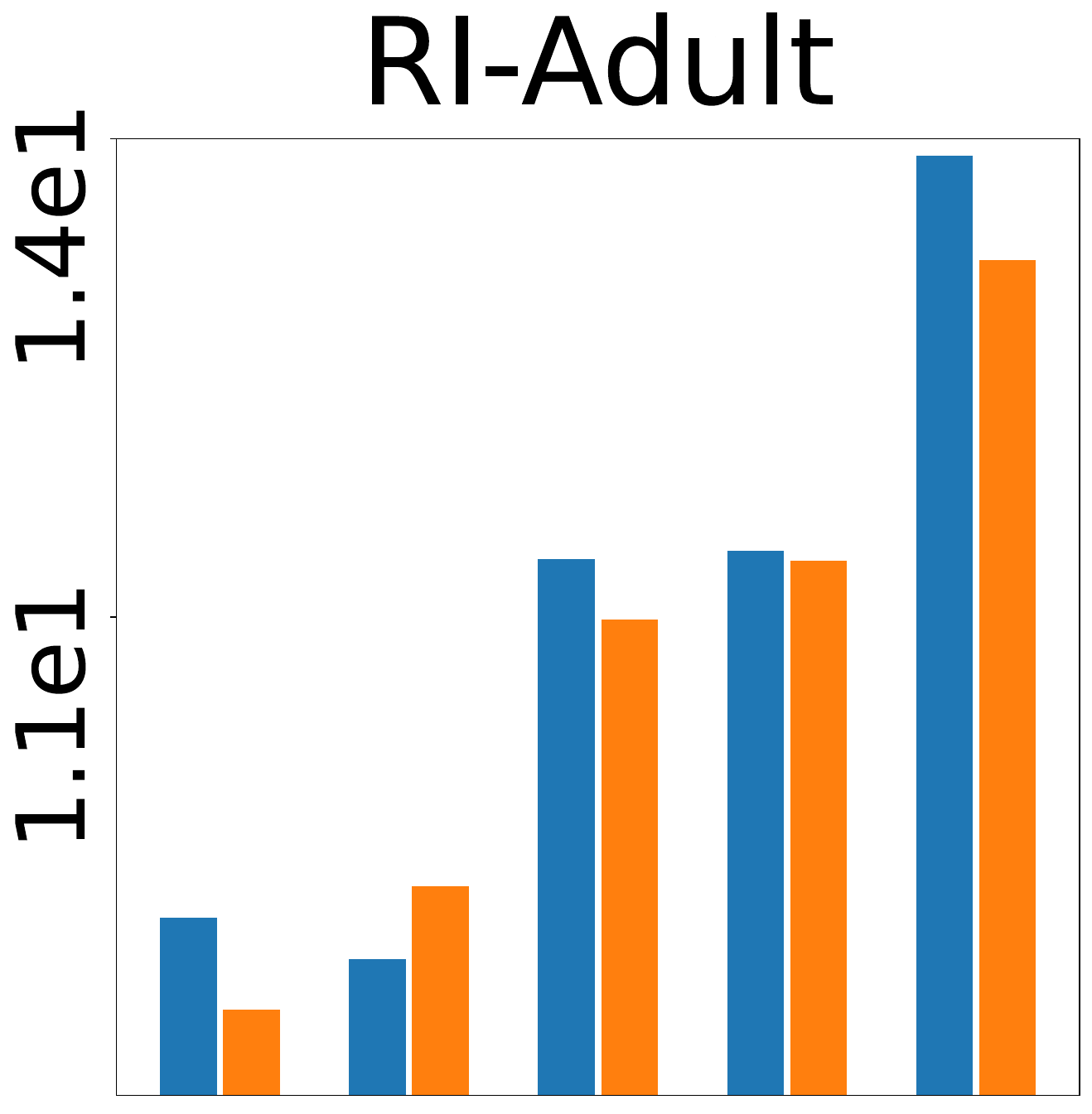}
        \includegraphics[width=0.2\columnwidth, height=0.16\columnwidth]{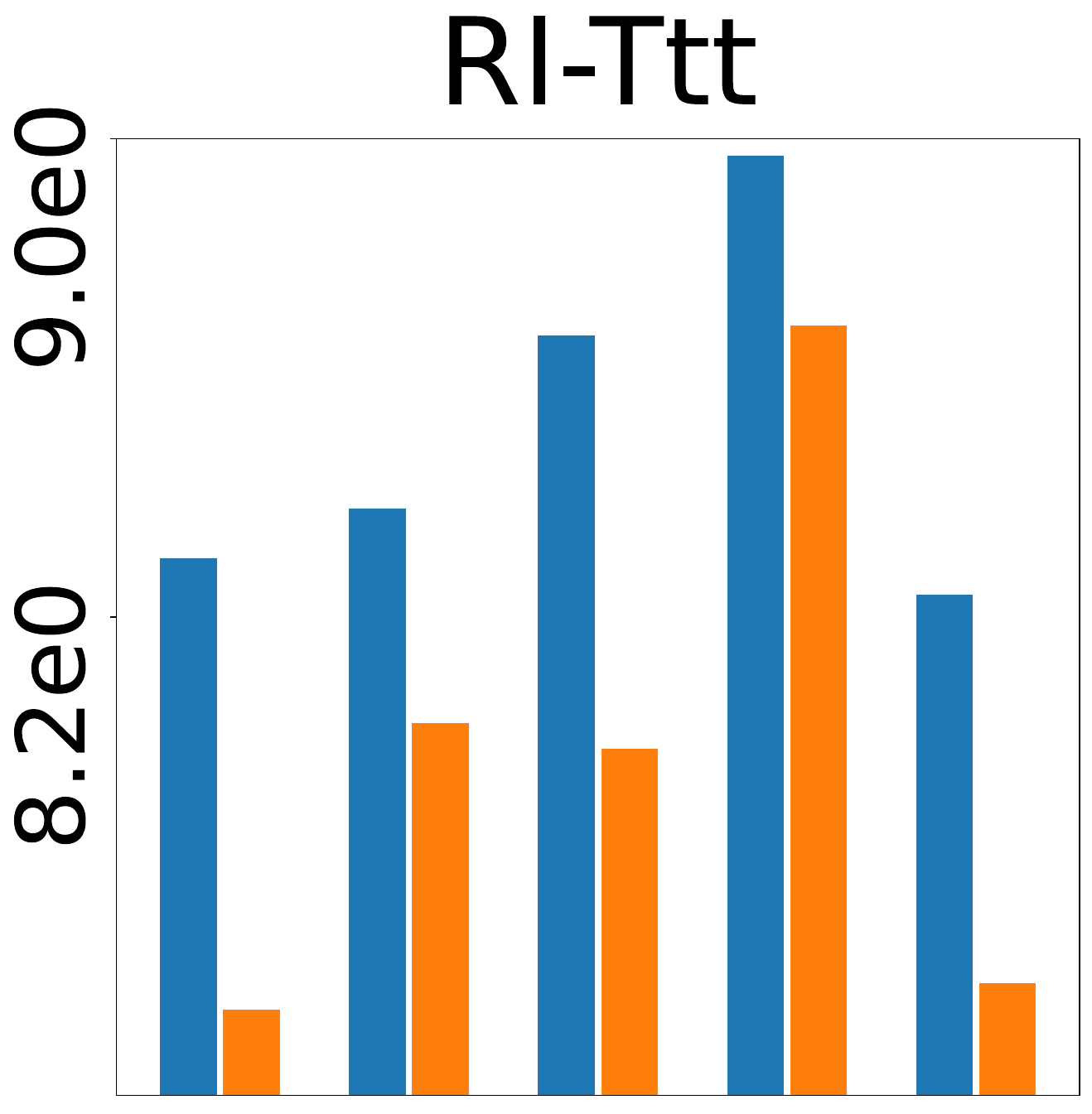}
        \includegraphics[width=0.2\columnwidth, height=0.16\columnwidth]{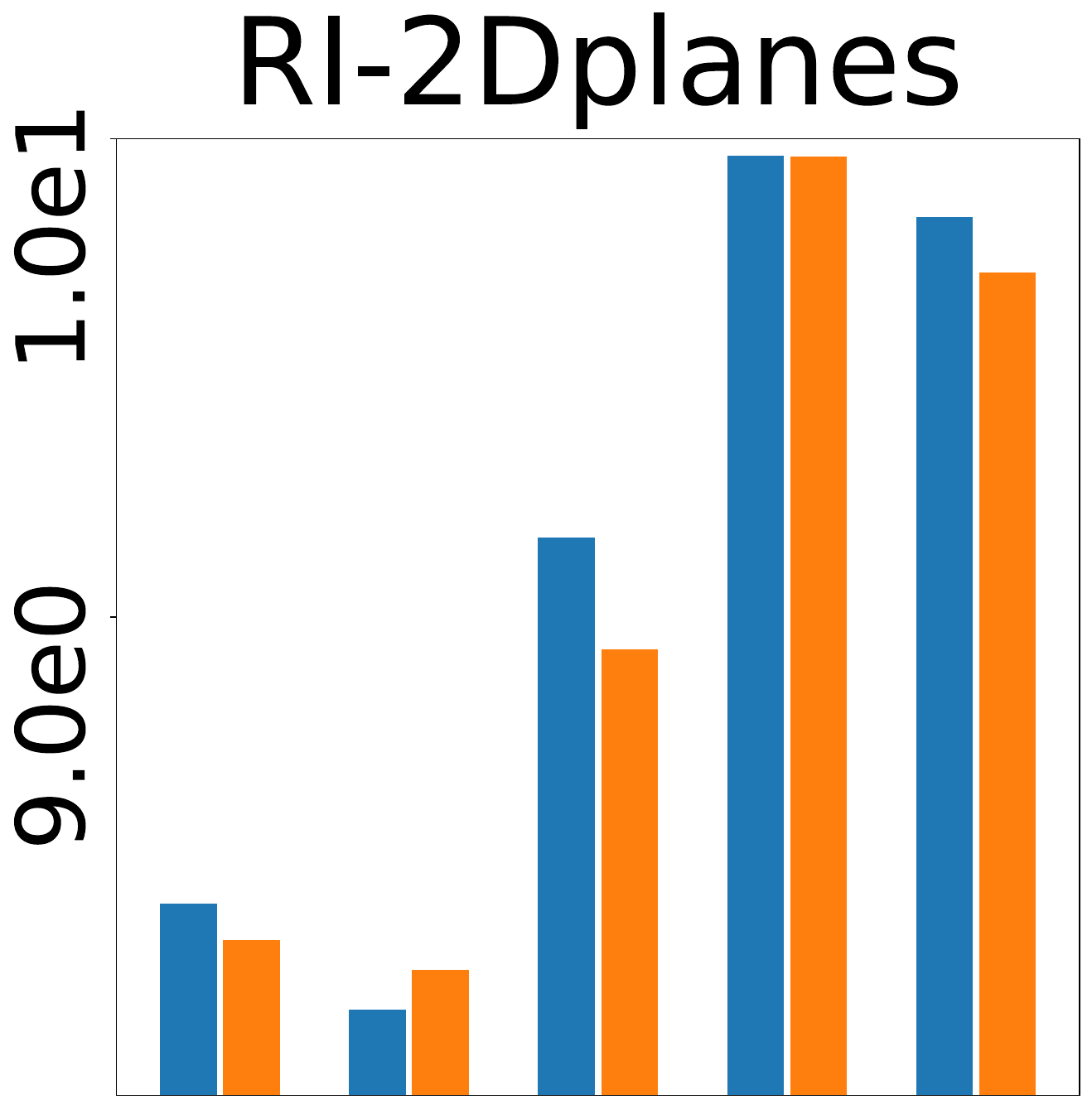}

        \includegraphics[width=0.2\columnwidth, height=0.16\columnwidth]{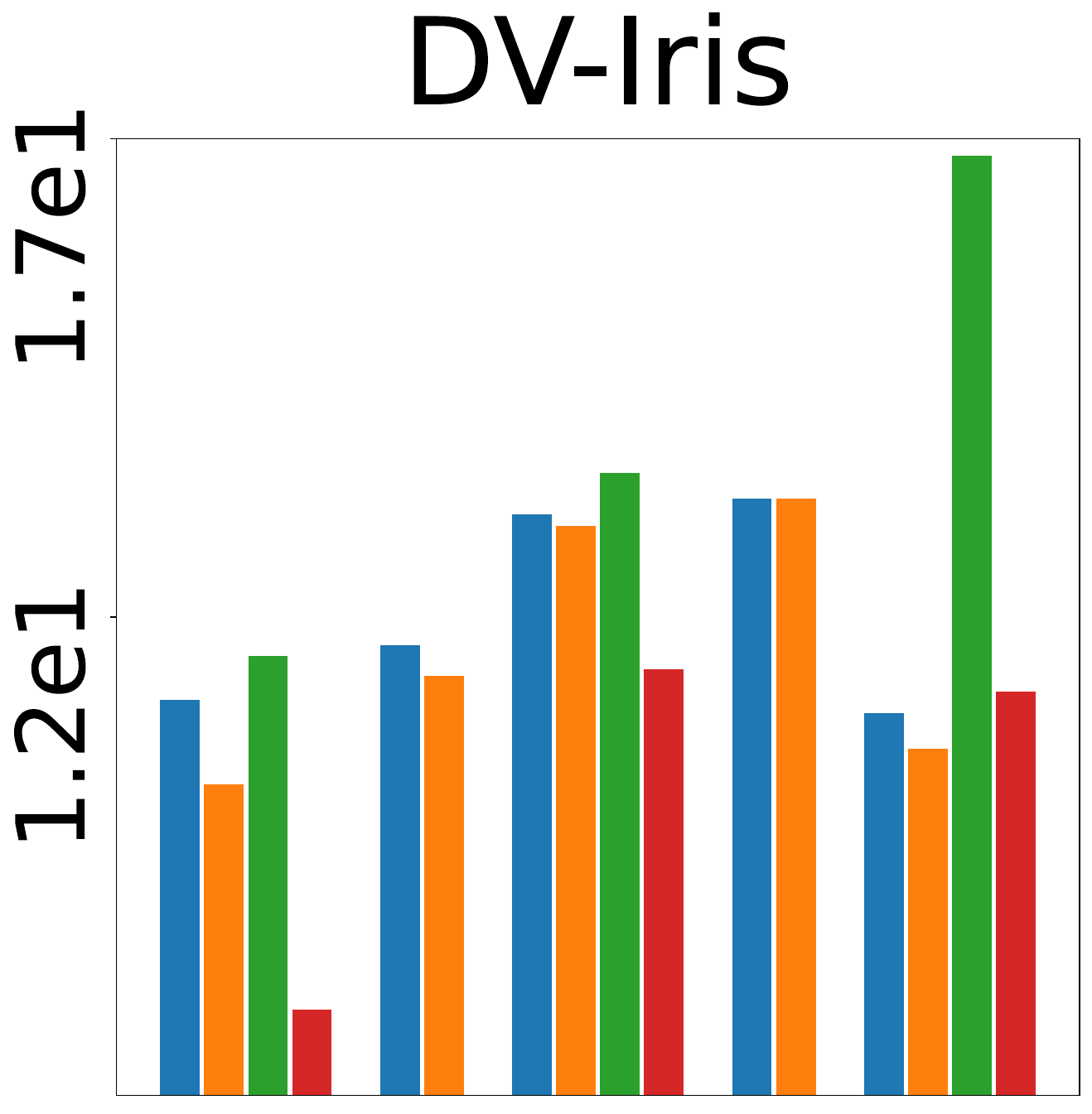}
        \includegraphics[width=0.2\columnwidth, height=0.16\columnwidth]{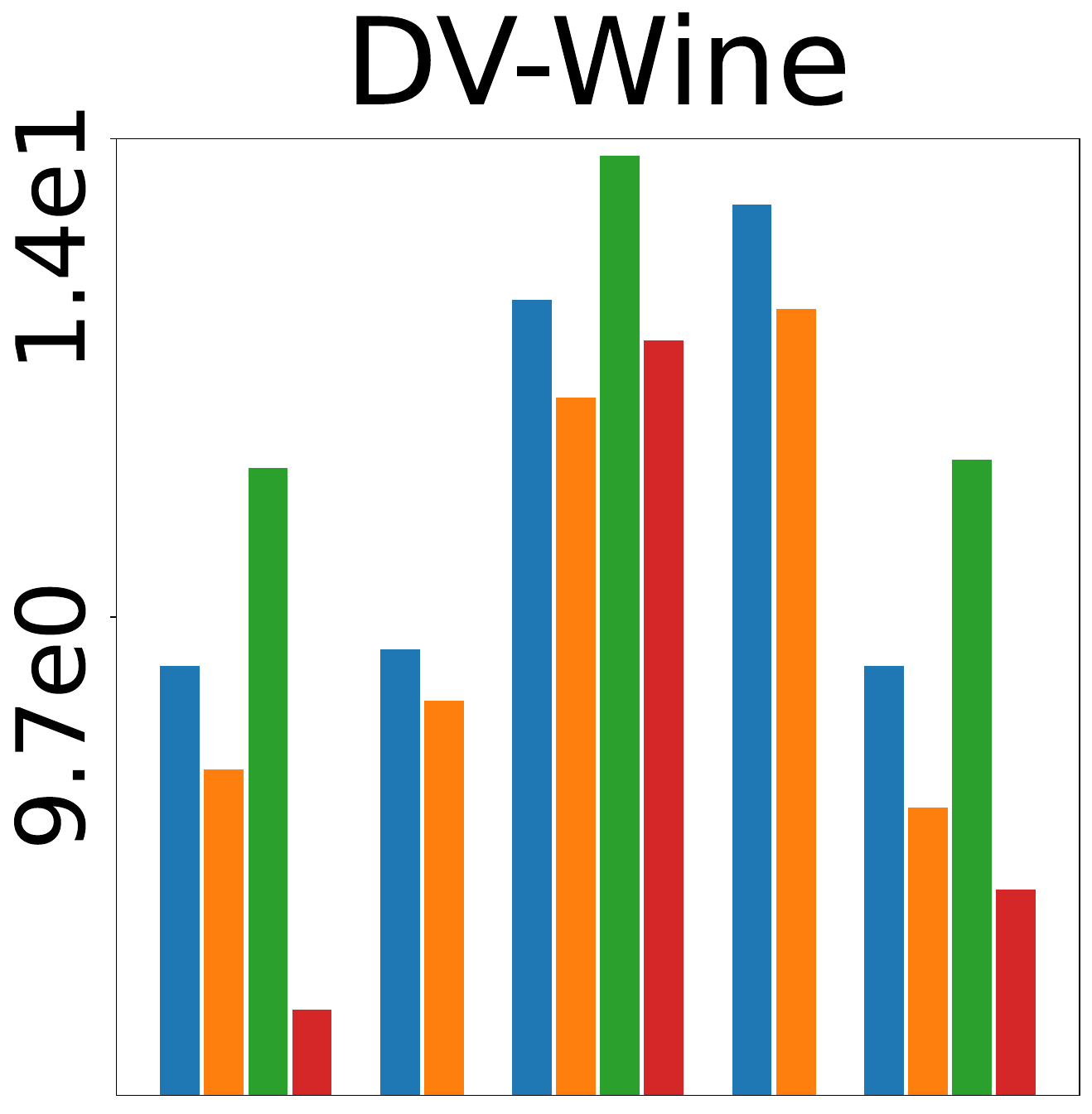}
        \includegraphics[width=0.2\columnwidth, height=0.16\columnwidth]{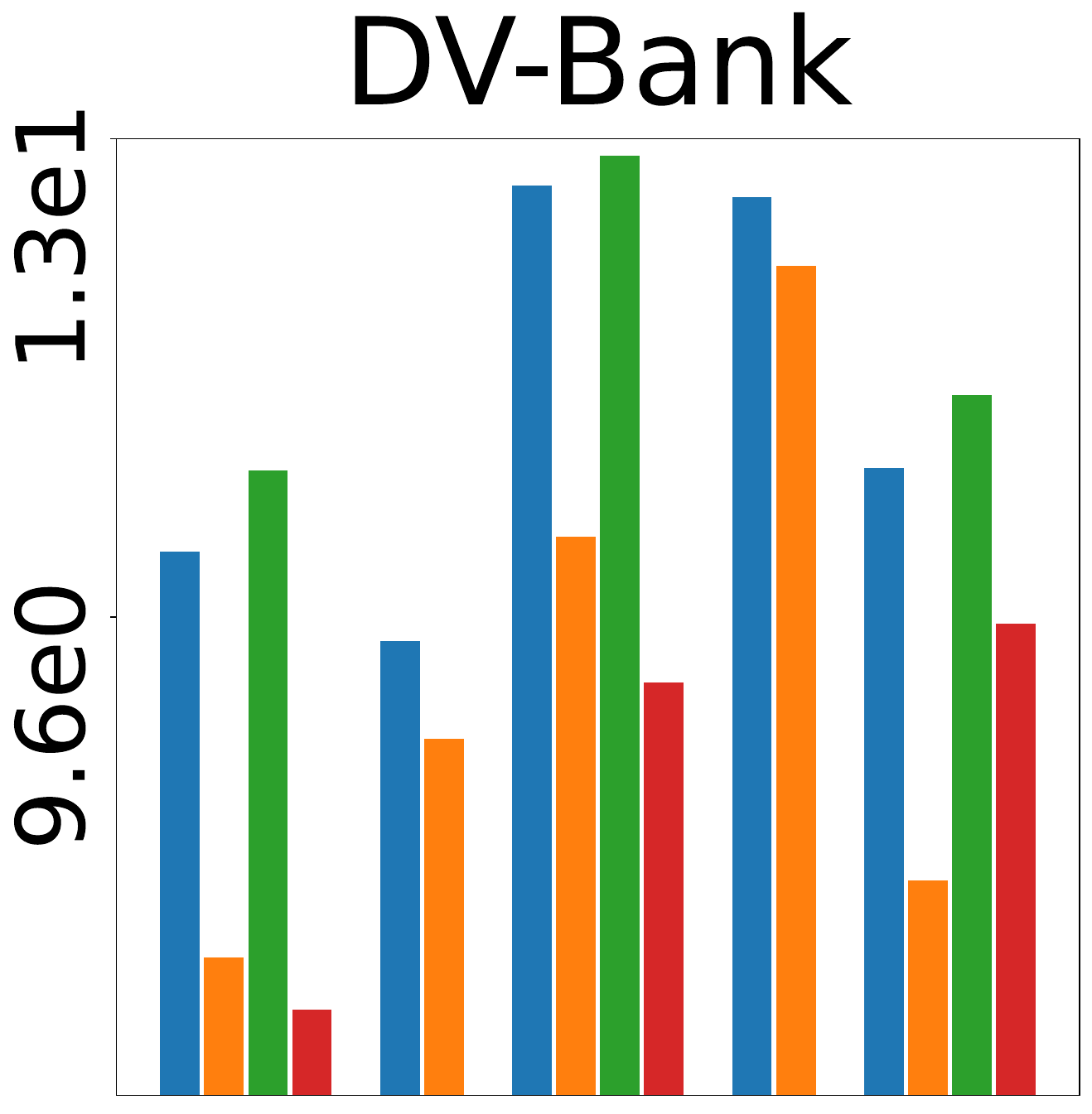}
        \includegraphics[width=0.2\columnwidth, height=0.16\columnwidth]{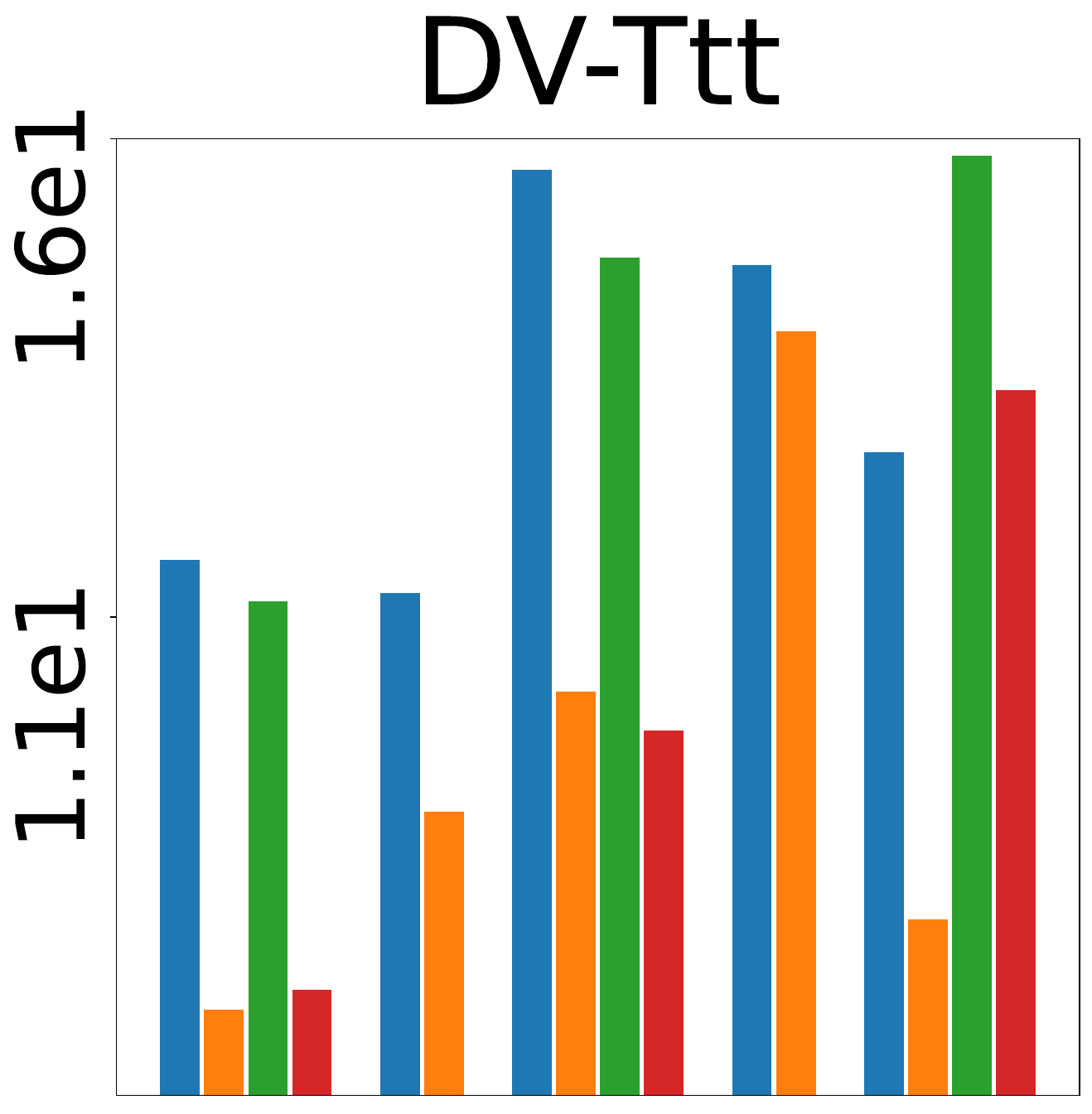}
        \includegraphics[width=0.2\columnwidth, height=0.16\columnwidth]{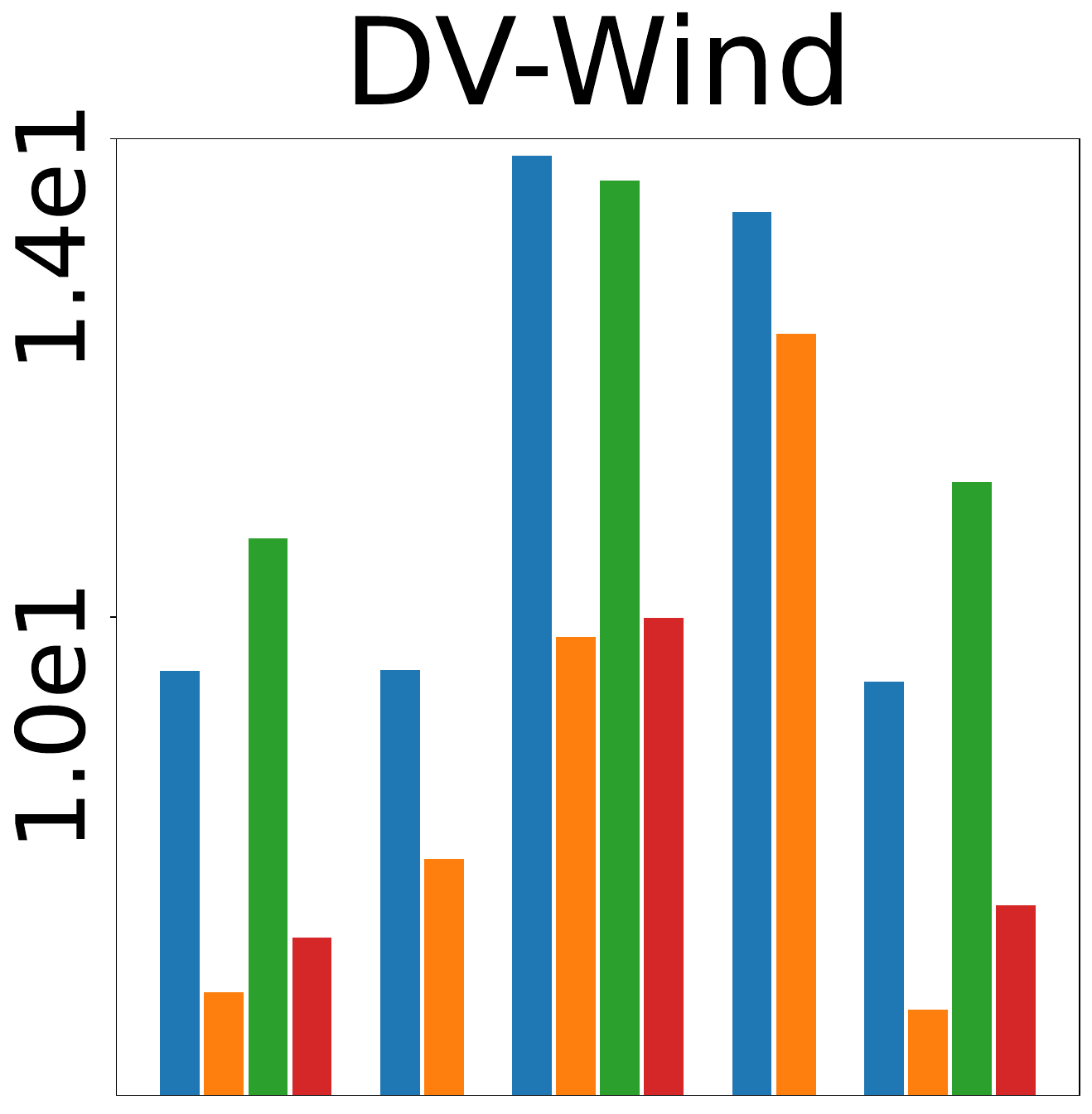}
    }
    \\
    
    \subfigure[The computation complexity, $N_{uc}$.] {
        \includegraphics[width=0.215\columnwidth, height=0.2\columnwidth]{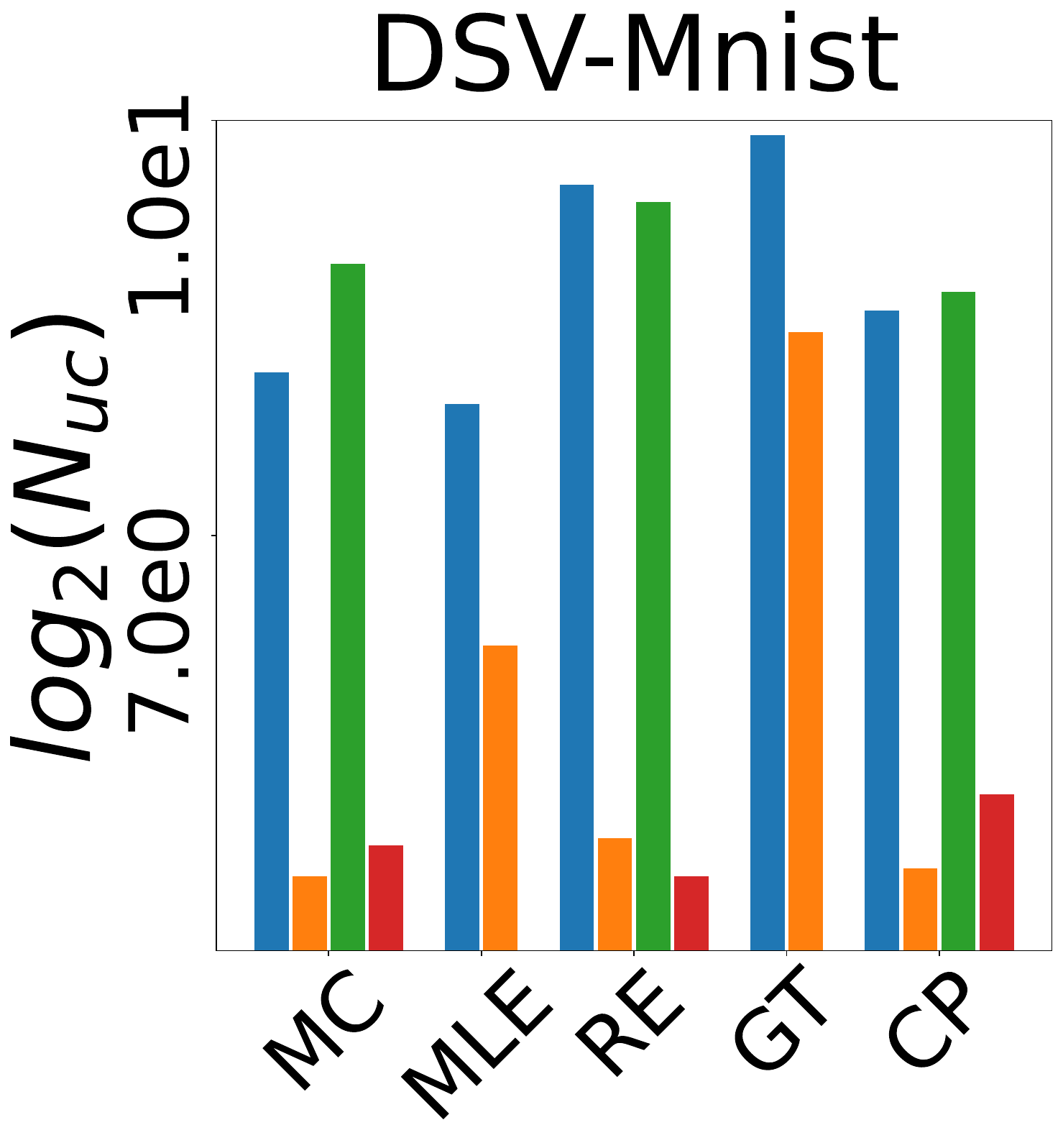}
        \includegraphics[width=0.2\columnwidth, height=0.2\columnwidth]{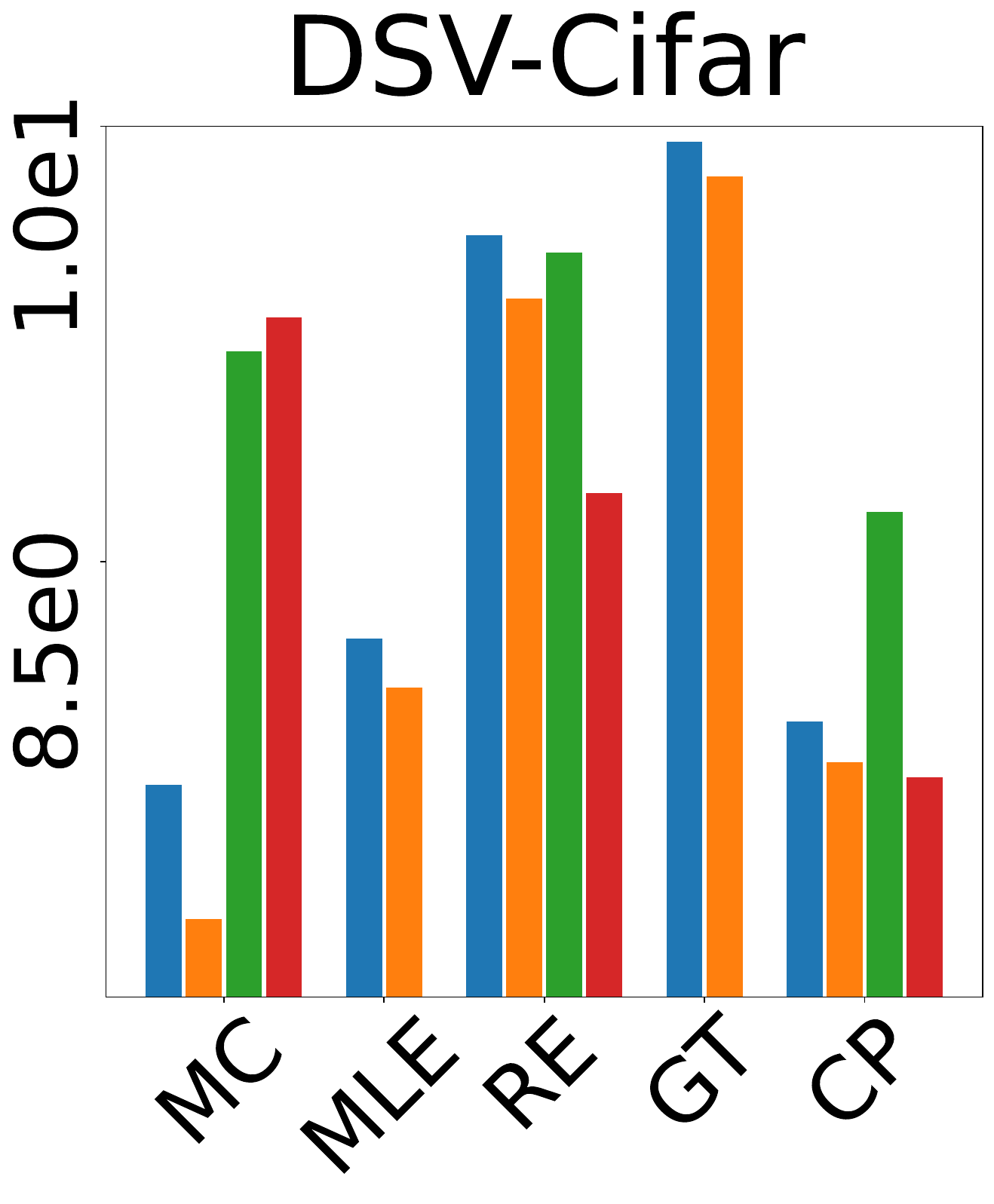}
        \includegraphics[width=0.2\columnwidth, height=0.2\columnwidth]{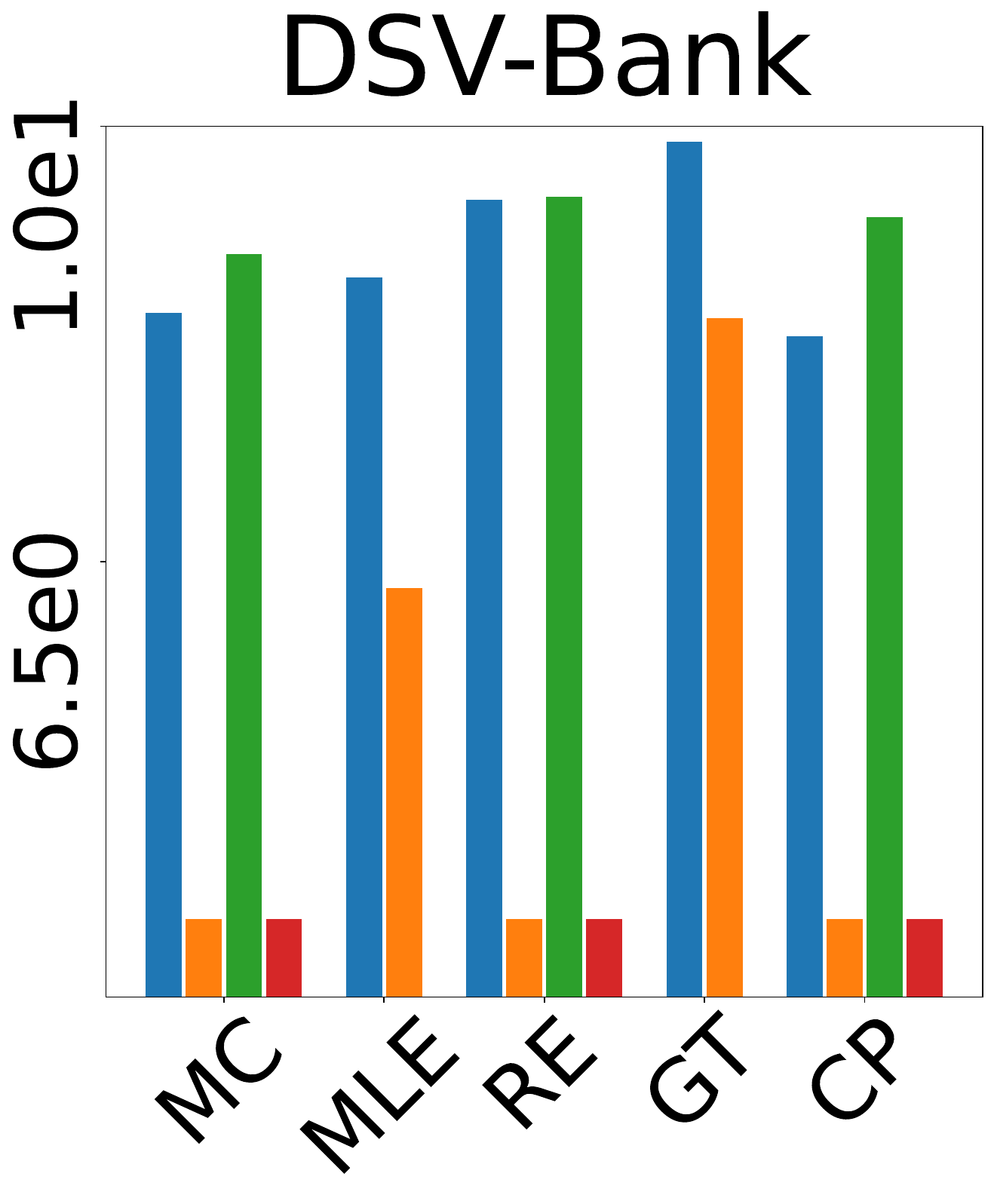}
        \includegraphics[width=0.2\columnwidth, height=0.2\columnwidth]{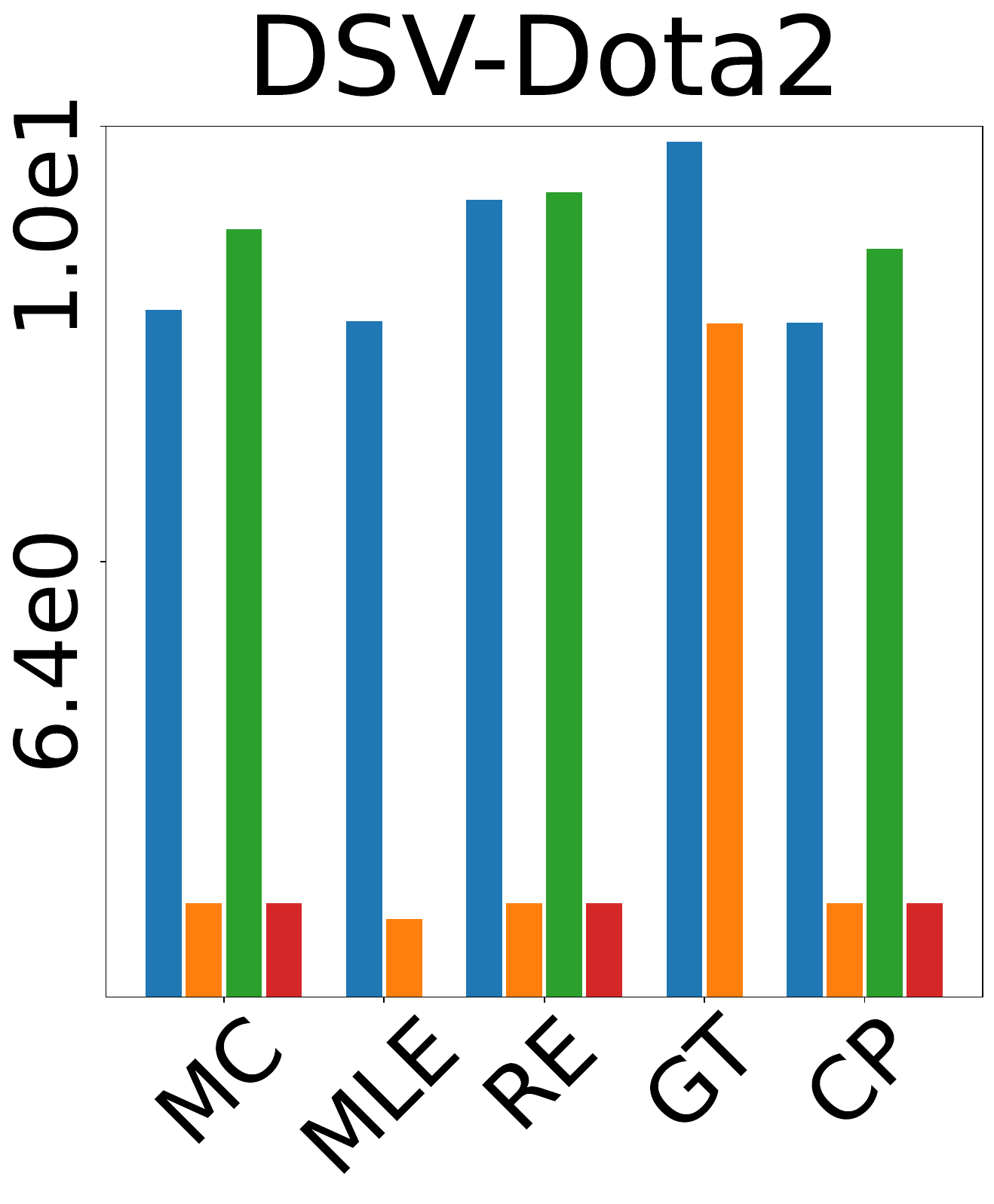}
        \includegraphics[width=0.2\columnwidth, height=0.2\columnwidth]{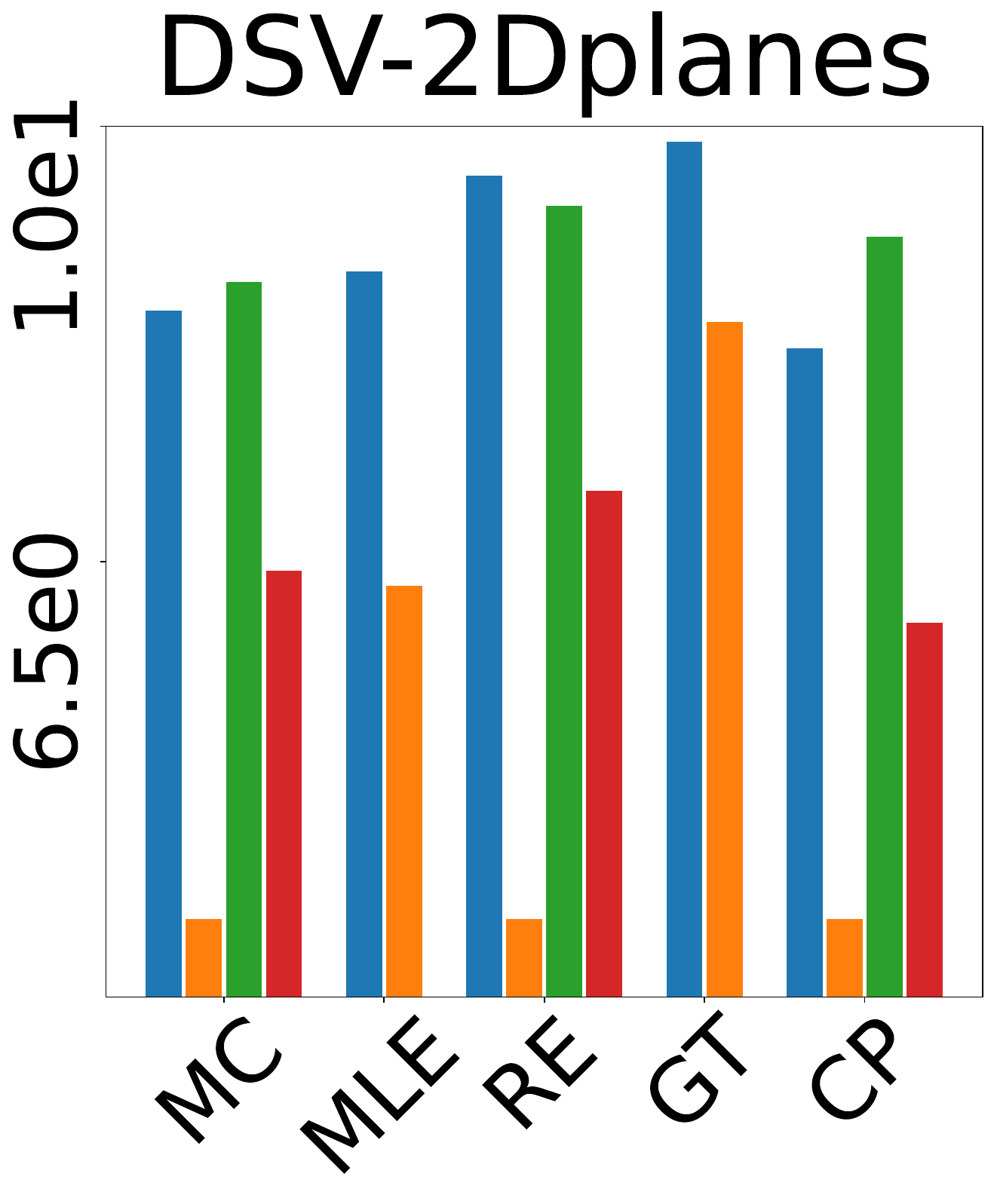}
        
        \includegraphics[width=0.2\columnwidth, height=0.2\columnwidth]{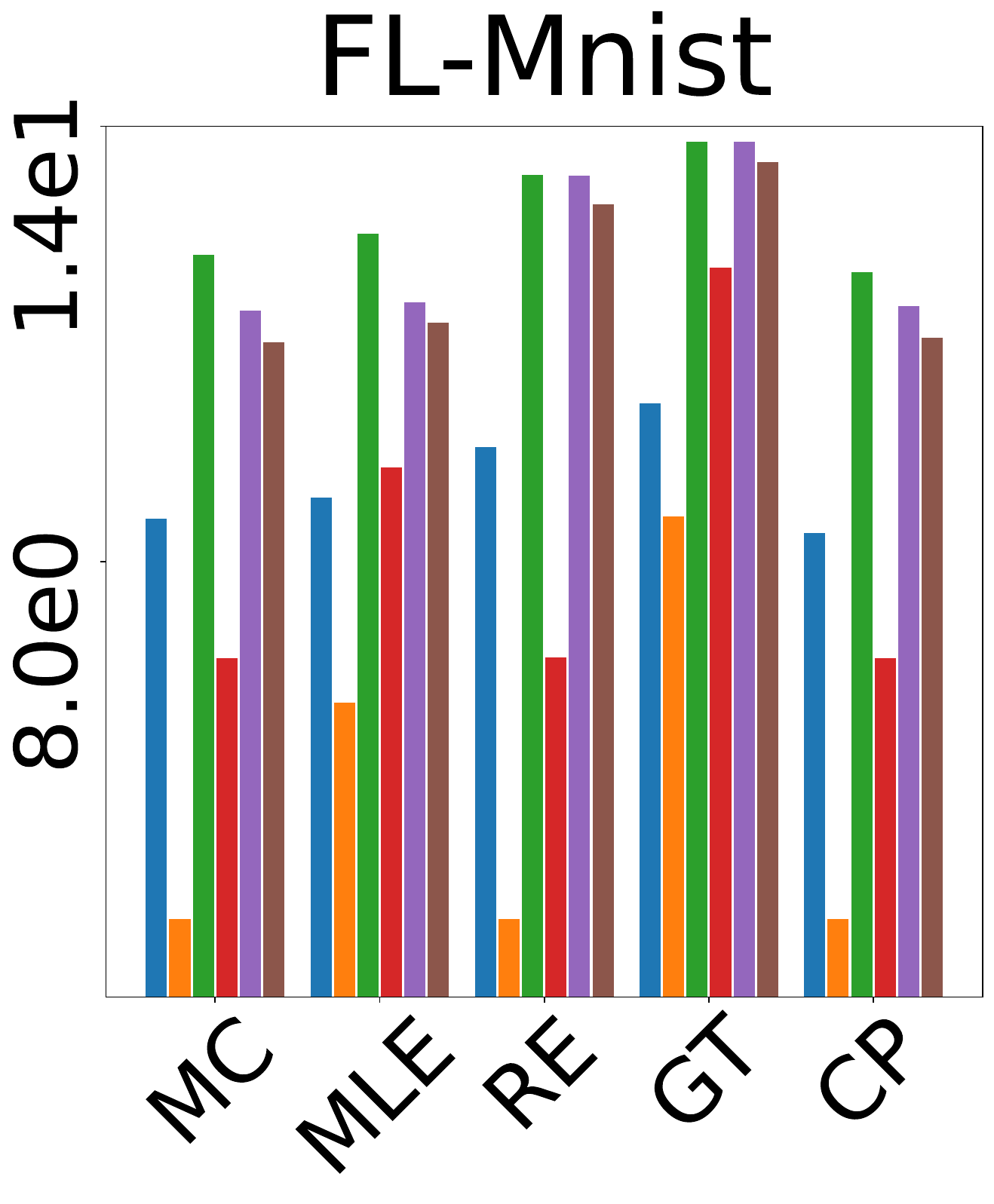}
        \includegraphics[width=0.2\columnwidth, height=0.2\columnwidth]{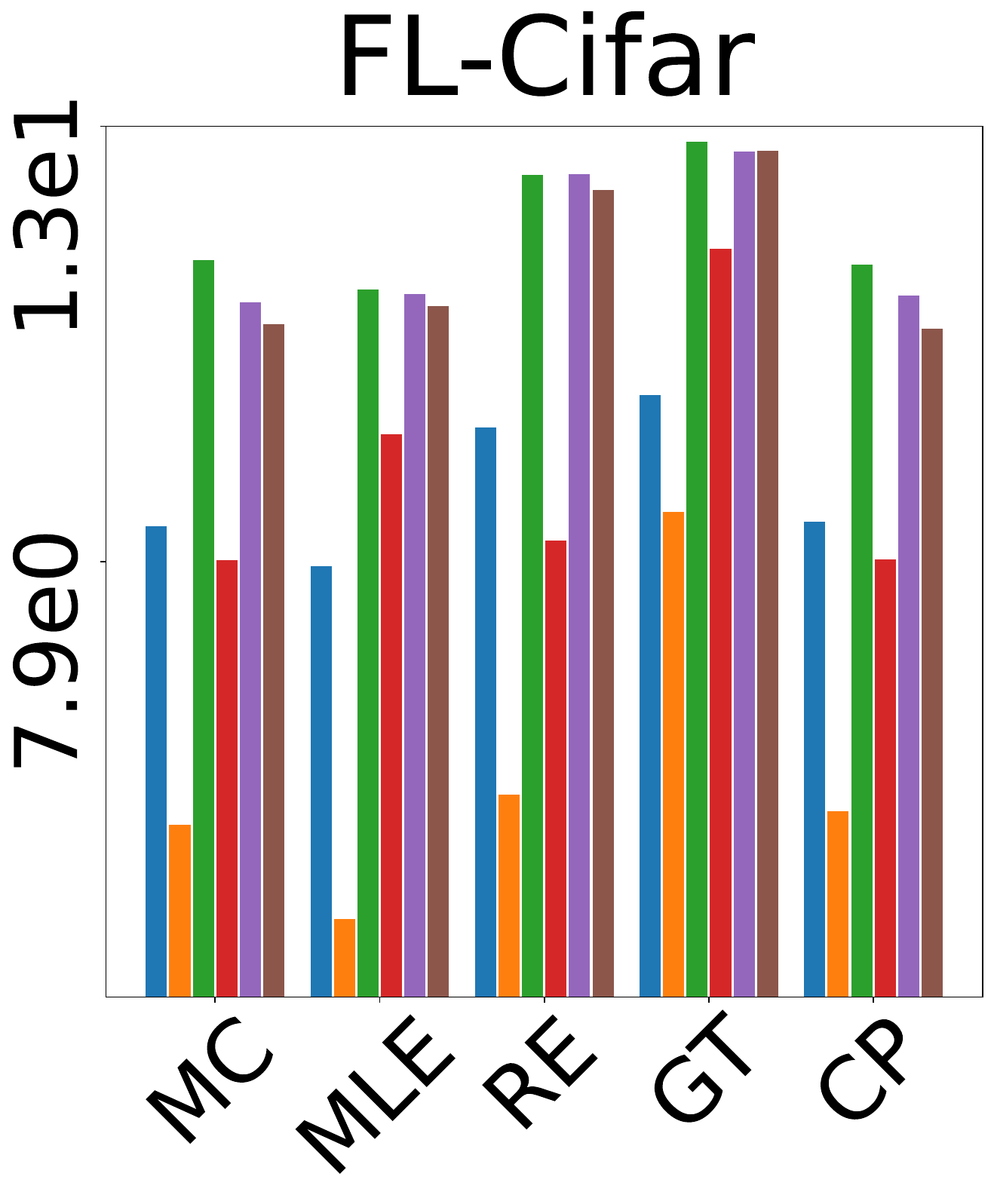}
        \includegraphics[width=0.2\columnwidth, height=0.2\columnwidth]{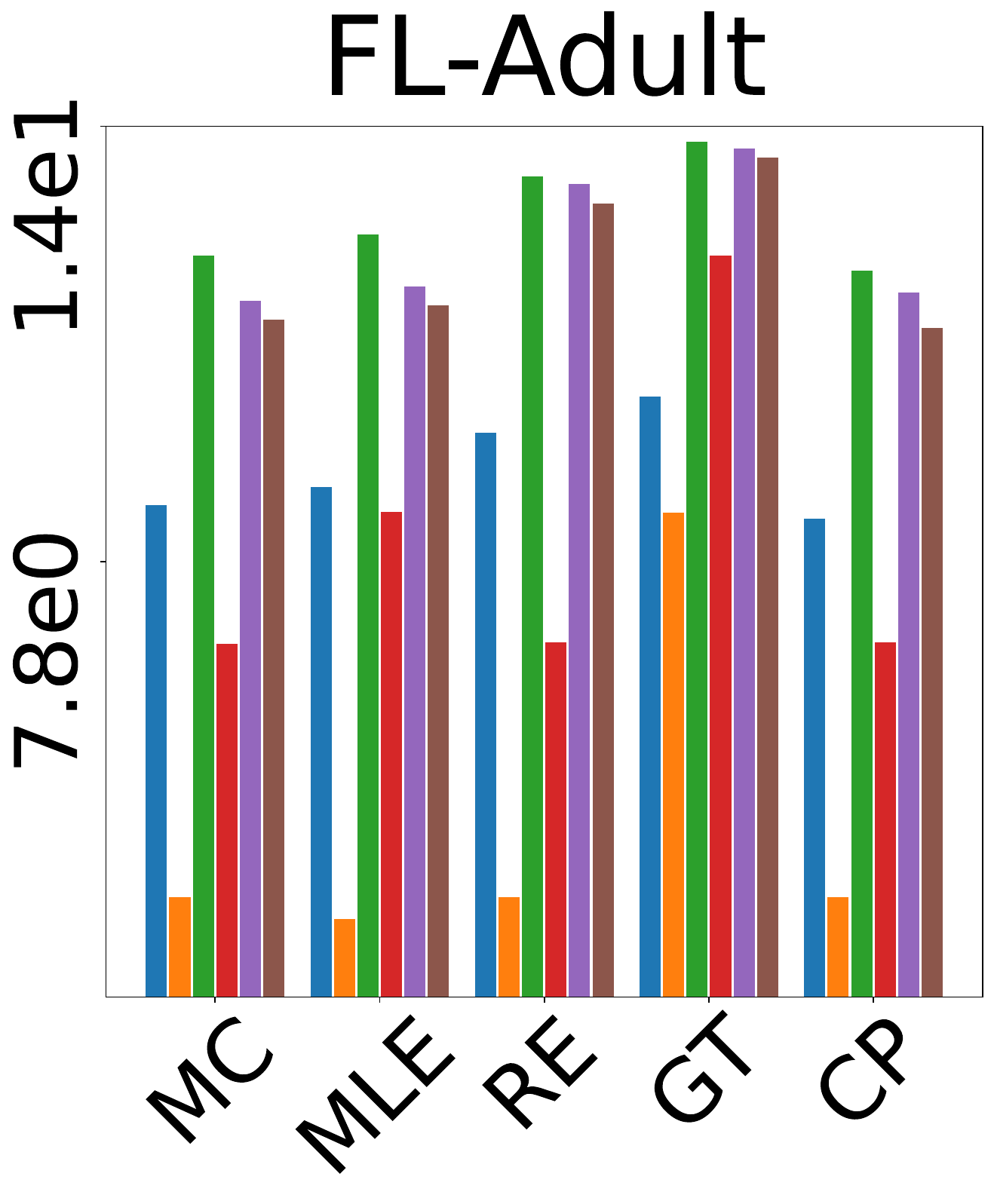}
        \includegraphics[width=0.2\columnwidth, height=0.2\columnwidth]{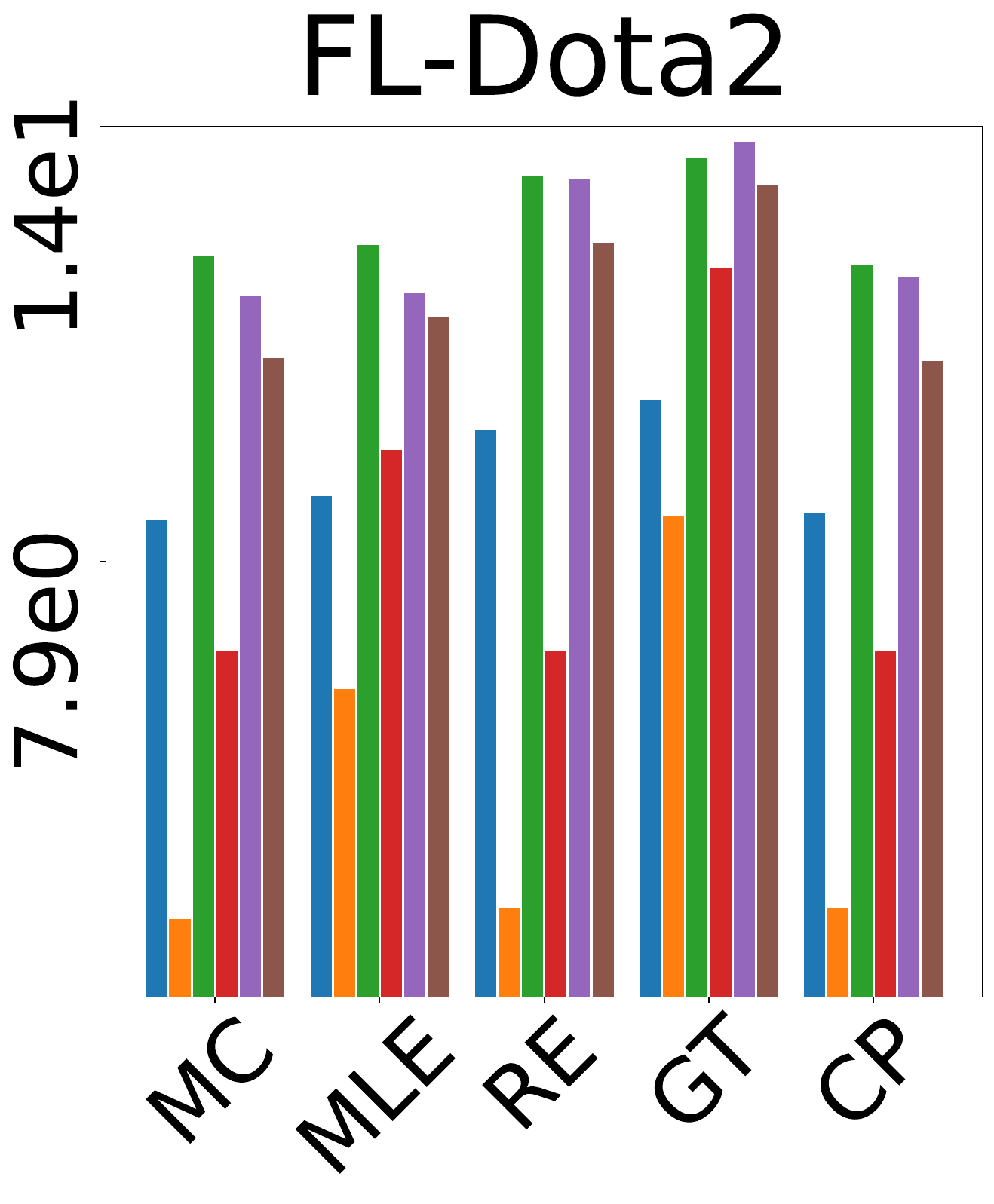}
        \includegraphics[width=0.2\columnwidth, height=0.2\columnwidth]{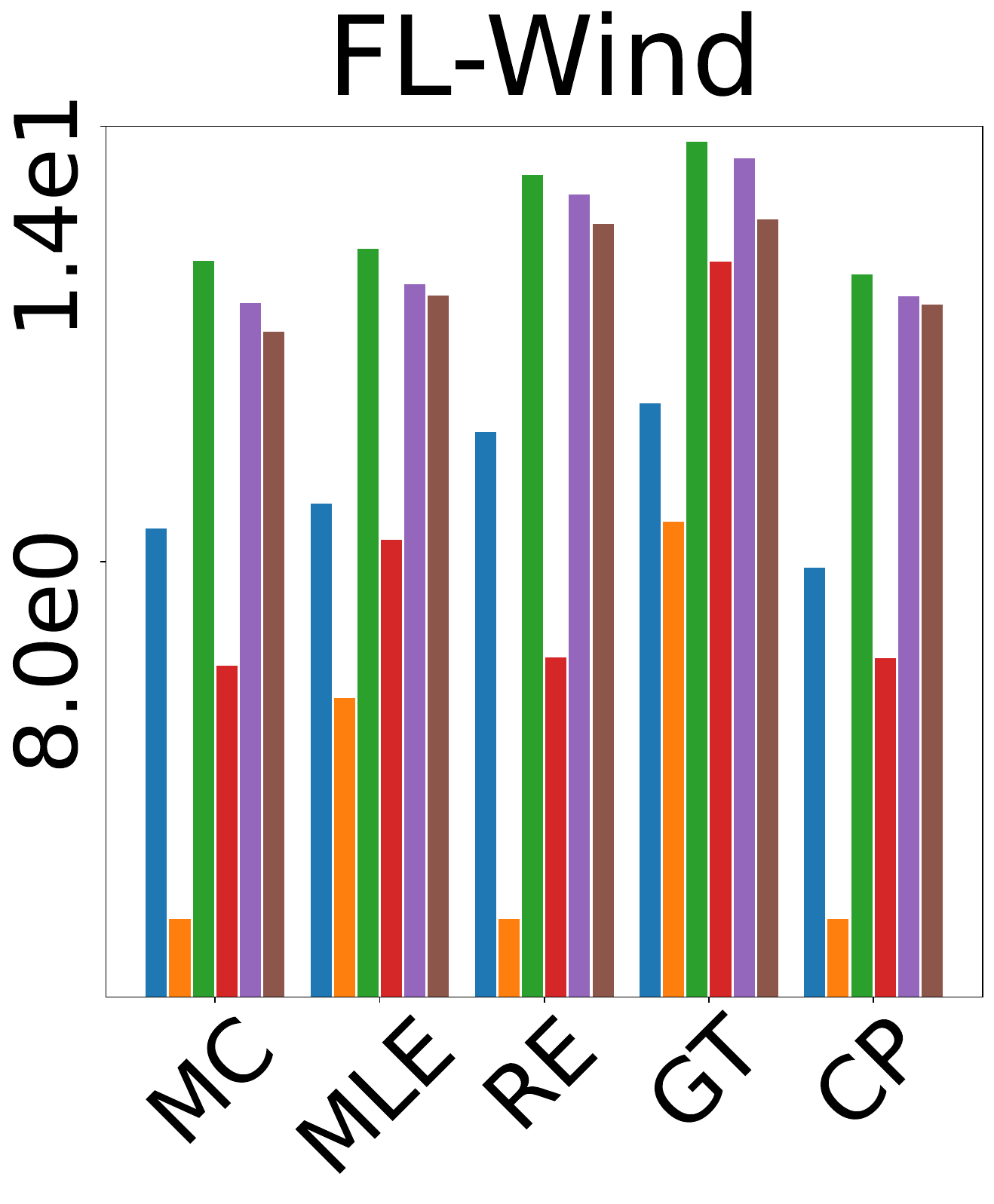}
    }  
    \\
    \mbox{
        \includegraphics[width=0.215\columnwidth, height=0.16\columnwidth]{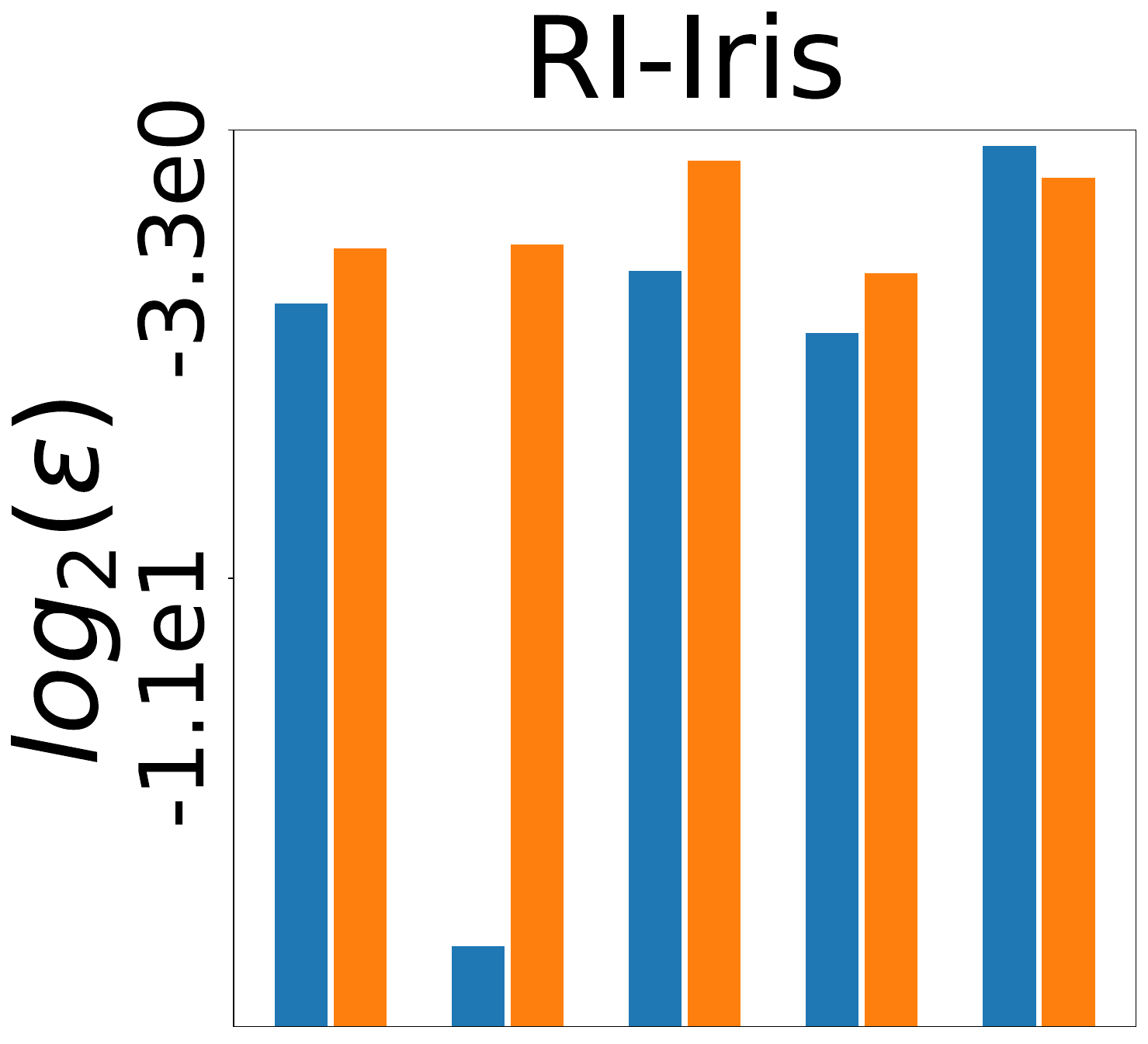}
        \includegraphics[width=0.2\columnwidth, height=0.16\columnwidth]{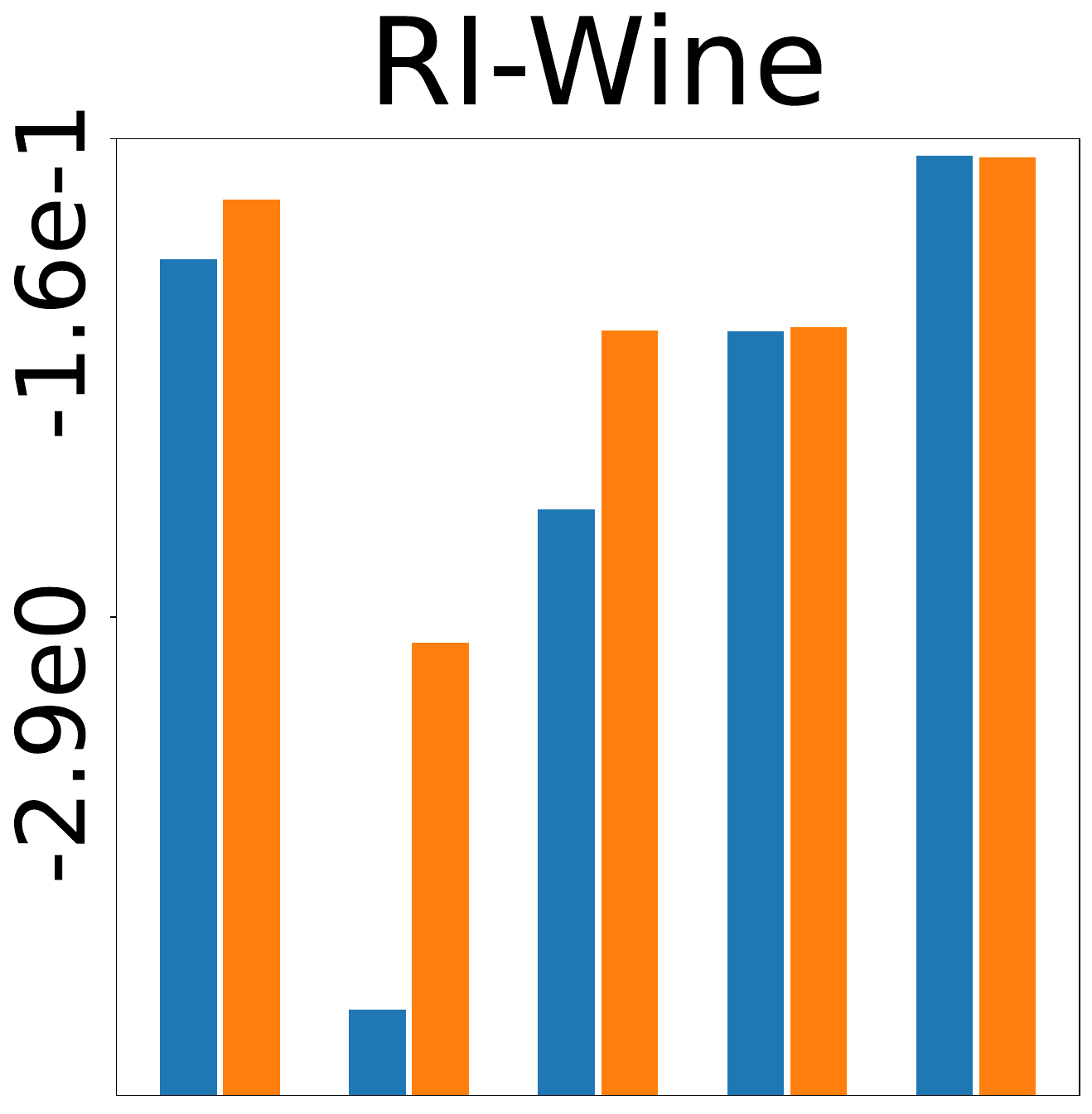}
        \includegraphics[width=0.2\columnwidth, height=0.16\columnwidth]{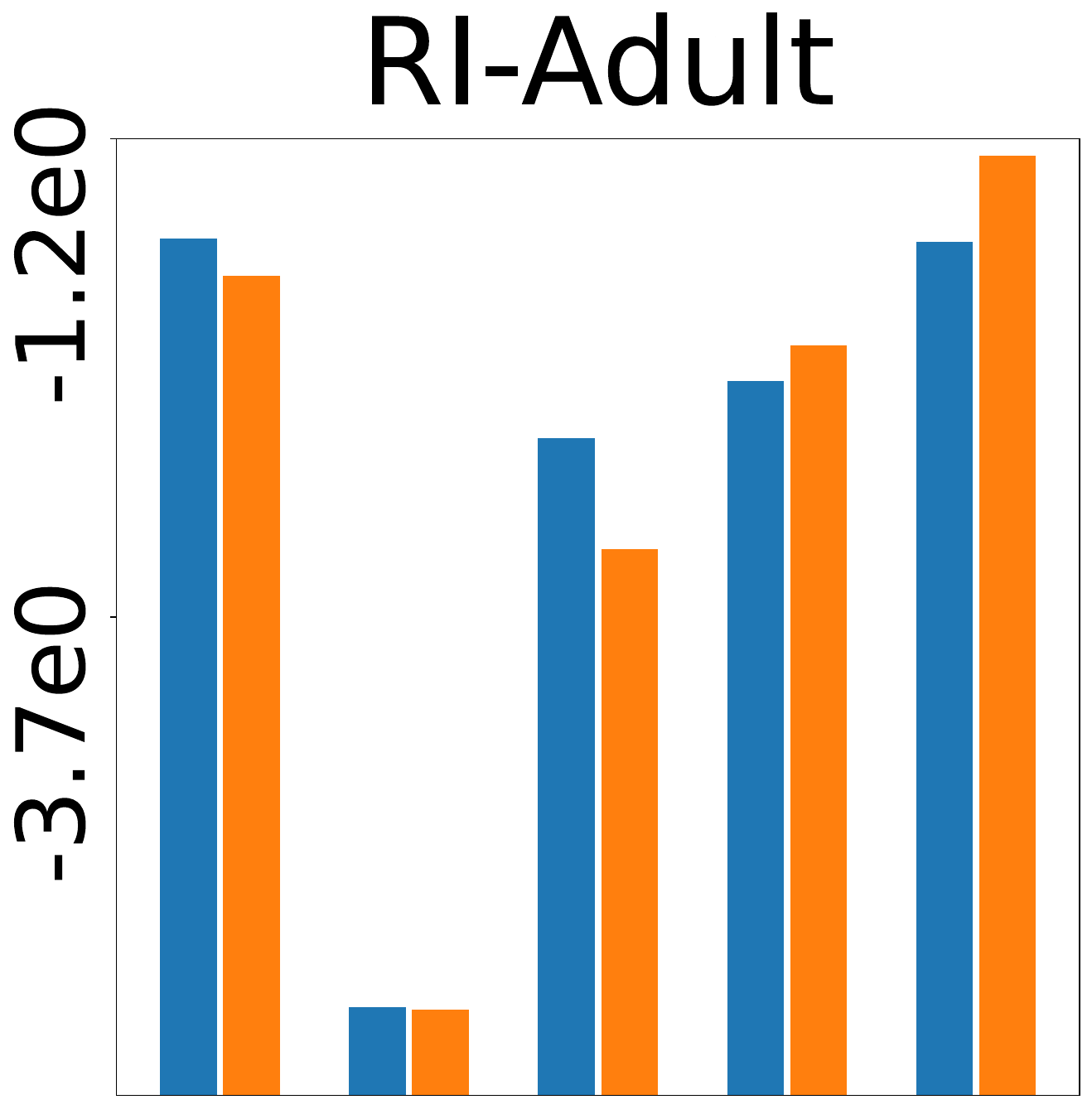}
        \includegraphics[width=0.2\columnwidth, height=0.16\columnwidth]{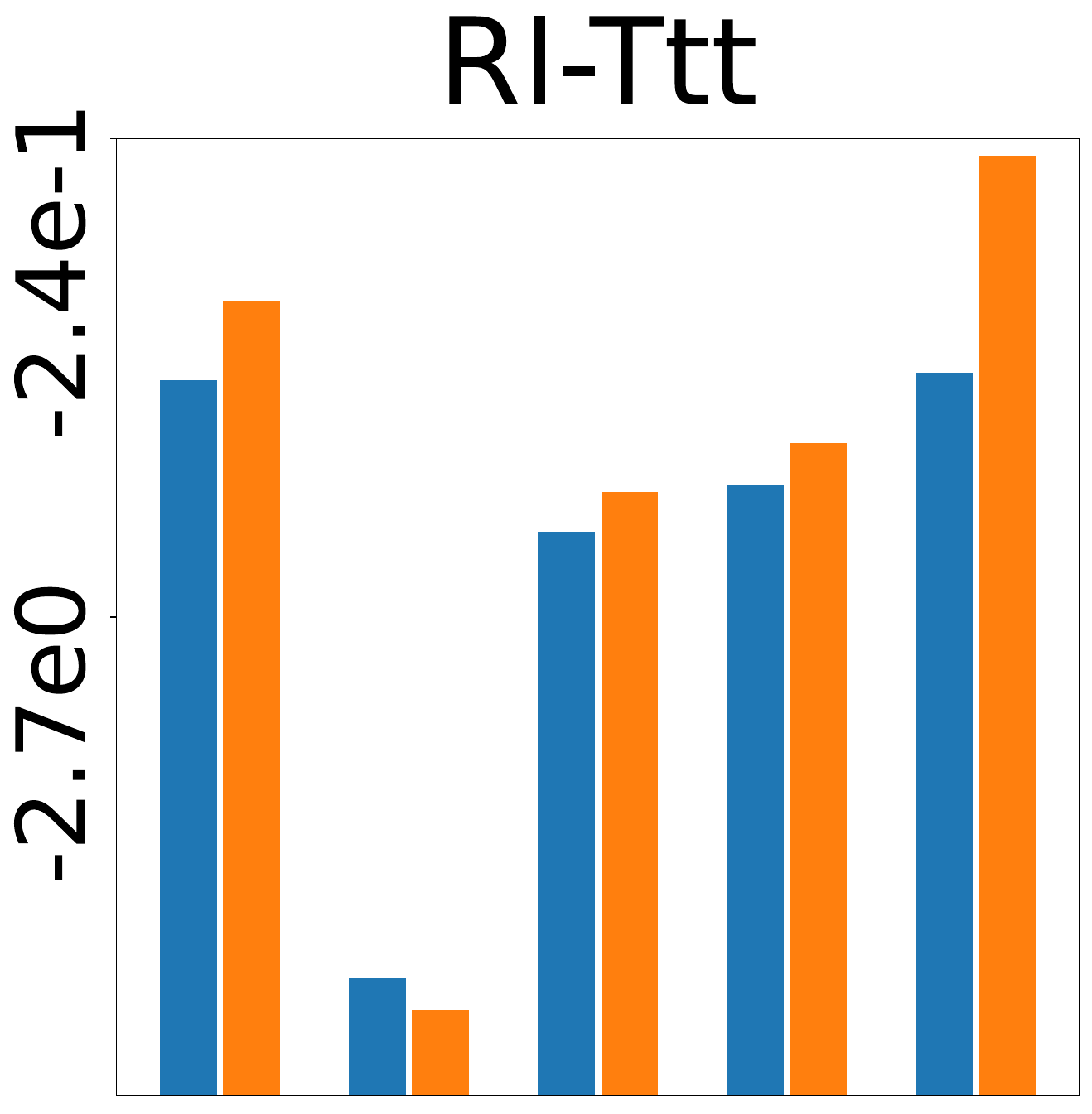}
        \includegraphics[width=0.2\columnwidth, height=0.16\columnwidth]{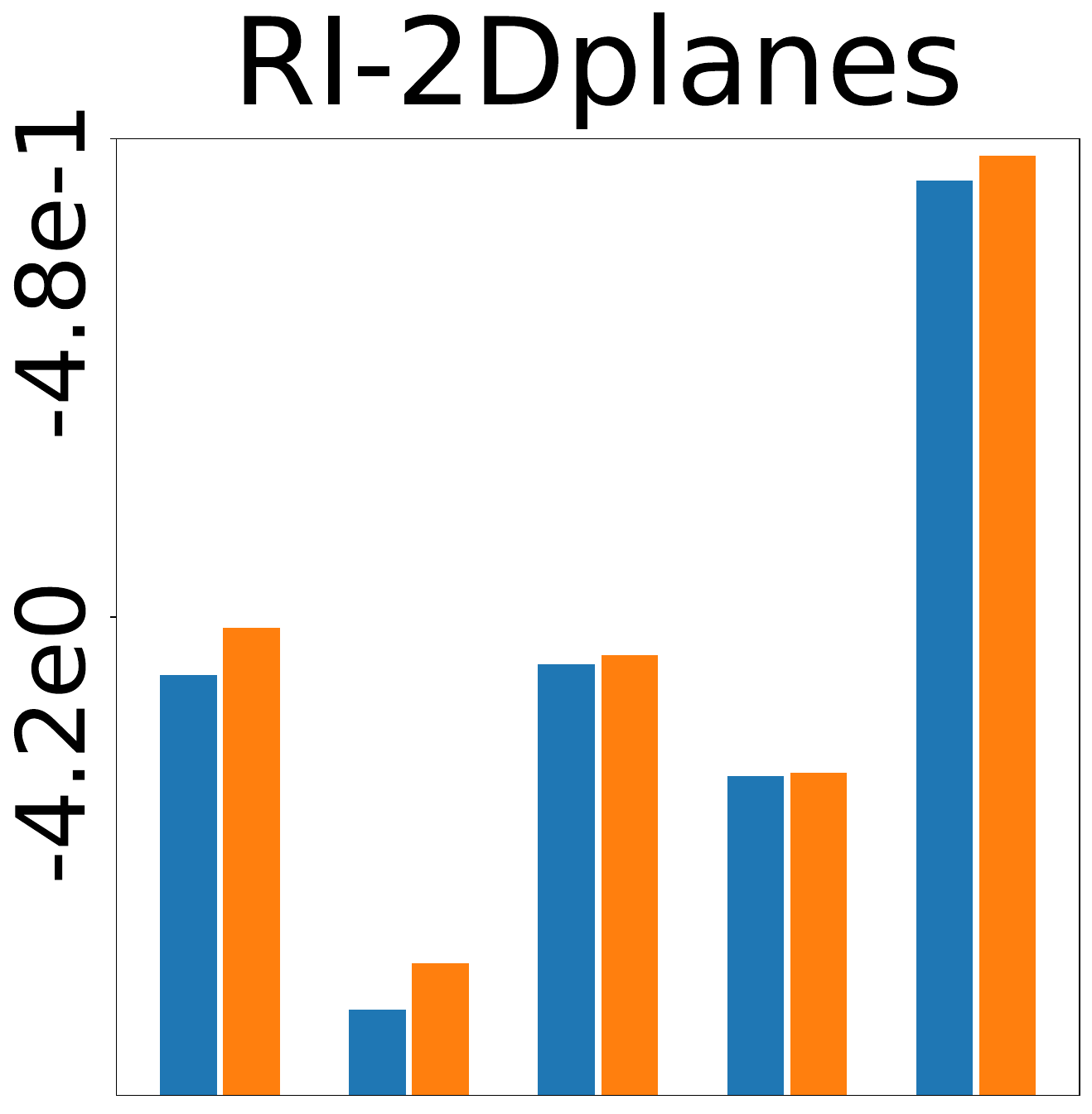}

        \includegraphics[width=0.2\columnwidth, height=0.16\columnwidth]{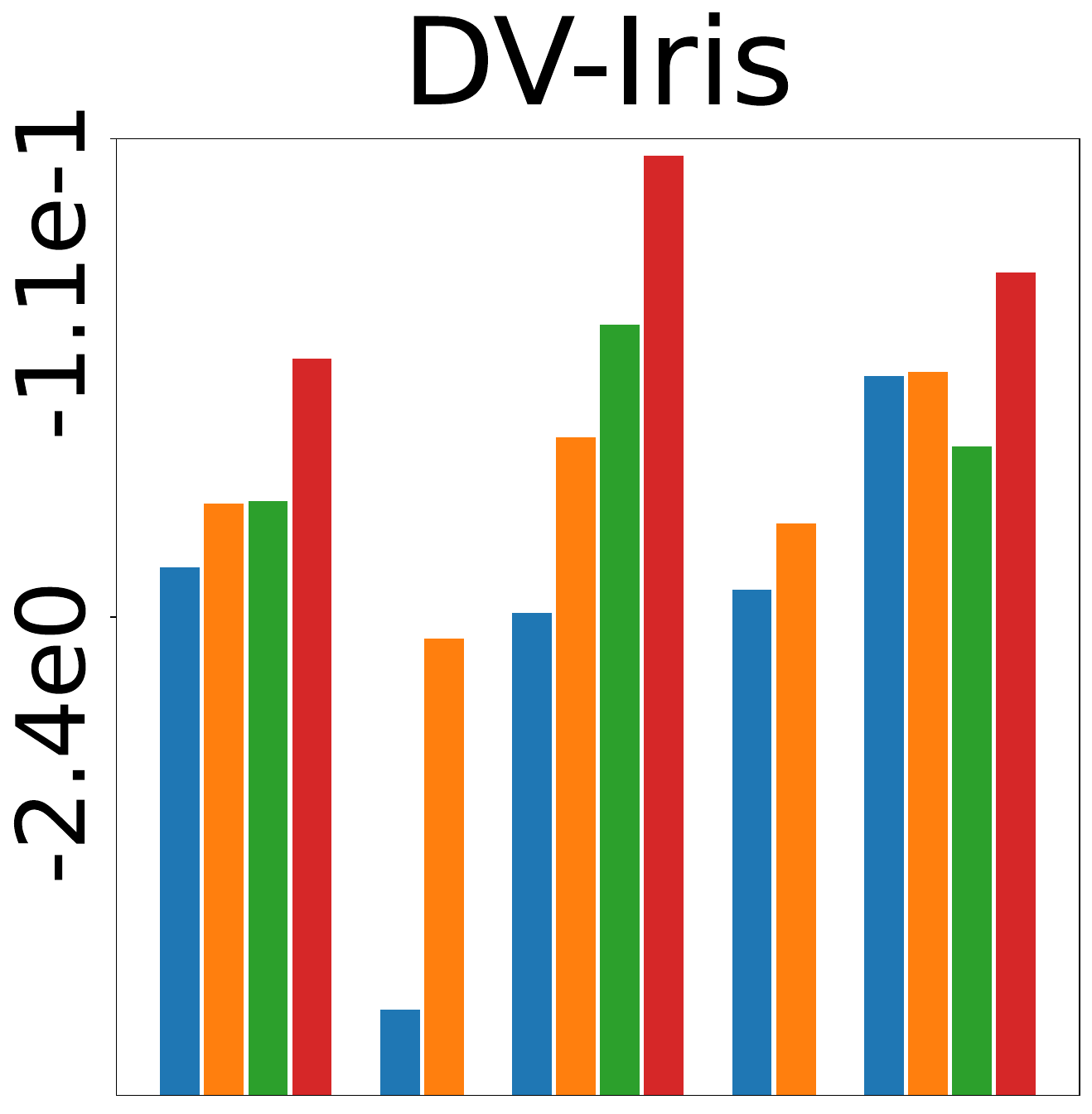}
        \includegraphics[width=0.2\columnwidth, height=0.16\columnwidth]{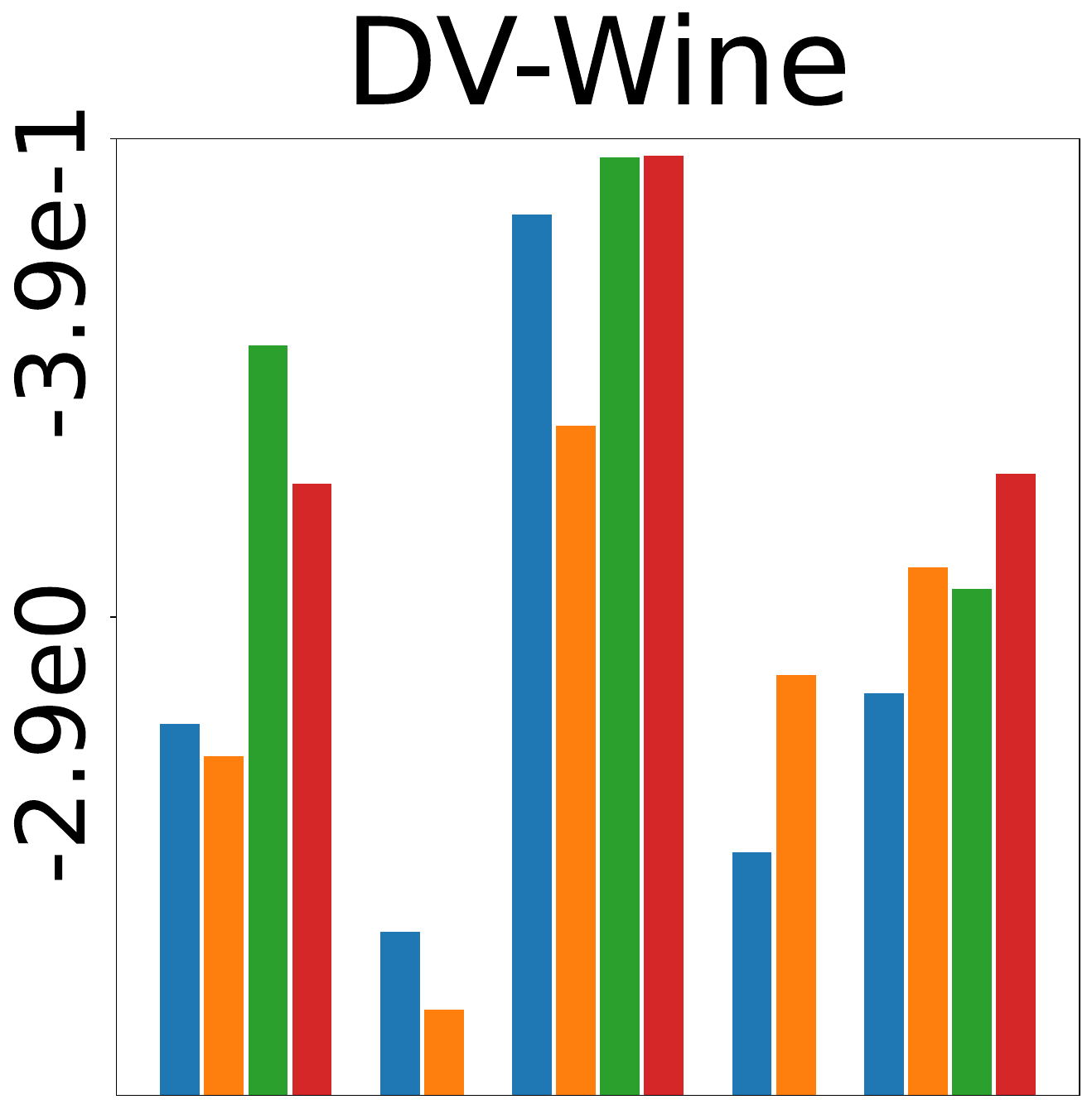}
        \includegraphics[width=0.2\columnwidth, height=0.16\columnwidth]{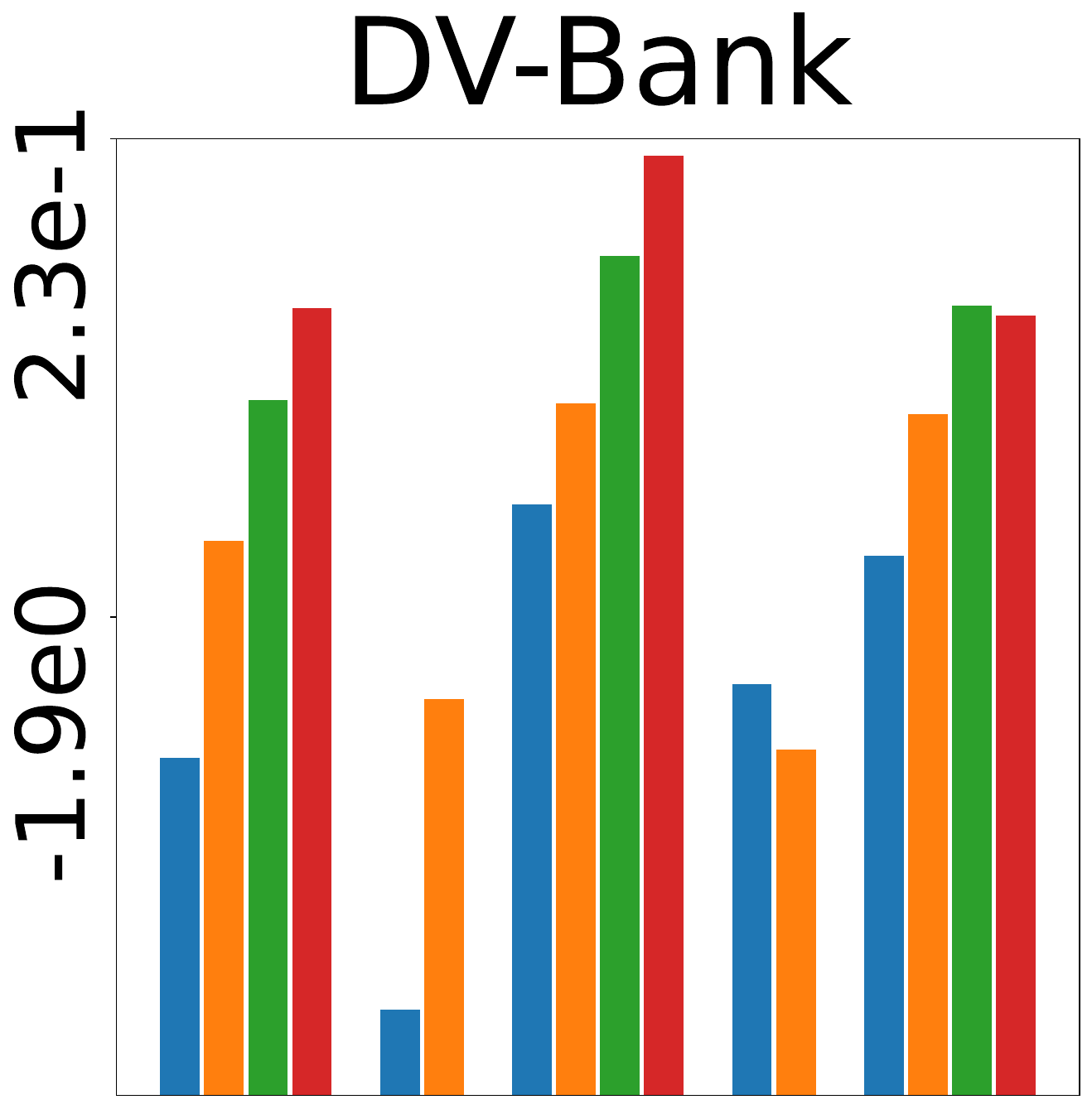}
        \includegraphics[width=0.2\columnwidth, height=0.16\columnwidth]{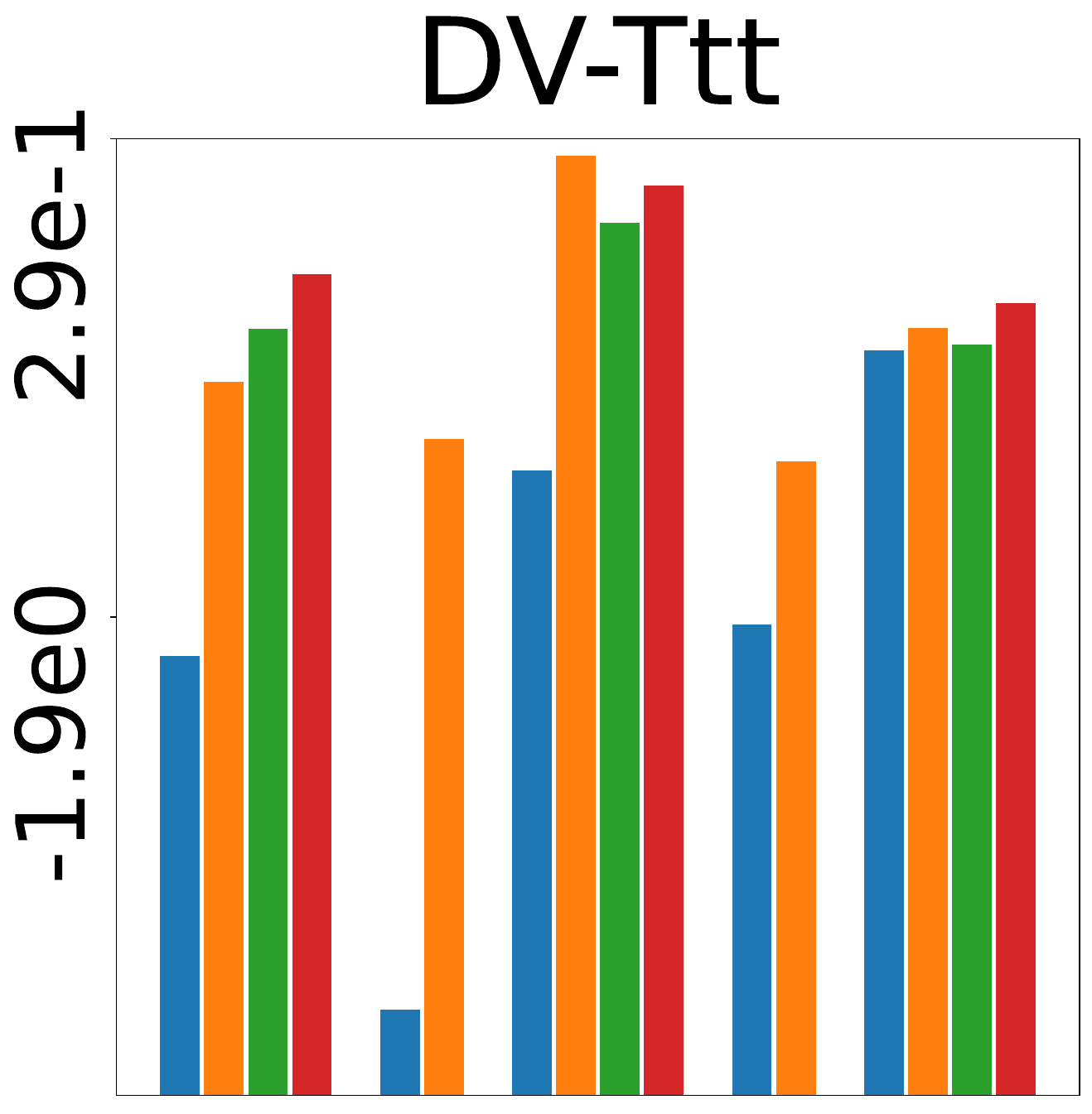}
        \includegraphics[width=0.2\columnwidth, height=0.16\columnwidth]{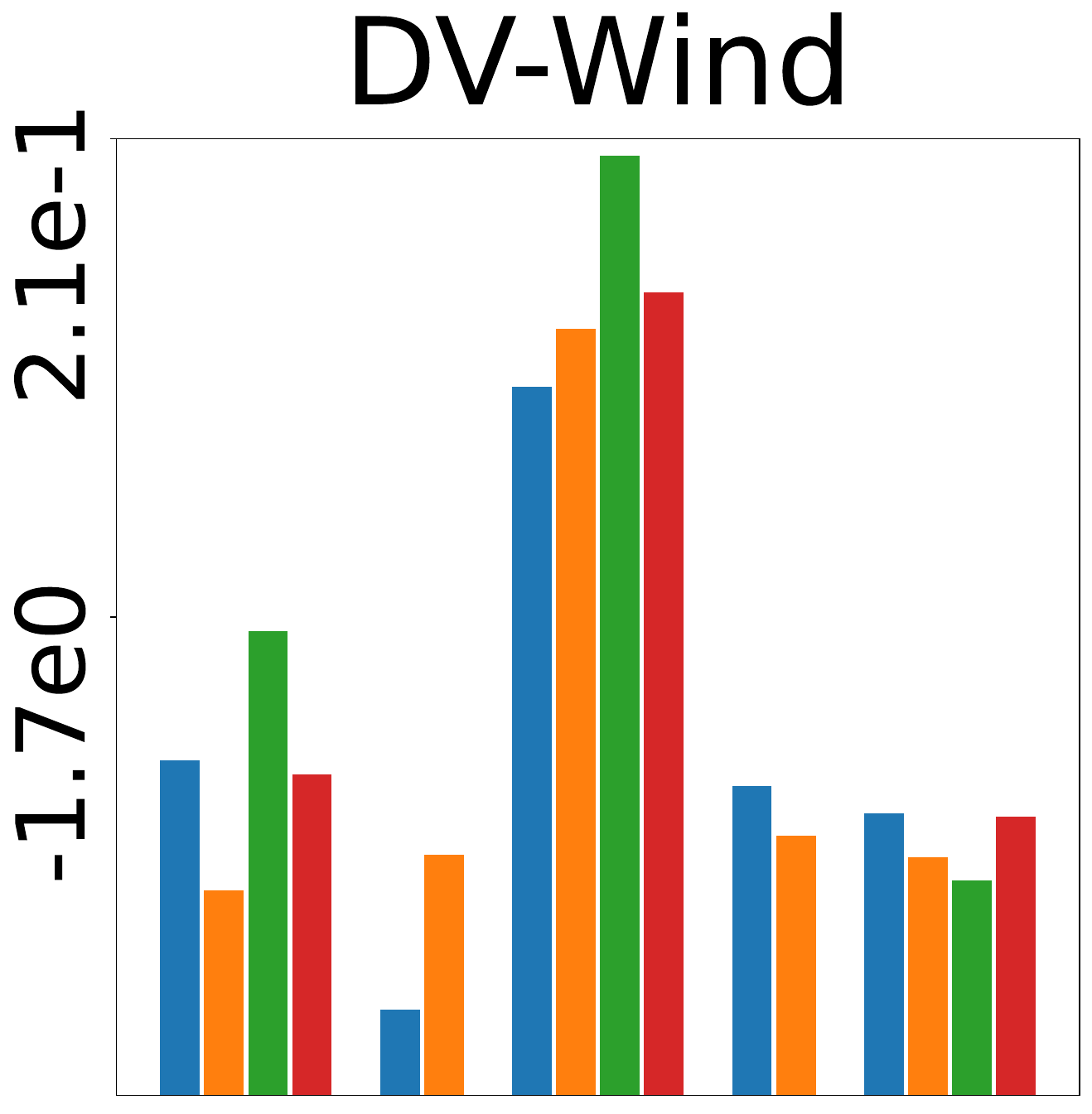}
    }
    \\
    
    \subfigure[The approximation error measured by $\epsilon$.] {
        \includegraphics[width=0.215\columnwidth, height=0.2\columnwidth]{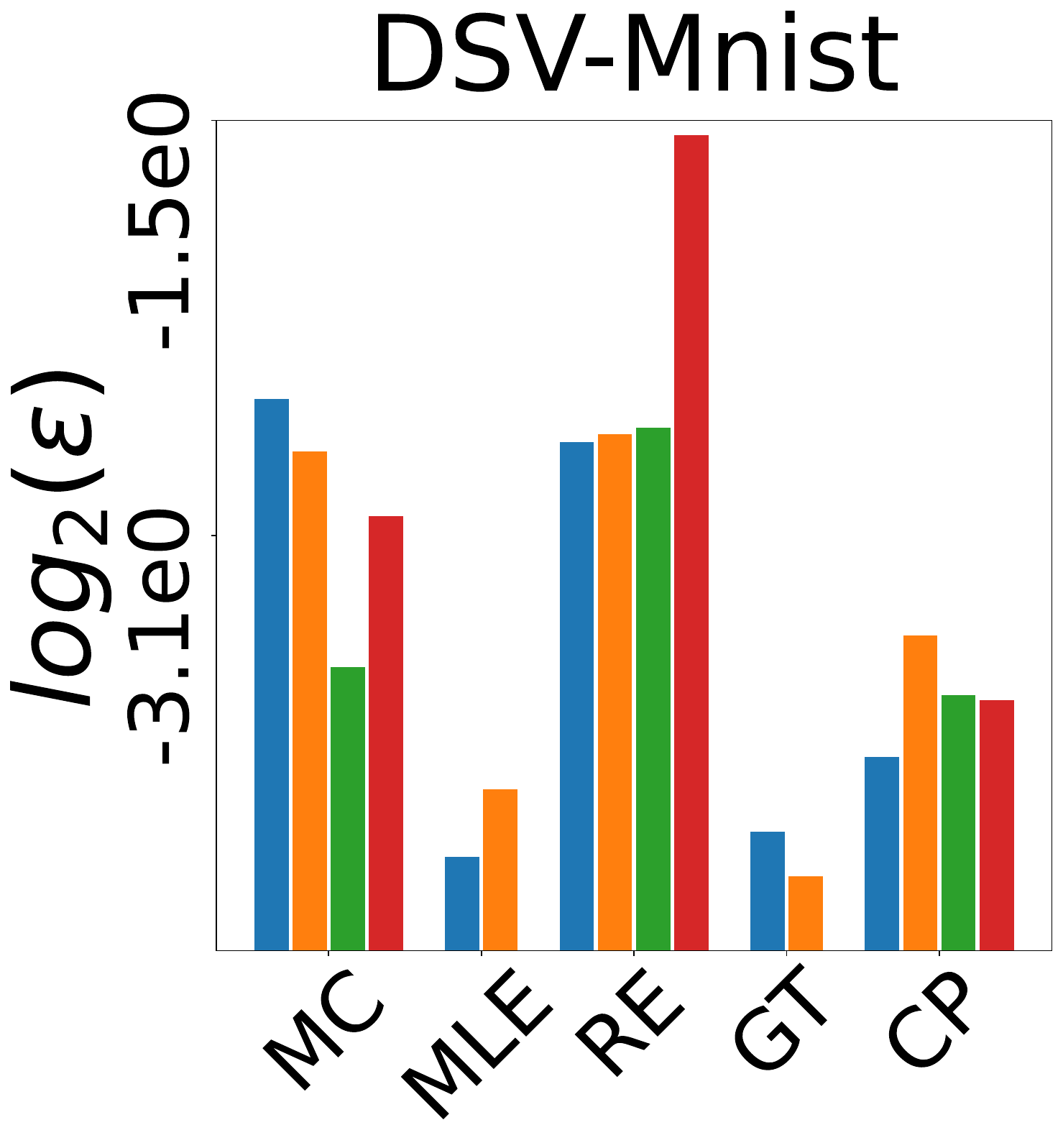}
        \includegraphics[width=0.2\columnwidth, height=0.2\columnwidth]{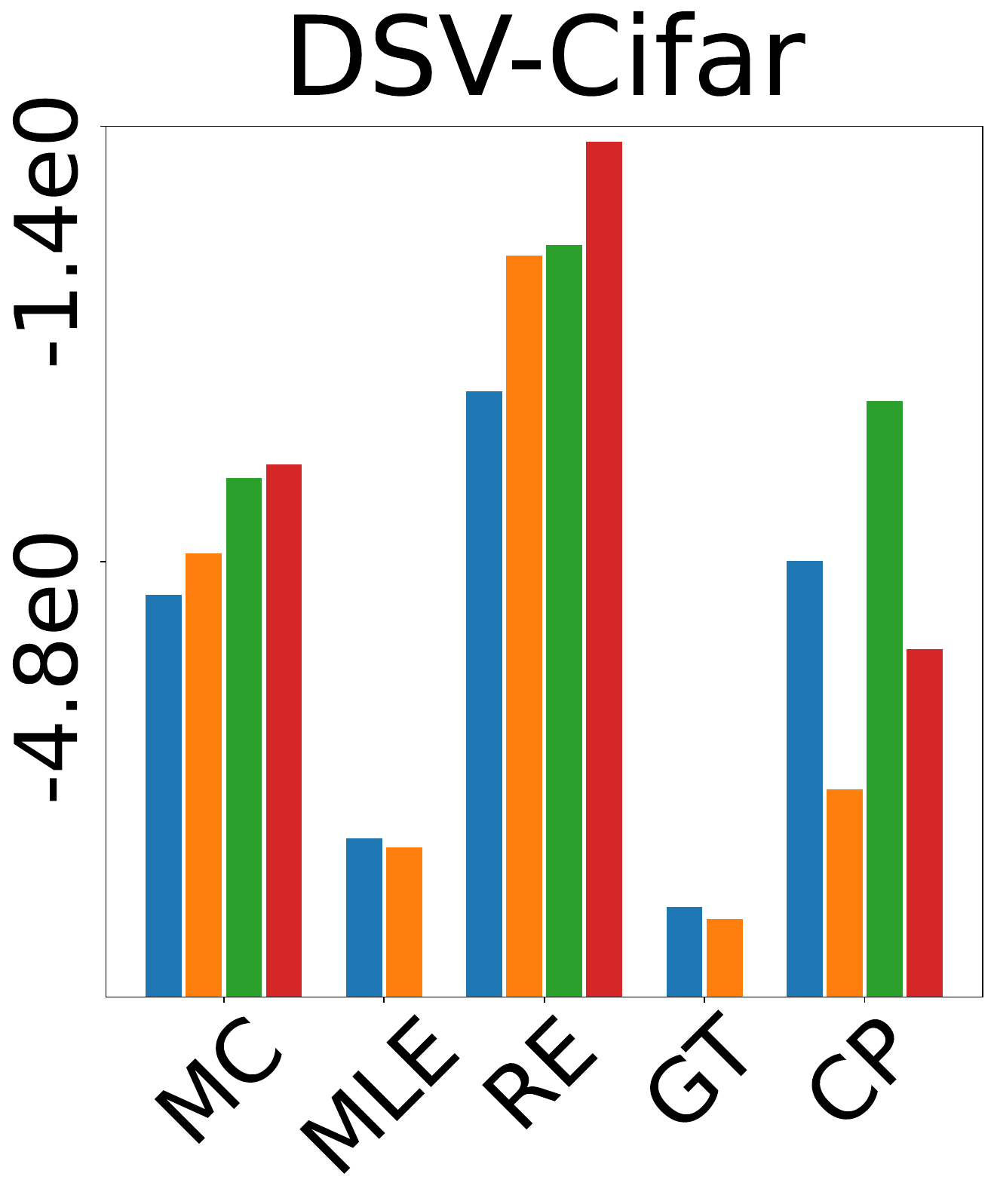}
        \includegraphics[width=0.2\columnwidth, height=0.2\columnwidth]{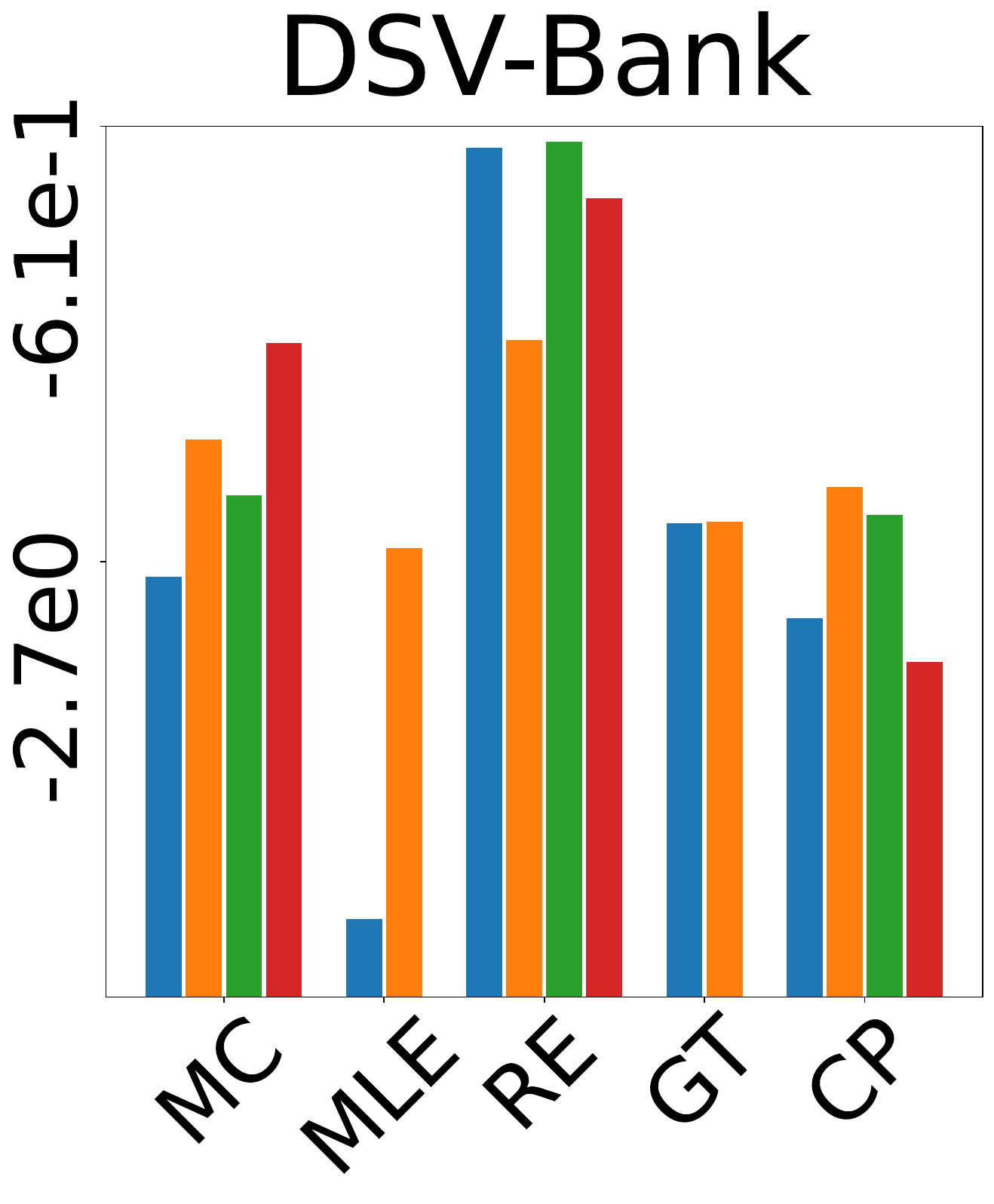}
        \includegraphics[width=0.2\columnwidth, height=0.2\columnwidth]{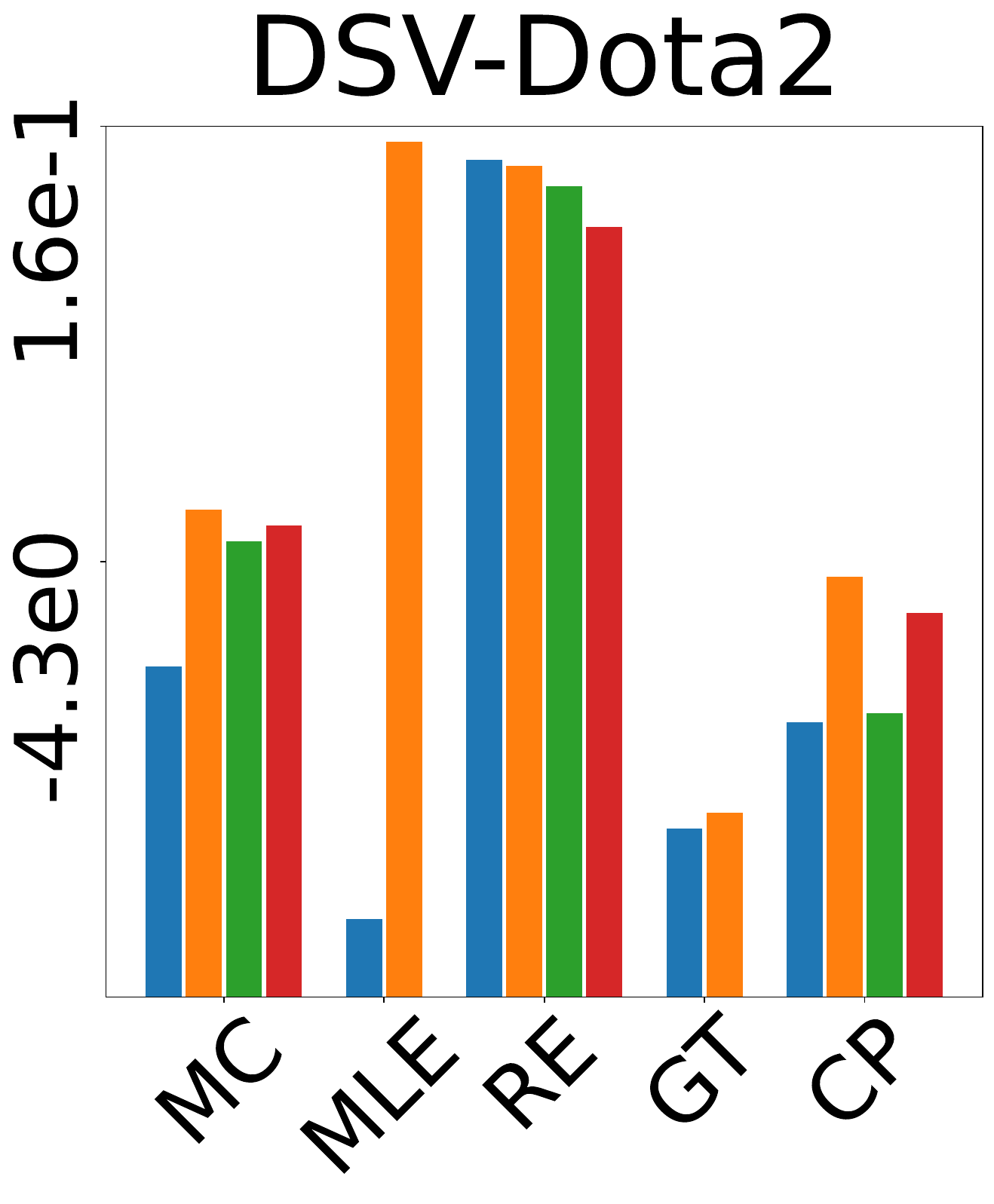}
        \includegraphics[width=0.2\columnwidth, height=0.2\columnwidth]{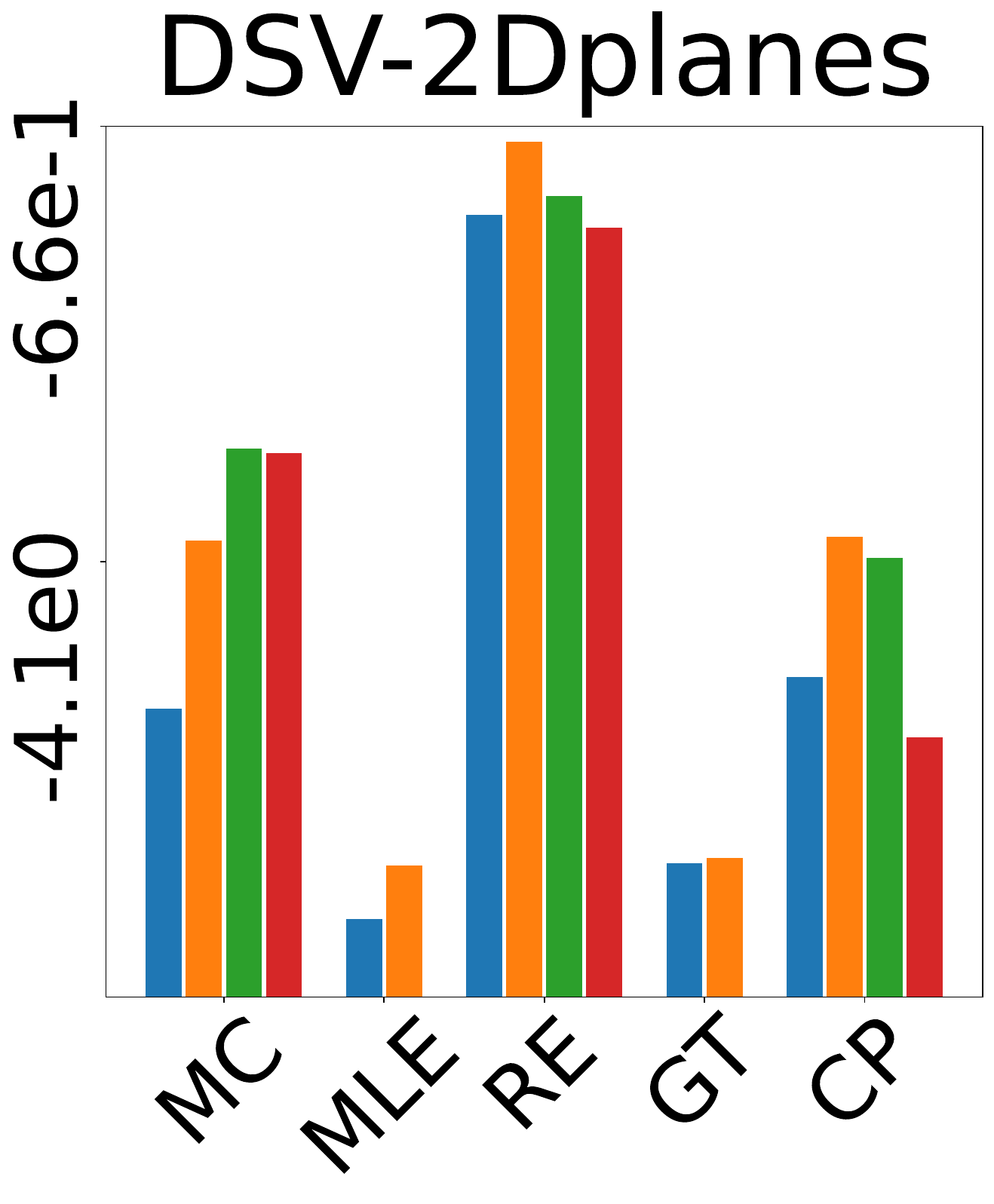}
        
        \includegraphics[width=0.2\columnwidth, height=0.2\columnwidth]{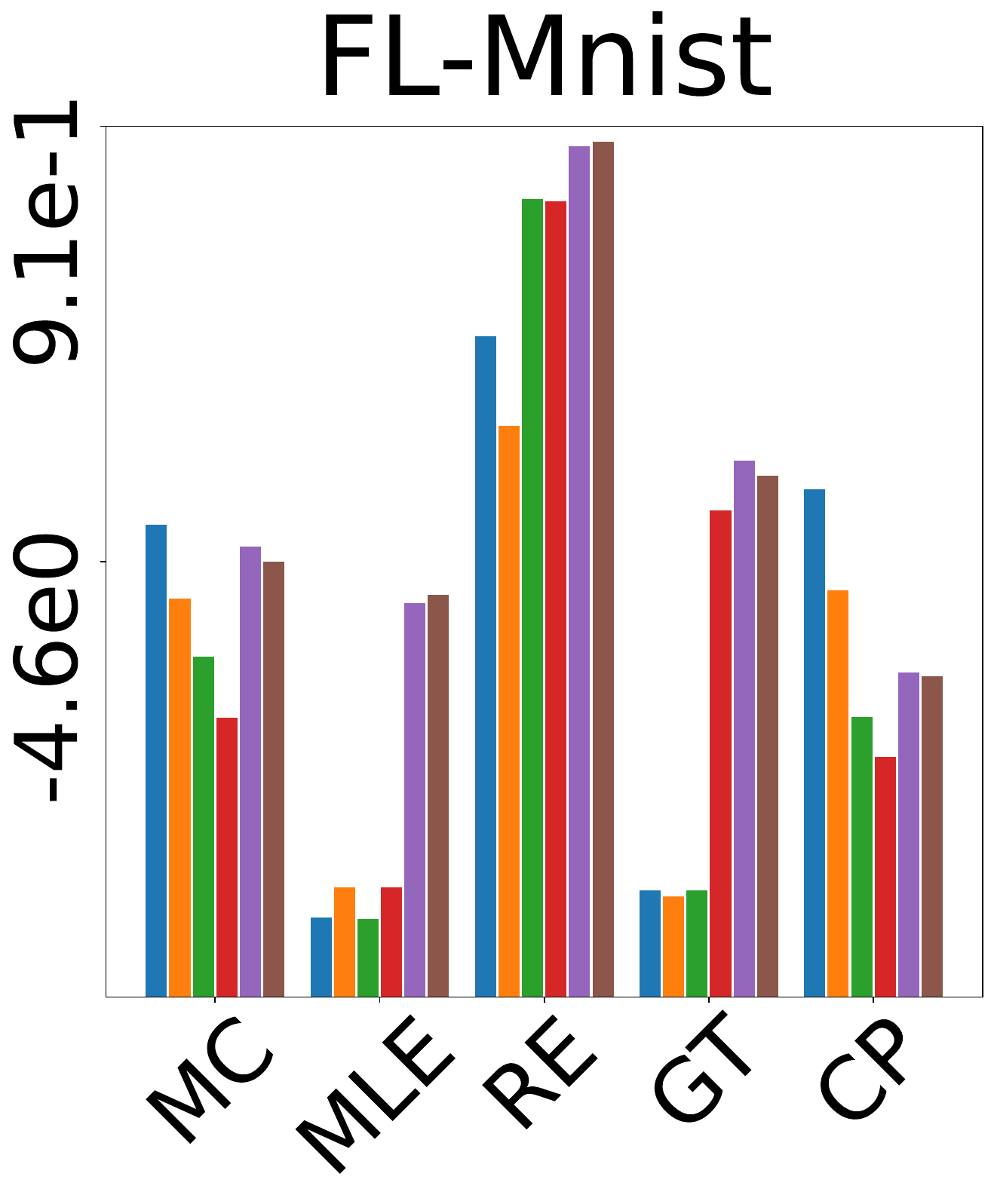}
        \includegraphics[width=0.2\columnwidth, height=0.2\columnwidth]{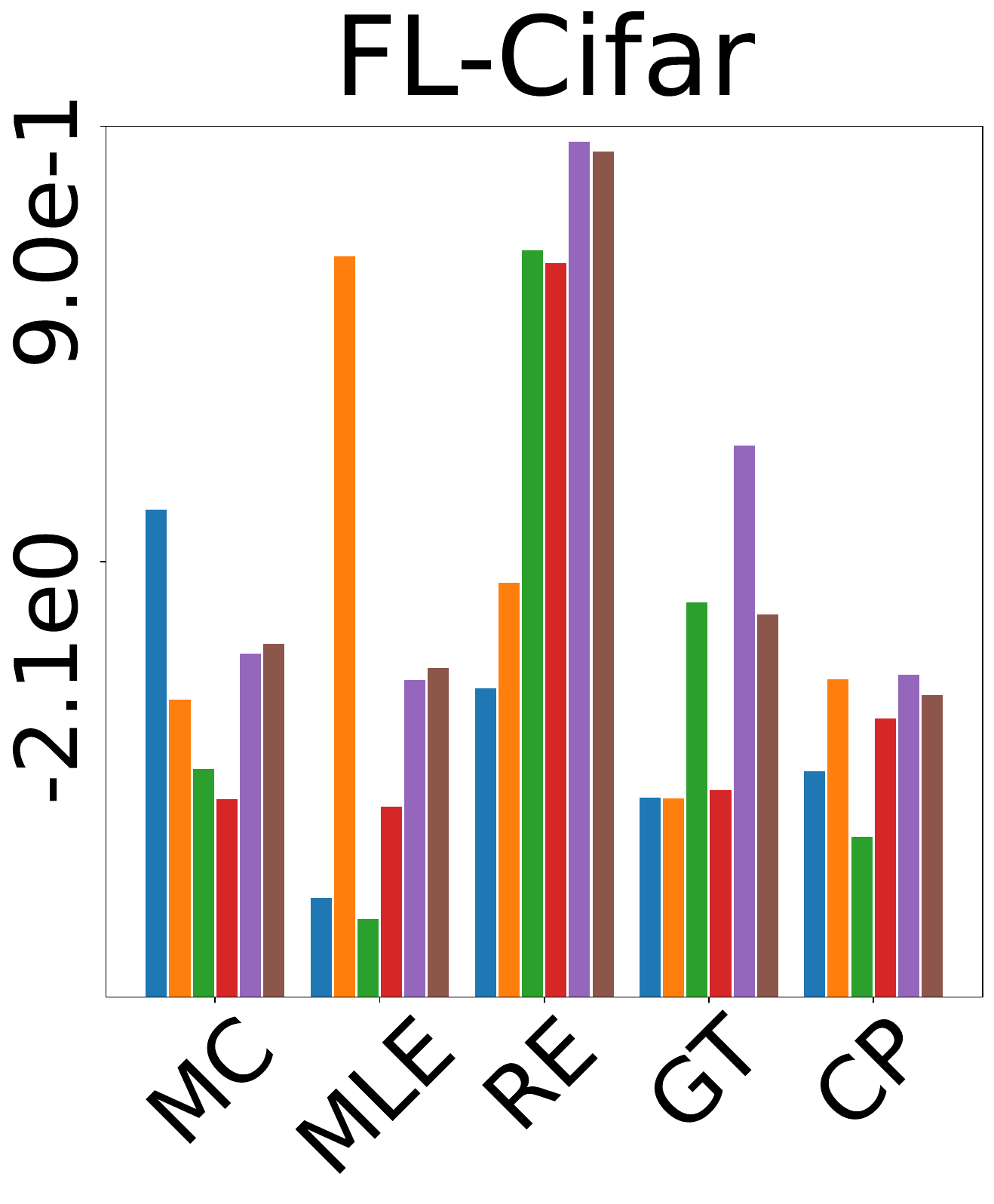}
        \includegraphics[width=0.2\columnwidth, height=0.2\columnwidth]{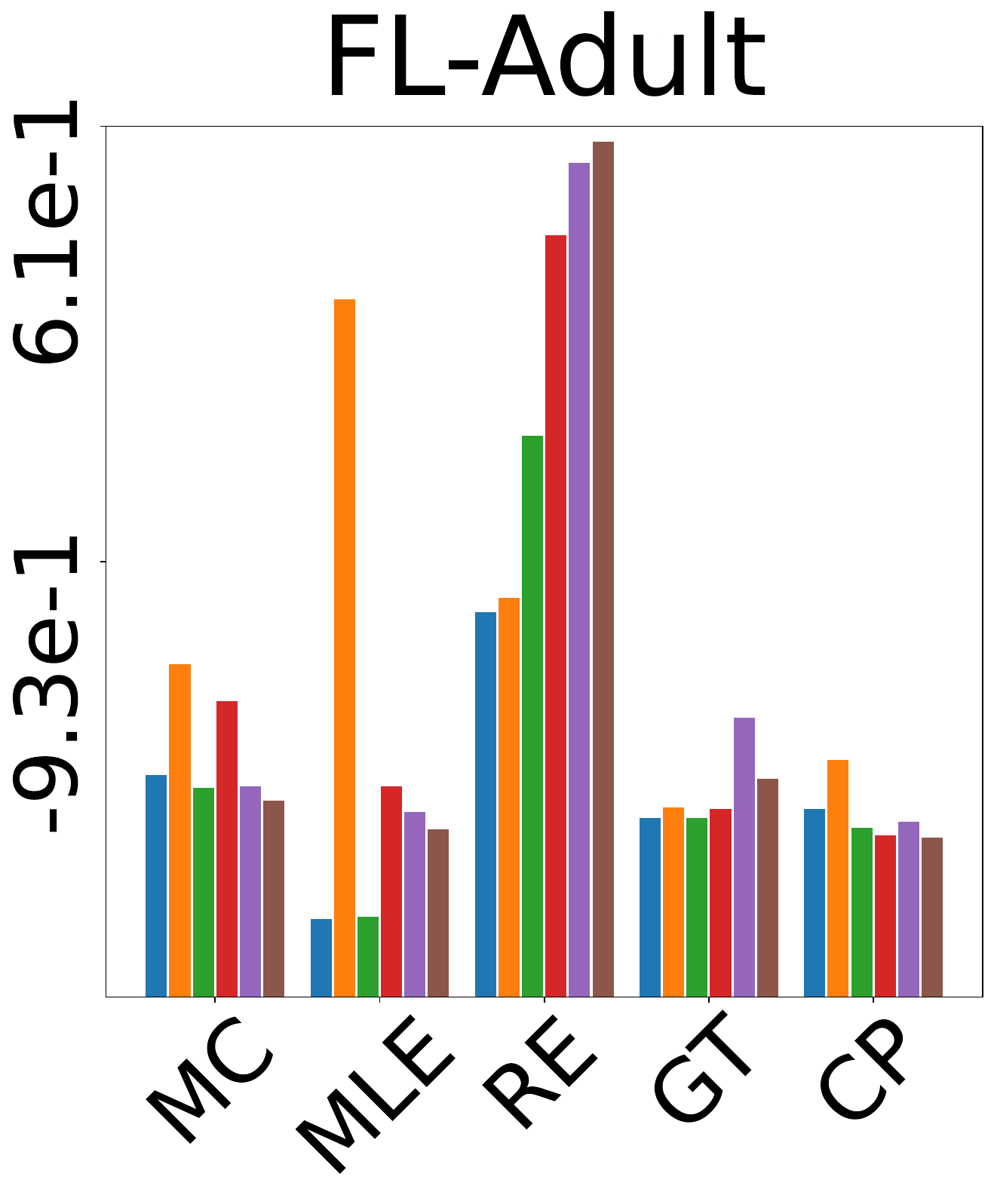}
        \includegraphics[width=0.2\columnwidth, height=0.2\columnwidth]{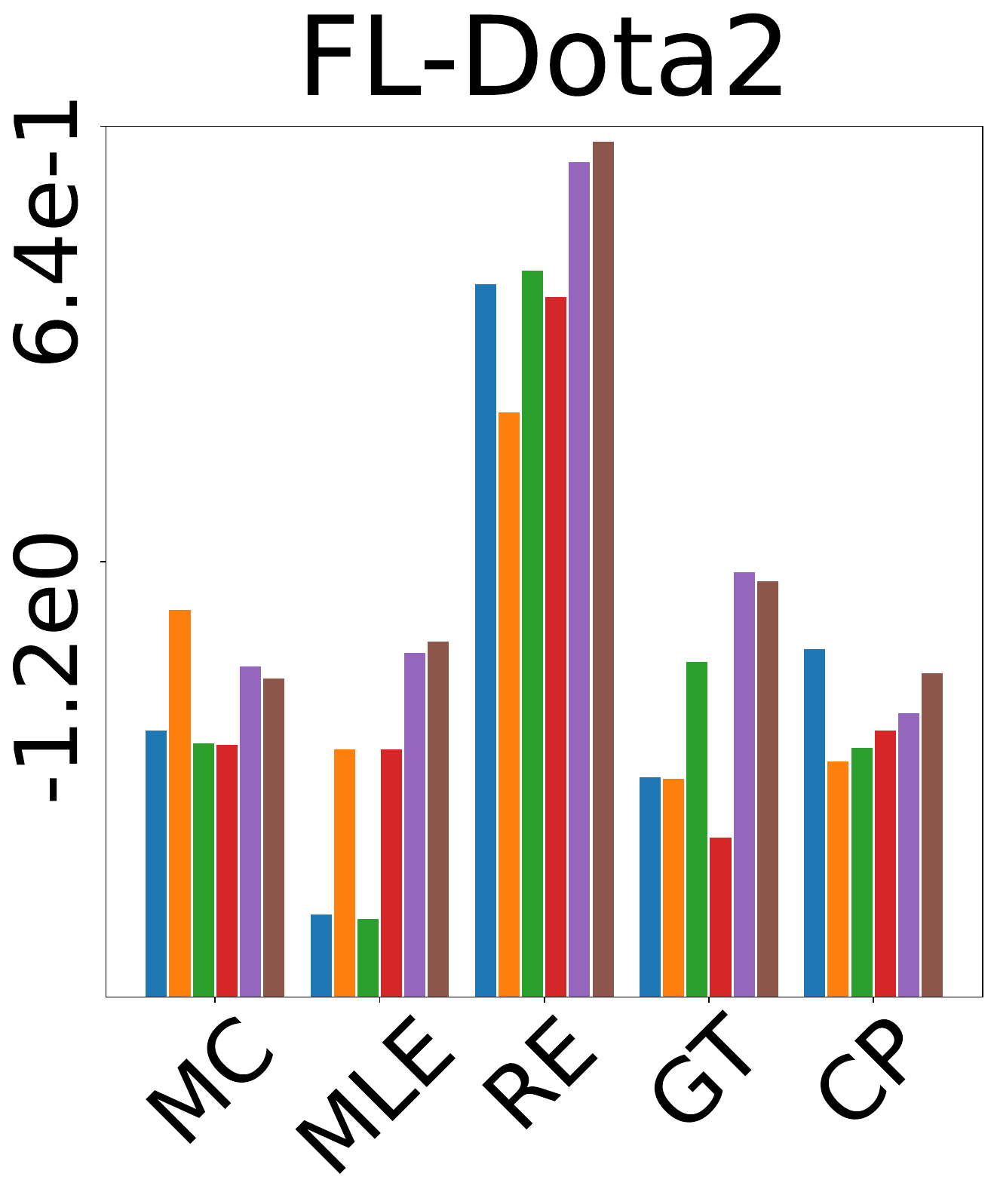}
        \includegraphics[width=0.2\columnwidth, height=0.2\columnwidth]{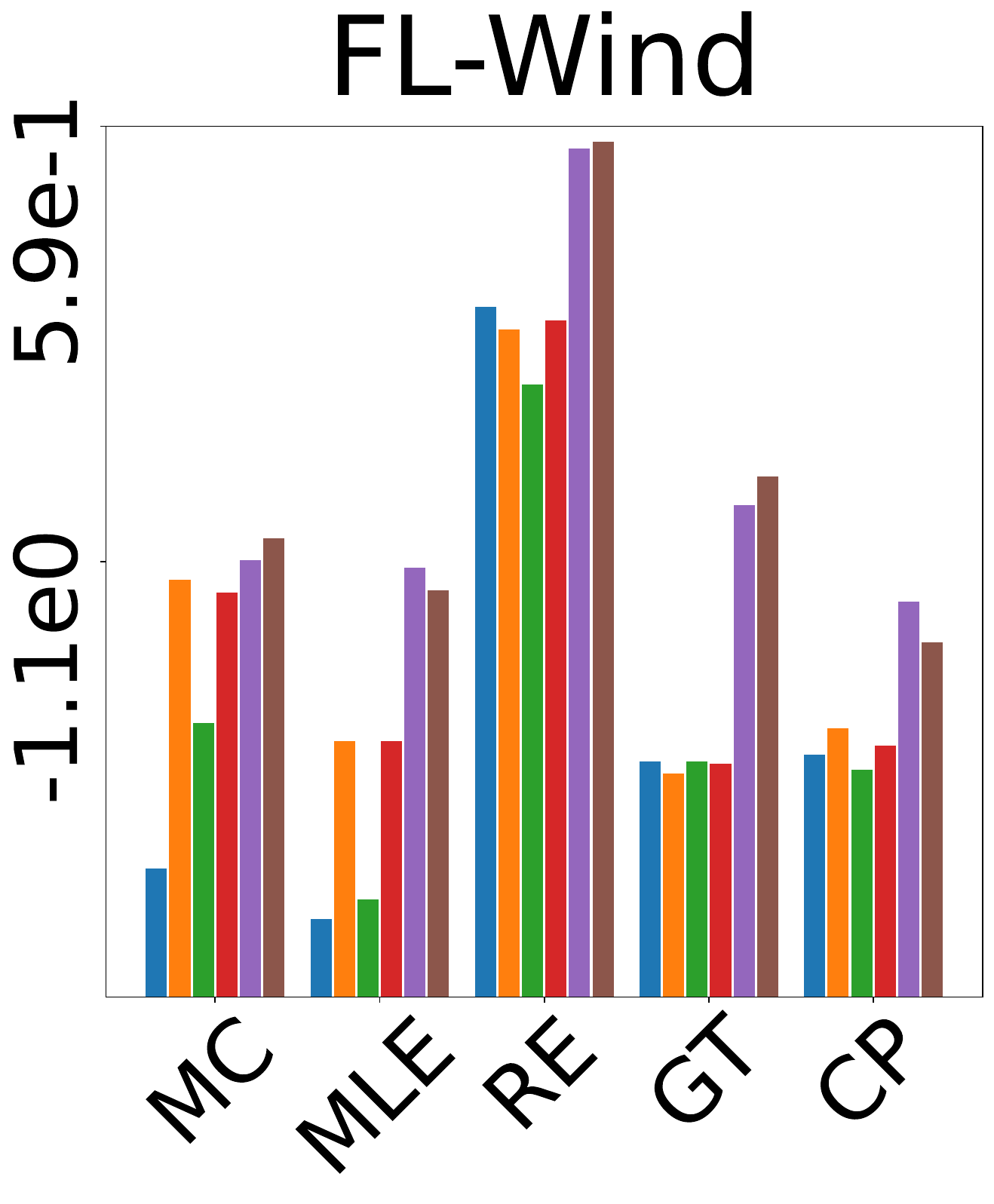}
    }  
    \caption{
    Efficiency performance (x-axis: the base SV computing algorithm). The smaller the three metrics, the more efficient and accurate the computation.
    }
    \label{fig:efficiency_results}
\end{figure*}
In this section, we use \textit{SVBench} to implement multiple SV computing algorithms, including both the base algorithms that have been studied by prior works and the hybrid algorithms with novel combinations of SV computing techniques. 
Using the implemented algorithms, we not only validate the usability, modularity, and flexibility of \textit{SVBench} but also study the following four sets of evaluations in order to answer the aforementioned questions:

\begin{itemize}
\item \S\ref{subsec: exp_efficiency} compares the efficiency of five base SV computing algorithms with several hybrid algorithms, answering the question highlighted at the end of \S\ref{sssec: computation_efficiency}.

\item \S\ref{subsec: exp_impactsOfError} investigates the relationship among the computation efficiency, approximation error, and the effectiveness of SV, solving the problem proposed at the end of \S\ref{sssec:approximation_error}.

\item \S\ref{subsec: exp_privacy} examines the effectiveness of existing measures for preventing SV-driven attacks and the impacts of the measures on the effectiveness of SV, tackling the problem left in \S\ref{sssec:privacy_challenge}.

\item \S\ref{subsec: exp_interpretability} explores the relationship between the SVs of the four types of players in DA and the overall utility of their associated tasks, offering insights to the question bold in \S\ref{sssec:interpretation_challenge}.
\end{itemize}

\Cref{tab:exp_settings_method} summarizes the detailed configurations specified for each algorithm. We note that this work is the first attempt to combine the $\spadesuit$-tagged base algorithm with the five optimization techniques. We use MLE as the base algorithm in §5.2–§5.4, since it generally achieves
better efficiency and accuracy performance than the other base algorithms in §5.1 and
varying base algorithms would not influence conclusions in those subsections.

For the generality of findings from experiments, we select 10 canonical datasets  from a large number of literature to conduct four types of DA tasks -- result interpretation (RI), data tuple valuation (DV), dataset valuation (DSV), and federated learning (FL), which enable evaluations of SV for different types of players defined in the current DA domain. 
\Cref{tab:exp_task_settings} summarizes the datasets, including Iris \cite{UCI_dataset}, Wine \cite{UCI_dataset}, Mnist \cite{lecun-mnisthandwrittendigit-2010}, Cifar-10 \cite{Krizhevsky2009LearningML}, Adult \cite{UCI_dataset}, Tic-Tac-Toe (Ttt) \cite{UCI_dataset}, Bank \cite{UCI_dataset}, Dota2 \cite{UCI_dataset}, Wind \cite{OpenML2013,OpenML2020}, 2Dplanes \cite{OpenML2013,OpenML2020}, and our task settings. More details of task settings and the code are available at GitHub\footnote{\url{https://github.com/zjuDBSystems/SVBench}. \label{fn:github_url}}.

\subsection{Computation Efficiency} \label{subsec: exp_efficiency}

This section investigates the efficiency performance of five base SV computing algorithms (MC, RE, MLE, GT, CP) and the hybrid algorithms, each of which selects an optimization strategy from \Cref{tab:configuration_parameters} and integrates this strategy with a base algorithm. 
To control the time cost of each experiment, we follow previous work \cite{liu2022gtg} to monitor the approximation stability by $\Delta \hat{\phi} = \frac{1}{5n} \sum_{m=1}^{5} \sum_{i=1}^{n} |\frac{\hat{\phi}^e_i-\hat{\phi}^{e-m \times n}_i}{\hat{\phi}^e_i}|$ and terminate the approximation when the convergence criterion ($\Delta \hat{\phi} < \tau$) is satisfied, where $\hat{\phi}^e_i$ is the approximate SV of player $p_i$ after $e$ times of utility computation, $\tau$ is the convergence threshold set to 0.05 for all tasks. 
We measure the efficiency performance by the total time cost $N_{uc}\times T_{uc}$, and show the computation complexity $N_{uc}$ and the approximation error $\epsilon=1-\frac{\sum_{i=1}^{n}\hat{\phi}_i \cdot \phi_i}{\sqrt{\sum_{i=1}^{n}\hat{\phi}_i^2 } \cdot \sqrt{\sum_{i=1}^{n}\phi_i^2} }$ \cite{liu2022gtg}, as a reference in the results.

\Cref{fig:efficiency_results} presents the efficiency results. The number of bars differs across tasks due to (1) no applicable GA/TSS techniques for RI tasks and no TSS technique applicable for DV tasks; (2) MLE/GT incompatible with the GA technique devised for DV tasks.

As denoted by blue bars, MLE generally achieves the leading efficiency performance (lower $T_{uc}\times N_{uc}$ and $N_{uc}$) among the five base algorithms. 
The major reason is that, unlike MLE, algorithms such as RE, GT, and CP must solve auxiliary optimization problems (e.g., weighted least squares in RE, feasibility constraints in GT, or convex objectives in CP) in the computation, introducing an extra cost.
In particular, when it is hard to solve the corresponding optimization problem for the DA task (e.g., RI-Adult, RI-2Dplanes, and DV-Ttt), prolonged convergence will appear. 
Although CP can converge faster than MLE in some tasks, e.g., RI-Iris, RI-Ttt, DV-Iris, it has a strong dependency on the sparse SV assumption and performs well only when that assumption is satisfied (to some extent). We also notice that MC can outperform MLE occasionally (e.g., in DSV-Cifar, FL-Dota2). 
However, MLE generally achieves higher accuracy (lower $\epsilon$) than MC under identical convergence criteria, due to its closed-form integral expression of SV, enabling more deterministic computation with high precision (e.g., via Gaussian quadrature).

By comparing bars of different colors, we note that integrating TC reduces the SV computation complexity ($N_{uc}$) and thus reduces the total time cost ($N_{uc}\times T_{uc}$), in most cases. 
The effects in RI-Iris is insignificant, since this task only has four players and all algorithms can quickly generate exact SVs without further efficiency optimization. 
TC can reduce $N_{uc}$ because it avoids computing the marginal contribution of new players who join a coalition unnecessarily, particularly when that coalition's utility $U(\mathcal{S})$ is close to the overall utility achieved by the grand coalition $U(\mathcal{N})$, for example $U(\mathcal{S})>90\% U(\mathcal{N})$. 
However, if few coalitions meet such a condition, TC would pose trivial positive impacts on SV computation efficiency.
To intensify TC's effects in reducing cost, setting a looser condition, e.g., $U(\mathcal{S})>80\% U(\mathcal{N})$, is a potential solution, but this may enlarge the approximation error. Therefore, a thorough tuning of the truncation condition is necessitated when adopting TC. 

By comparing \Cref{fig:efficiency_results}(a) and
\Cref{fig:efficiency_results}(b) vertically, we can tell that GA and TSS techniques, though they may contribute negatively to reducing $N_{uc}$, are very helpful in reducing the average time cost of utility computation ($T_{uc}$). 
GA and TSS can reduce $T_{uc}$ because they use fewer training batches and fewer test data samples in marginal contribution estimation, respectively.  
However, these operations may enlarge the variance of approximate SVs in different iterations of utility computations, leading to an increasing number of utility computations ($N_{uc}$) for convergence.
When using GA or TSS together with TC, their negative impacts on computation complexity can be mitigated.

\textbf{Research Direction 1: Exploration on innovative hybrid SV computing algorithms.} 
Our evaluations show that the hybrid SV computing algorithms integrating multiple efficiency optimization techniques achieve better performance than the algorithms using only one of the integrated techniques in most cases. 
We highly recommend \textit{combining TC with a base SV computing algorithm} in all DA tasks and \textit{activating GA and TSS} when the utility computation (e.g., in FL) needs costly model training and evaluation. For the base algorithm, it is recommended to choose MLE (or MC). 
In summary, innovative hybrid SV computing algorithms with extensive experiments are anticipated for efficient SV computation in the literature.

\subsection{Approximation Error} \label{subsec: exp_impactsOfError}

This section explores the relationship among the approximation error, computation efficiency, and the effectiveness of SV in different DA tasks. We compare the performance of approximating SV using three widespread sampling techniques, \textit{random sampling}, \textit{stratified sampling}, and \textit{antithetic sampling}. 
The computation complexity is still measured by $N_{uc}$. 
The impact of approximation error on the effectiveness of SV is quantified by the score $\sum_{i=1}^n (\frac{\hat{\phi}_i}{\sum_{i=1}^n\hat{\phi}_i} - \frac{\phi_i}{\sum_{i=1}^n \phi_i})$. 
We also report $\Delta \hat{\phi}$, the stability of approximate SVs defined in the previous section, as a reference. 


\begin{figure}[t]
    \centering
    \begin{minipage}{0.75\columnwidth}
        \centering \includegraphics[width=\columnwidth]{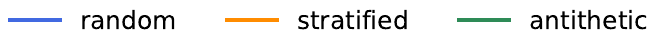}
    \end{minipage}
    \\
    
    \hspace{-15pt}
    \begin{minipage}{0.95\columnwidth}
    \mbox{
        \includegraphics[width=0.215\columnwidth, height=0.2\columnwidth]{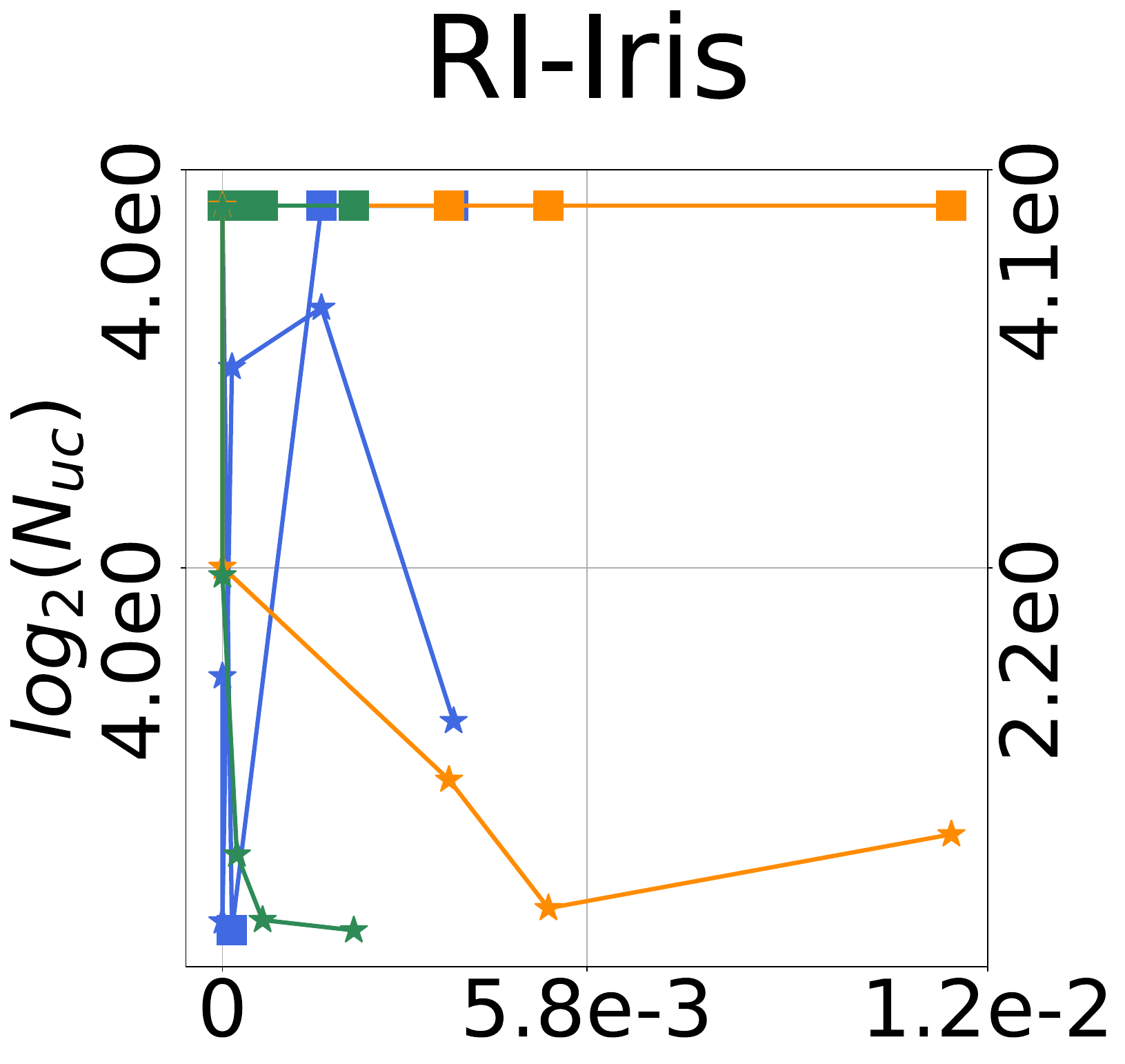}
        \includegraphics[width=0.2\columnwidth, height=0.2\columnwidth]{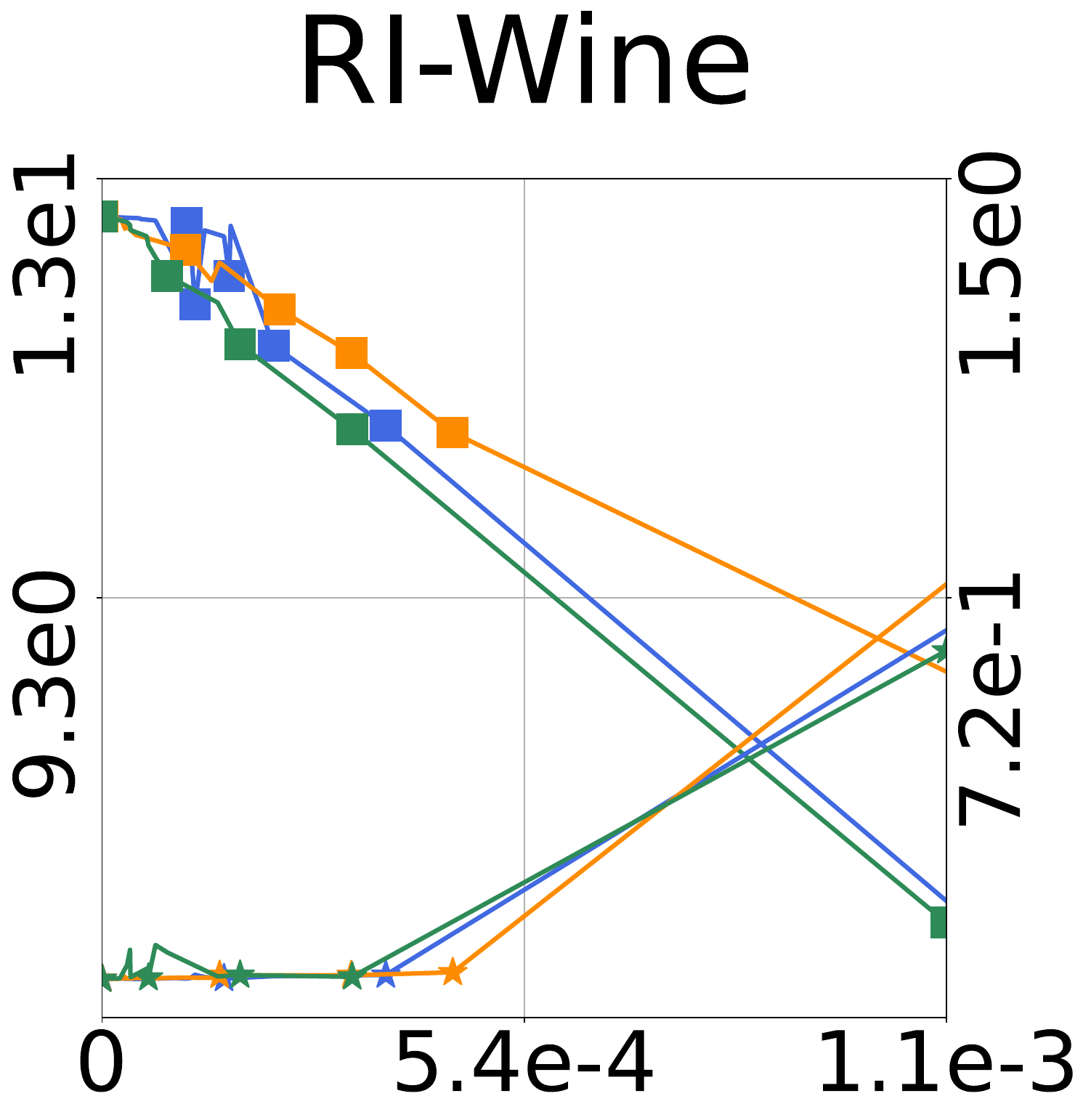}
        \includegraphics[width=0.2\columnwidth, height=0.2\columnwidth]{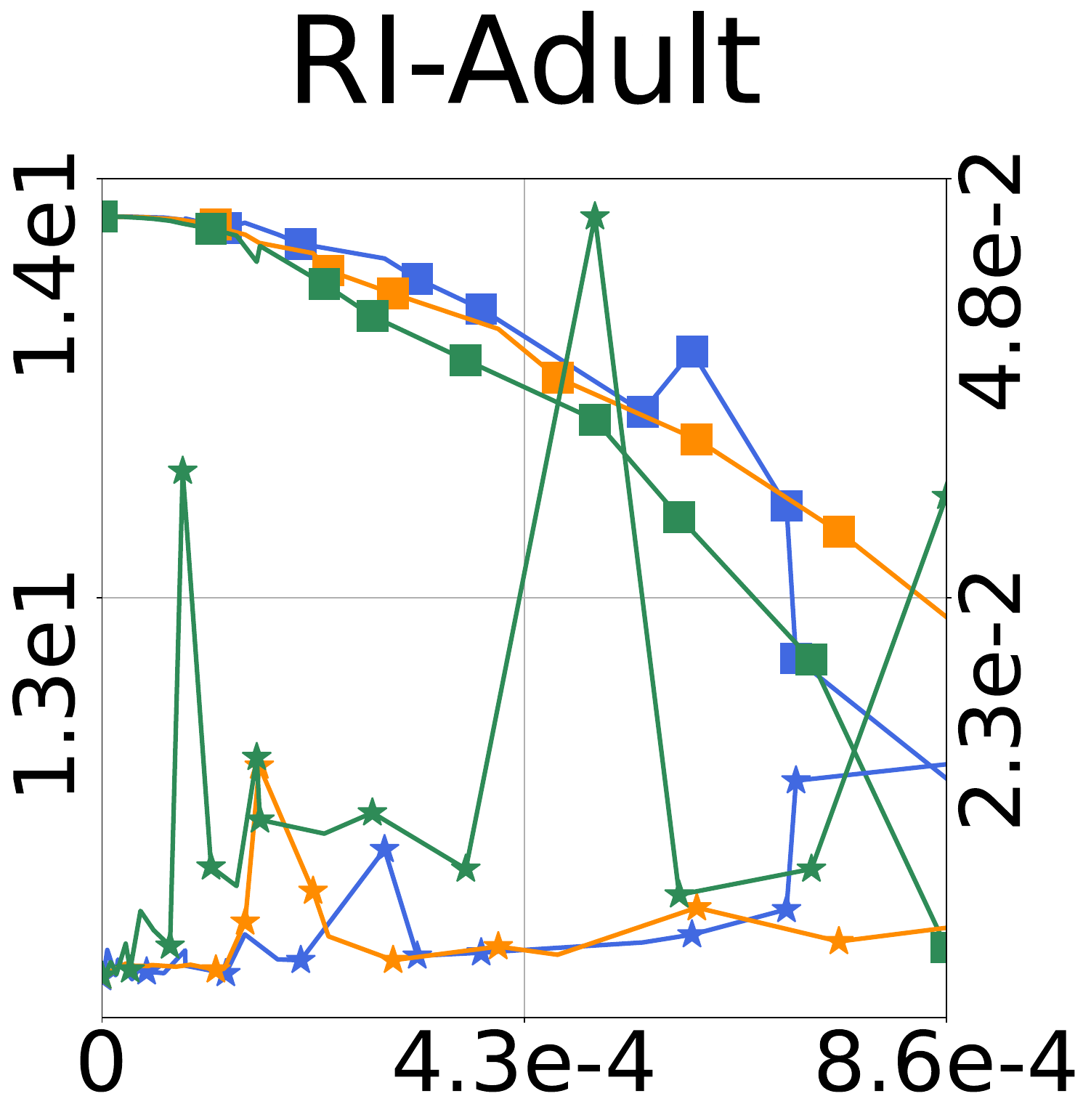}
        \includegraphics[width=0.2\columnwidth, height=0.2\columnwidth]{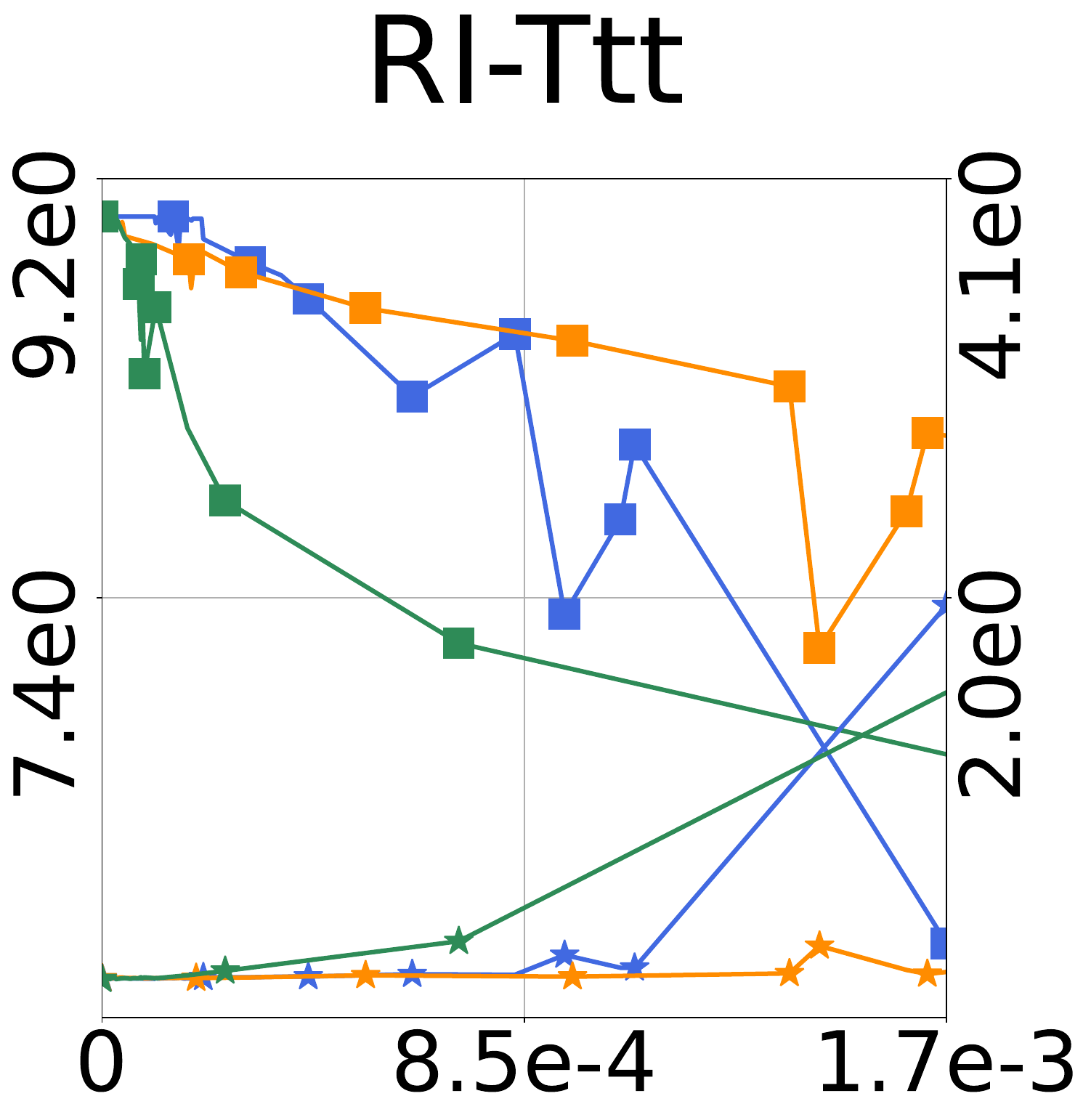}
        \includegraphics[width=0.215\columnwidth, height=0.2\columnwidth]{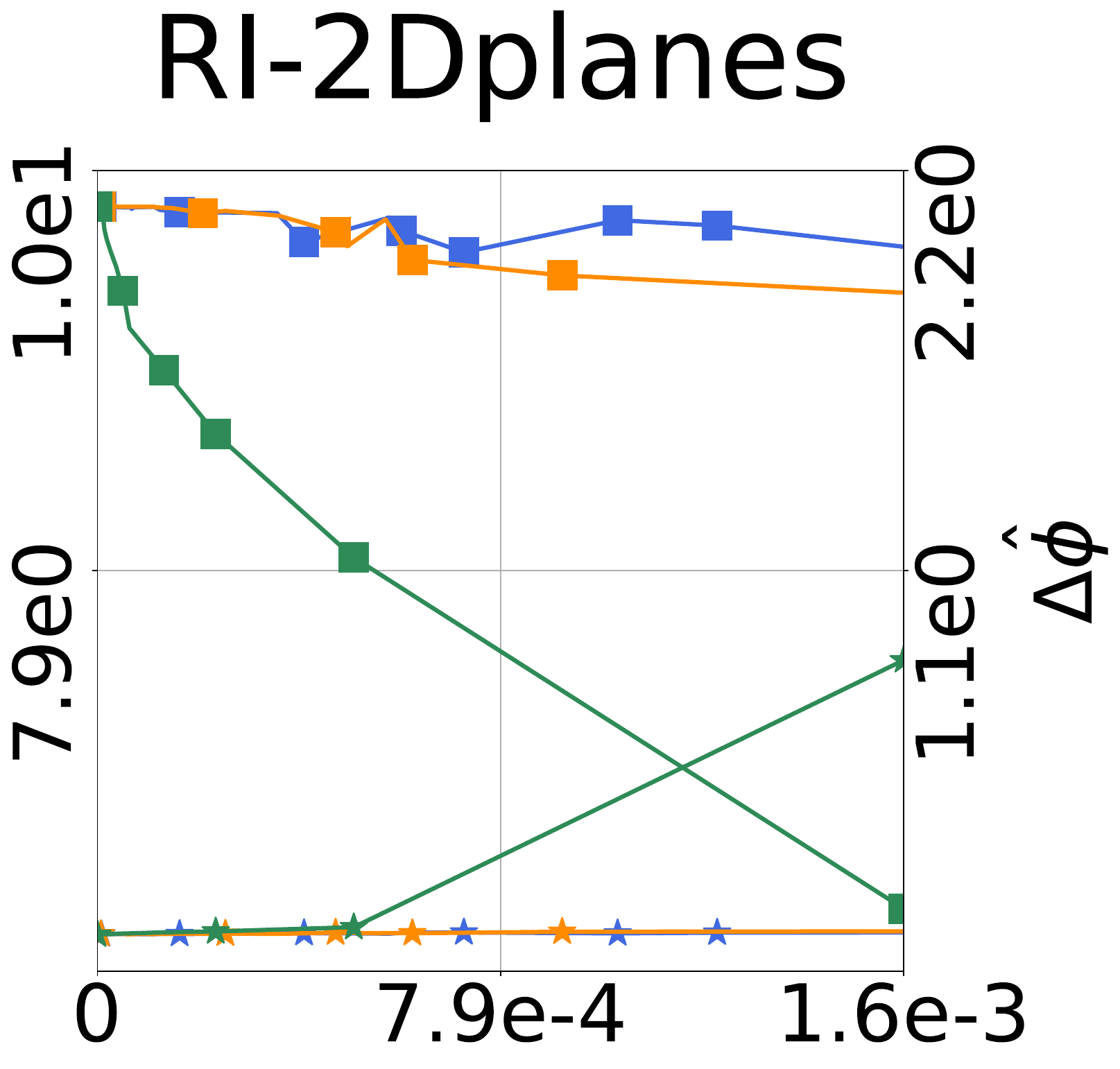}
    }
    \\
    \mbox{
        
        \includegraphics[width=0.215\columnwidth, height=0.2\columnwidth]{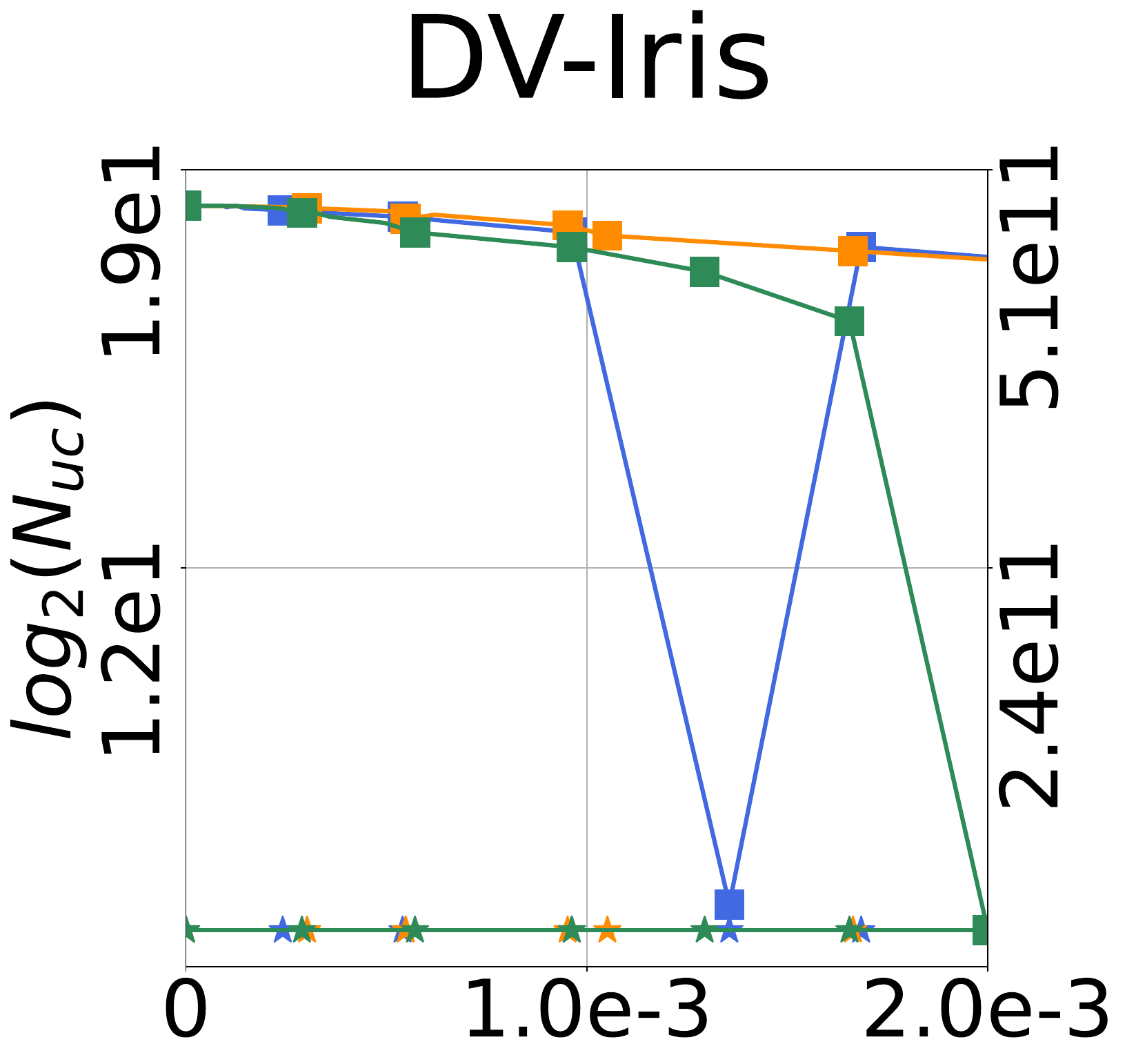}
        \includegraphics[width=0.2\columnwidth, height=0.2\columnwidth]{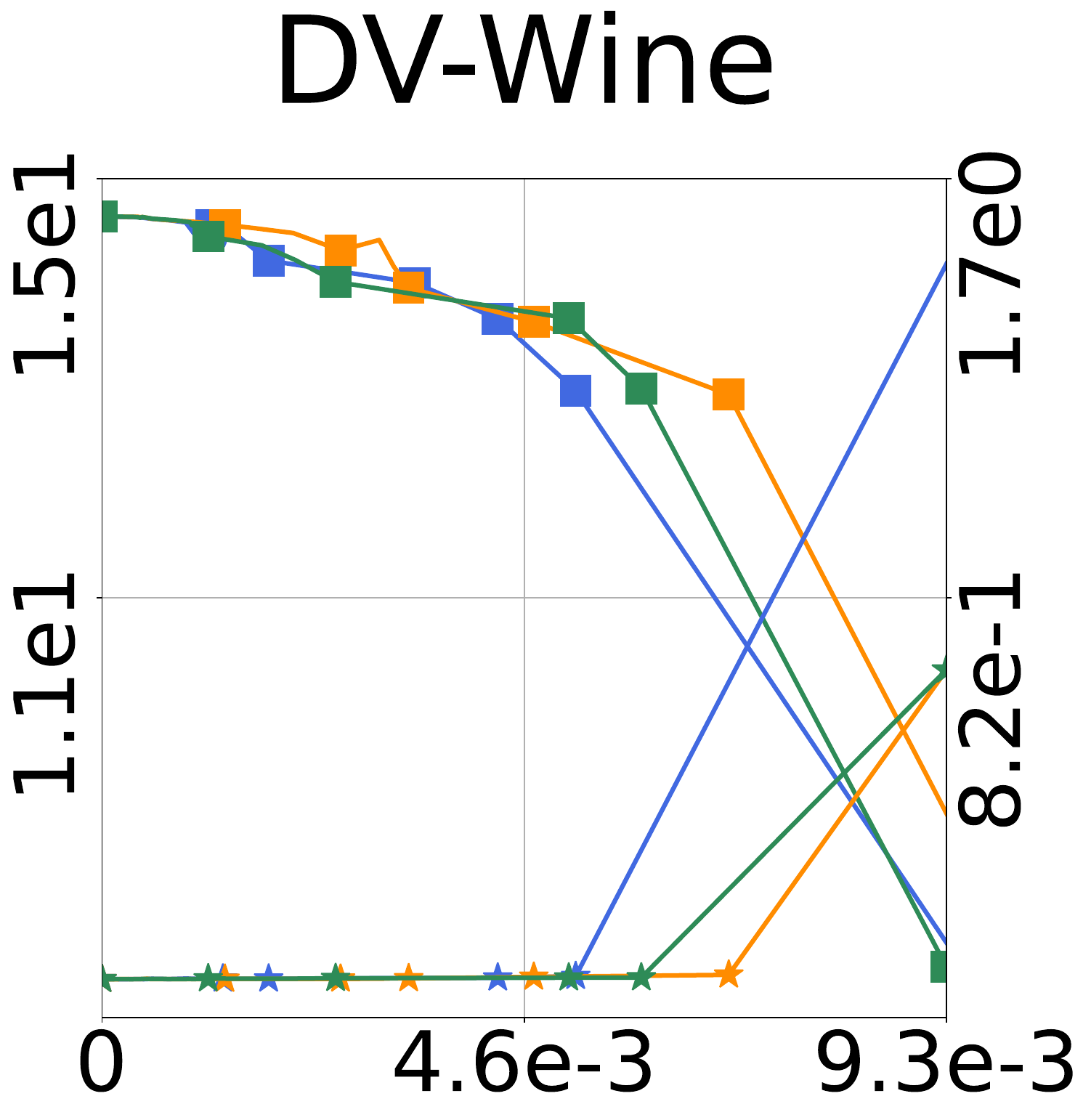}
        \includegraphics[width=0.2\columnwidth, height=0.2\columnwidth]{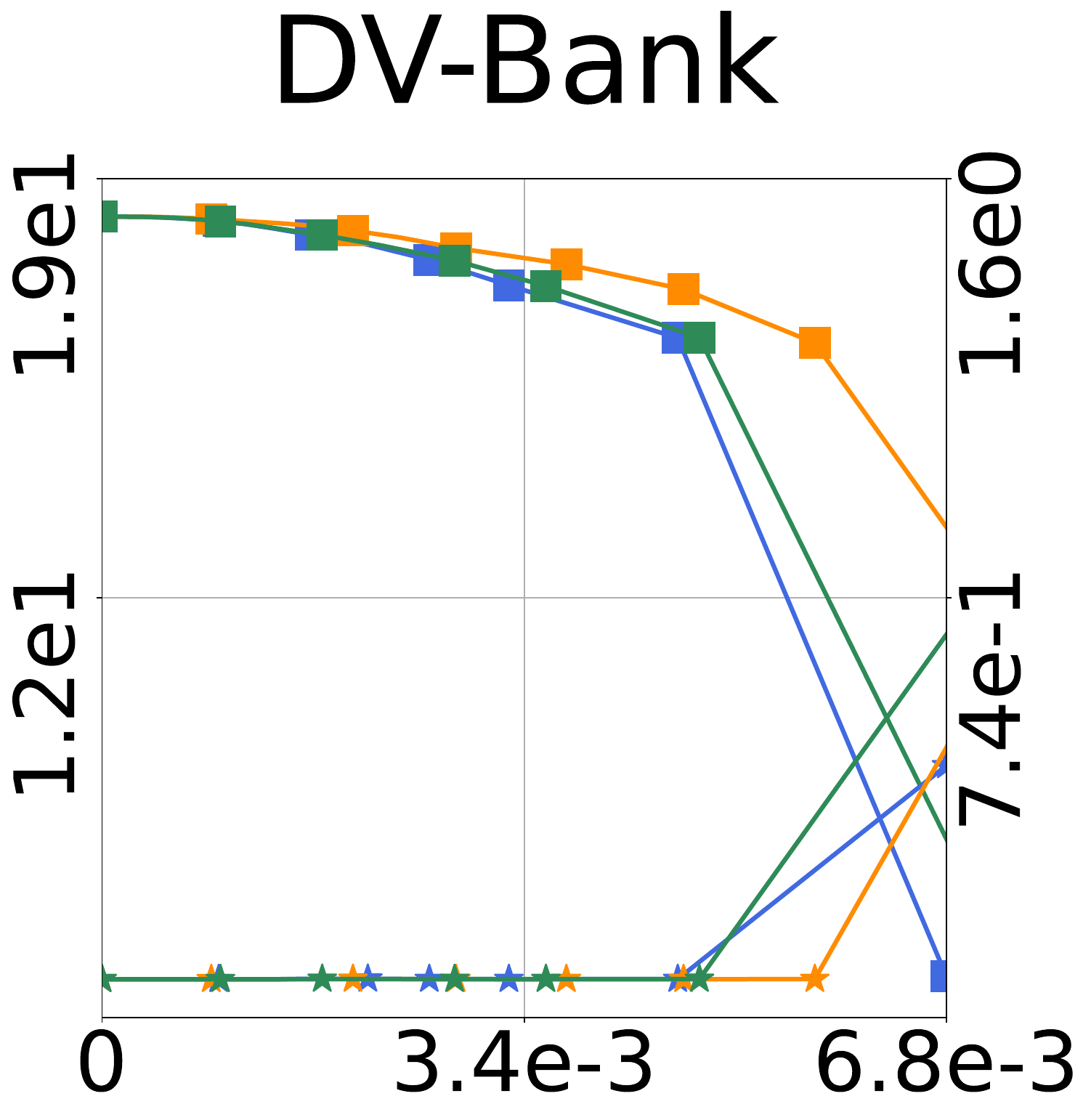}
        \includegraphics[width=0.2\columnwidth, height=0.2\columnwidth]{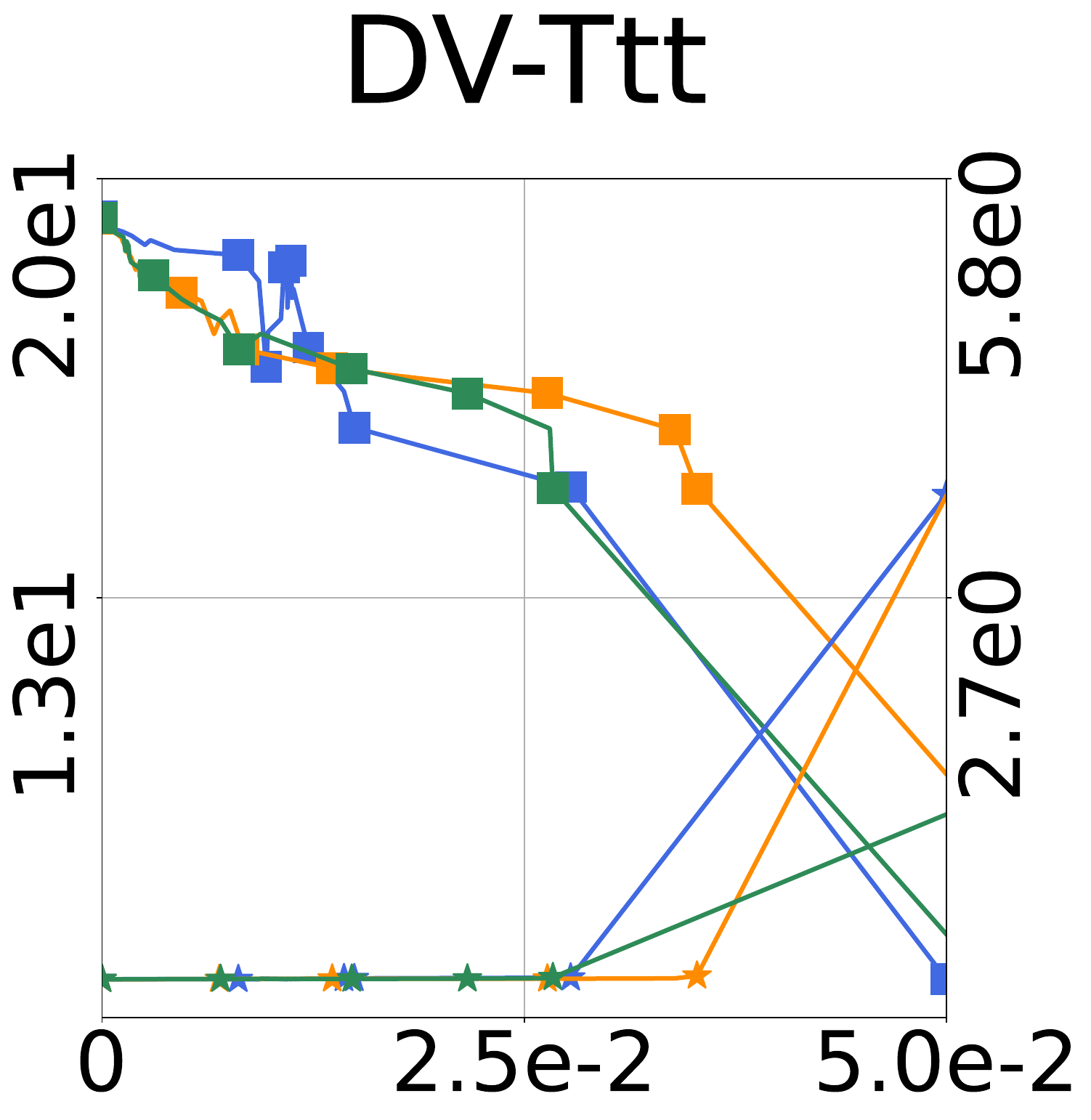}
        \includegraphics[width=0.215\columnwidth, height=0.2\columnwidth]{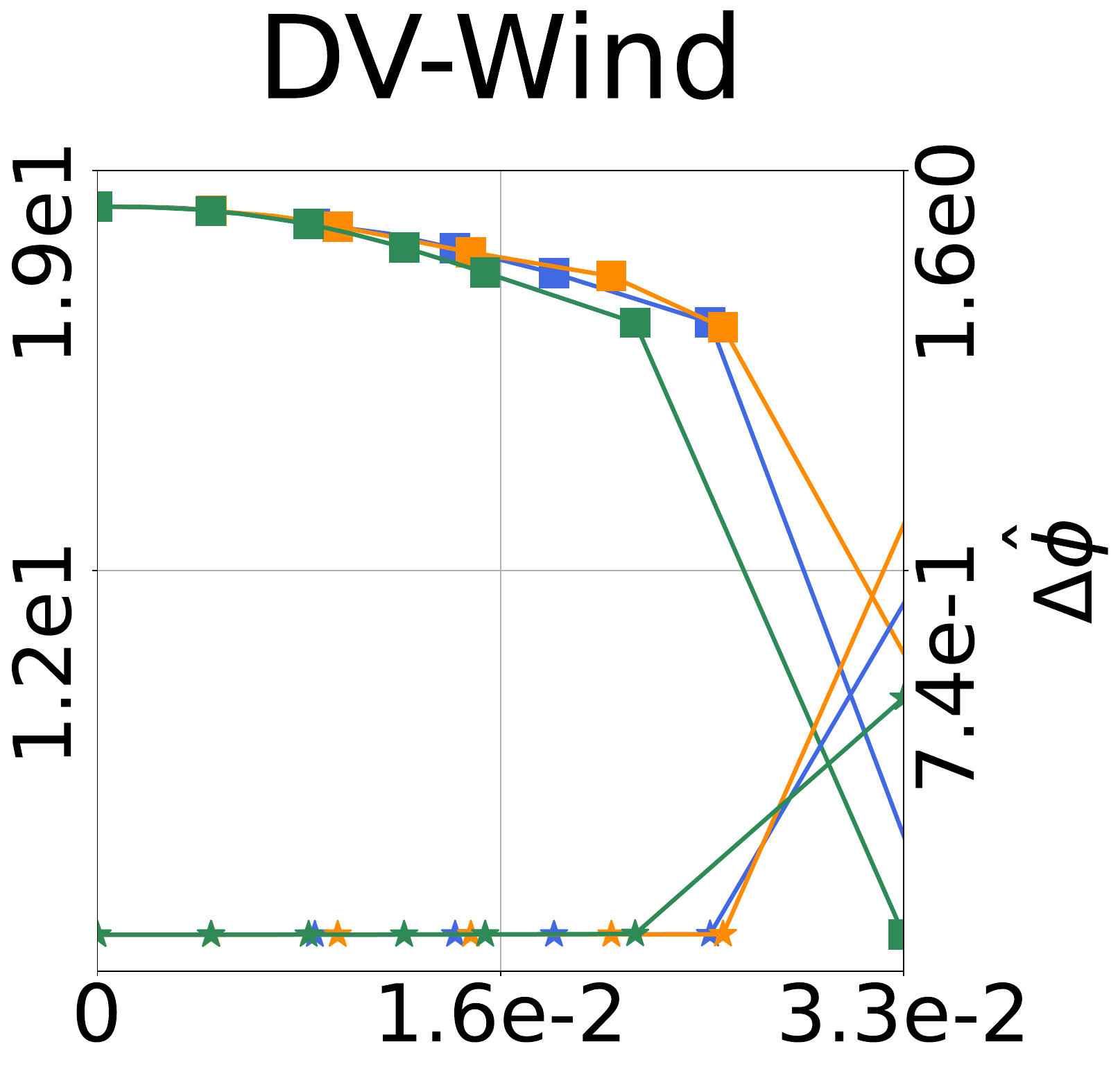}
    }
    \\
    \mbox{
        
        \includegraphics[width=0.215\columnwidth, height=0.2\columnwidth]{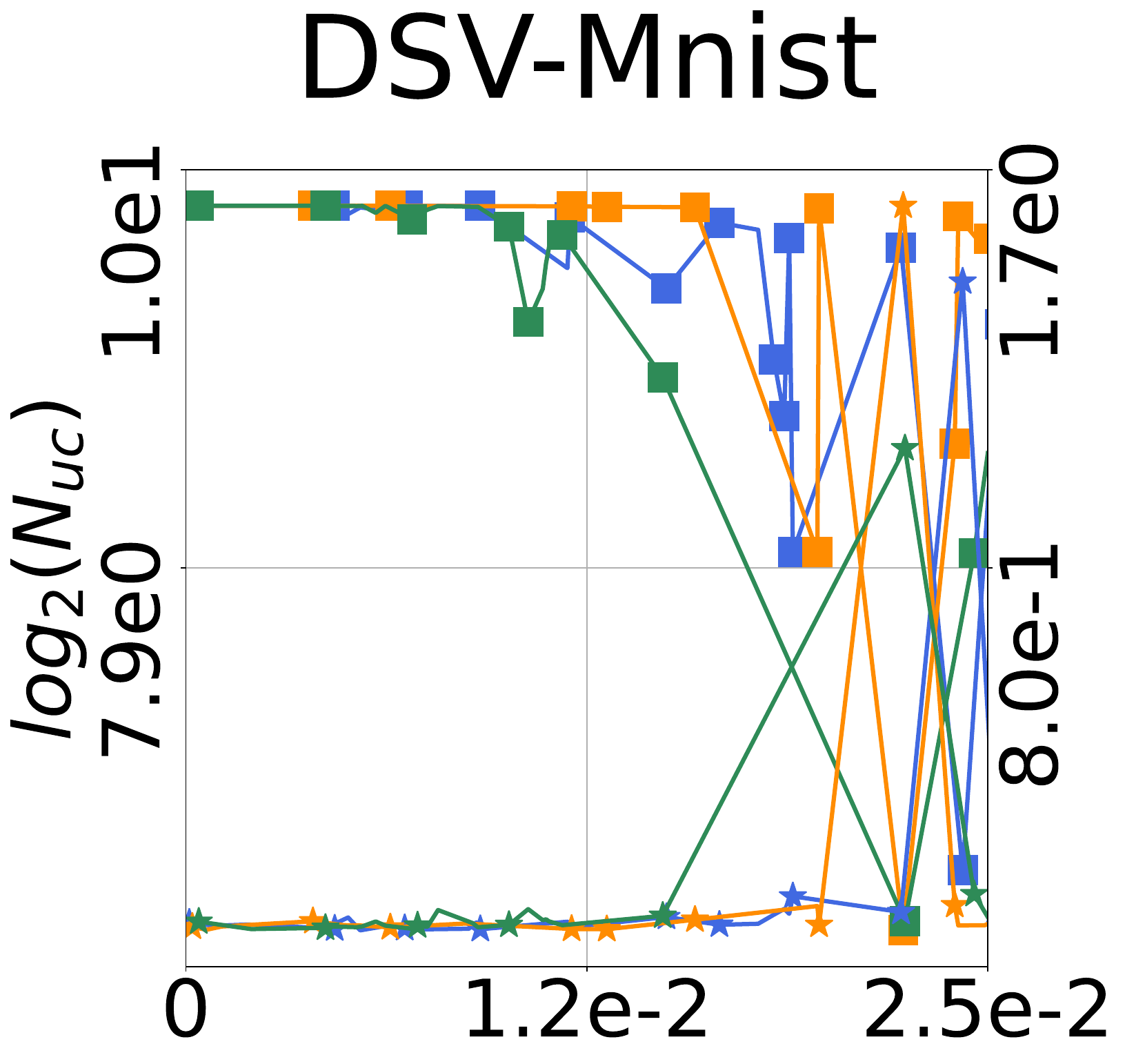}
        \includegraphics[width=0.2\columnwidth, height=0.2\columnwidth]{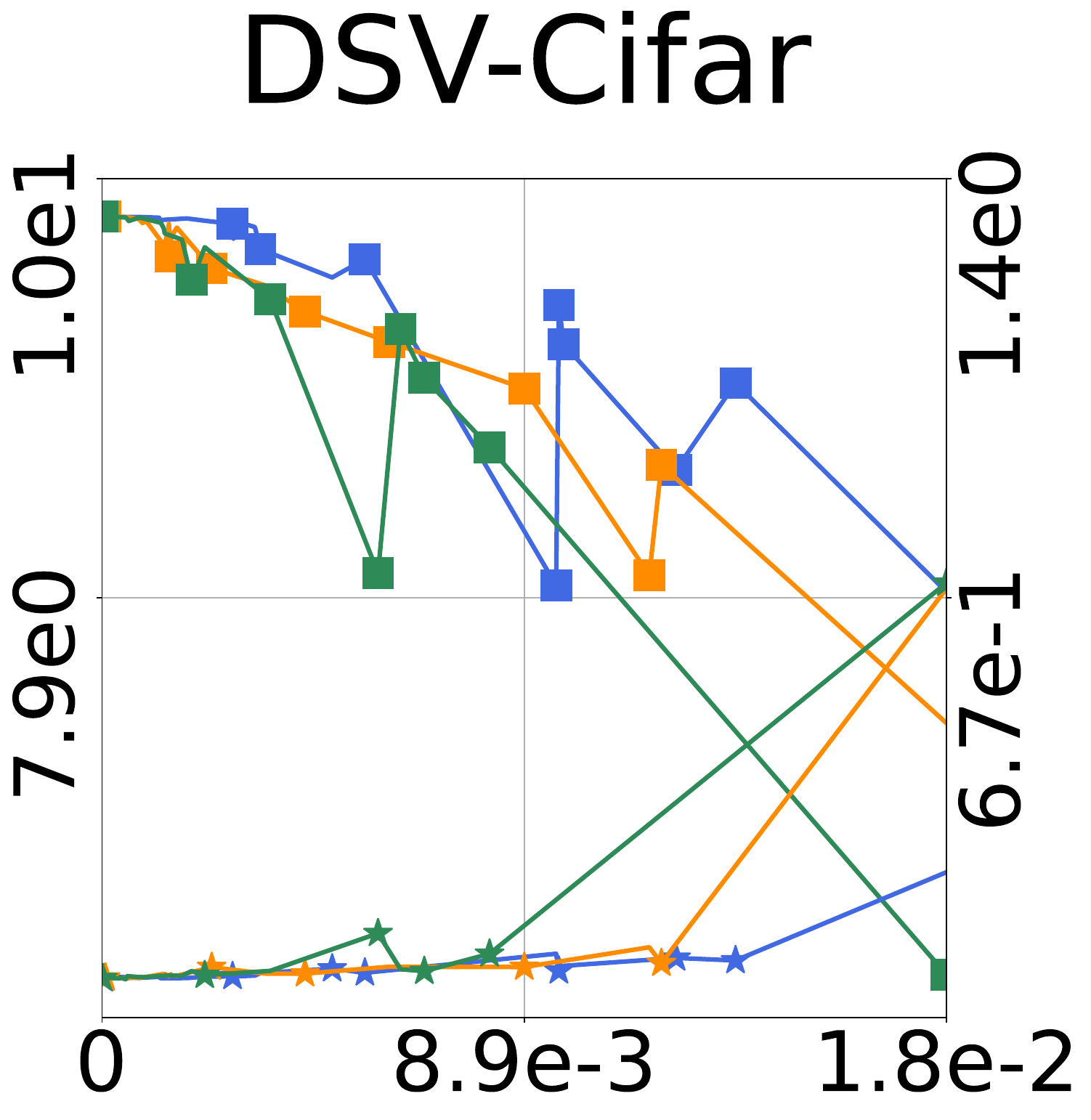}
        \includegraphics[width=0.2\columnwidth, height=0.2\columnwidth]{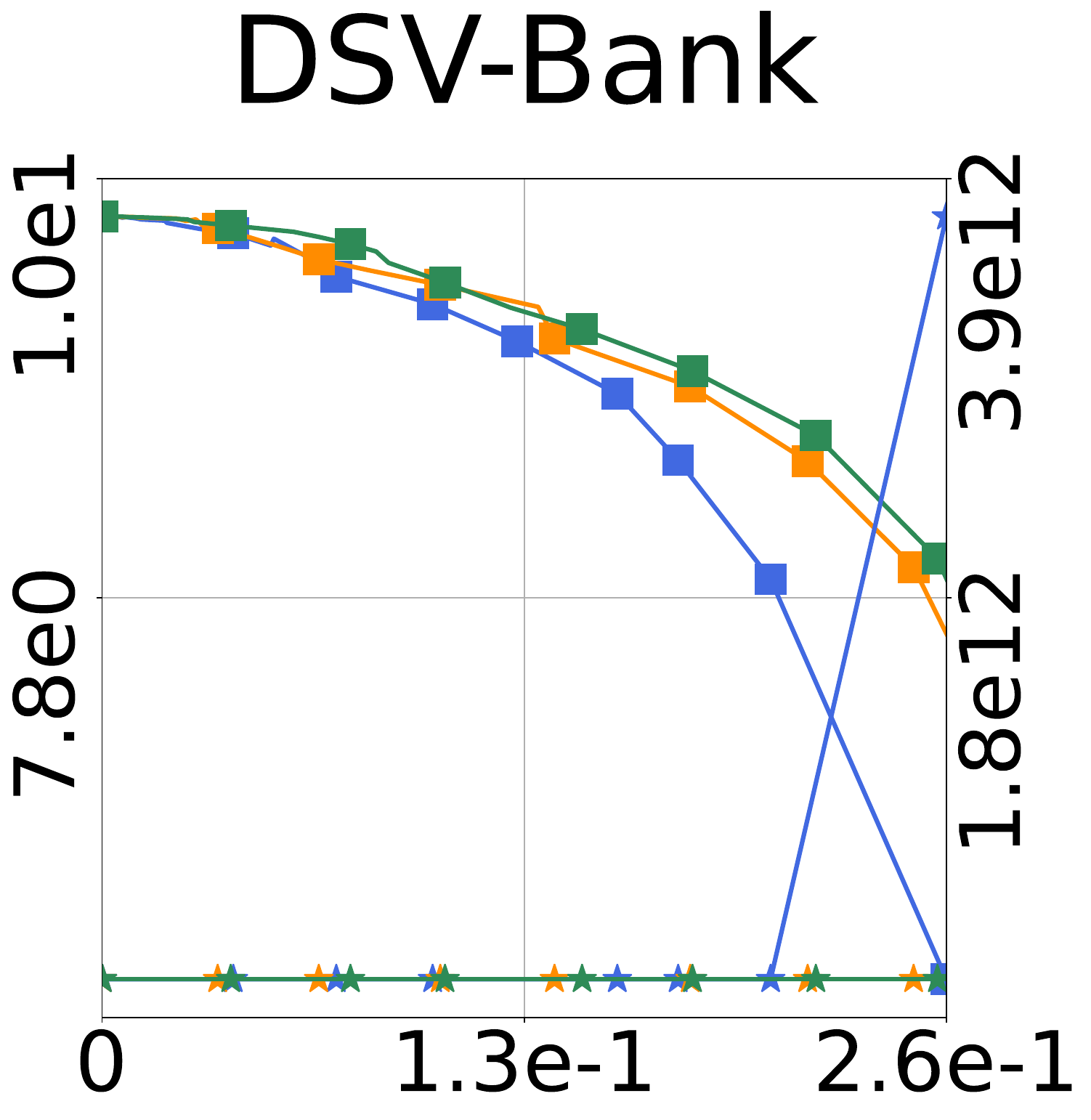}
        \includegraphics[width=0.2\columnwidth, height=0.2\columnwidth]{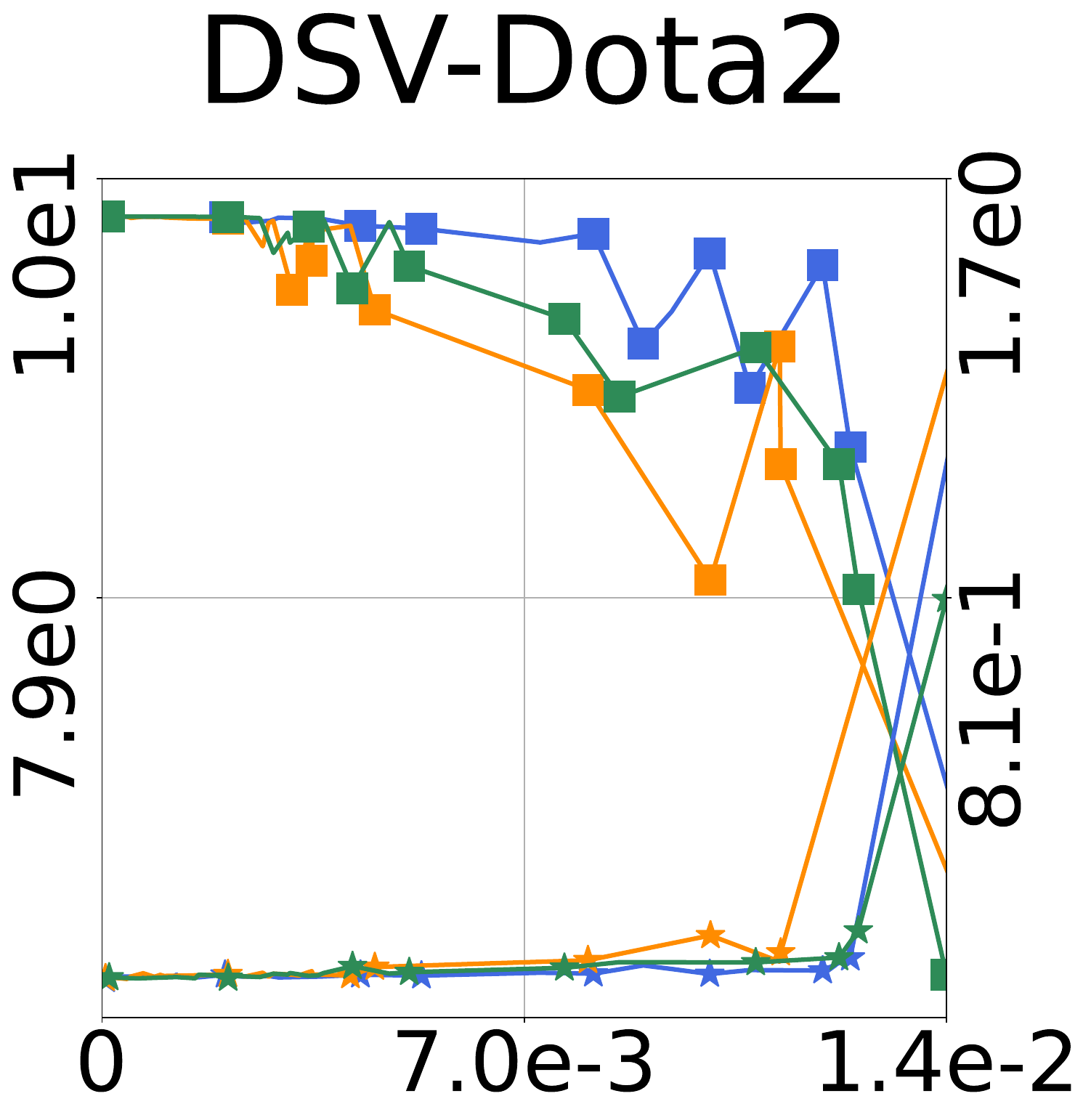}
        \includegraphics[width=0.215\columnwidth, height=0.2\columnwidth]{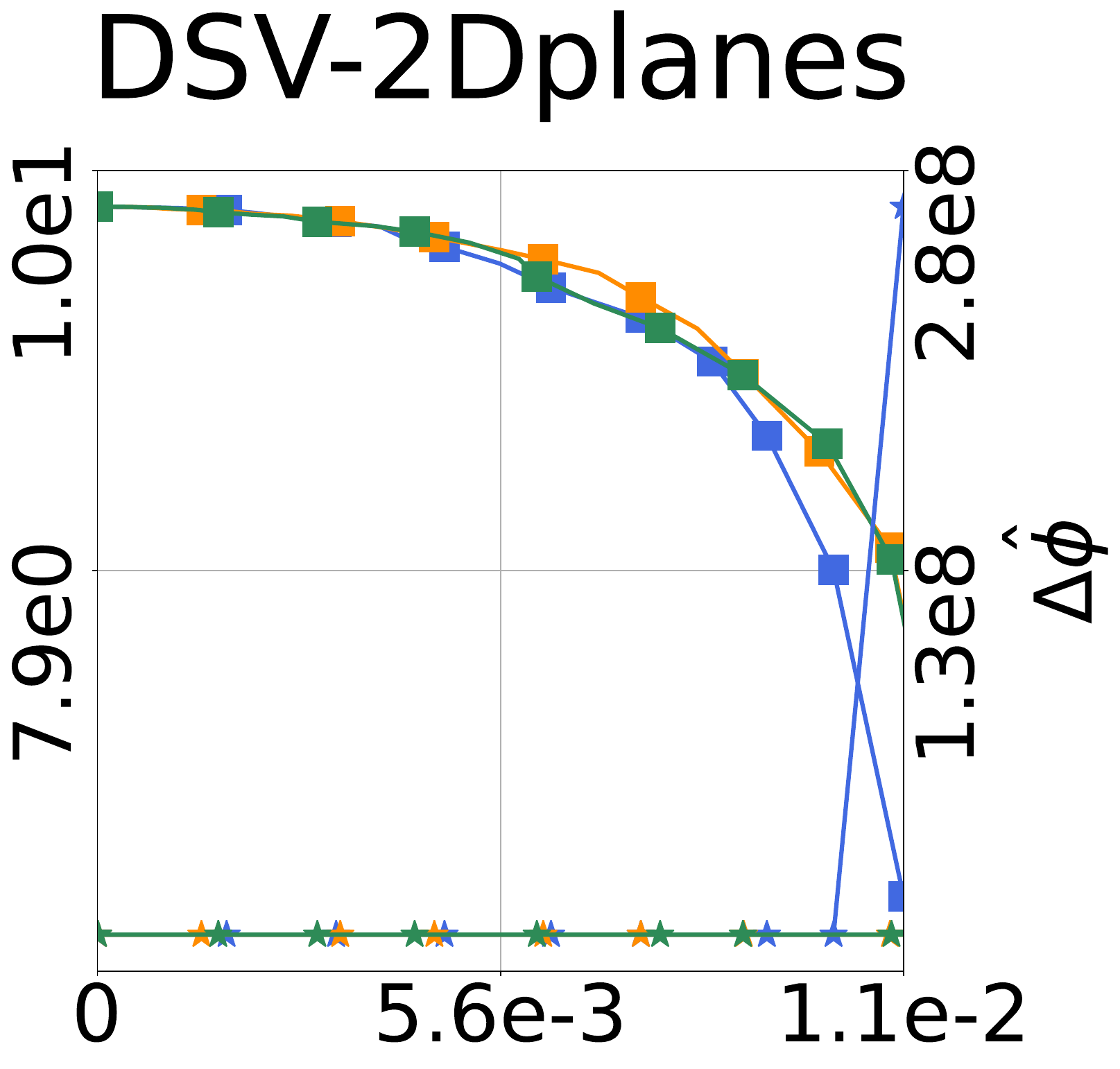}

    }  
    \\
    \subfigure[Left y-axis and square-tagged lines report $N_{uc}$. 
    Right y-axis and star-tagged lines report $\Delta\hat{\phi}$. 
    The lower the two values, the more efficient and stable the SV approximation. 
    ] {\label{fig:exp_EfficiencyAndApproximation:complexity}
        \includegraphics[width=0.215\columnwidth, height=0.2\columnwidth]{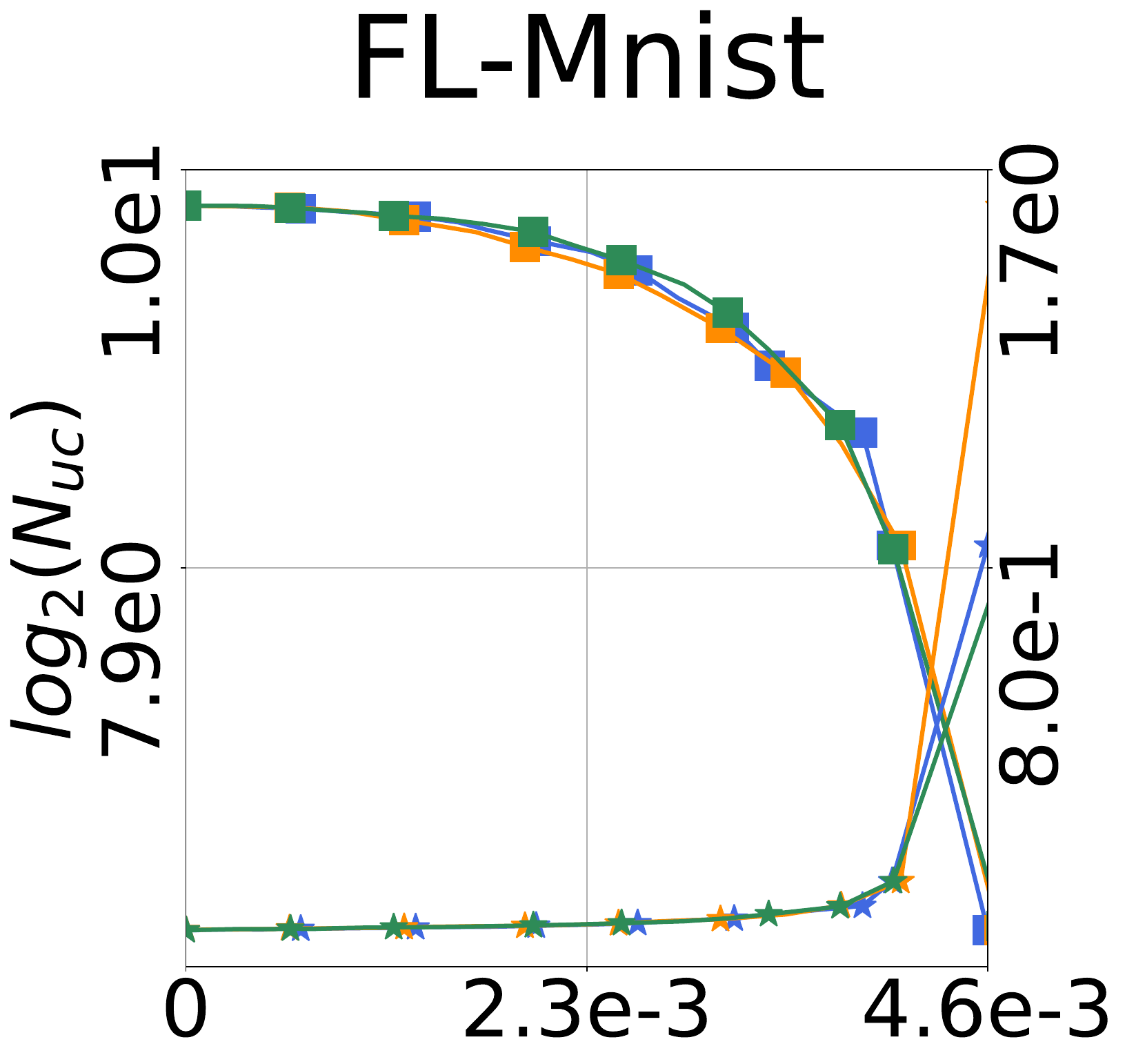}
        \includegraphics[width=0.2\columnwidth, height=0.2\columnwidth]{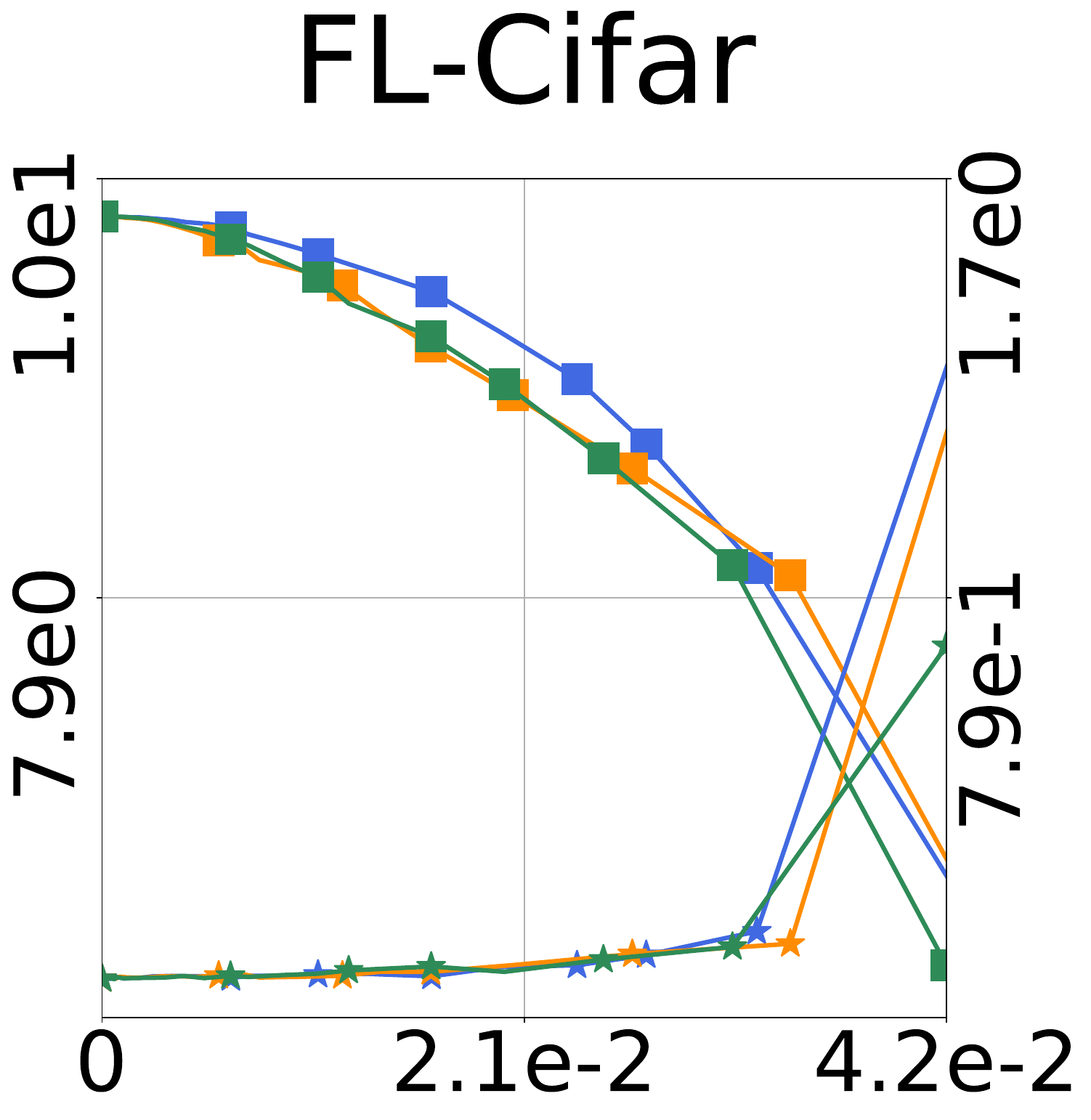}
        \includegraphics[width=0.2\columnwidth, height=0.2\columnwidth]{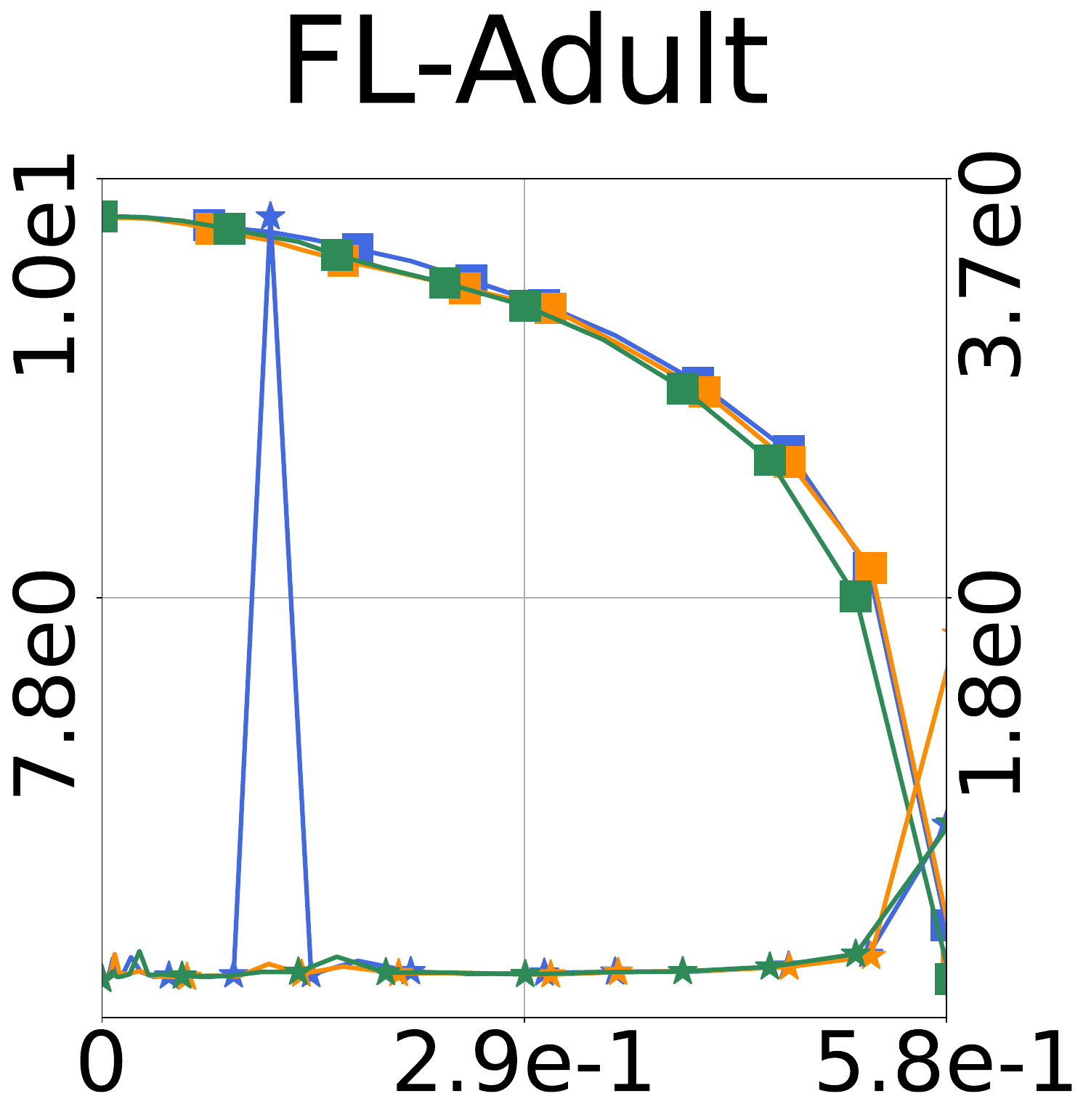}
        \includegraphics[width=0.2\columnwidth, height=0.2\columnwidth]{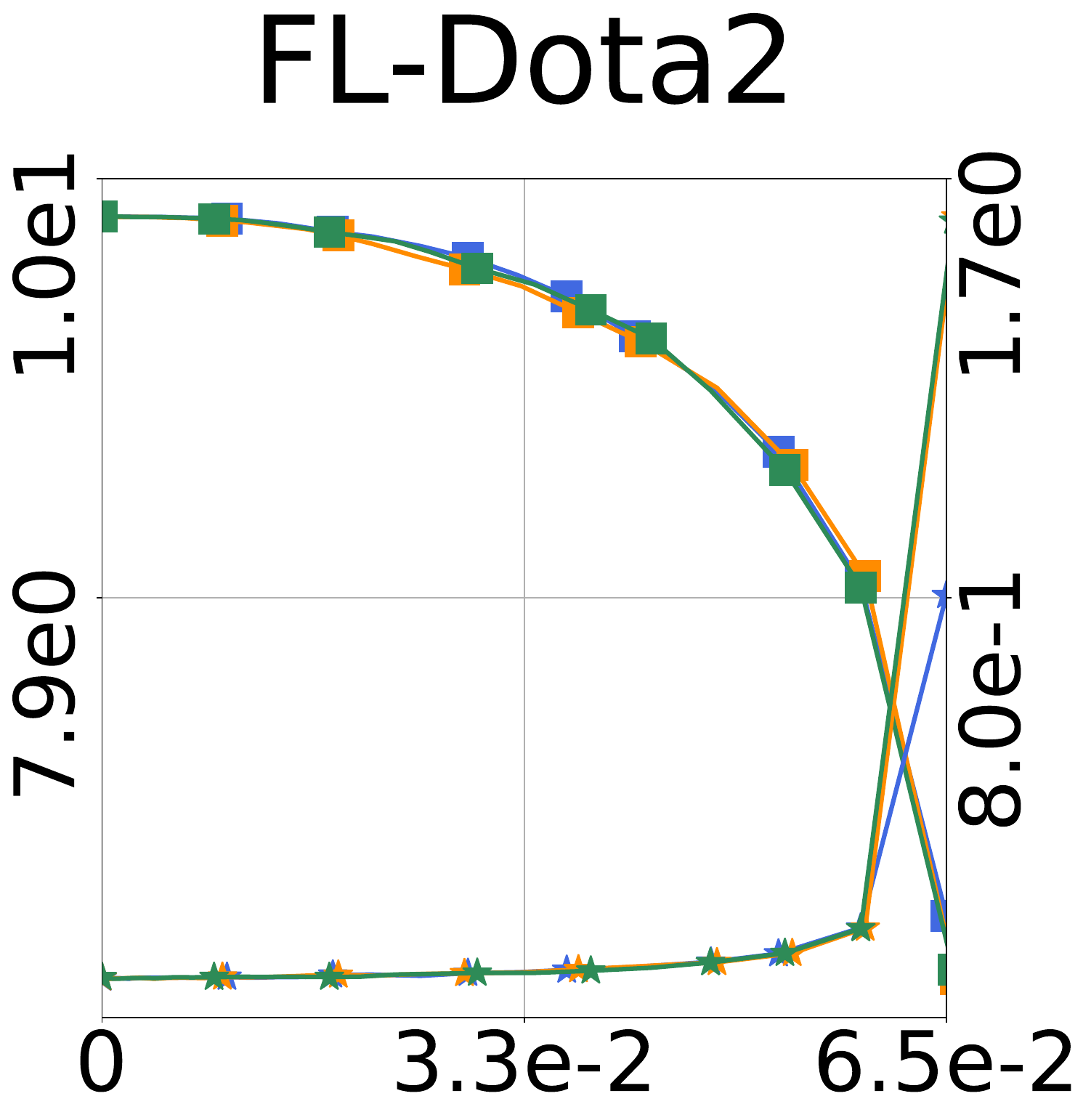}
        \includegraphics[width=0.215\columnwidth, height=0.2\columnwidth]{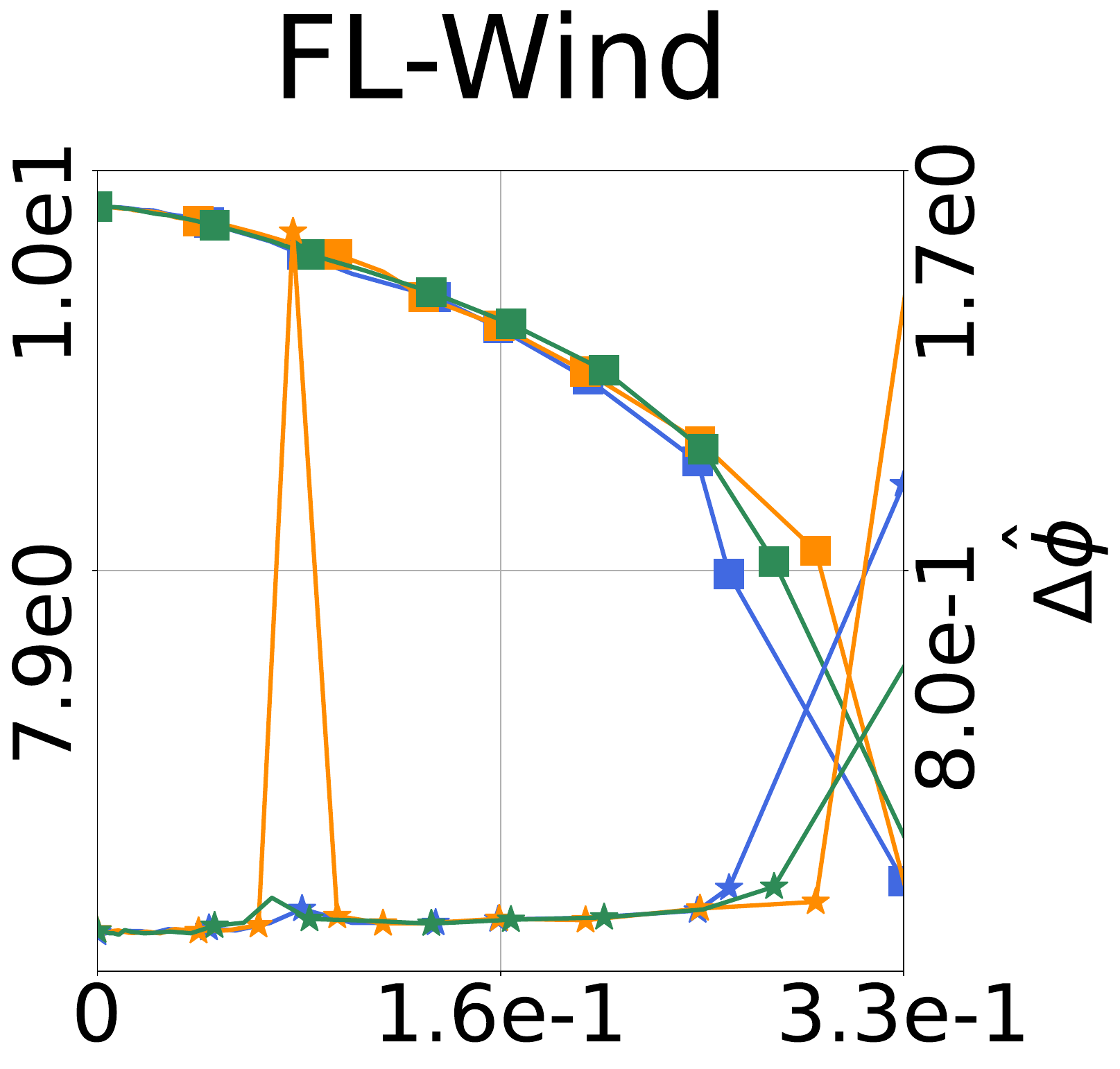}
    }  
    \end{minipage}
    \\
    \hspace{-15pt}
    \begin{minipage}{0.95\columnwidth}
    \mbox{
        \includegraphics[width=0.215\columnwidth, height=0.2\columnwidth]{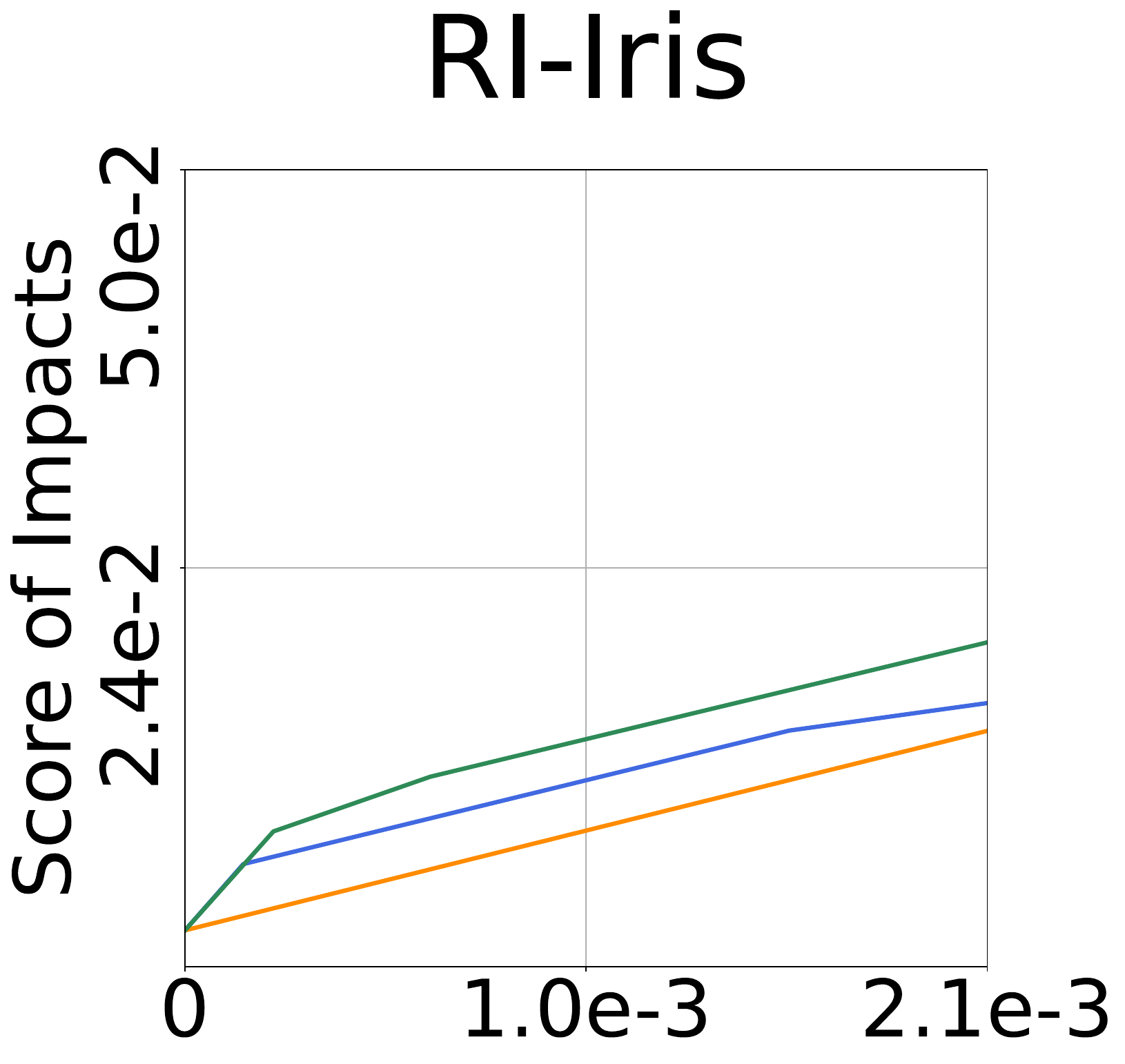}
        \includegraphics[width=0.2\columnwidth, height=0.2\columnwidth]{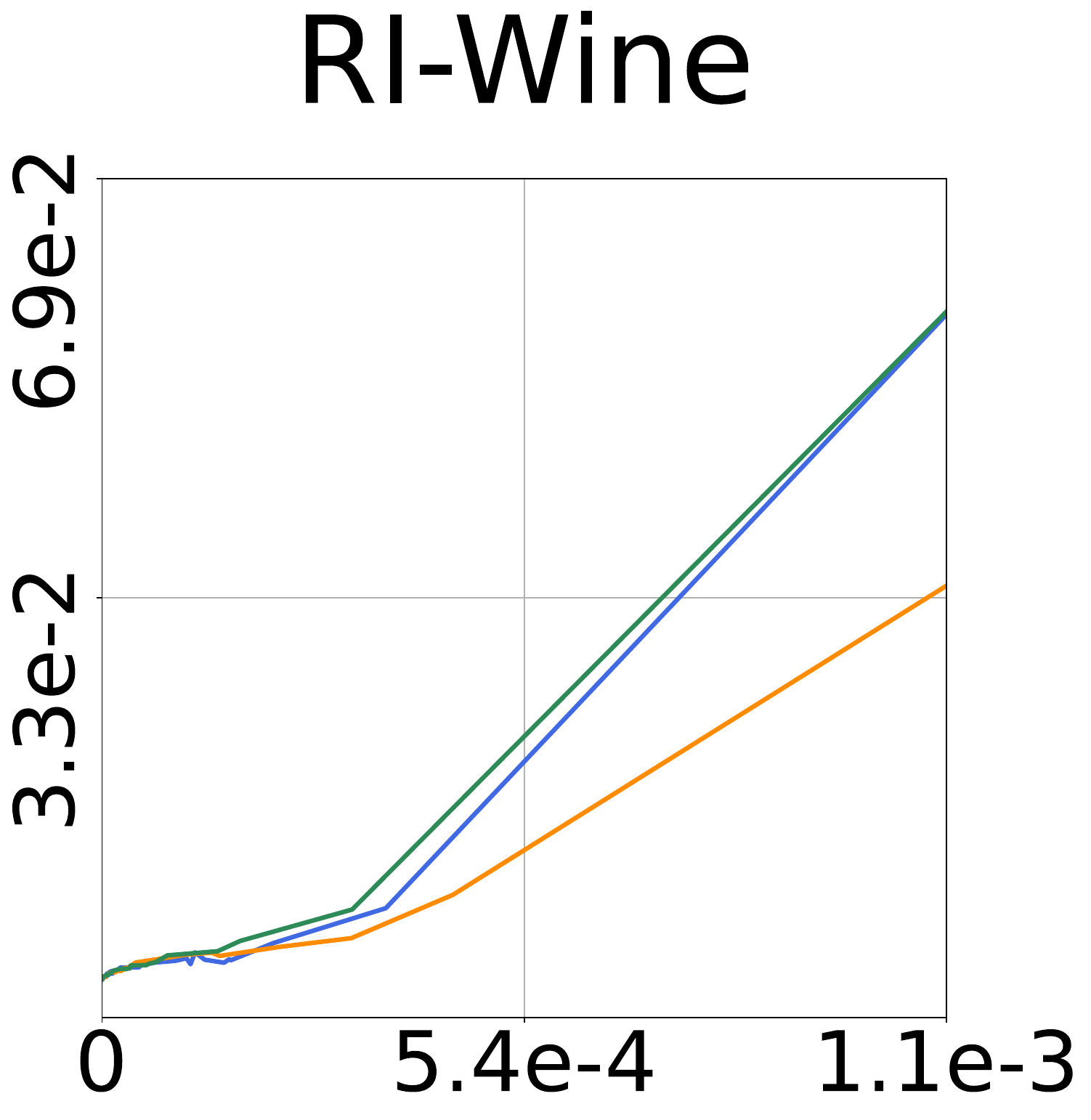}
        \includegraphics[width=0.2\columnwidth, height=0.2\columnwidth]{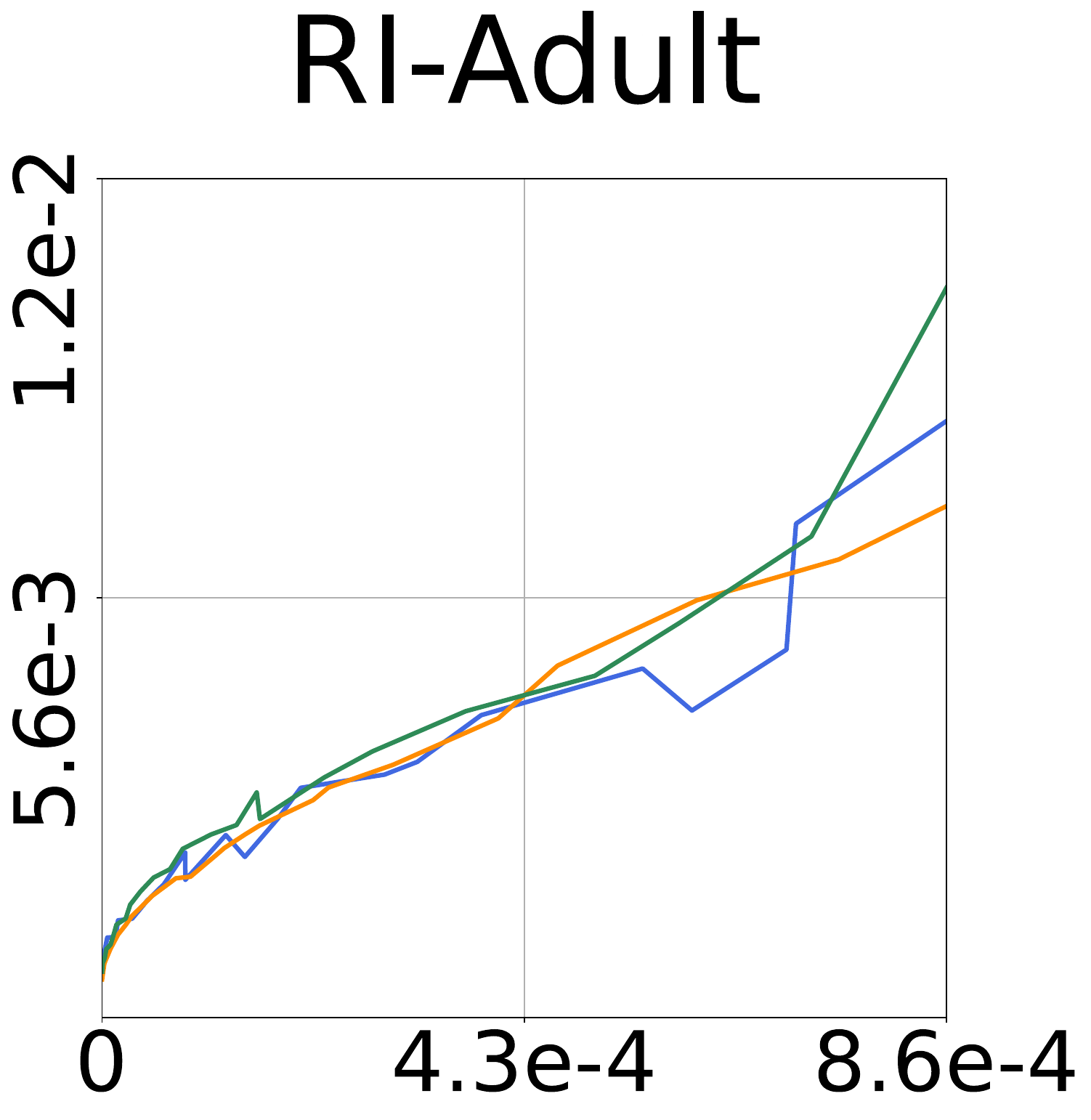}
        \includegraphics[width=0.2\columnwidth, height=0.2\columnwidth]{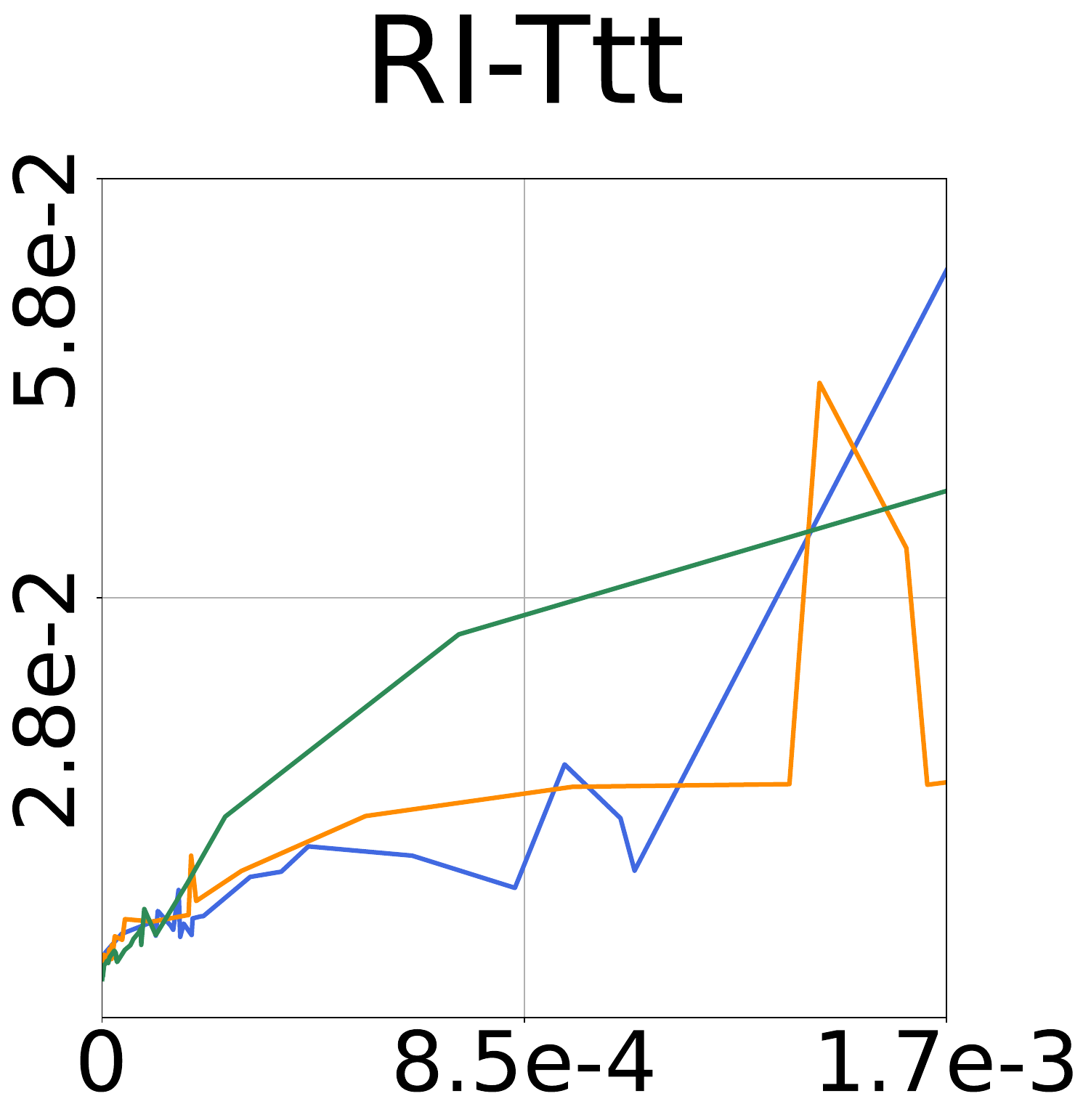}
        \includegraphics[width=0.2\columnwidth, height=0.2\columnwidth]{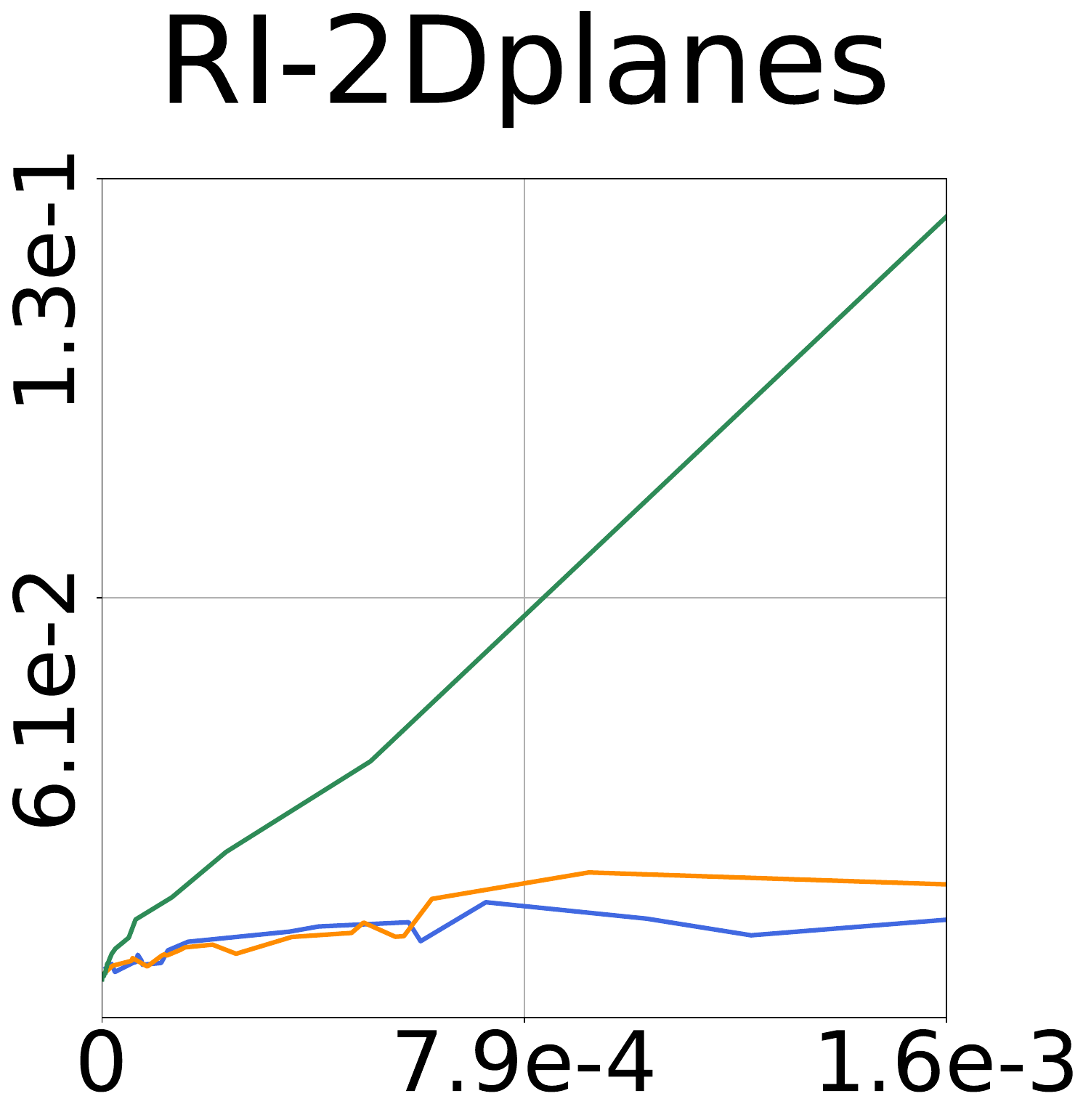}
        
    }
    \\
    \mbox{
        
        \includegraphics[width=0.215\columnwidth, height=0.2\columnwidth]{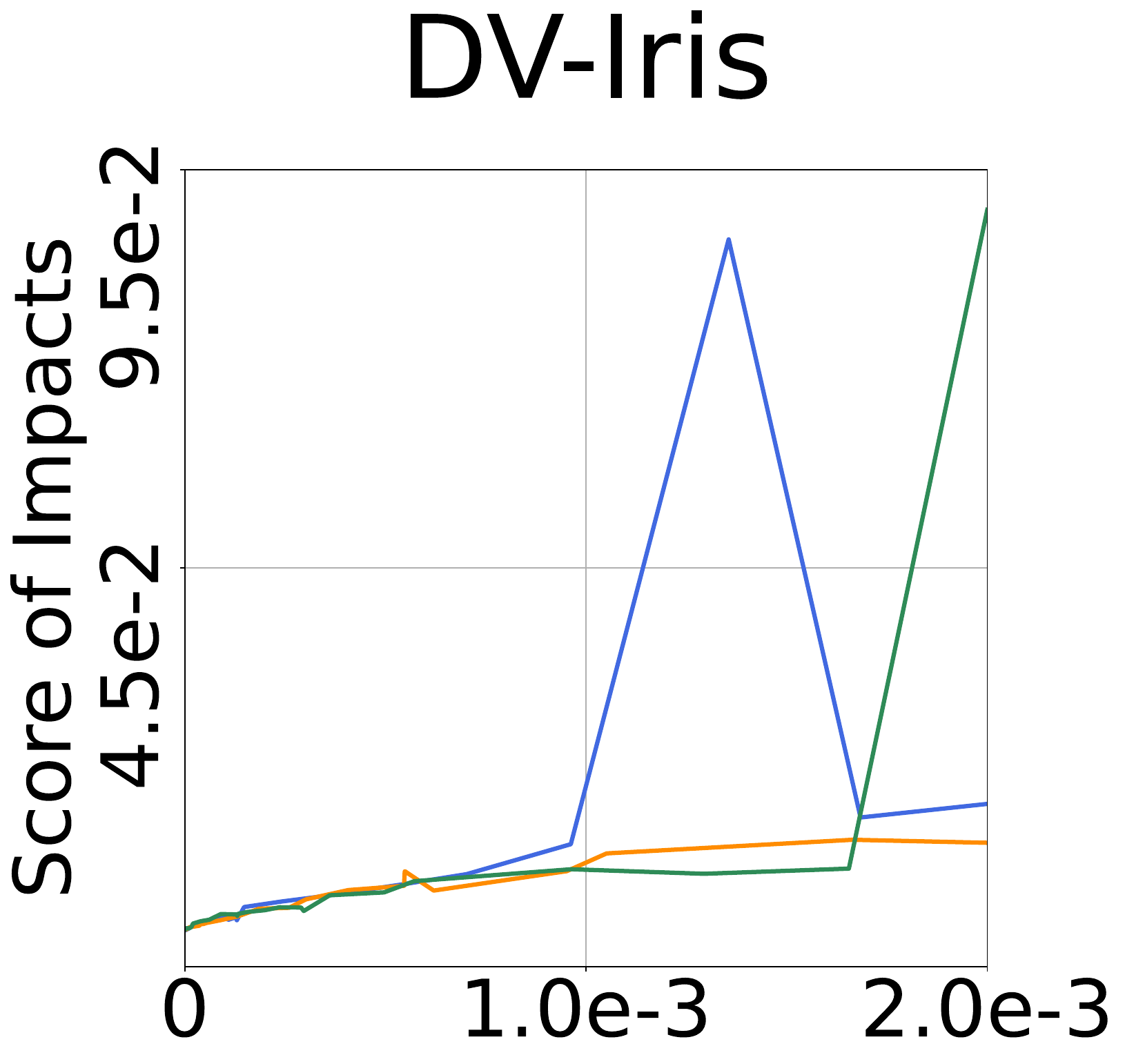}
        \includegraphics[width=0.2\columnwidth, height=0.2\columnwidth]{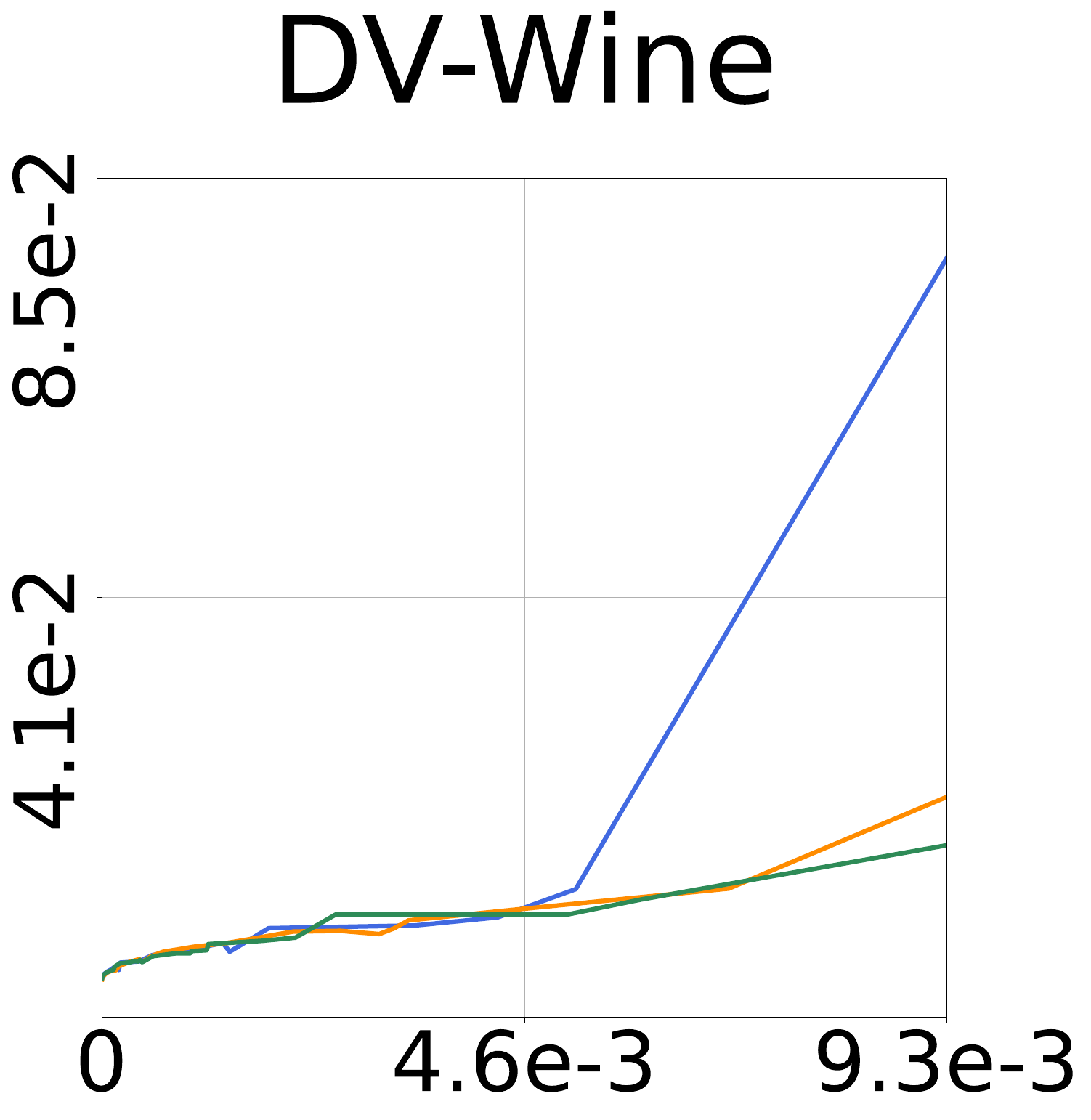}
        \includegraphics[width=0.2\columnwidth, height=0.2\columnwidth]{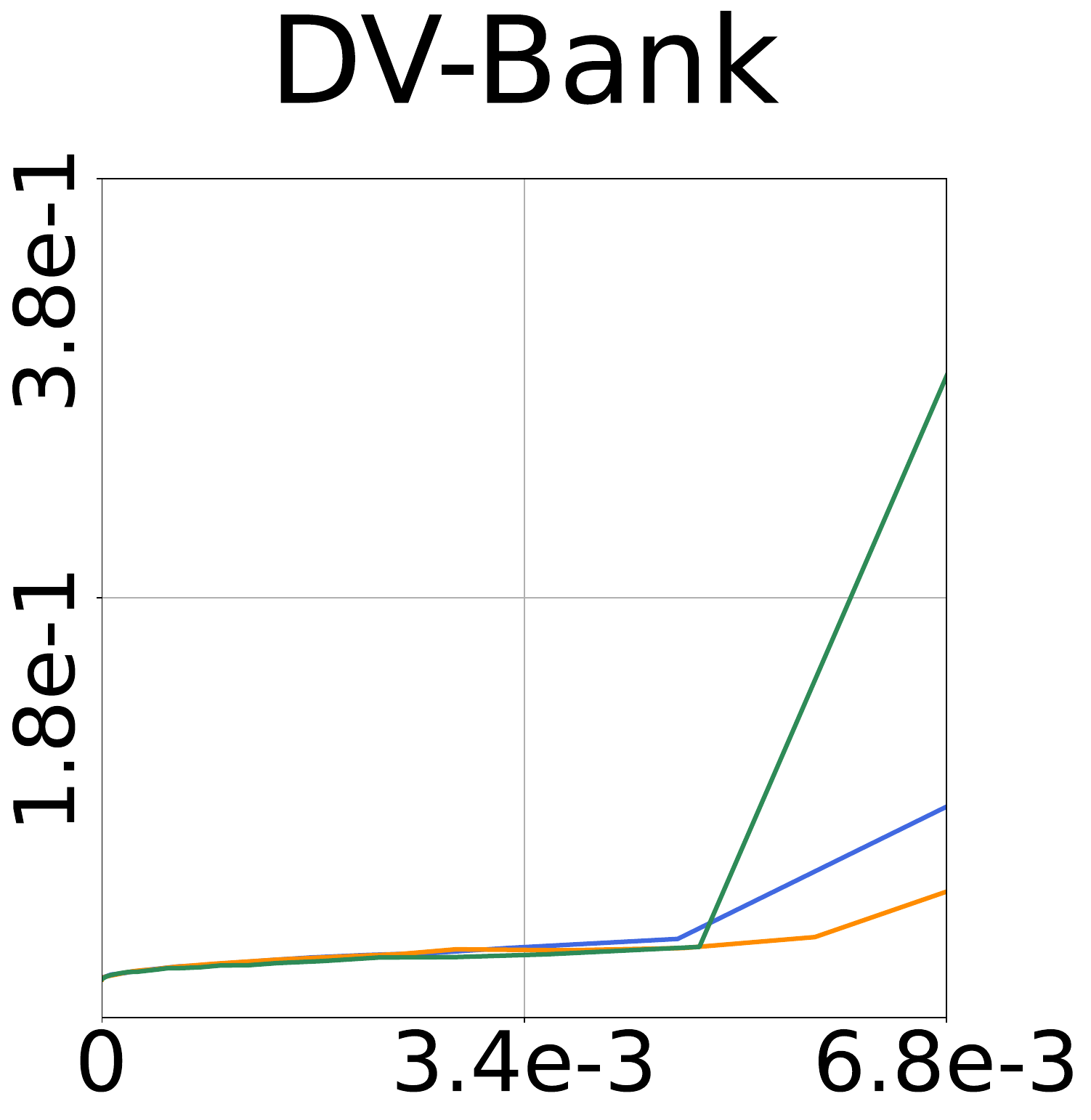}
        \includegraphics[width=0.2\columnwidth, height=0.2\columnwidth]{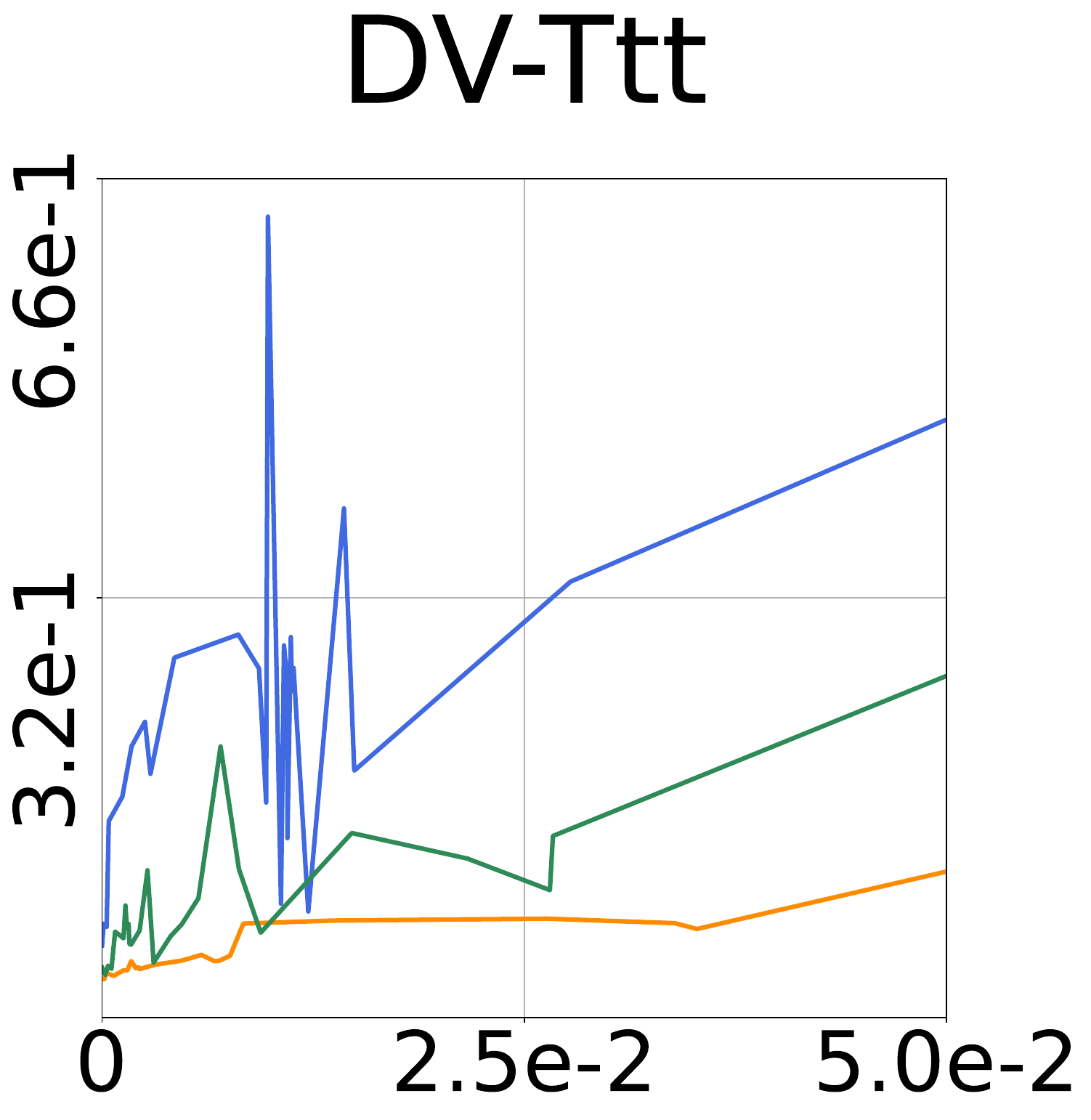}
        \includegraphics[width=0.2\columnwidth, height=0.2\columnwidth]{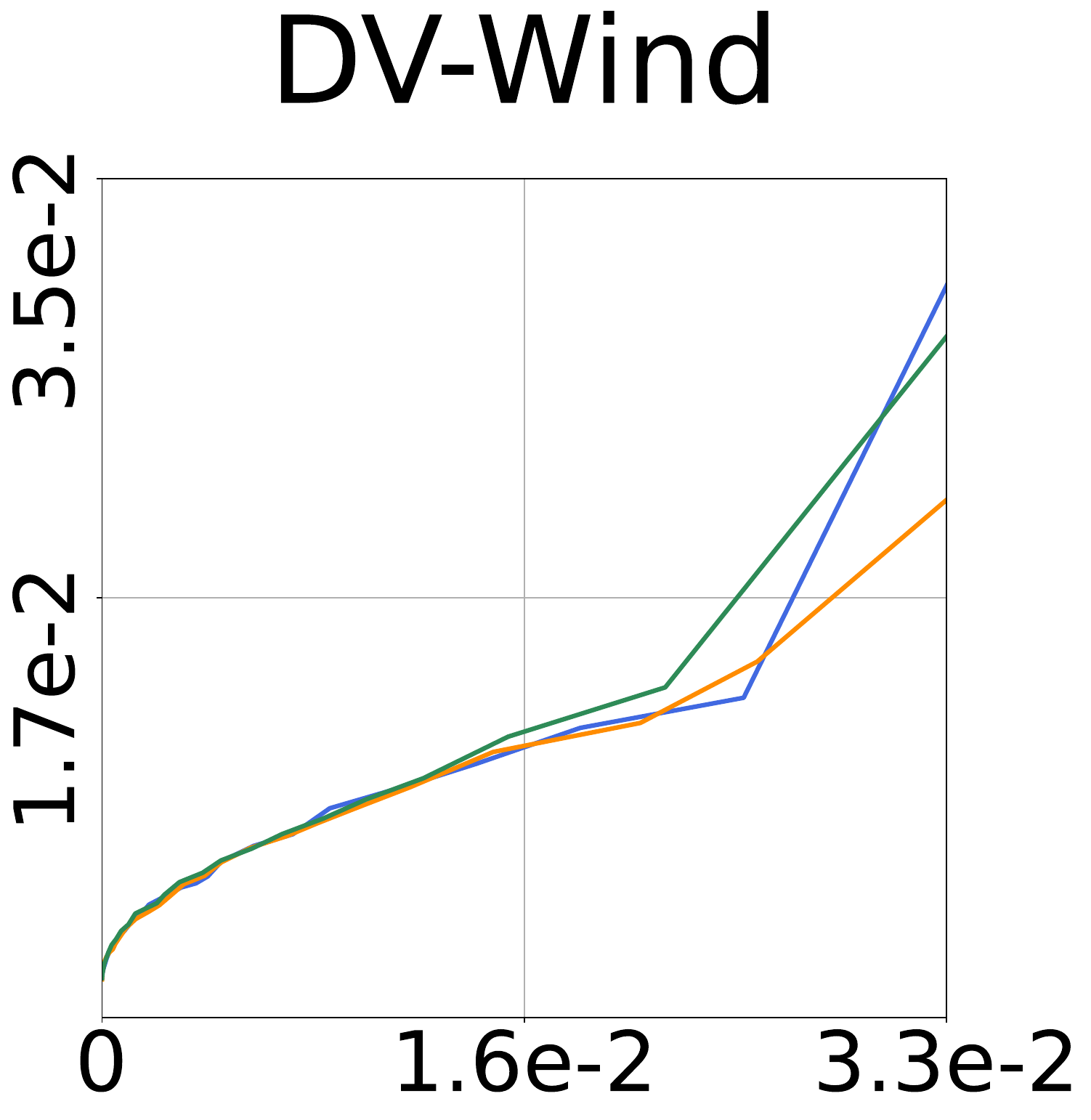}
    }
    \\
    \mbox{
        
        \includegraphics[width=0.215\columnwidth, height=0.2\columnwidth]{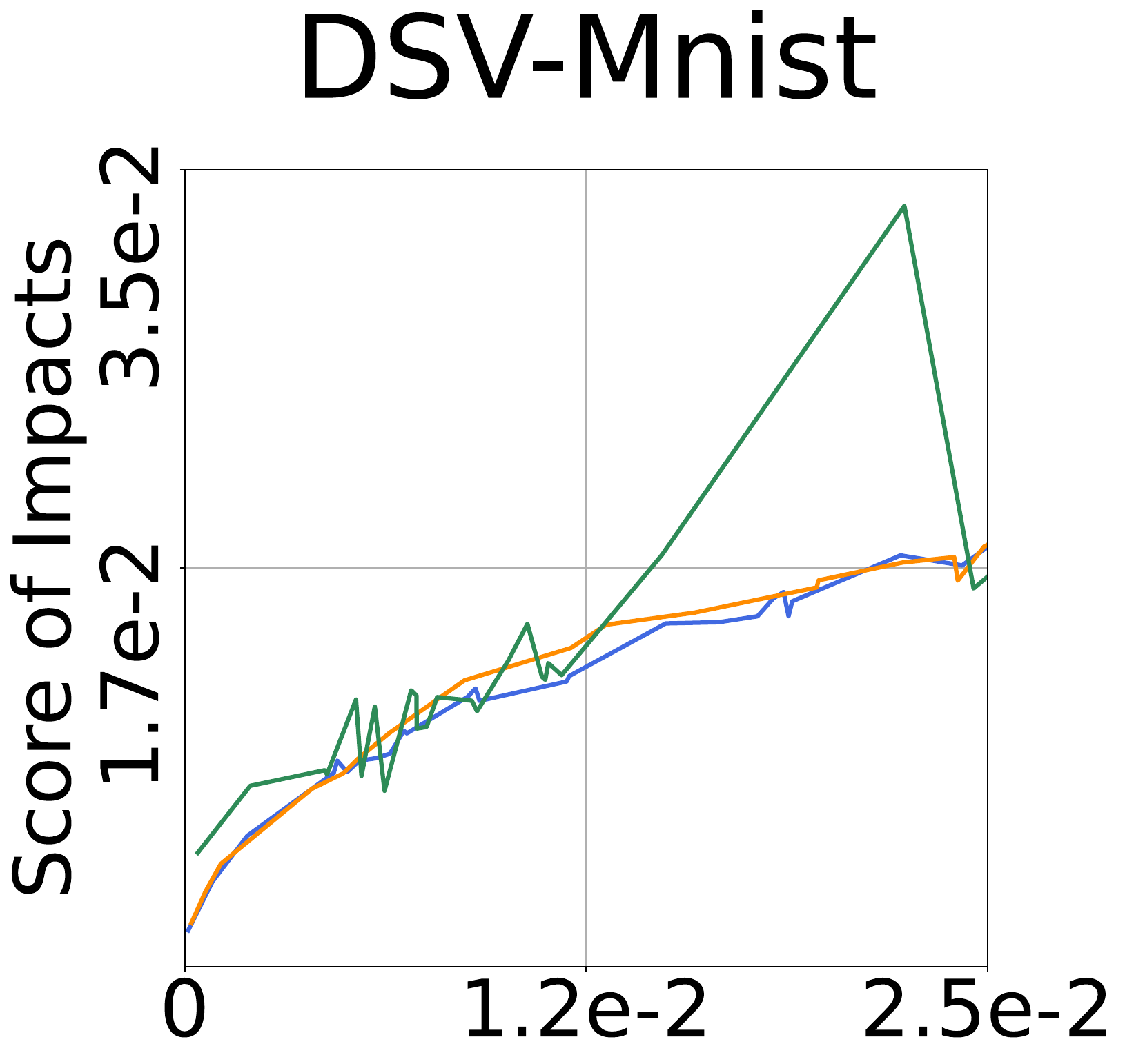}
        \includegraphics[width=0.2\columnwidth, height=0.2\columnwidth]{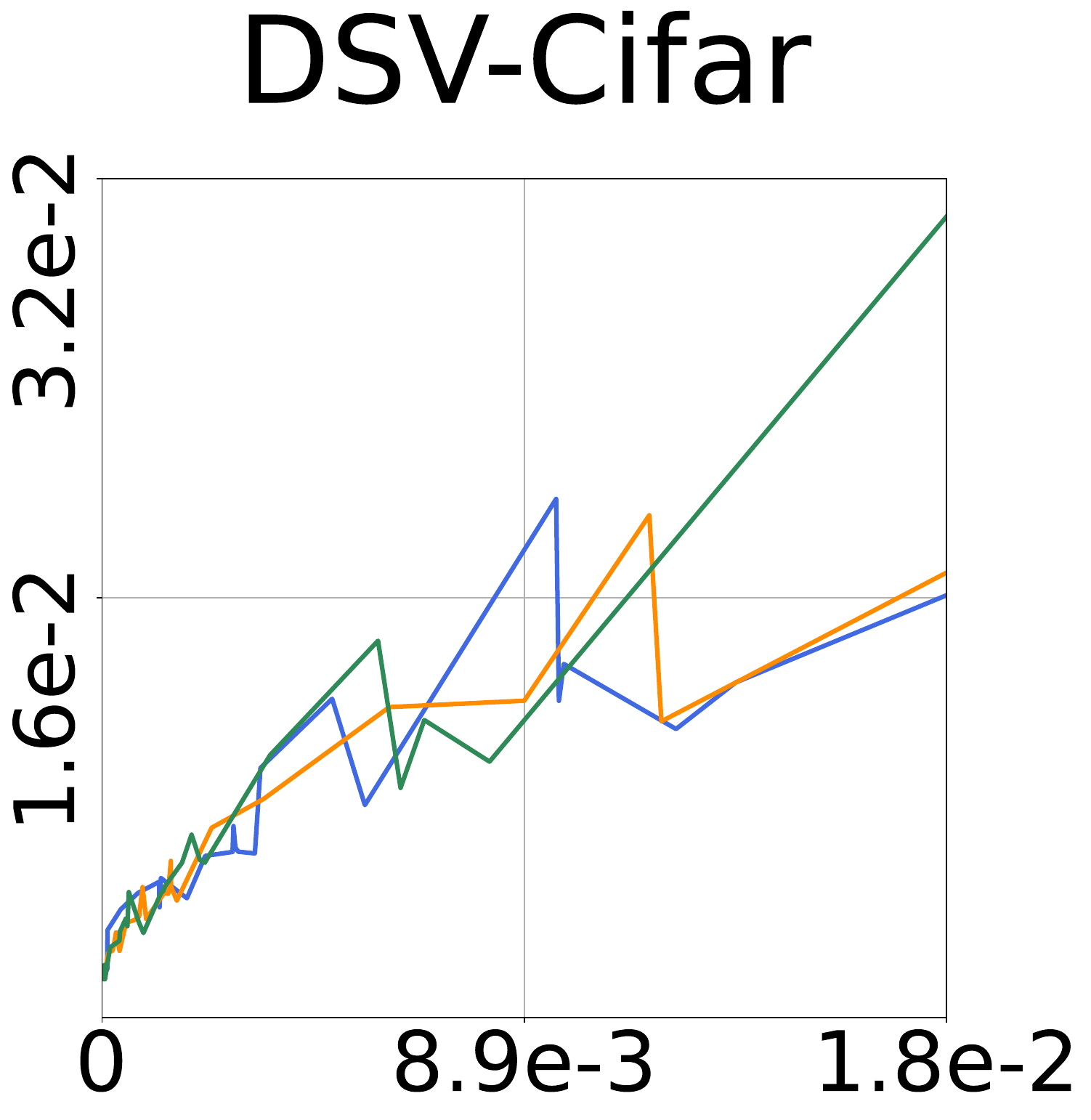}
        \includegraphics[width=0.2\columnwidth, height=0.2\columnwidth]{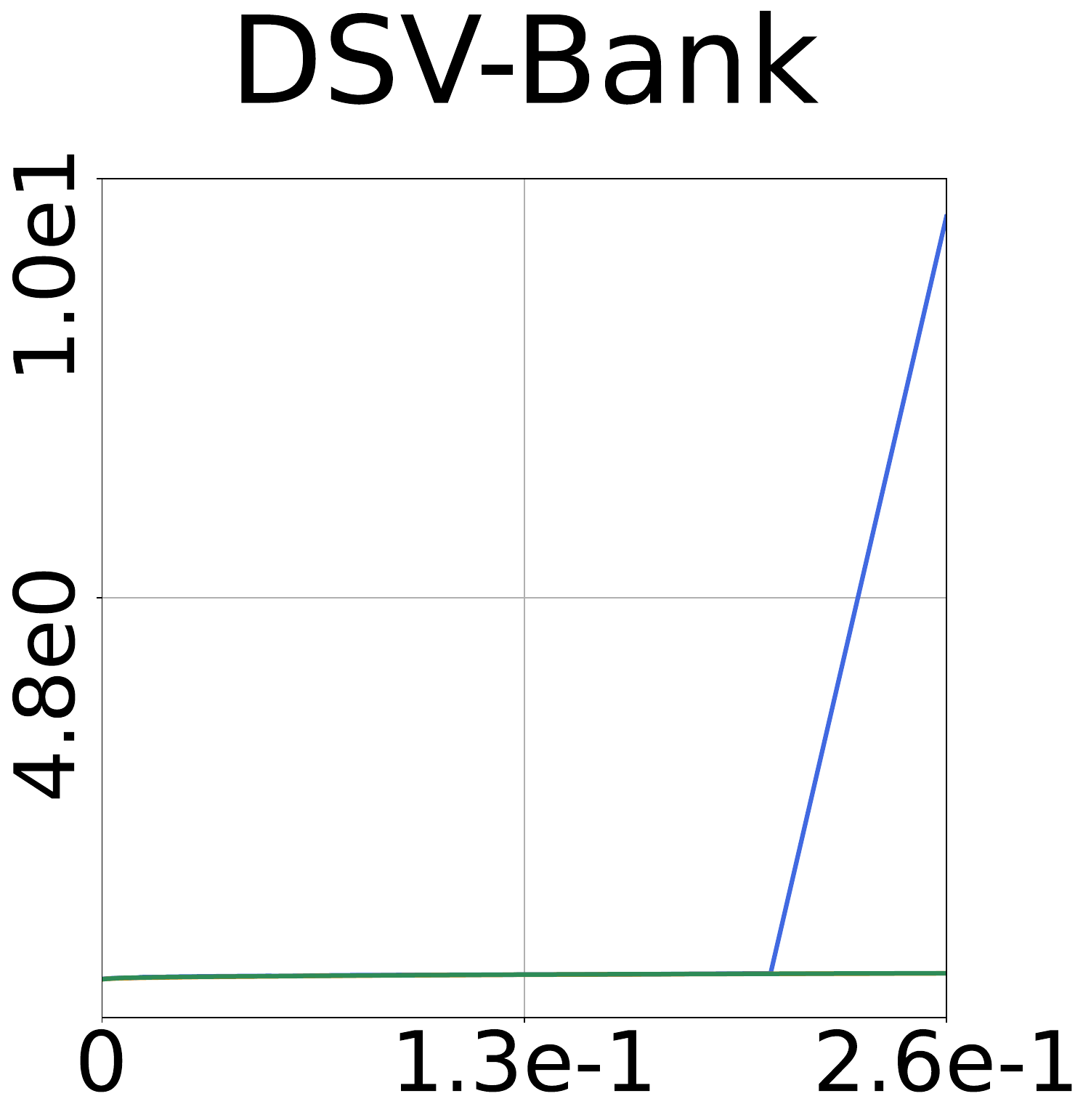}
        \includegraphics[width=0.2\columnwidth, height=0.2\columnwidth]{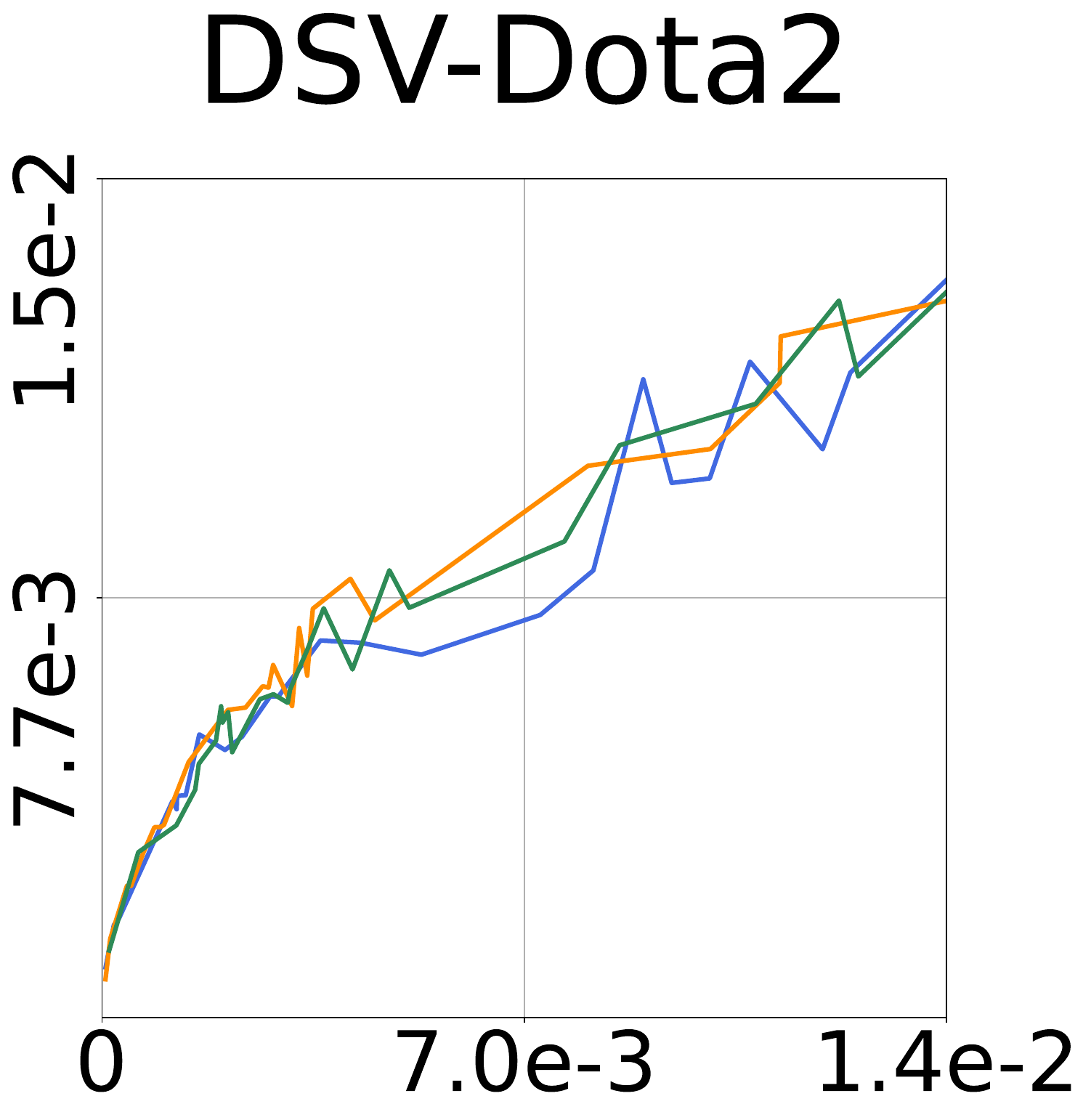}
        \includegraphics[width=0.2\columnwidth, height=0.2\columnwidth]{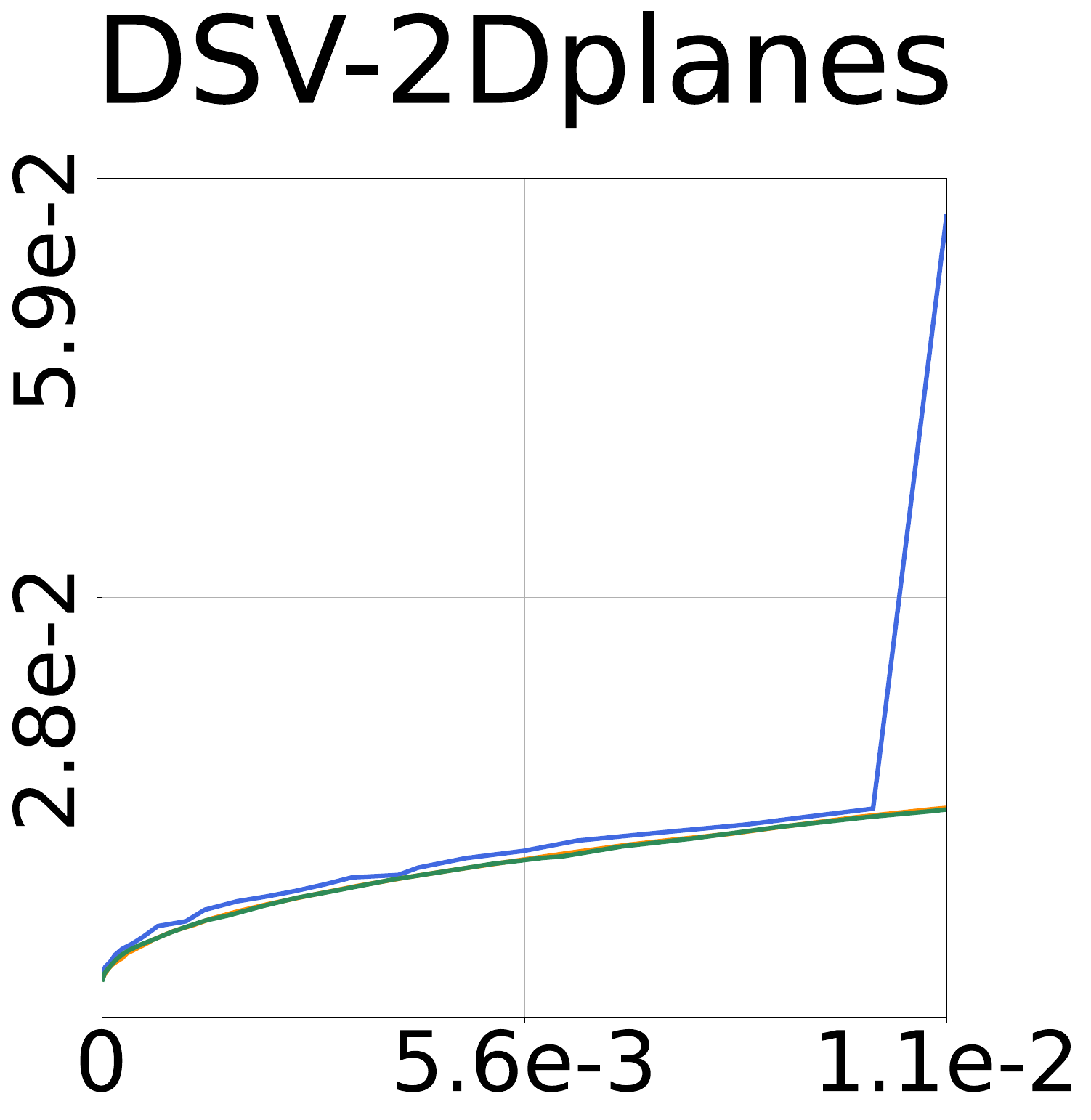}

    }  
    \\
    \subfigure[The score of impacts on SV effectiveness. The smaller the score, the better effectiveness the approximate SV has.] {\label{fig:exp_EfficiencyAndApproximation:ranking}
        \includegraphics[width=0.215\columnwidth, height=0.2\columnwidth]{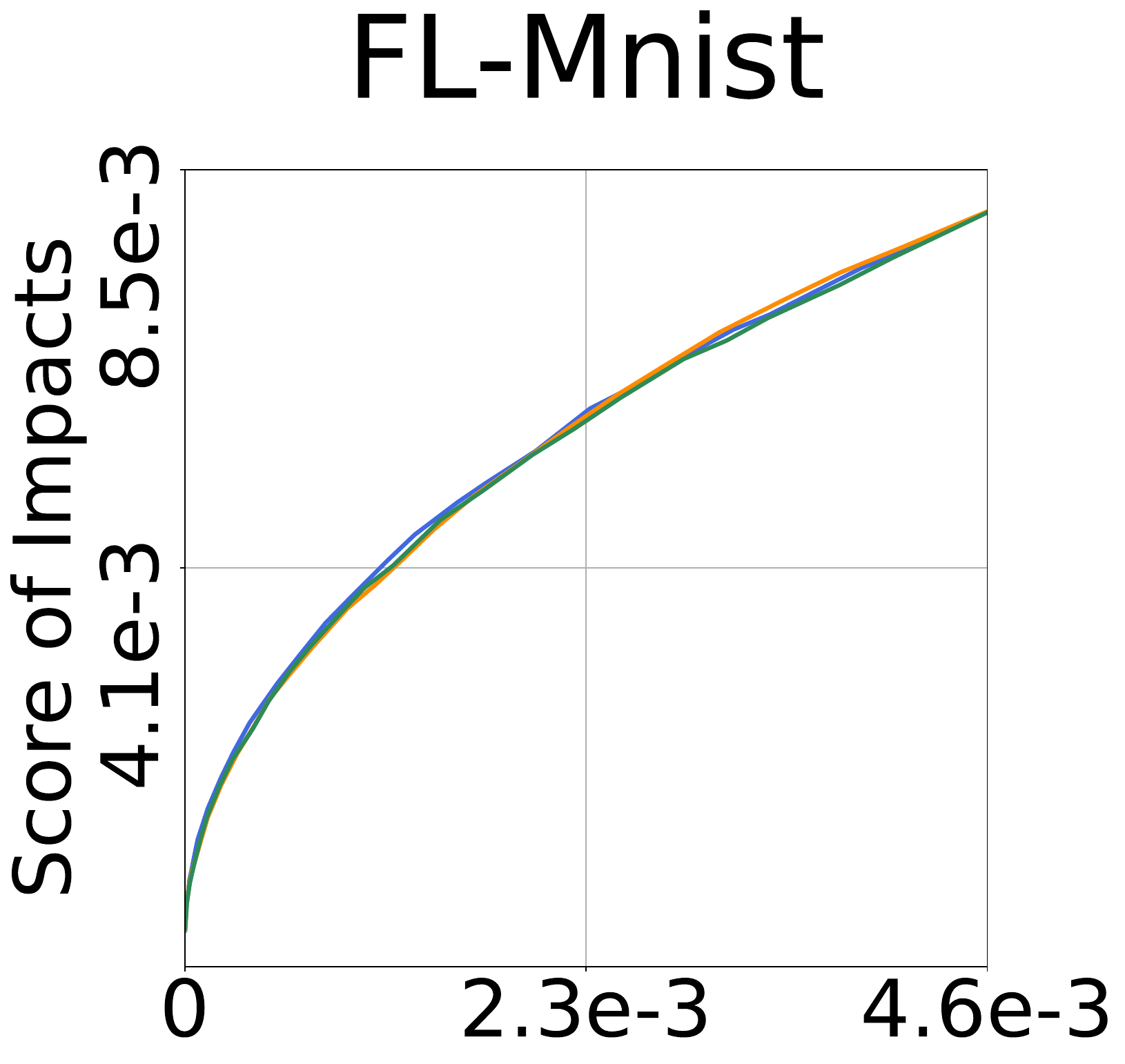}
        \includegraphics[width=0.2\columnwidth, height=0.2\columnwidth]{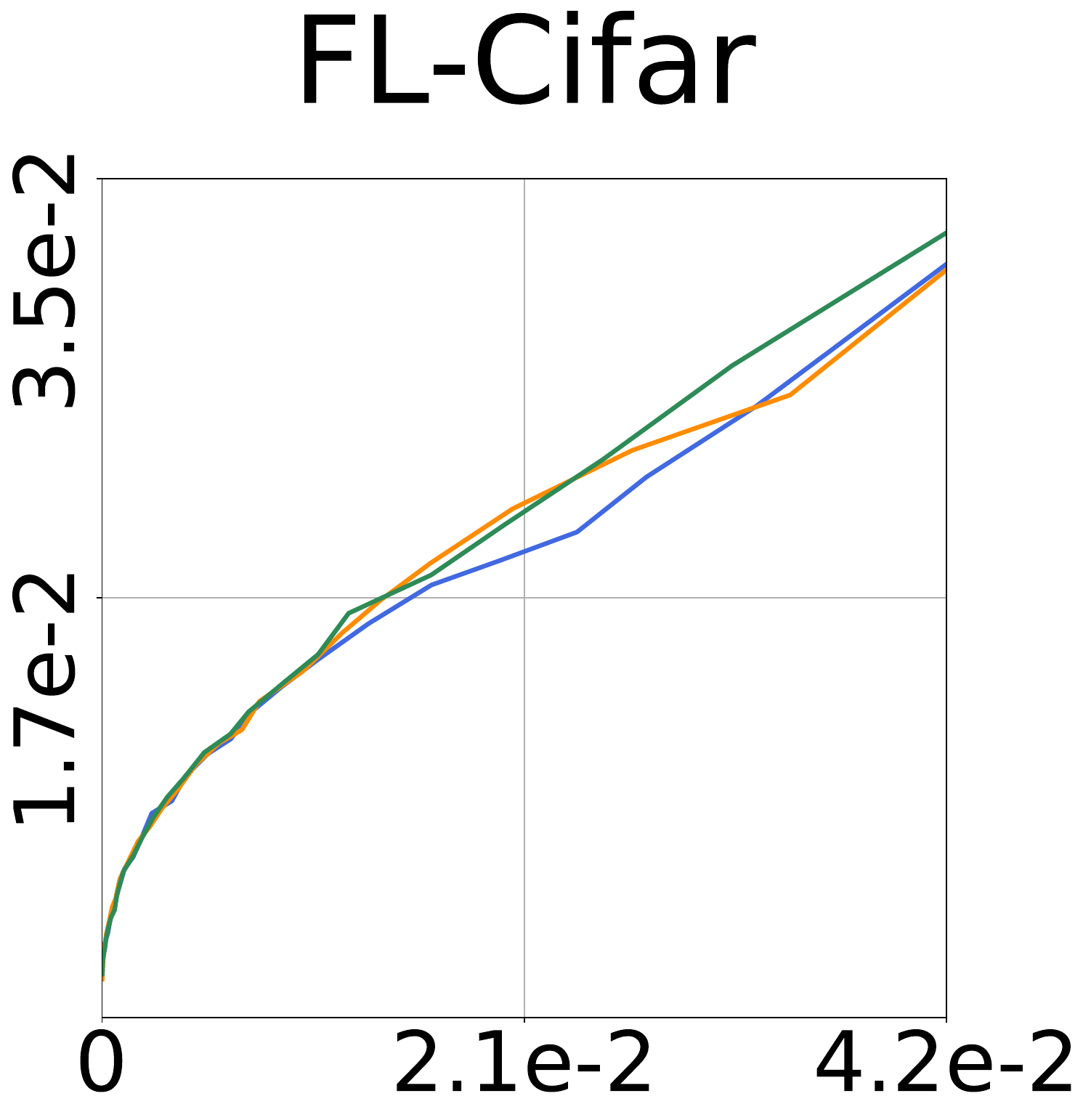}
        \includegraphics[width=0.2\columnwidth, height=0.2\columnwidth]{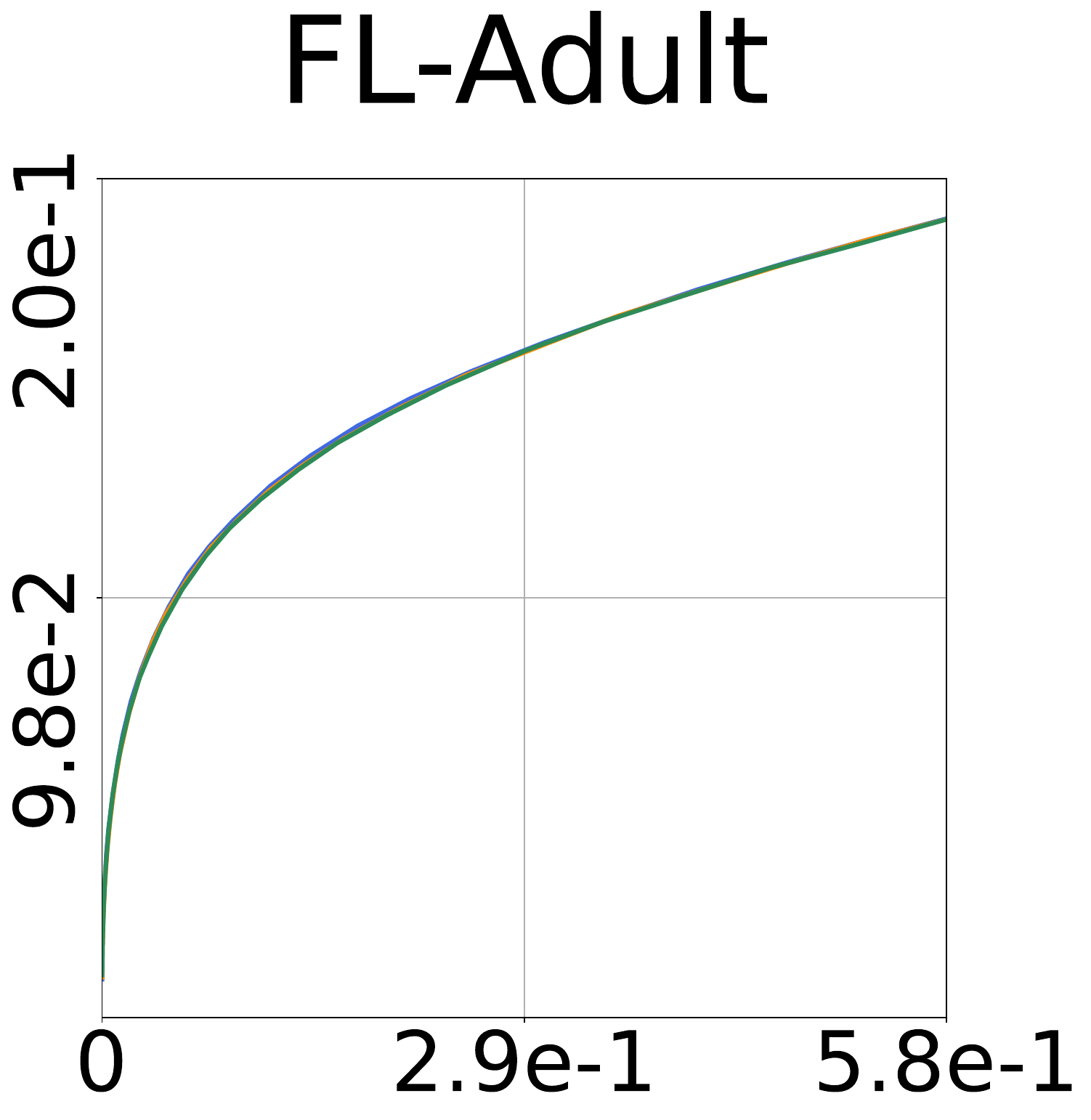}
        \includegraphics[width=0.2\columnwidth, height=0.2\columnwidth]{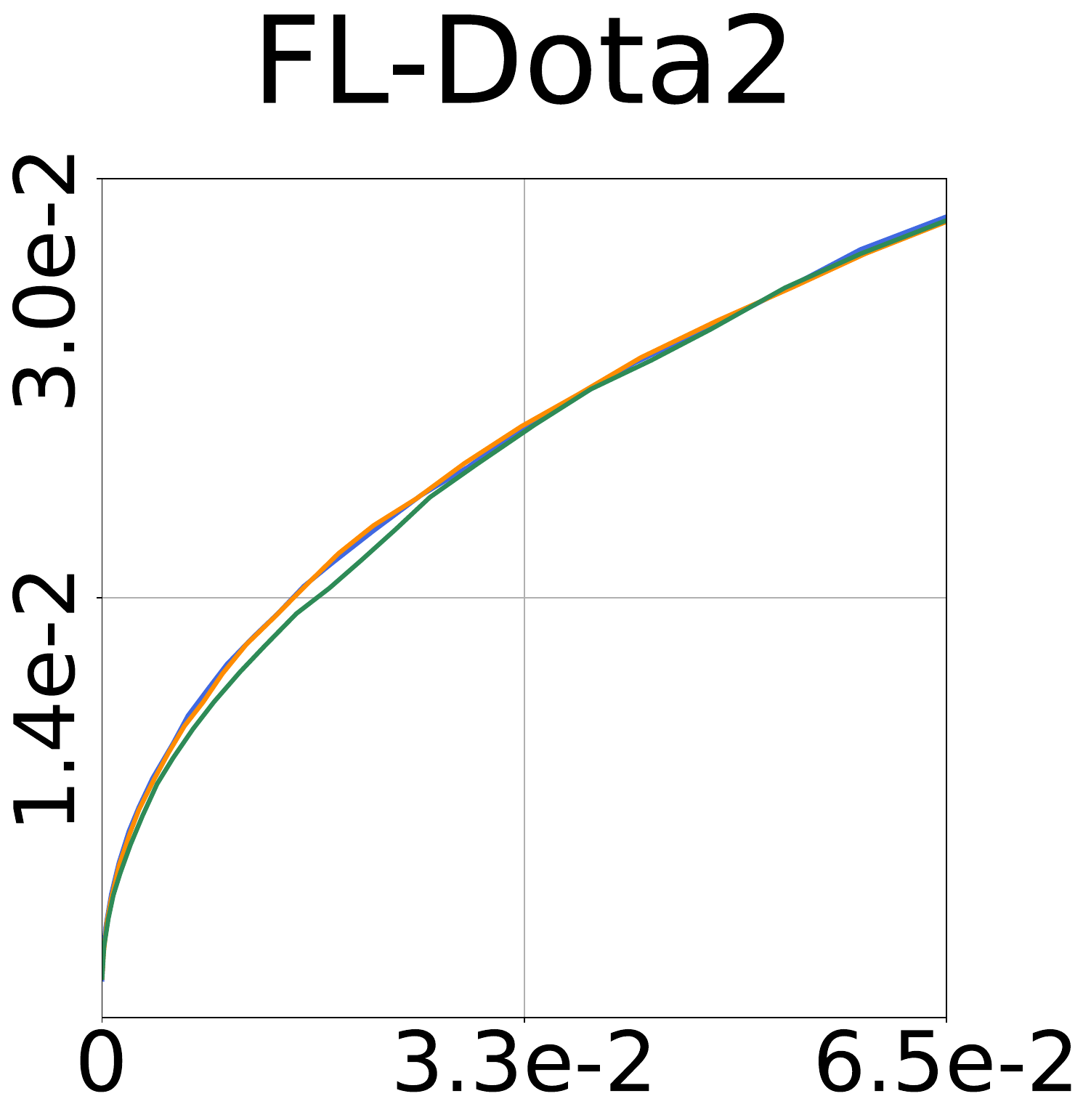}
        \includegraphics[width=0.2\columnwidth, height=0.2\columnwidth]{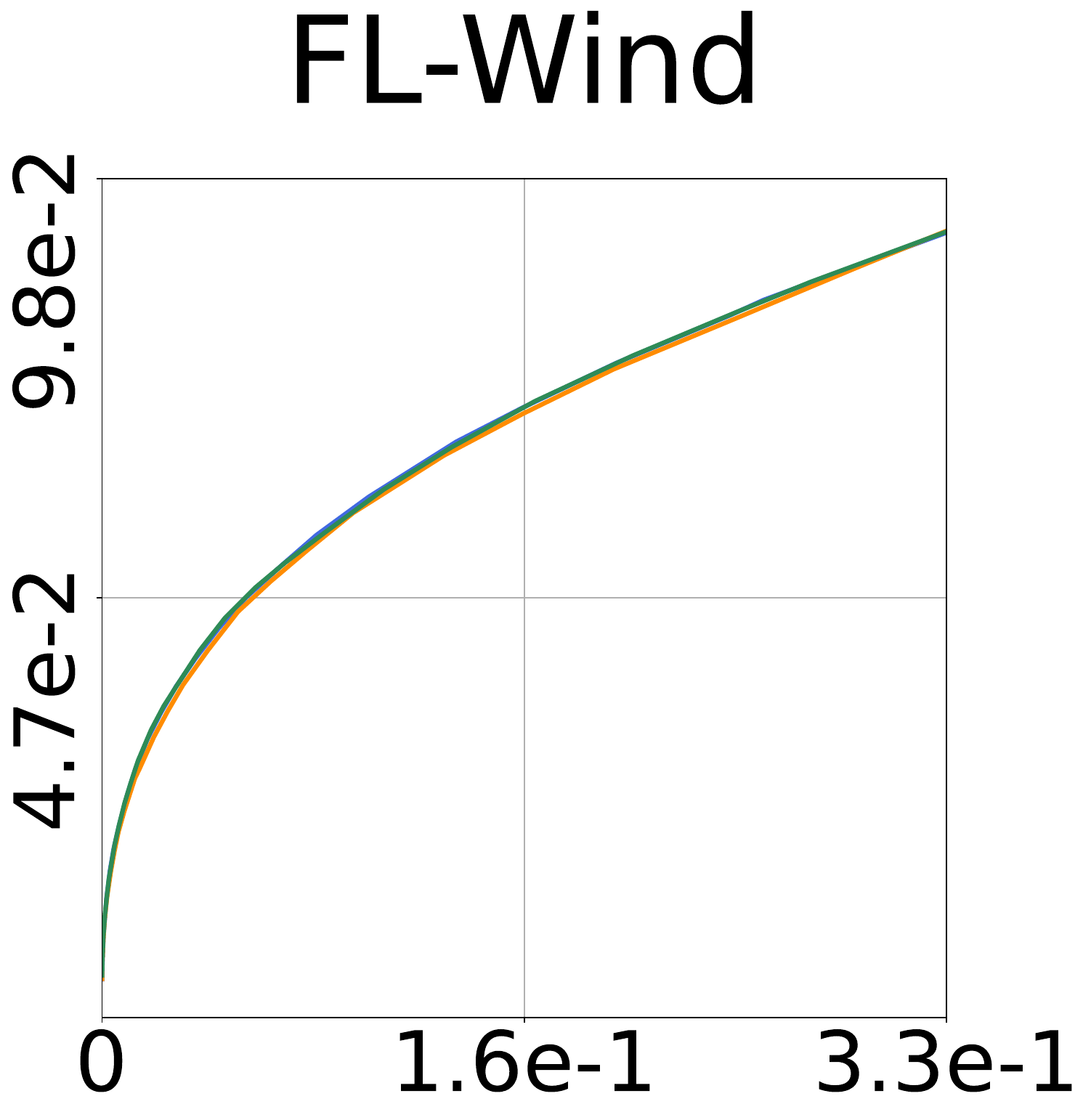}
    }  

    \end{minipage}
    
    \caption{
    Impacts of approximation error (x-axis: the $\epsilon$ value).
    }
    \label{fig:exp_EfficiencyAndApproximation}
\end{figure}

\Cref{fig:exp_EfficiencyAndApproximation:complexity} presents the relationship between the approximation error of three SV computing algorithms and the computation efficiency and stability, while \Cref{fig:exp_EfficiencyAndApproximation:ranking} presents the impacts of approximation error on the effectiveness of SV. 
The square-tagged lines in \Cref{fig:exp_EfficiencyAndApproximation:complexity} show that, in most tasks, the larger the SV approximation error, the smaller the computation complexity. 
This is due to far fewer utility values needing to be computed when a larger tolerance is given to the approximation error, regardless of the strategy for sampling coalitions, as discussed in \S\ref{sssec:approximation_error}. 
However, as shown in the star-tagged lines in \Cref{fig:exp_EfficiencyAndApproximation:complexity}, the approximate SVs with a large error tend to be generated when the computation is far from an ideal convergence status (where $\Delta \hat{\phi}$ goes below a trivial threshold value). Those SVs may lead to a larger error in the scaled SV results, as shown in \Cref{fig:exp_EfficiencyAndApproximation:ranking}, indicating the potential ineffectiveness of using approximate SV for pricing, selection, weighting, and attribution in DA. 
We notice that the RI\_Iris task's results in Figure~\ref{fig:exp_EfficiencyAndApproximation:complexity} show a different pattern from the other tasks. 
This is for the same reason as the performance of this task in \S\ref{subsec: exp_efficiency}, that only four players in this task make all algorithms complete full sampling before converging. 
In conclusion, the above results consolidate the necessity to strike a balance among the approximation error, computation complexity, and the effectiveness of SV. To achieve such a balance, we highlight the following future directions. 

\textbf{Research Direction 2: Investigation on the runtime dynamic tuning of convergence criterion when using sampling-based approximation techniques.} 
Take the criterion $\Delta \hat{\phi} < \tau$ and the DV task in \Cref{fig:SV_in_DA} as an example.
A small threshold (e.g.,$\tau=10^{-5}$) can be set at the start of SV approximation. 
Then, the task can \textit{periodically check} whether 
image samples ranked in the top 50\% based on the latest approximate SV produce more accurate classification models than those produced by images ranked in the bottom. Once satisfied, the approximation can be terminated to save computation cost.

\textbf{Research Direction 3: Exploration on lightweight fitting-based approximation techniques.} The fitting-based methods search for a mathematical relationship $\hat{\phi}_i = G(E(p_i))$, where $E(\cdot)$ encodes the player $p_i$ into a characteristic vector, e.g., encoding the image in \Cref{fig:SV_in_DA} into a vector composed of the image's pixel values and label indexes, and $G(\cdot)$ generates the unbiased approximate SV, e.g., the SV of the image. Once $G(E(p_i))$ is determined in a DA task, the approximate SV can be generated by $\textit{O}(1)$ complexity for new players in that task, significantly mitigating the conflicts between SV computation efficiency and approximation error. 
The major obstacle against the feasibility of learning $G(E(\cdot))$ at an affordable cost is \textit{the collection of a sufficient amount of ground-truth SVs (or high-quality surrogates of the ground-truth)}. 
This problem is highly expected to be solved with the surge in real-world SV applications.

\begin{figure*}[t]
    \centering
    \begin{minipage}{0.7\columnwidth}
        \hspace{-6pt}
        \centering \includegraphics[width=\columnwidth]{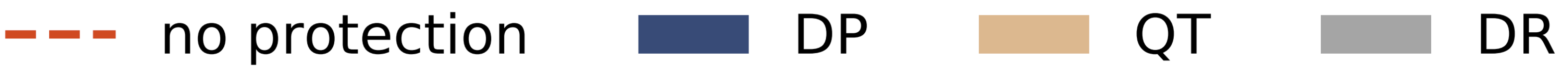}
    \end{minipage}
    \\
    \subfigure[The larger the MAE, the less privacy the SV exposes.] {
        \hspace{-8pt}
        \includegraphics[width=0.215\columnwidth, height=0.2\columnwidth]{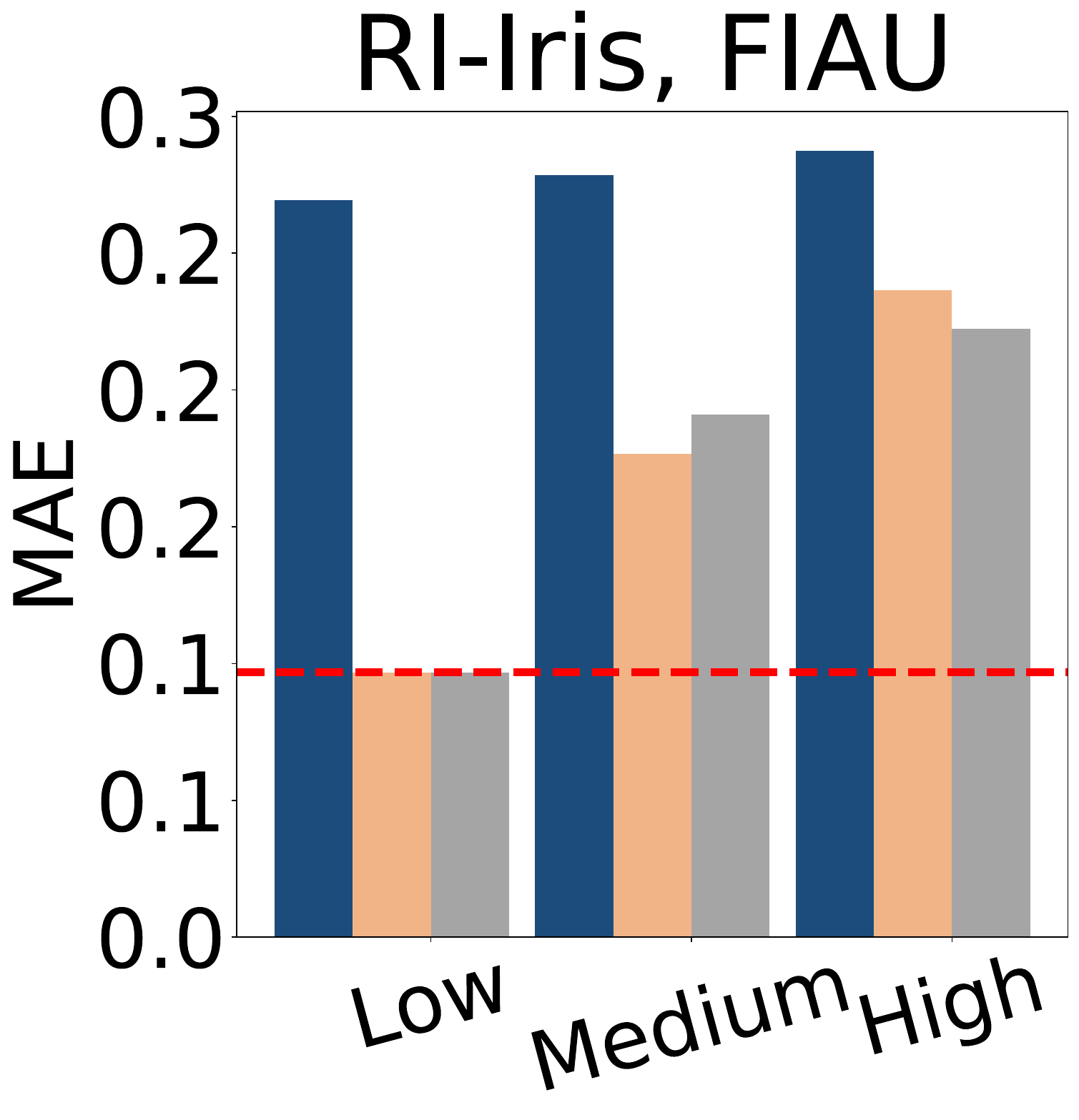}
        \includegraphics[width=0.2\columnwidth, height=0.2\columnwidth]{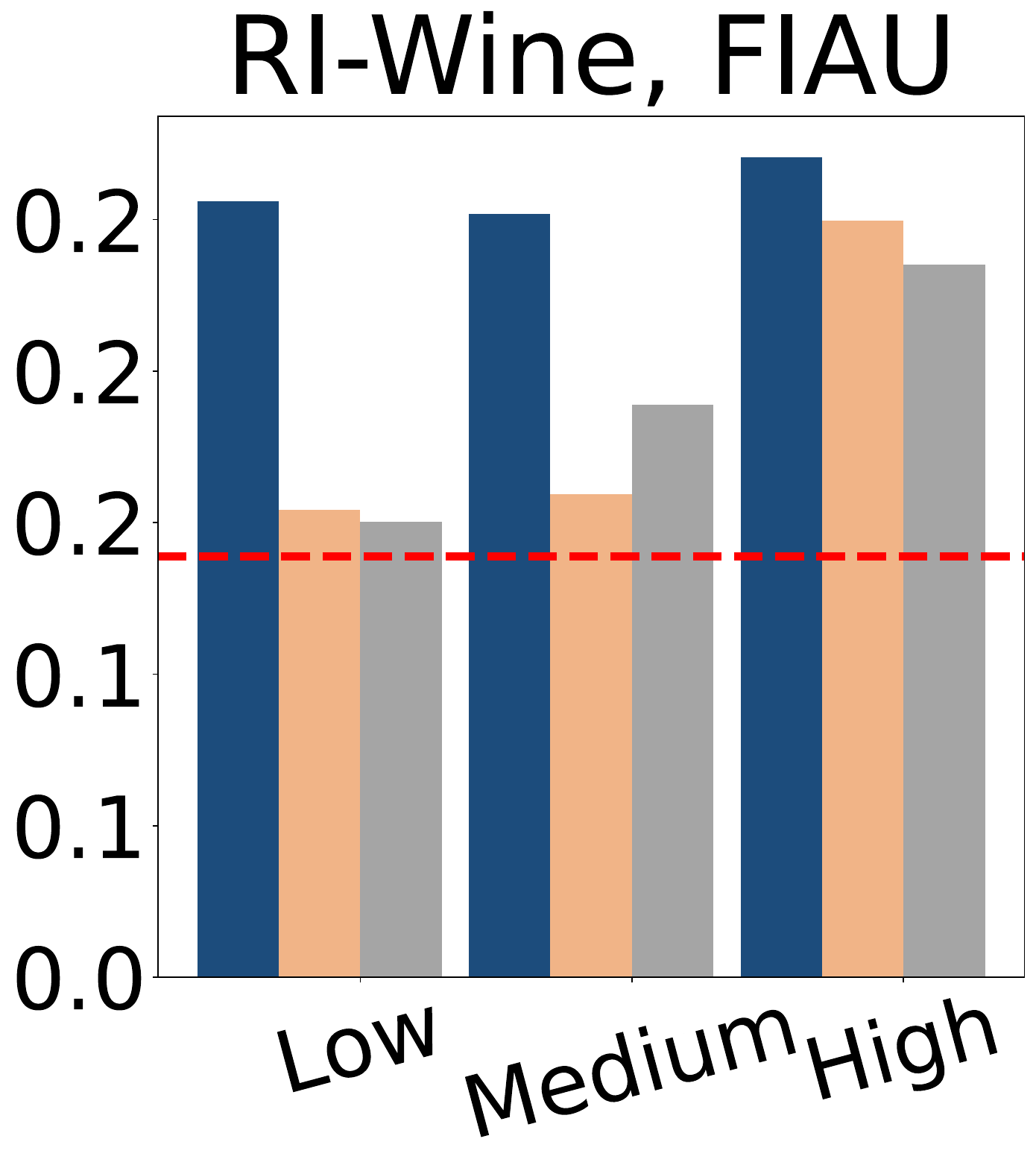}
        \includegraphics[width=0.2\columnwidth, height=0.2\columnwidth]{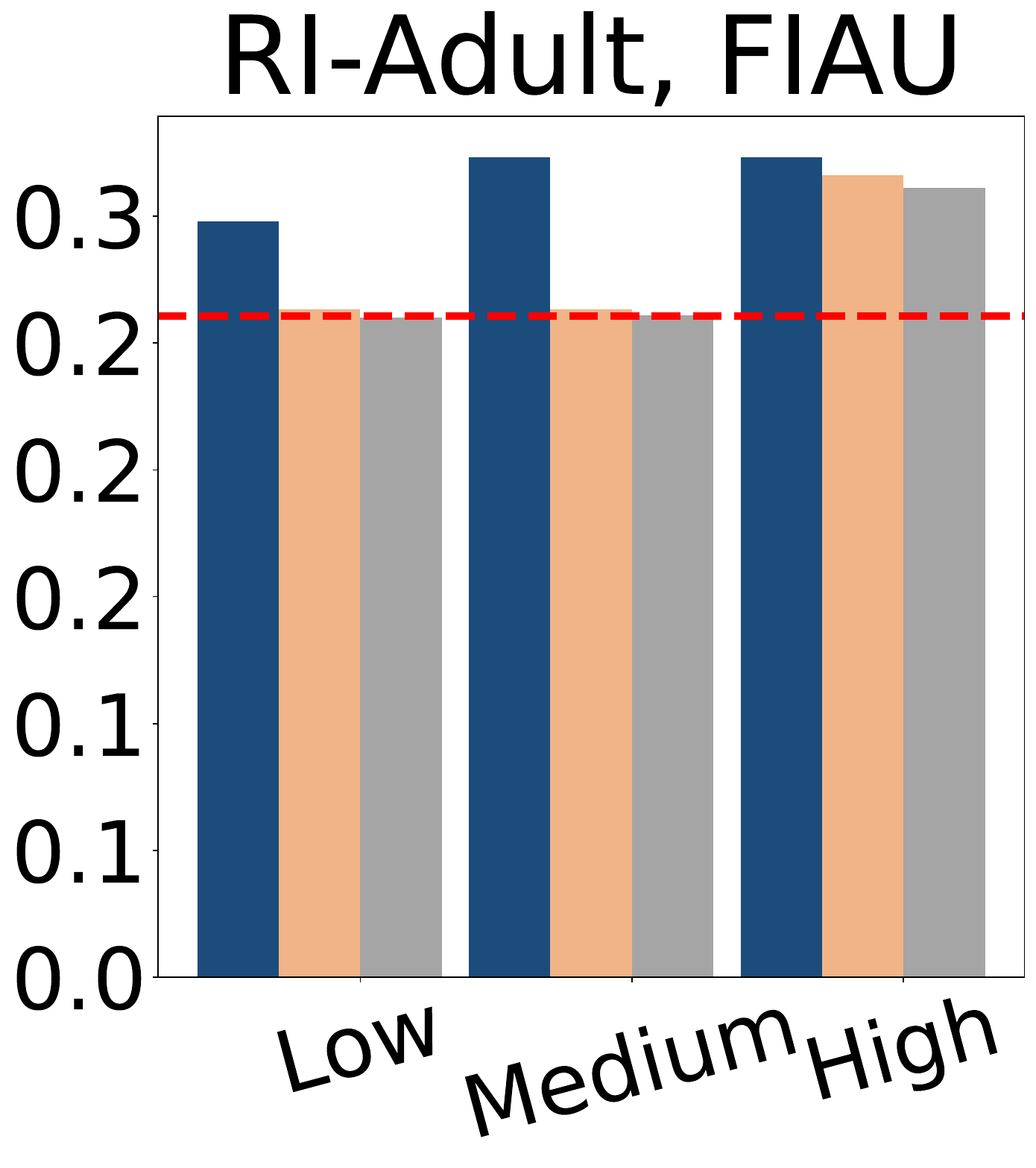}
        \includegraphics[width=0.2\columnwidth, height=0.2\columnwidth]{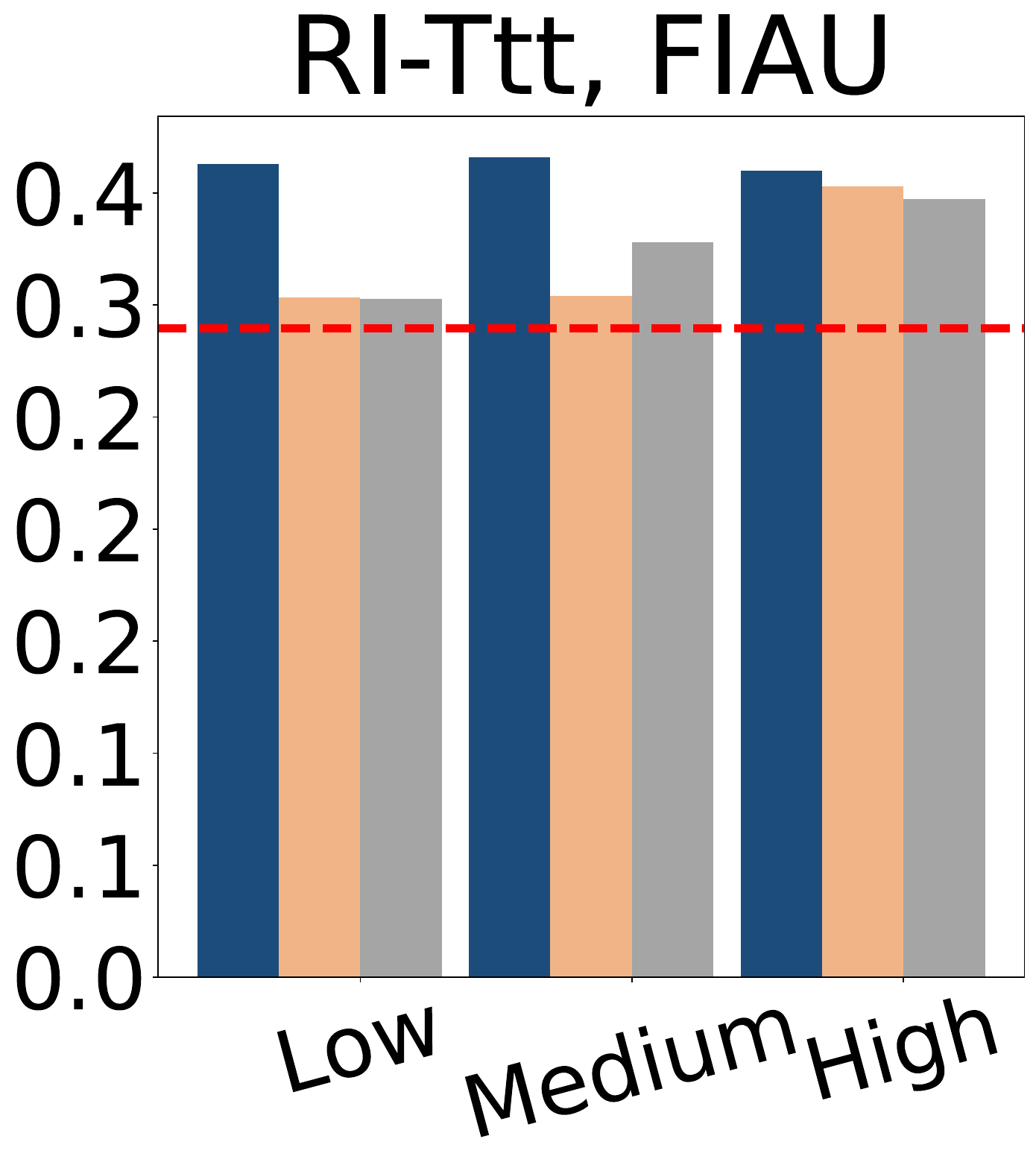}
        \includegraphics[width=0.2\columnwidth, height=0.2\columnwidth]{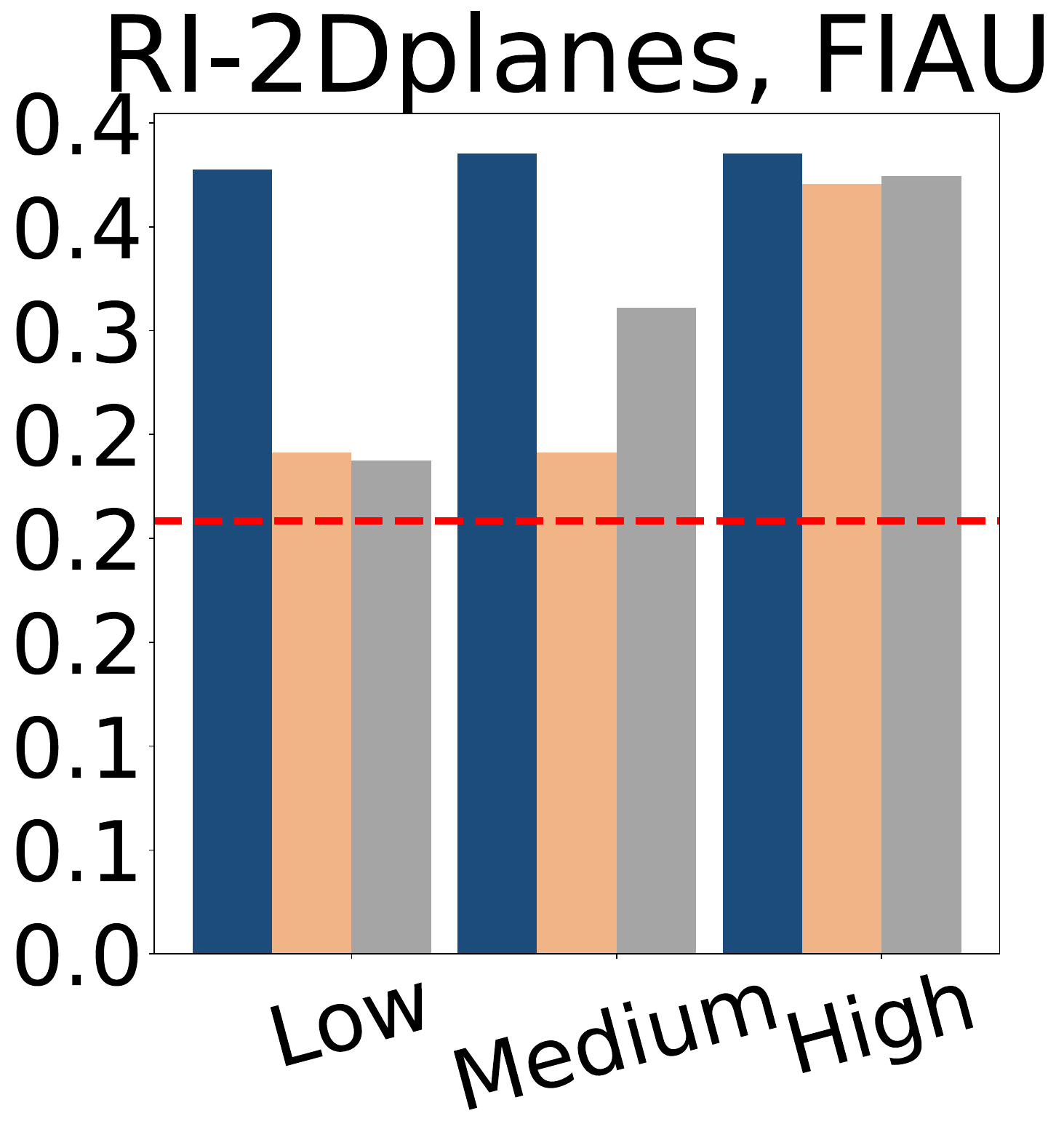}
        \hspace{5pt}
        \includegraphics[width=0.215\columnwidth, height=0.2\columnwidth]{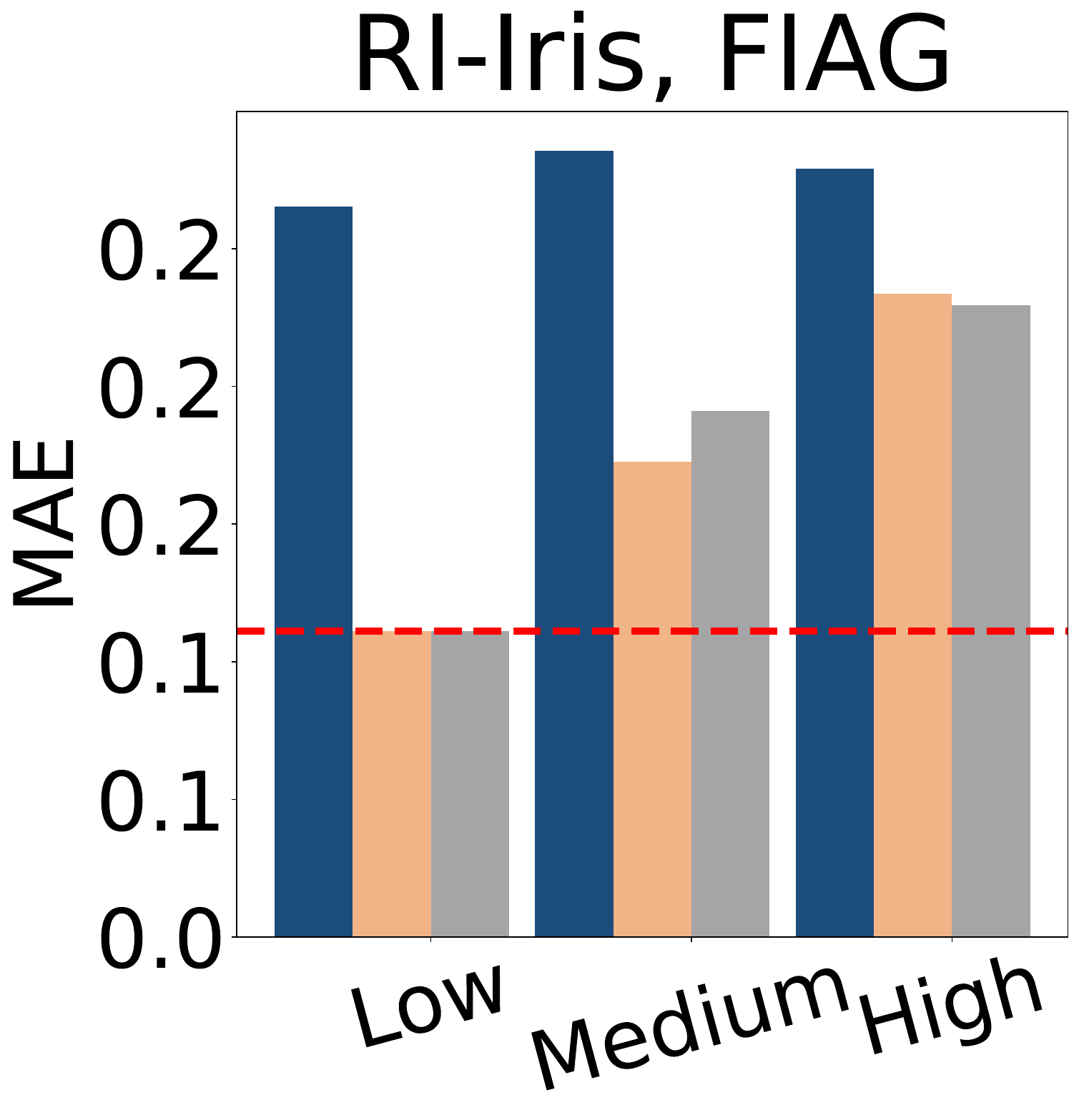}
        \includegraphics[width=0.2\columnwidth, height=0.2\columnwidth]{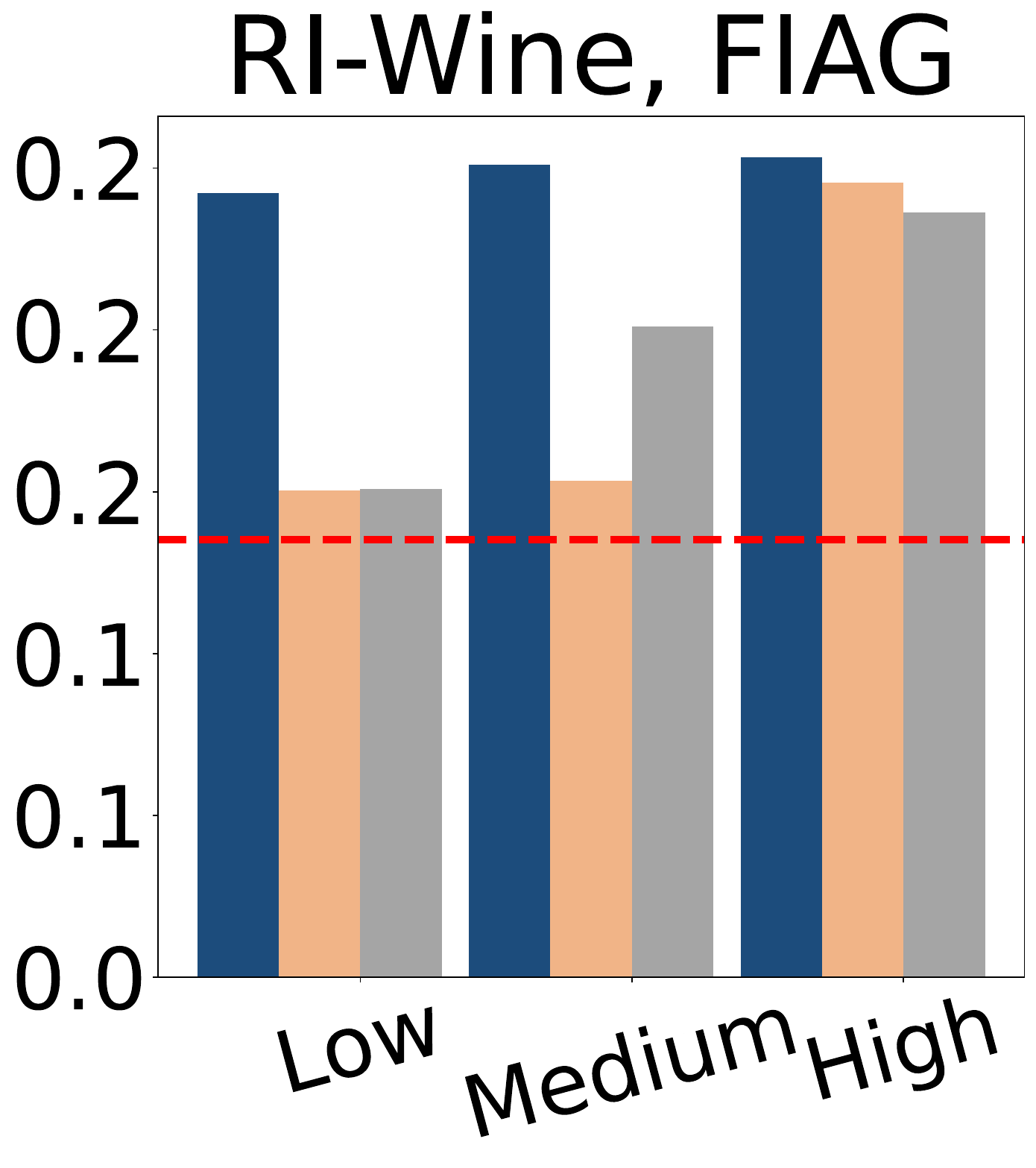}
        \includegraphics[width=0.2\columnwidth, height=0.2\columnwidth]{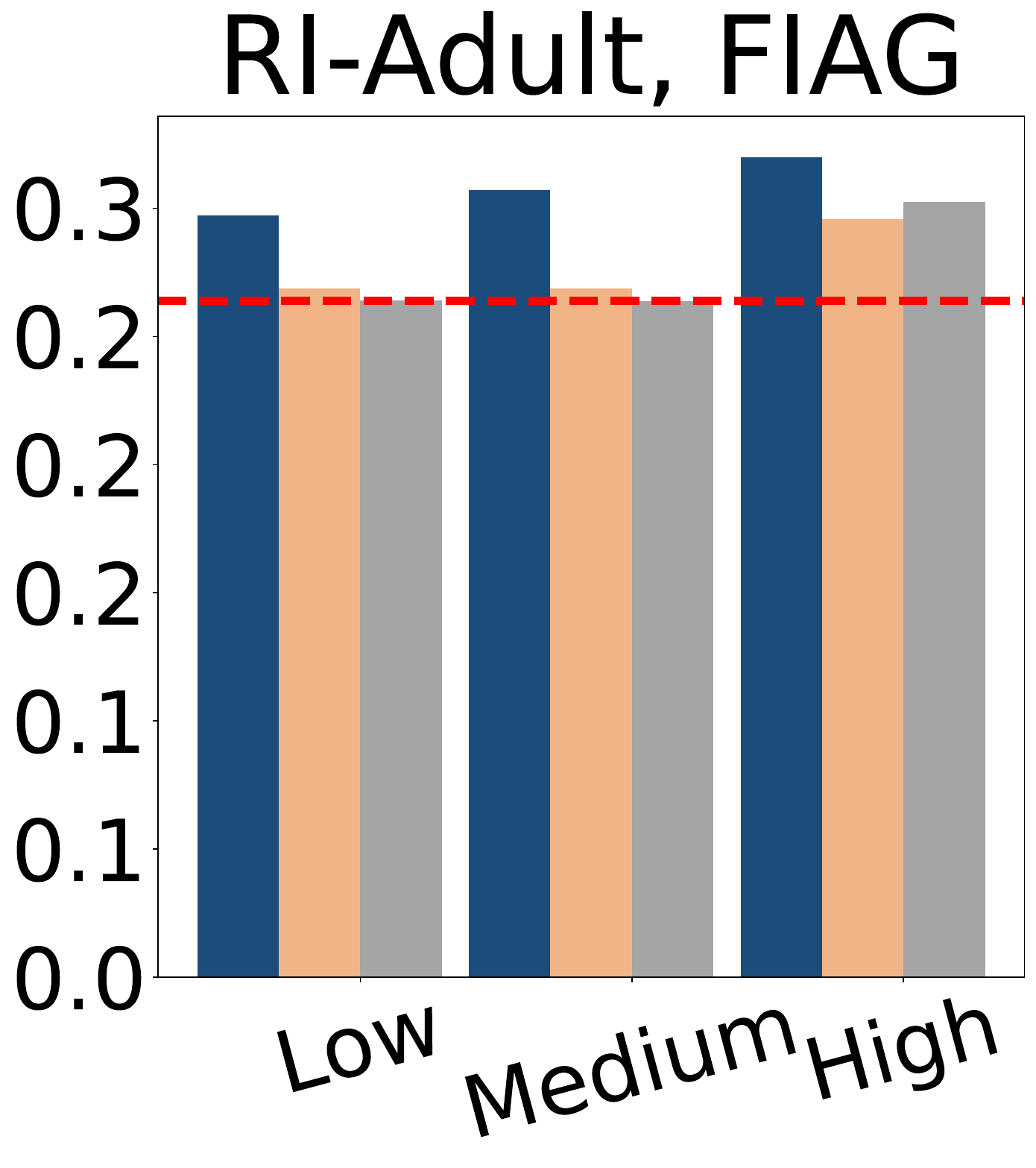}
        \includegraphics[width=0.2\columnwidth, height=0.2\columnwidth]{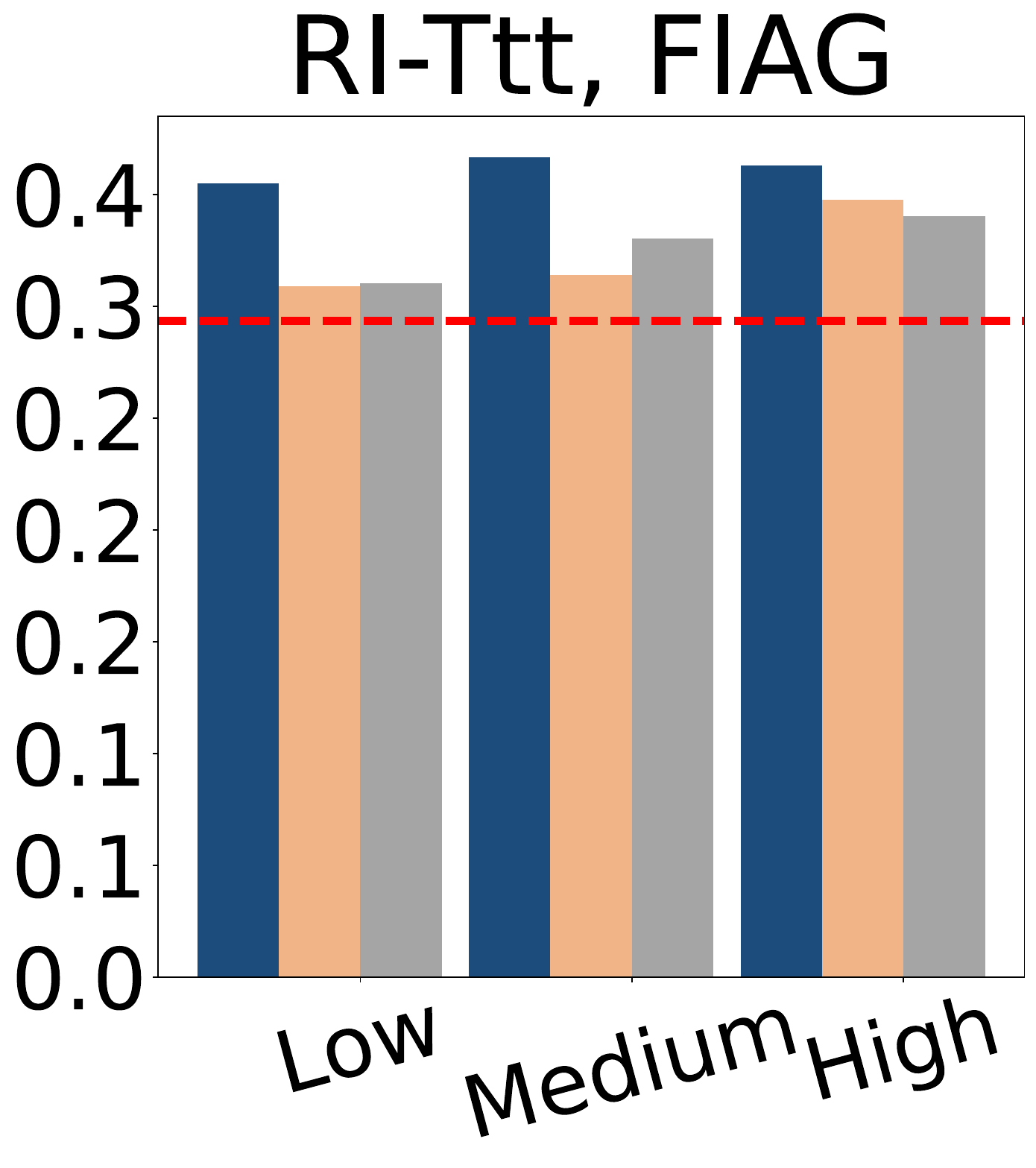}
        \includegraphics[width=0.2\columnwidth, height=0.2\columnwidth]{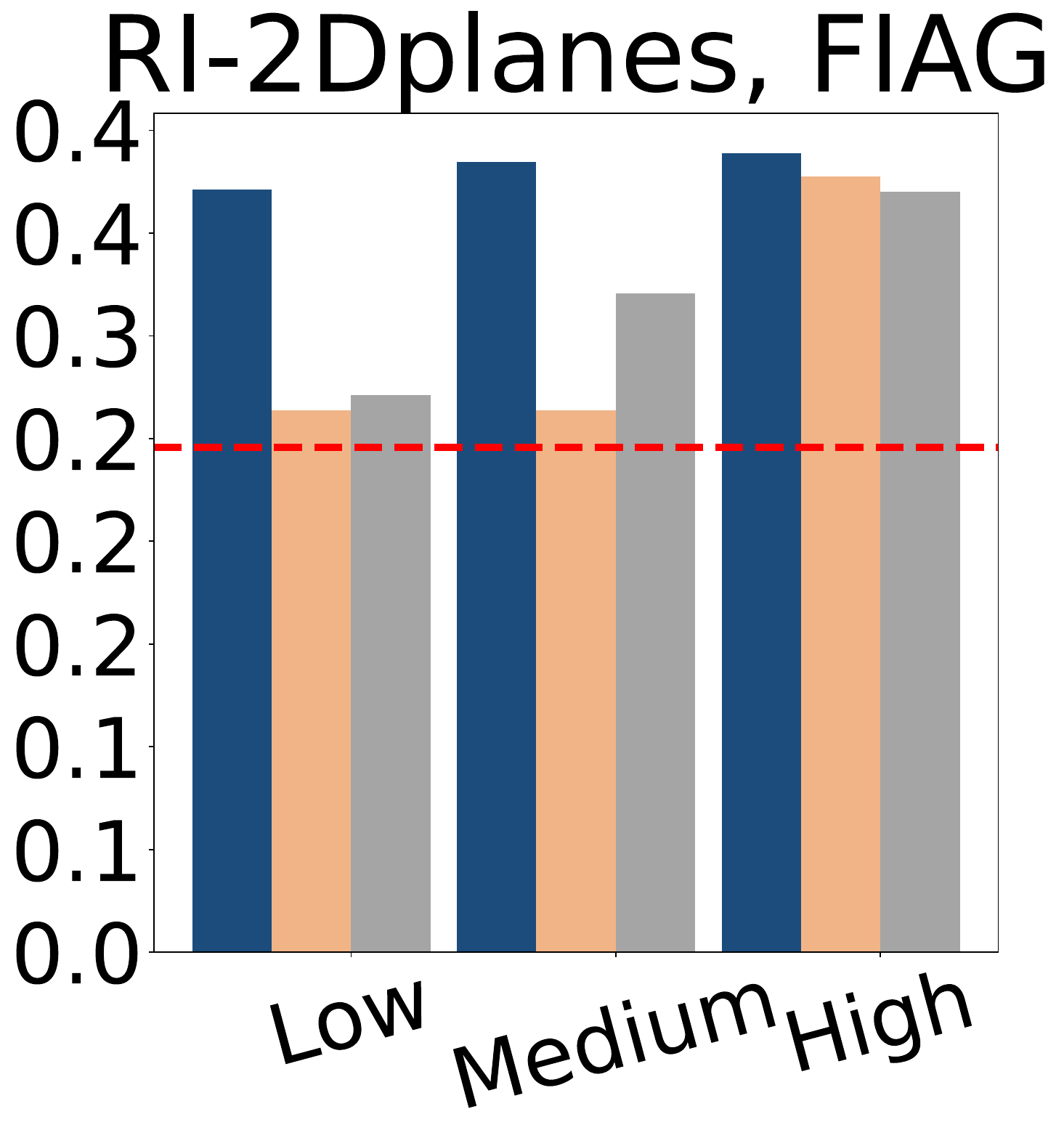}
    }  
    \\
    
    \subfigure[The smaller the variance, the less the impact on SV effectiveness.] {
        \hspace{-8pt}
        \includegraphics[width=0.215\columnwidth, height=0.2\columnwidth]{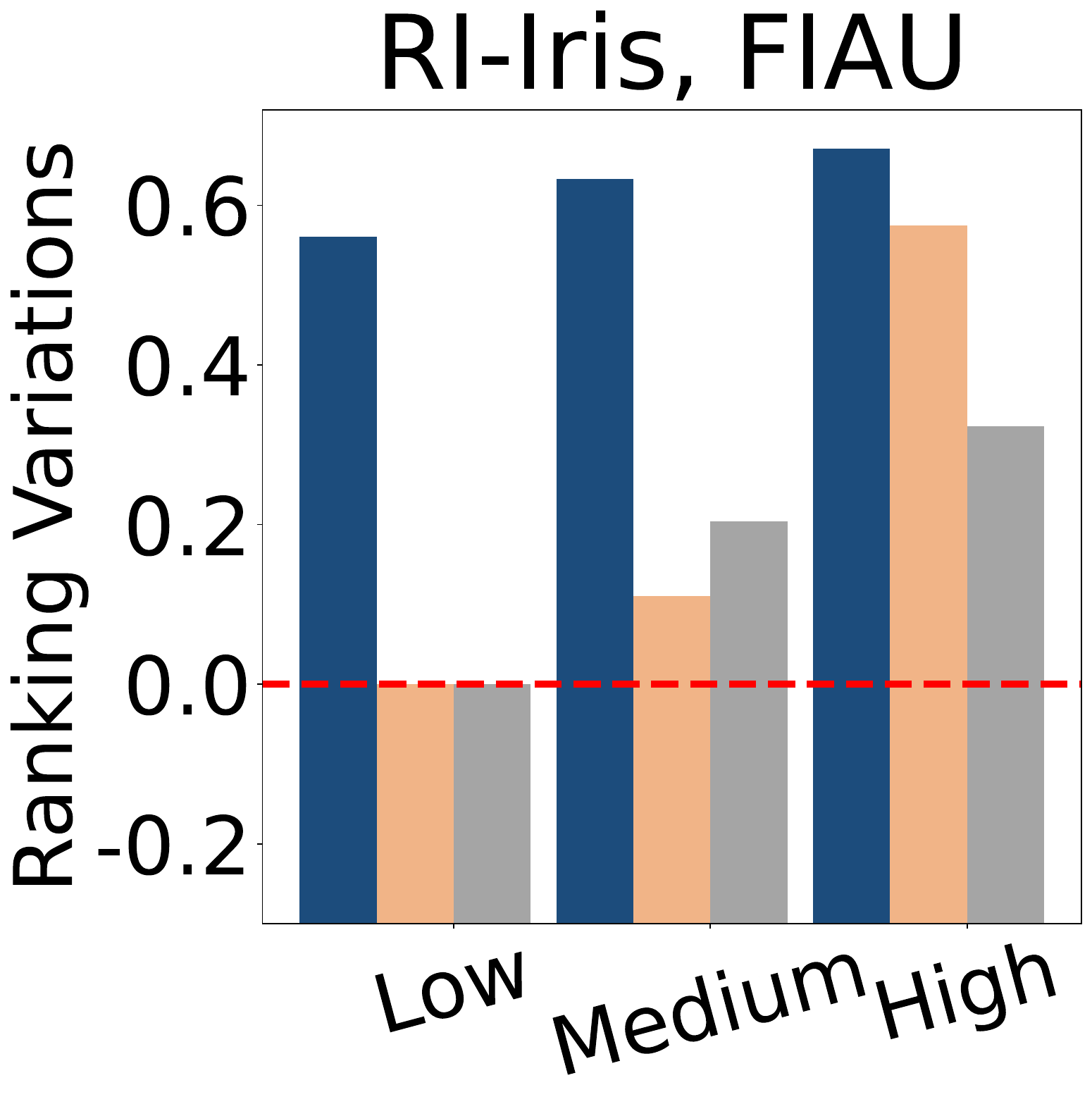}
        \includegraphics[width=0.2\columnwidth, height=0.2\columnwidth]{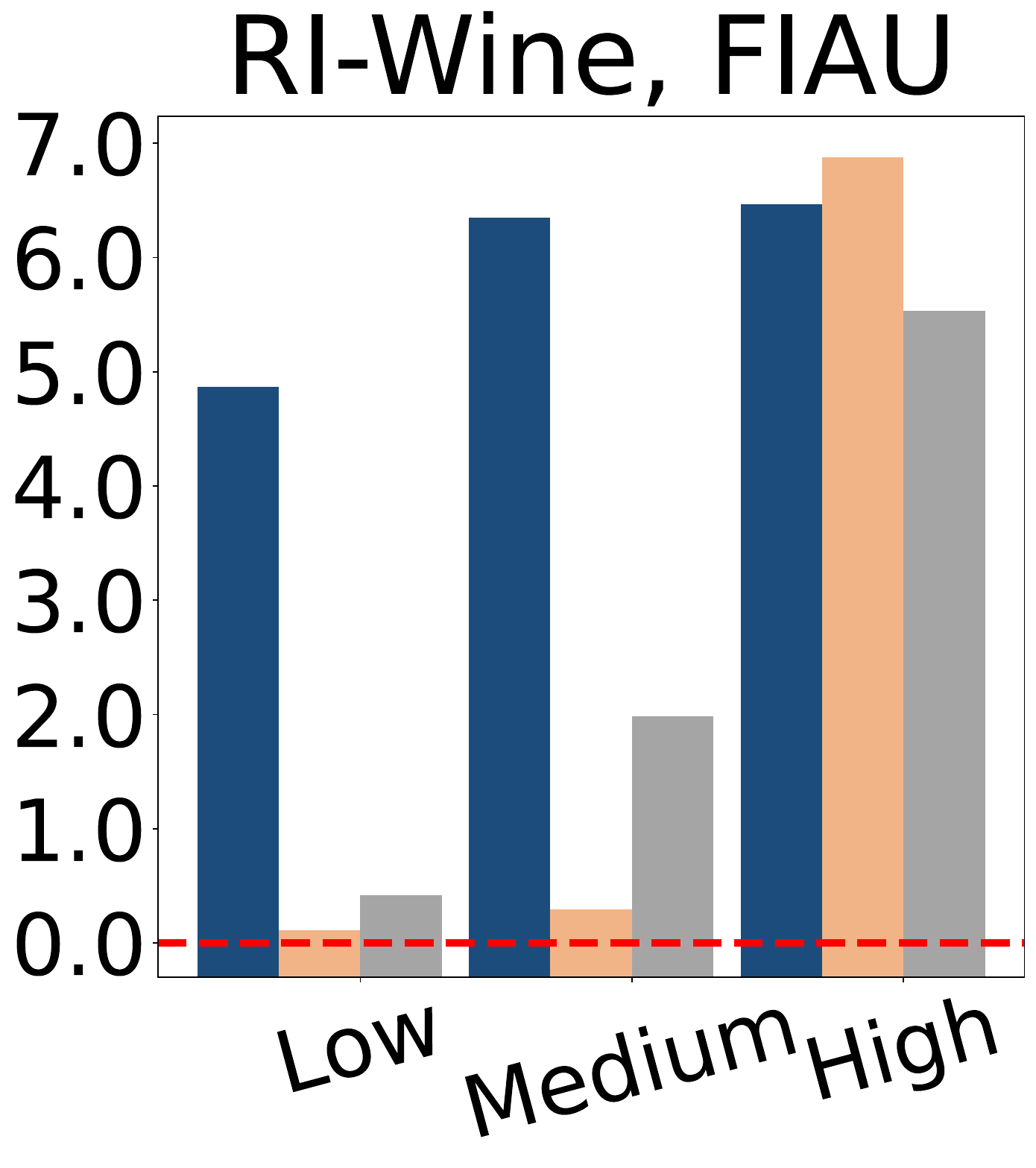}
        \includegraphics[width=0.2\columnwidth, height=0.2\columnwidth]{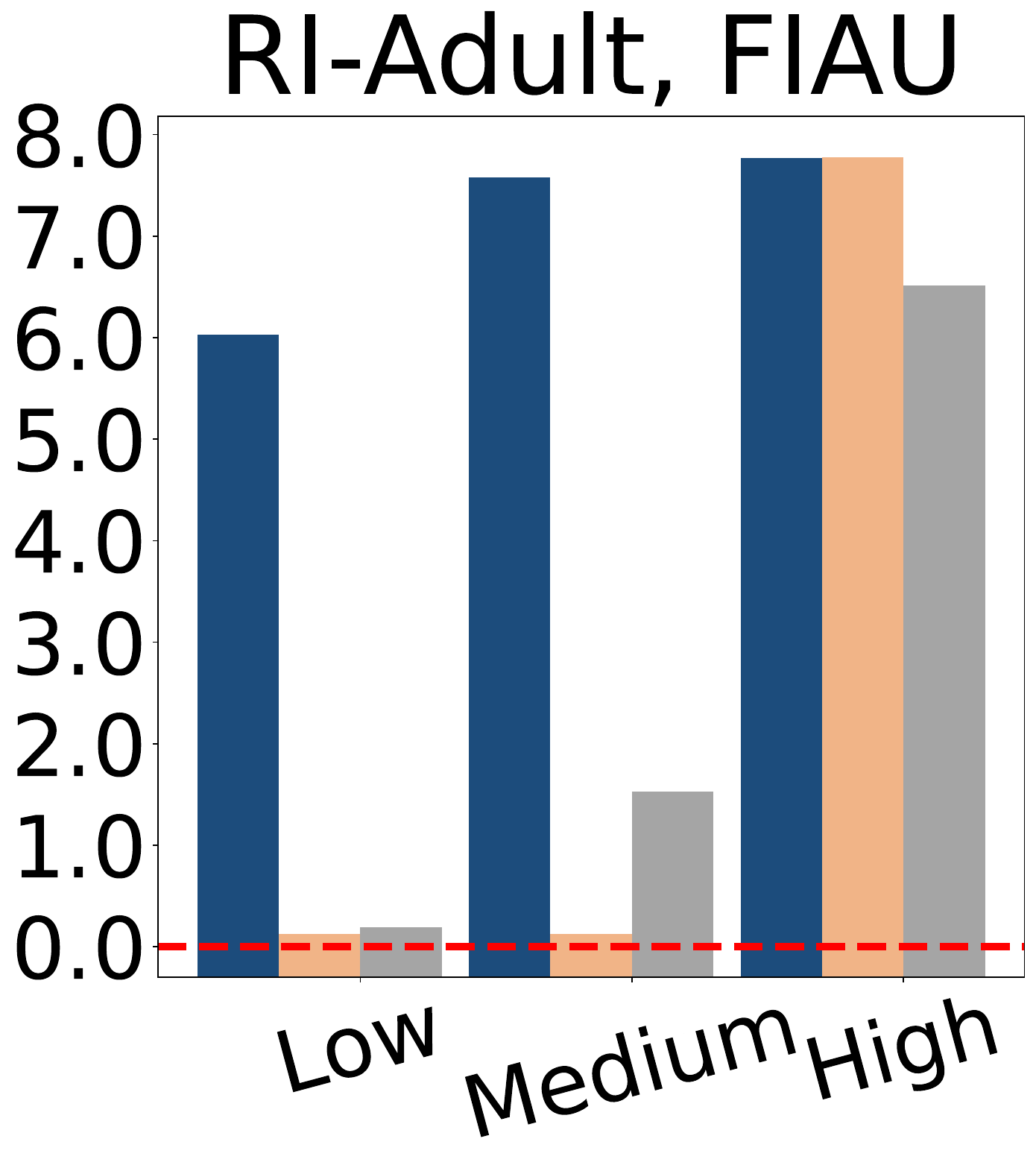}
        \includegraphics[width=0.2\columnwidth, height=0.2\columnwidth]{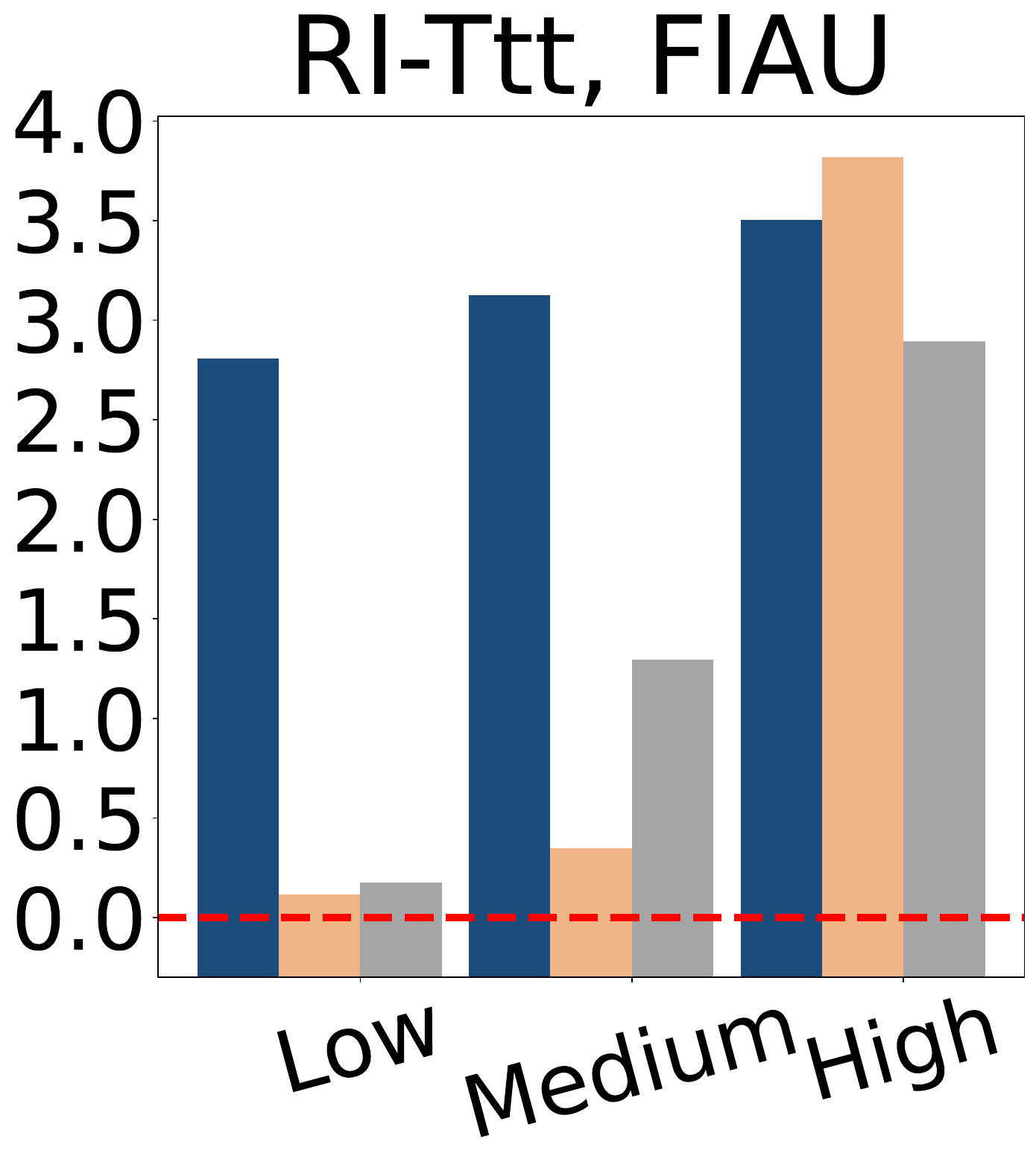}
        \includegraphics[width=0.2\columnwidth, height=0.2\columnwidth]{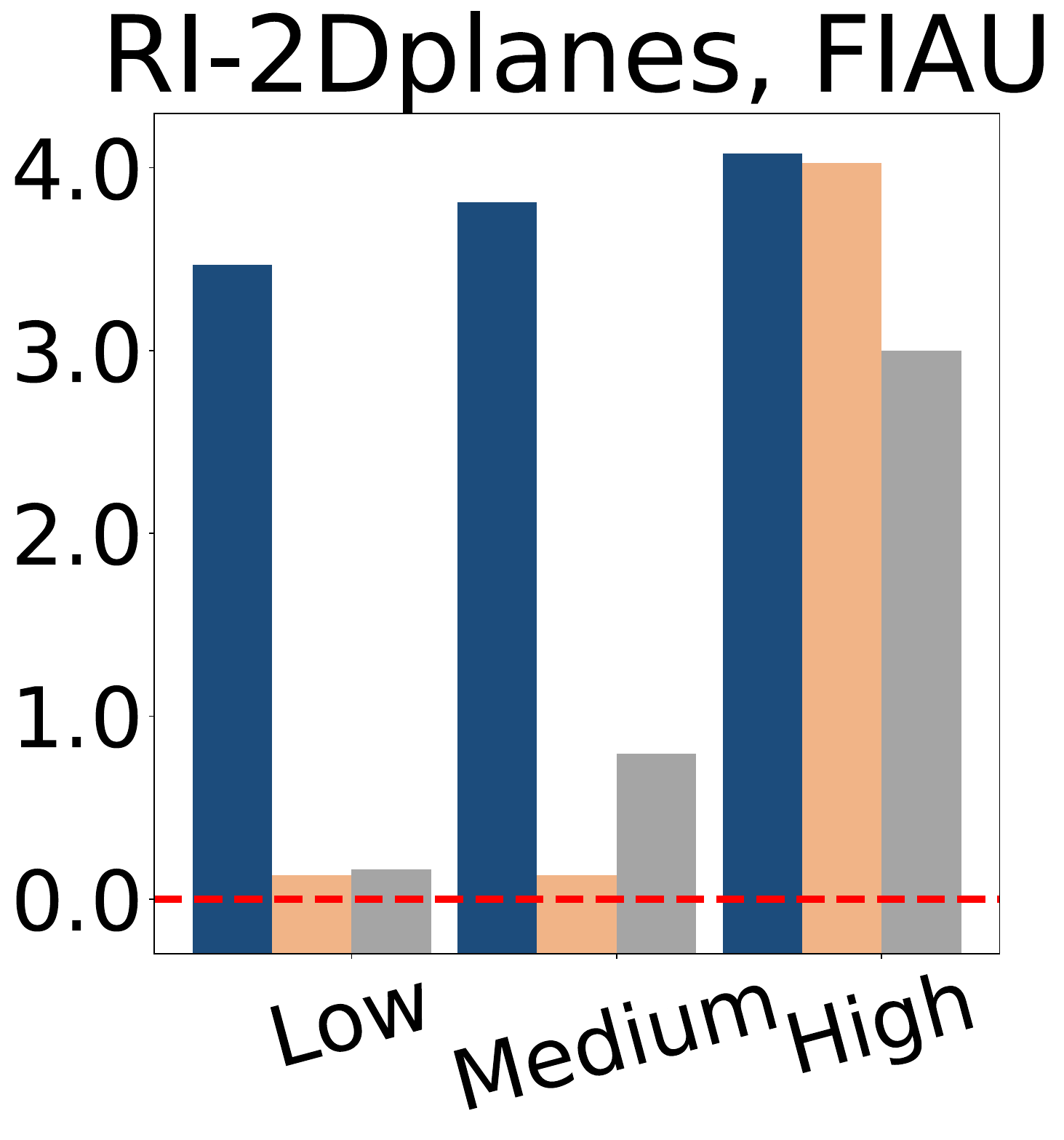}
        \hspace{5pt}
        \includegraphics[width=0.215\columnwidth, height=0.2\columnwidth]{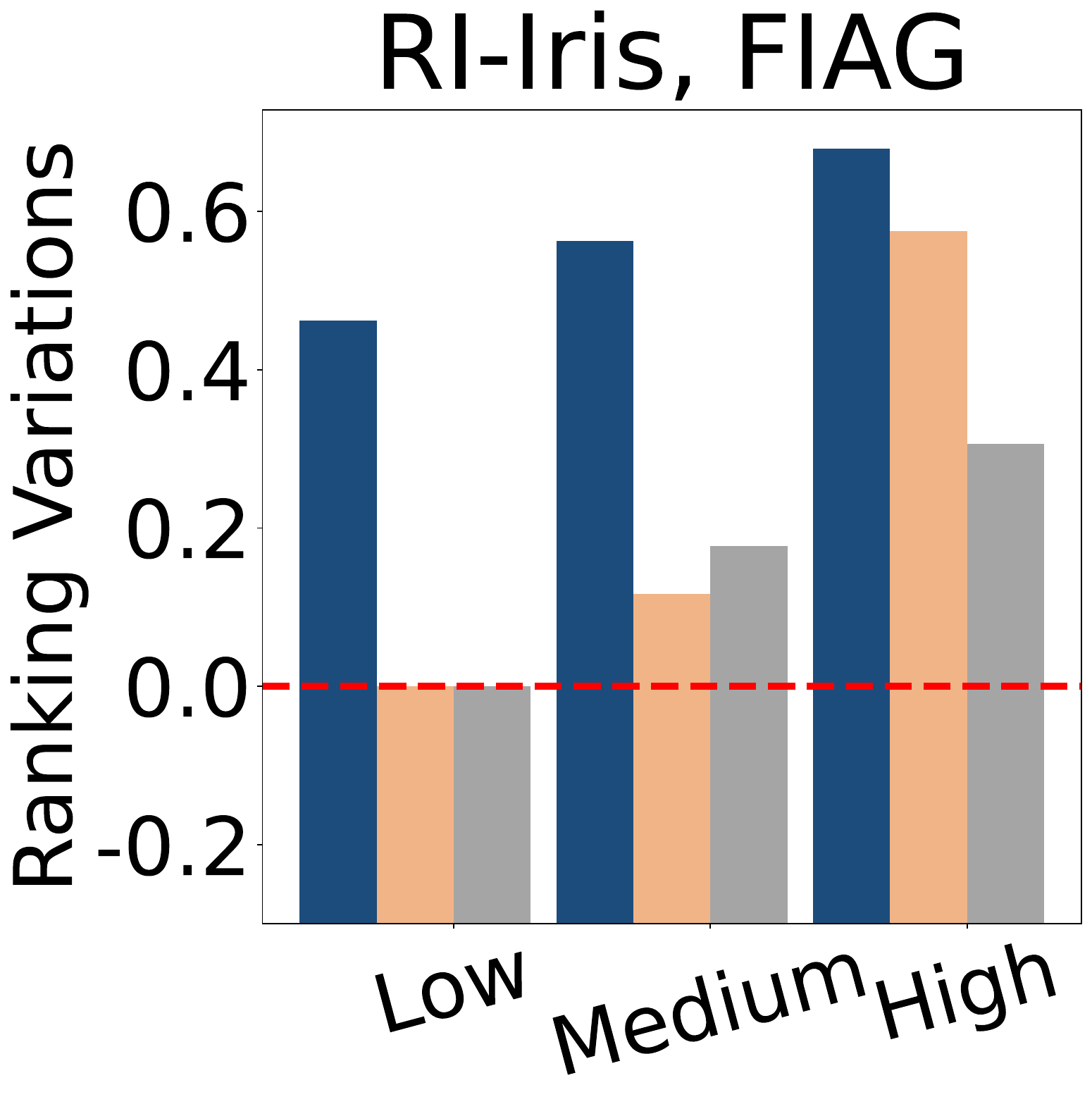}
        \includegraphics[width=0.2\columnwidth, height=0.2\columnwidth]{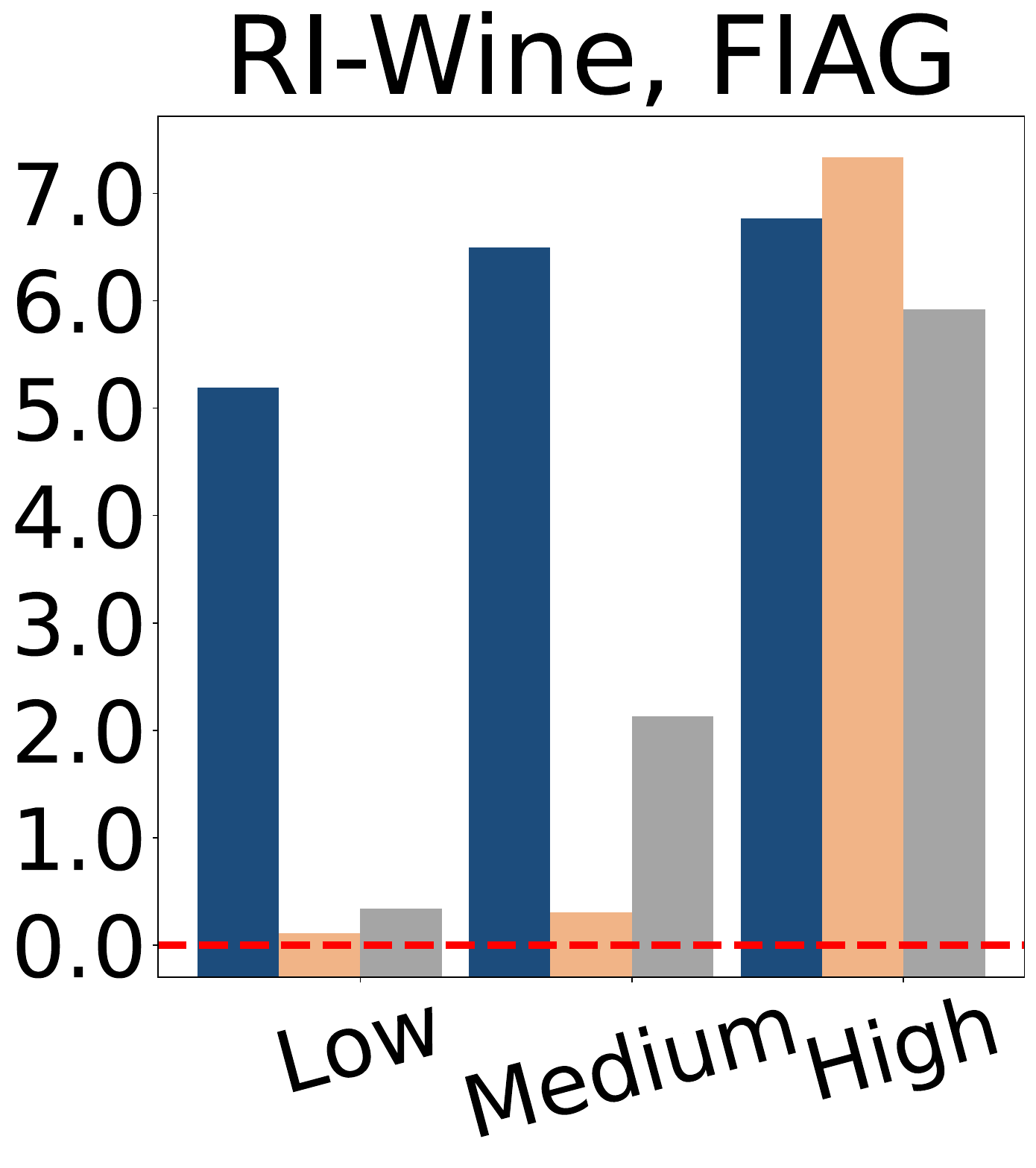}
        \includegraphics[width=0.2\columnwidth, height=0.2\columnwidth]{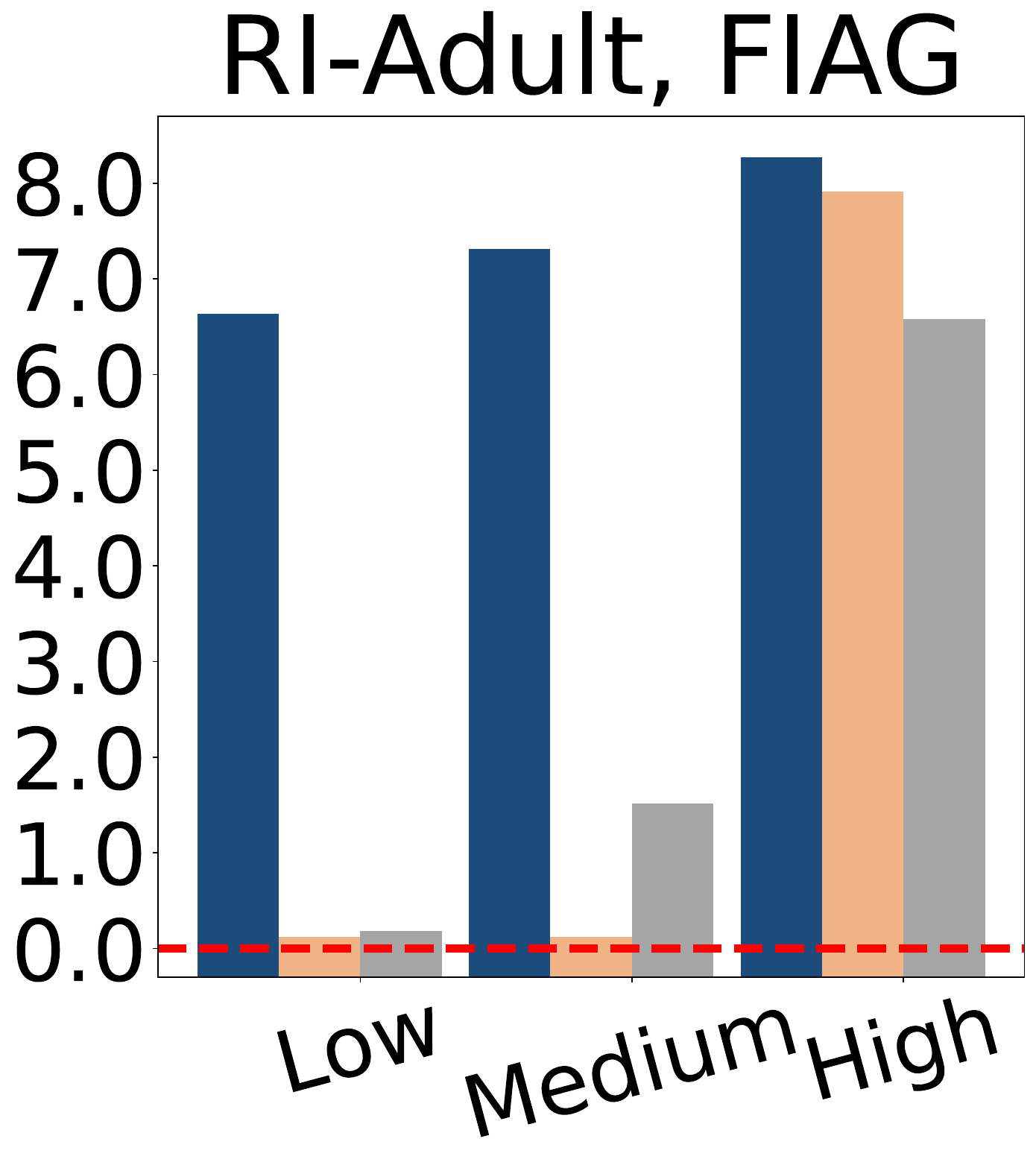}
        \includegraphics[width=0.2\columnwidth, height=0.2\columnwidth]{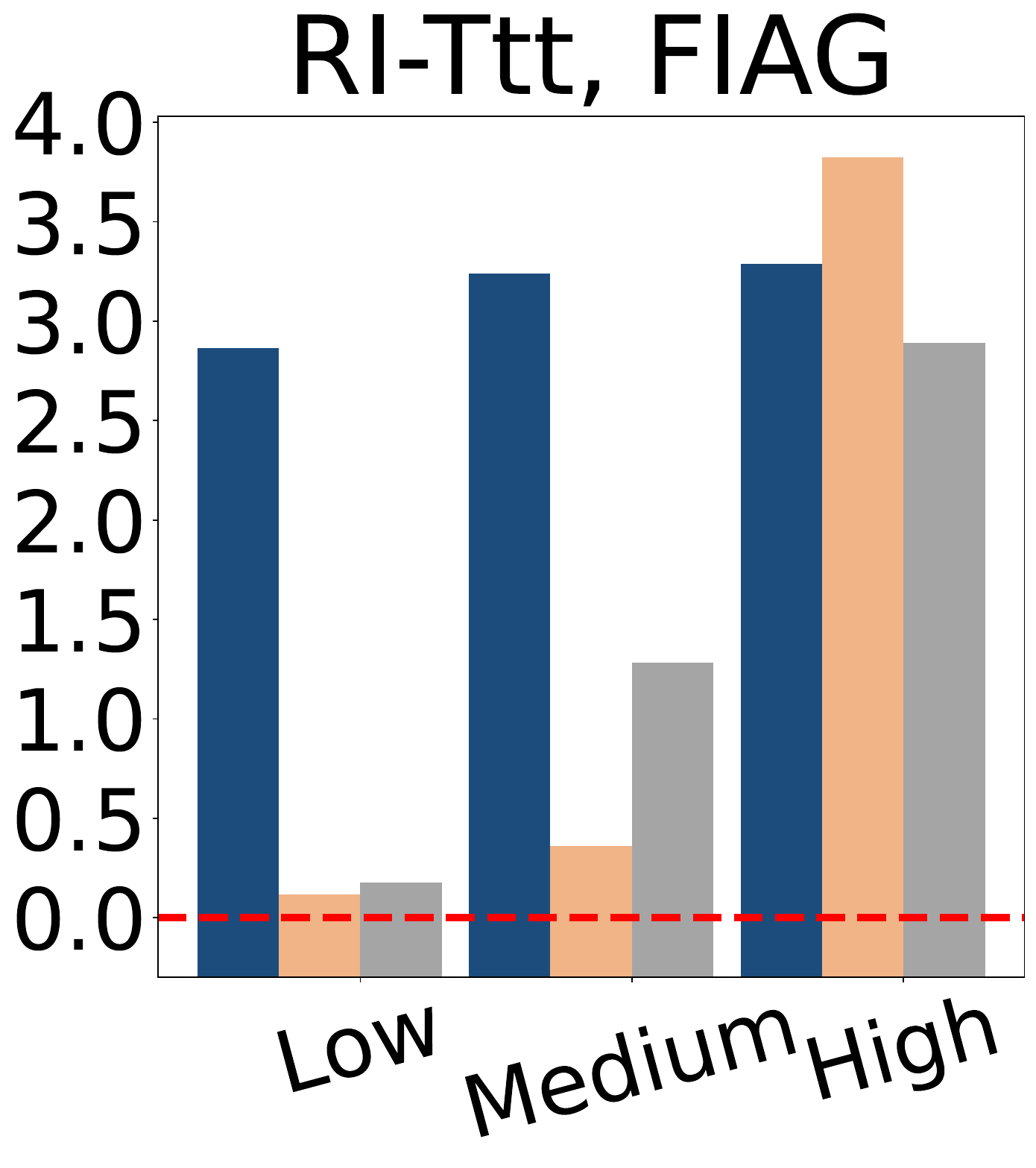}
        \includegraphics[width=0.2\columnwidth, height=0.2\columnwidth]{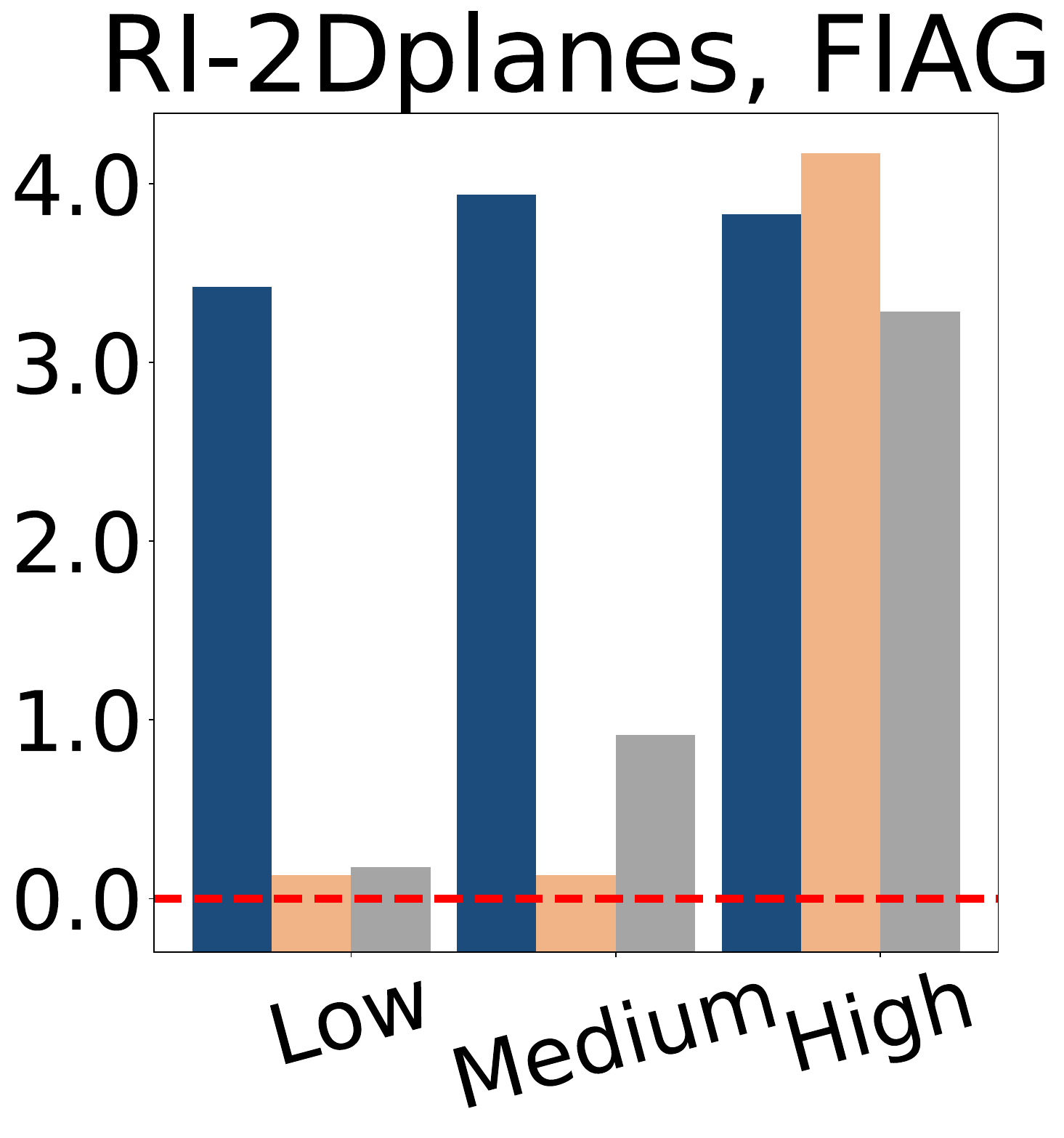}
    }  
    \\
    \mbox{
    \hspace{-10pt}
    \subfigure[The smaller the AUROC, the less privacy the SV exposes. ] {
         \includegraphics[width=0.215\columnwidth, height=0.2\columnwidth]{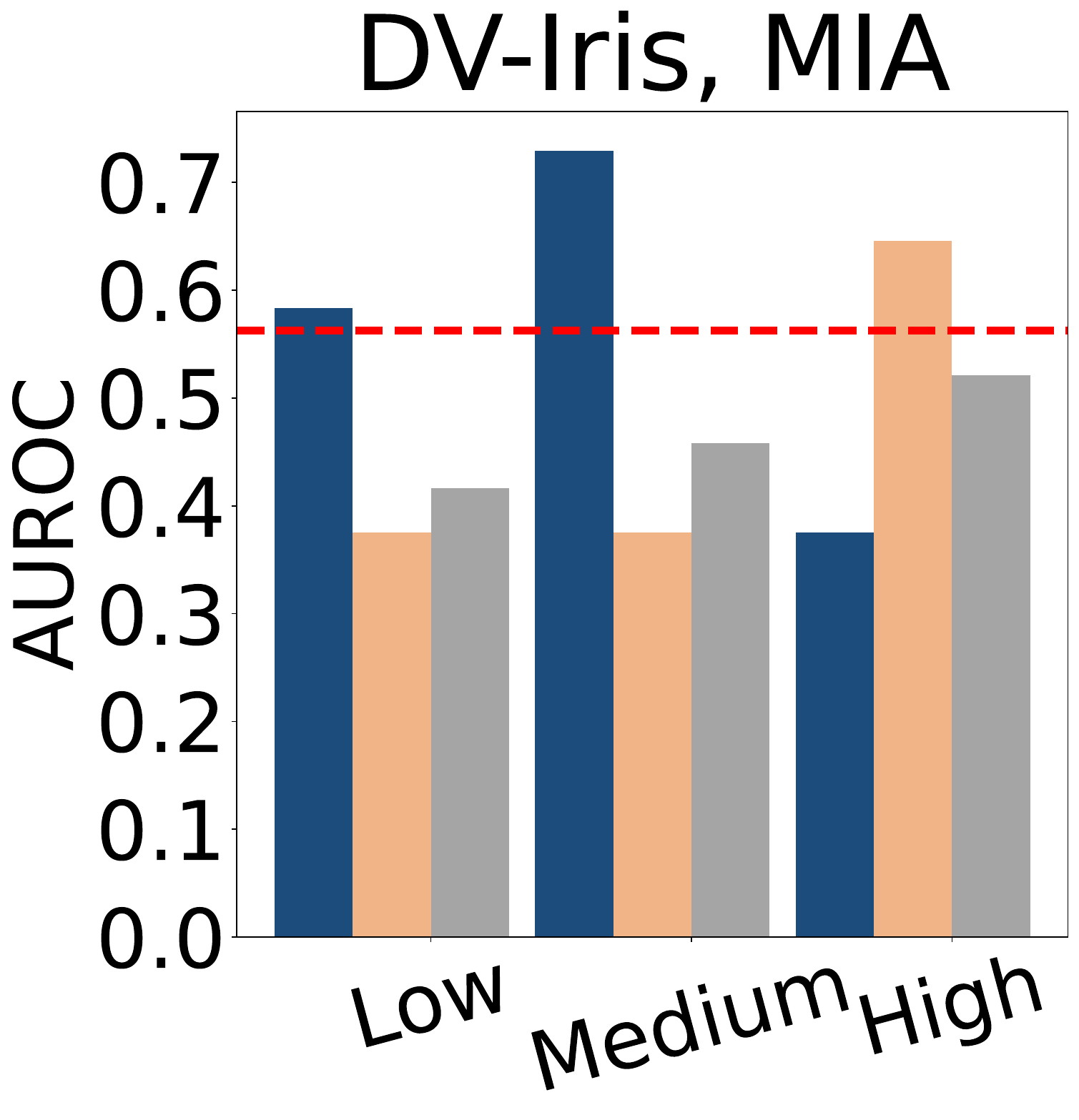}
        \includegraphics[width=0.2\columnwidth, height=0.2\columnwidth]{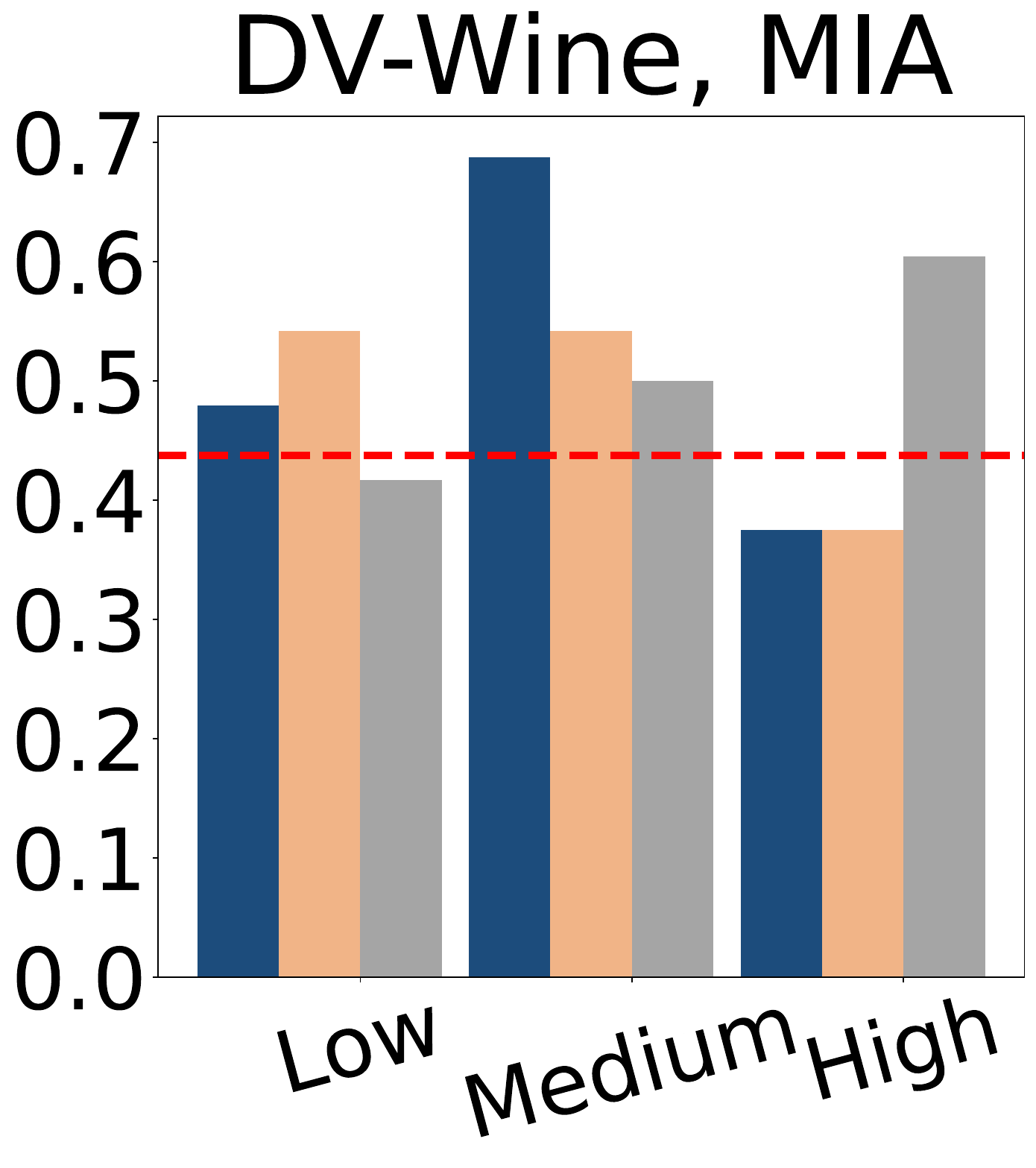}
        \includegraphics[width=0.2\columnwidth, height=0.2\columnwidth]{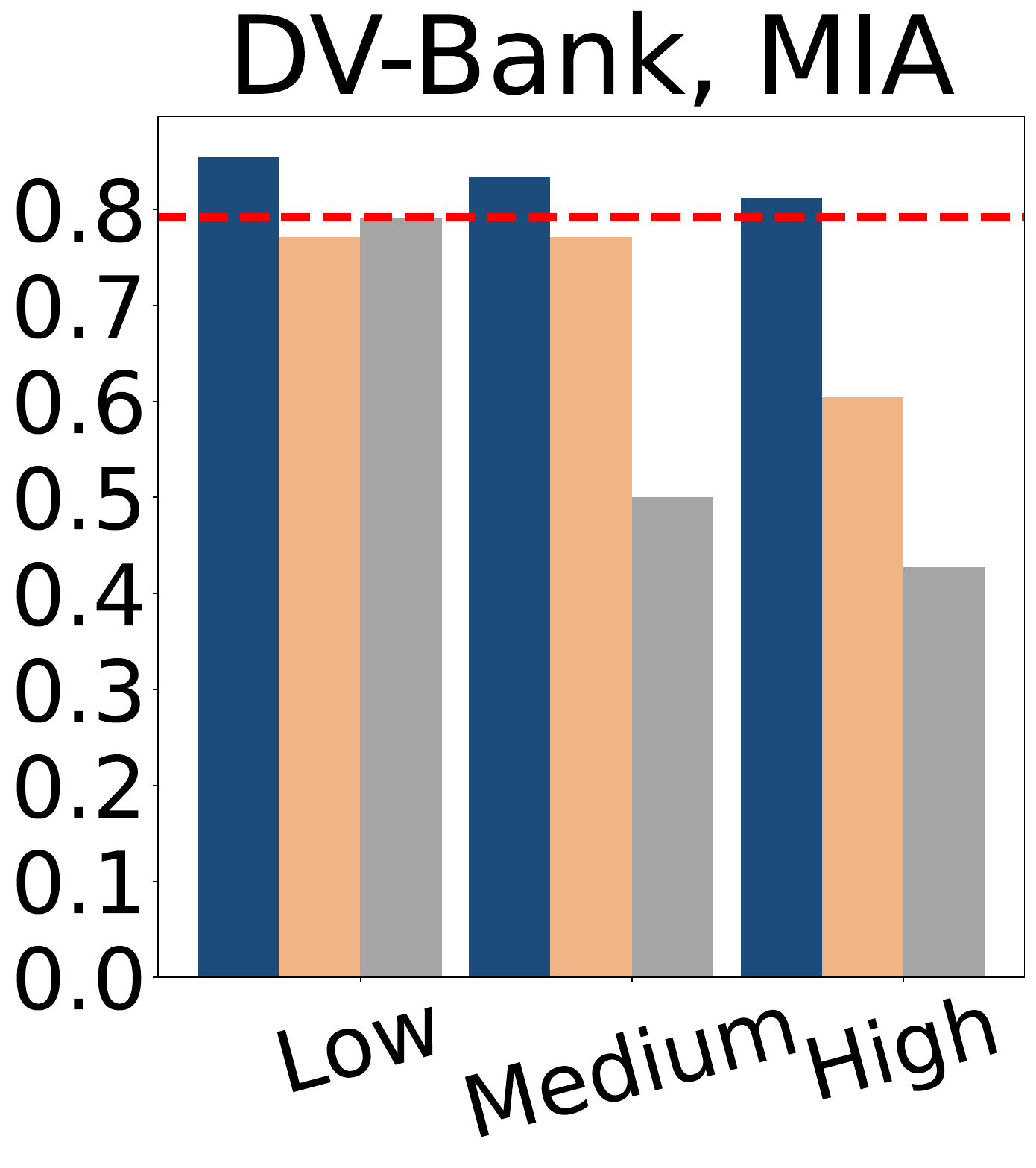}
        \includegraphics[width=0.2\columnwidth, height=0.2\columnwidth]{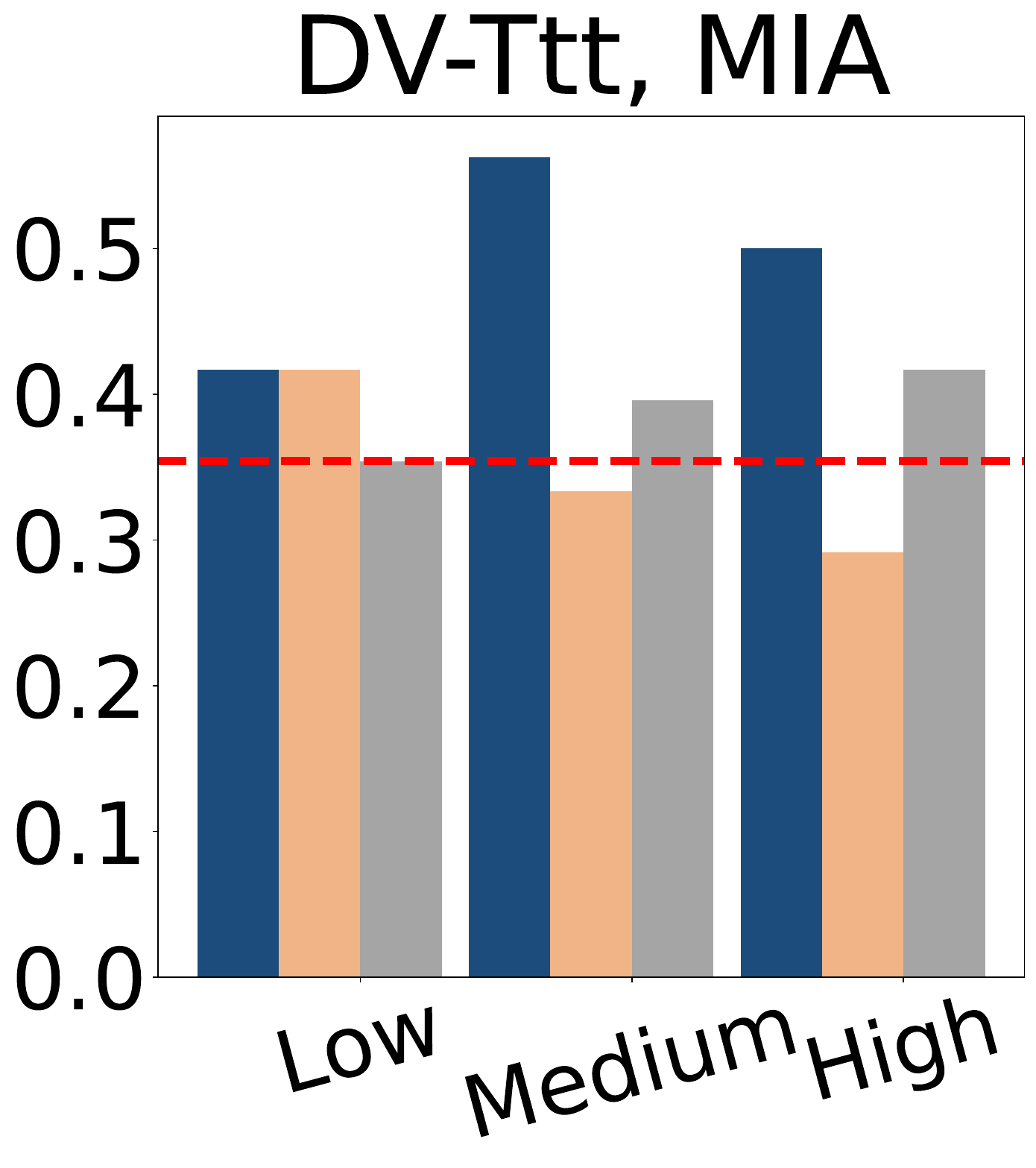}
        \includegraphics[width=0.2\columnwidth, height=0.2\columnwidth]{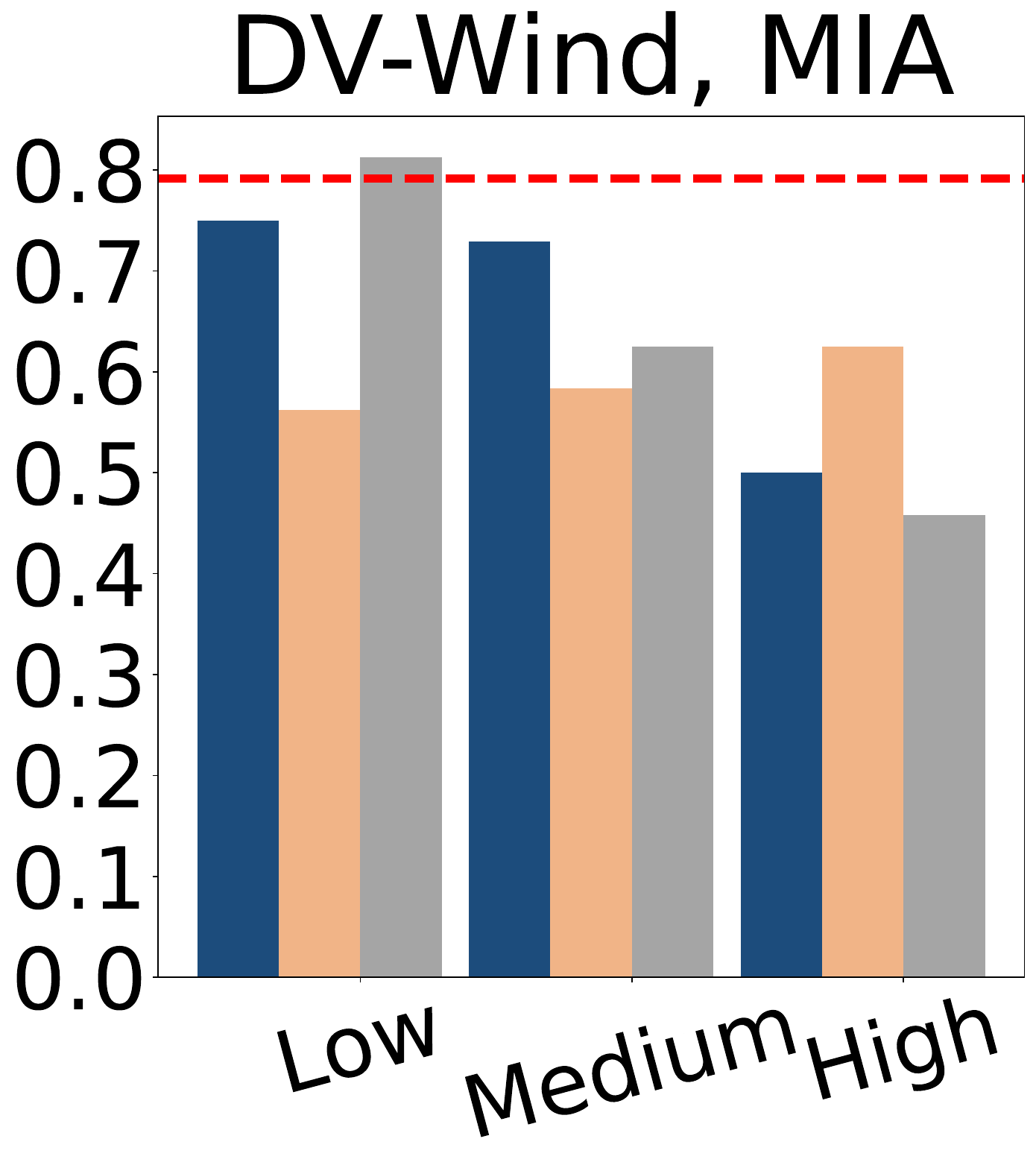}

    }  
    \subfigure[The smaller the variance, the less the impact on SV effectiveness.] {
        \includegraphics[width=0.215\columnwidth, height=0.2\columnwidth]{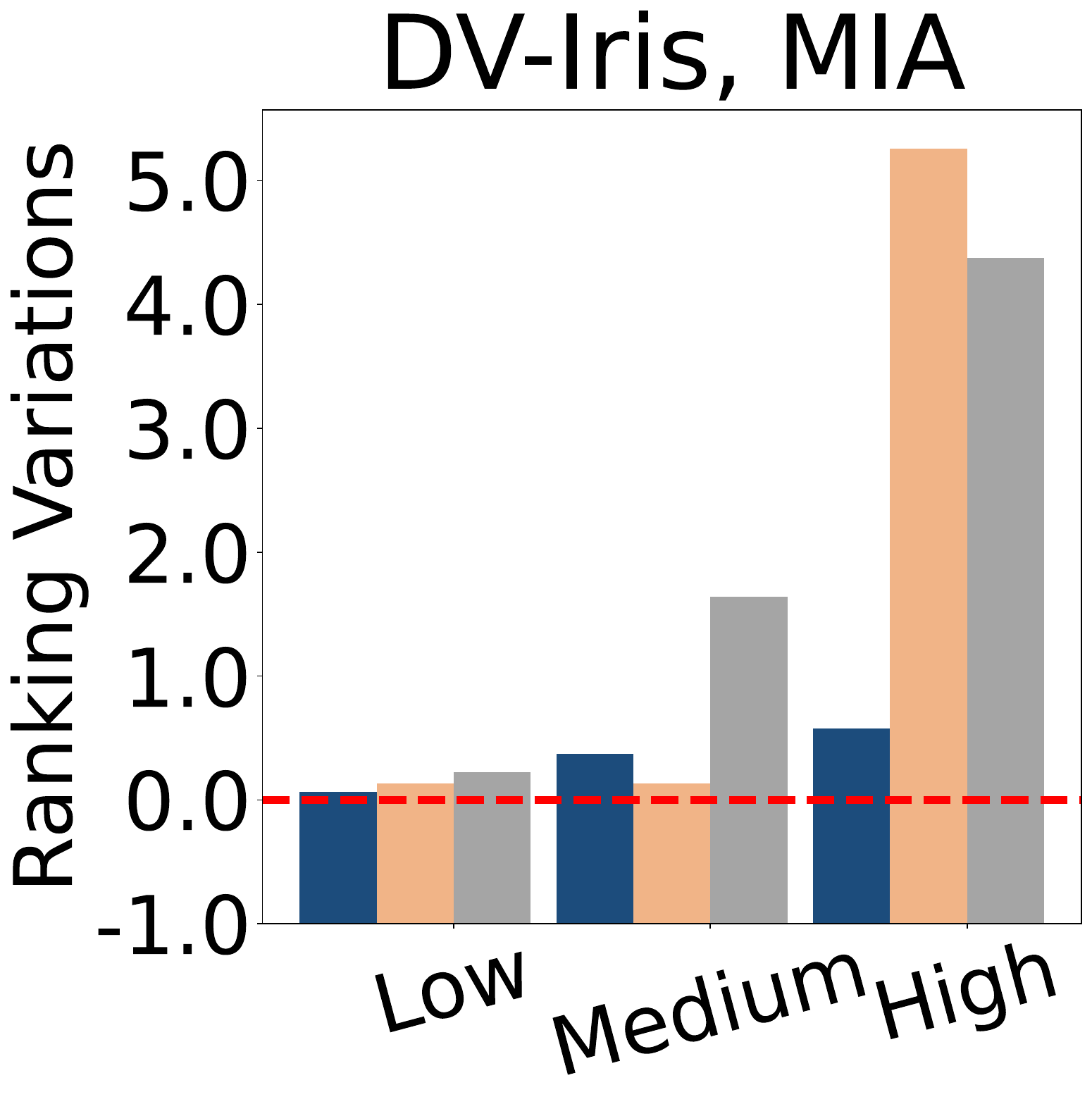}
        \includegraphics[width=0.2\columnwidth, height=0.2\columnwidth]{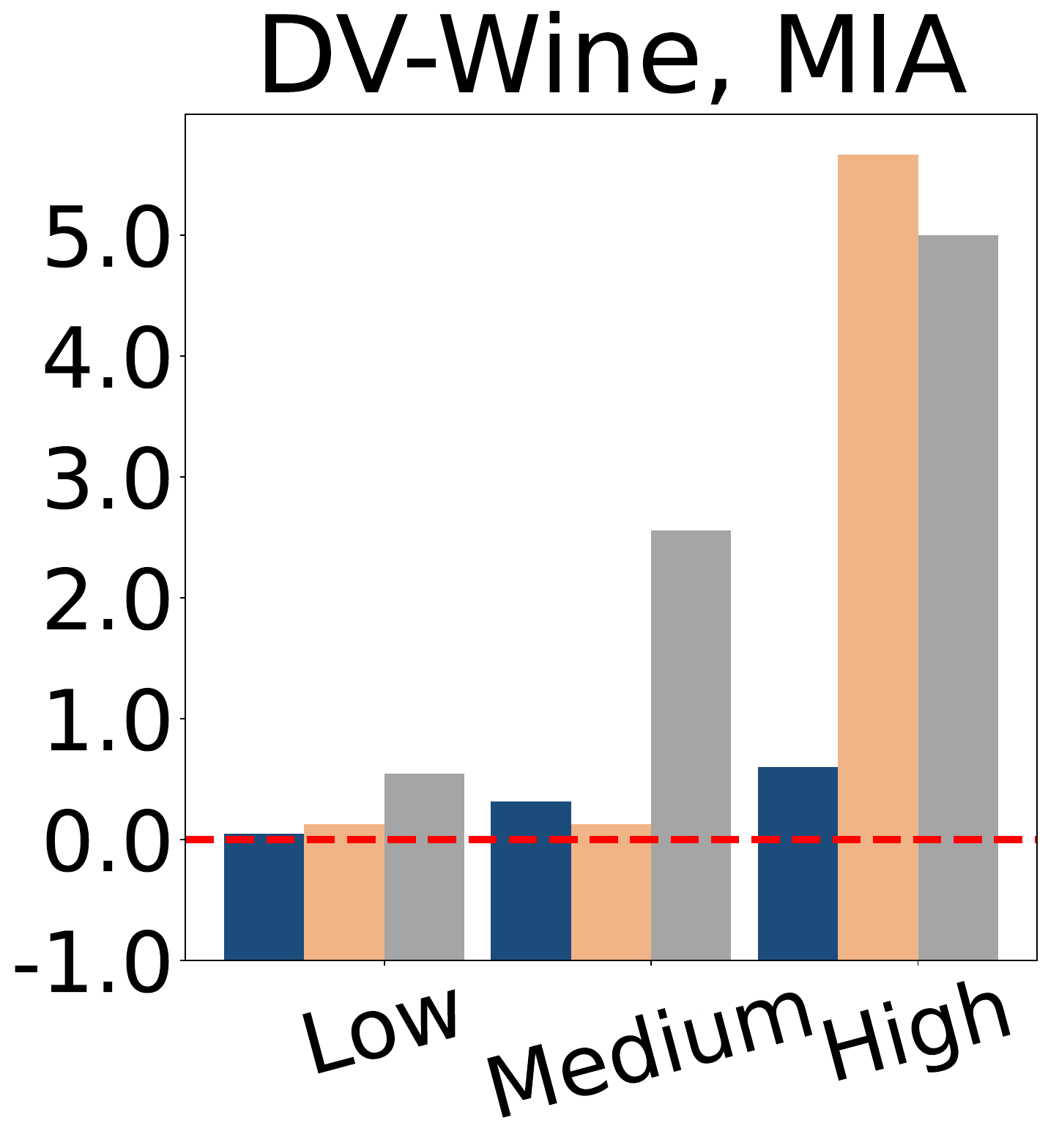}
        \includegraphics[width=0.2\columnwidth, height=0.2\columnwidth]{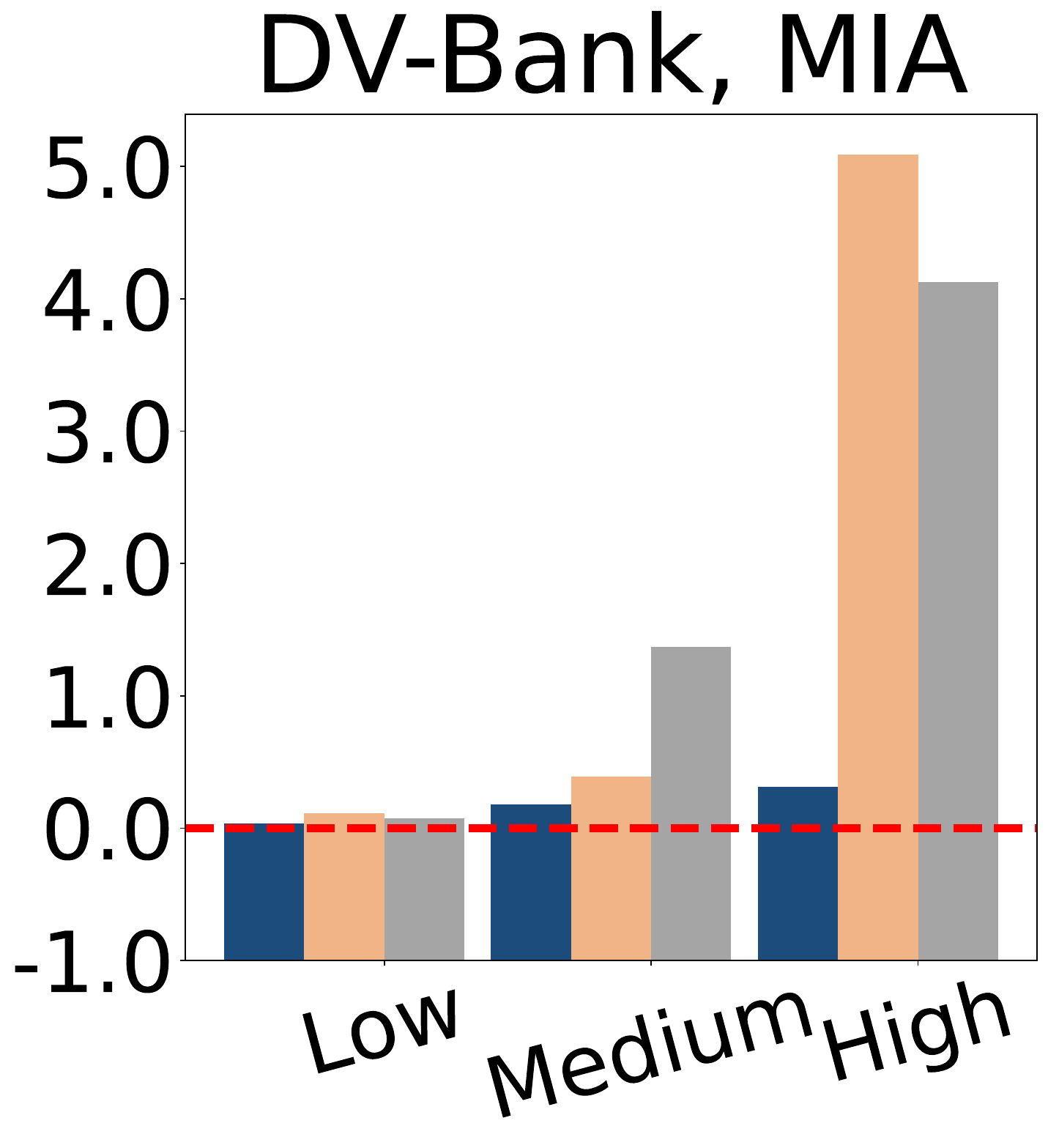}
        \includegraphics[width=0.2\columnwidth, height=0.2\columnwidth]{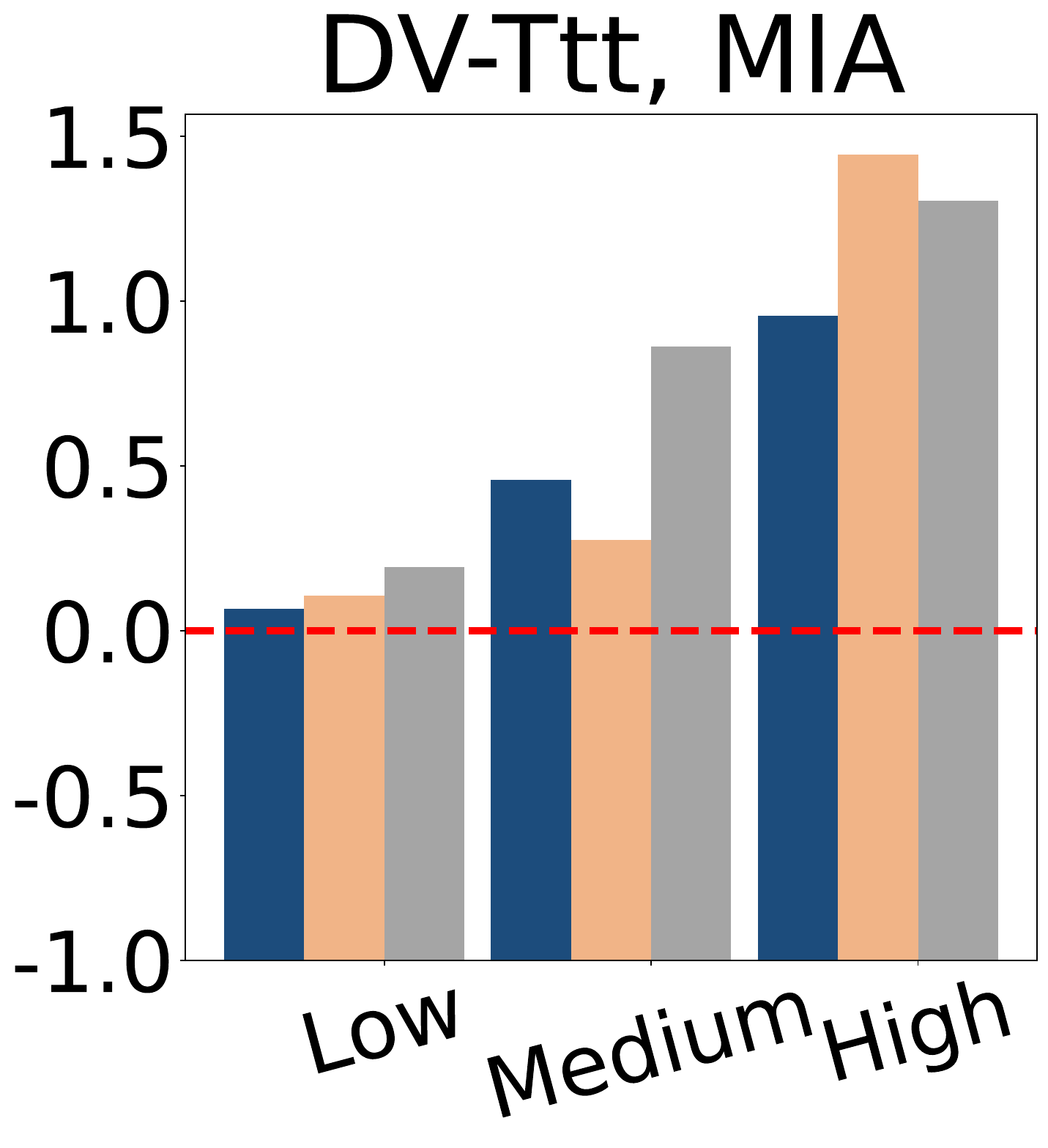}
        \includegraphics[width=0.2\columnwidth, height=0.2\columnwidth]{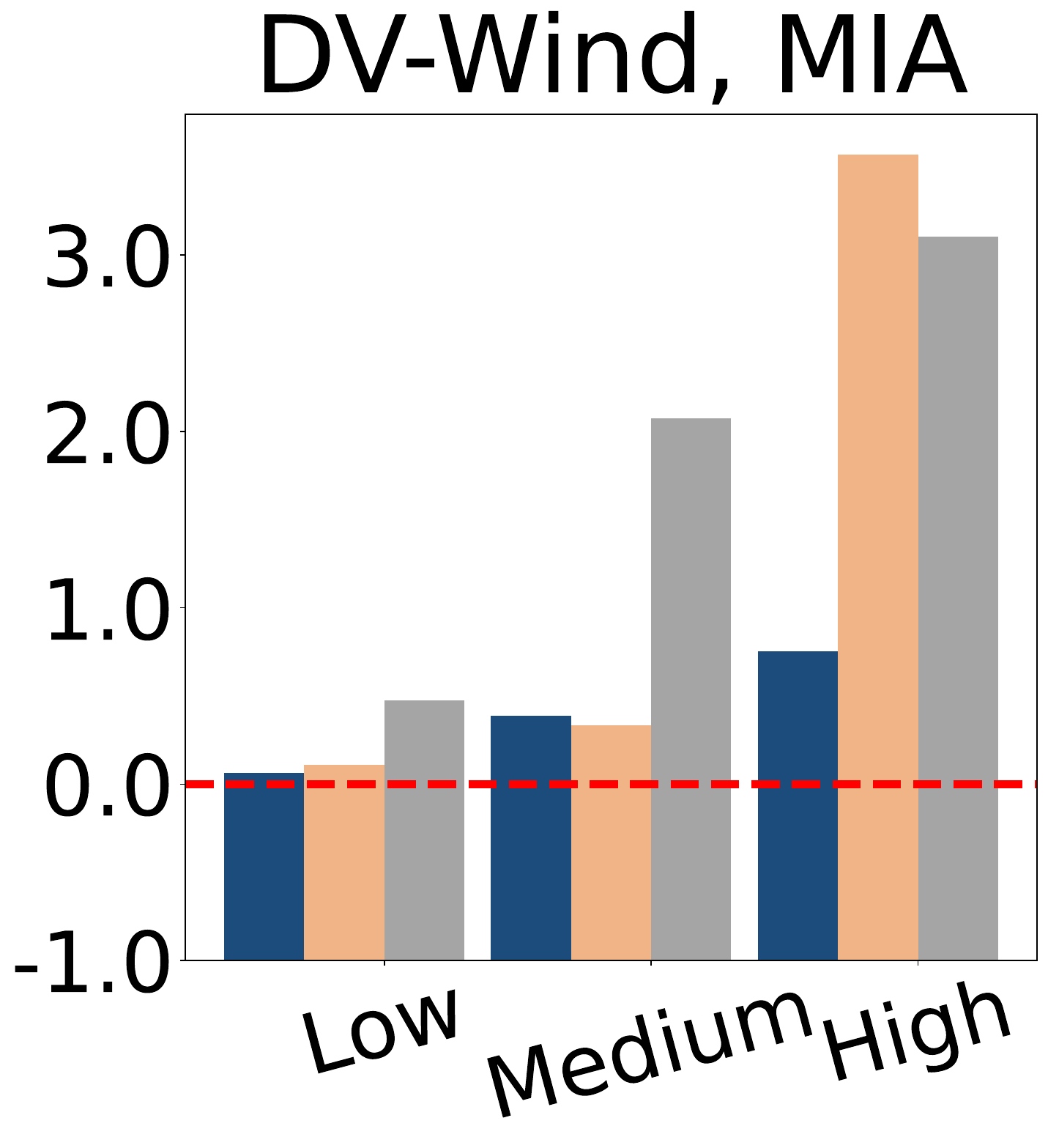}

    }  
    }
    \caption{Preventing SV-driven FIA and MIA (x-axis:  privacy protection strength). 
    }
    \label{fig:privacy_protection}
\end{figure*}

\subsection{Privacy Protection} \label{subsec: exp_privacy}

This section studies the effectiveness of three popular privacy protection techniques (\textit{DP}, \textit{QT}, \textit{DR}) for preventing SV-driven attacks, including \textit{FIA} \cite{Luo2022FeatureIA}, which may occur in RI tasks, and \textit{MIA} \cite{wang2023threshold}, which may occur in DV tasks. We also study the impacts of the three techniques on SV's effectiveness. 

The implementation of two attacks and three countermeasures is based on previous papers \cite{Luo2022FeatureIA, wang2023threshold}. 
We follow \citet{Luo2022FeatureIA} to use two random distributions, Uniform(0,1) and Gaussian(0.5,$0.25^2$), to perform FIA (called FIAU and FIAG respectively), and follow  \citet{wang2023threshold} to implement MIA. 
We tune the strength of privacy protection from low to high by varying (1) the standard deviation of noise generated by DP from 0.1, 0.5 to 0.9, (2) the number of distinctive discrete Shapley values produced by QT from $0.9n$, $0.5n$ to $0.1n$, and (3) the number of dimensions reduced for $\vec{\phi}$ by DR from $0.1n$, $0.5n$ to $0.9n$. We measure the performance in preventing SV-driven FIA by the MAE metric used by \citet{Luo2022FeatureIA} and measure the effectiveness of preventing SV-driven MIA by the AUROC score used by \citet{wang2023threshold}. 
As both RI and DV tasks rely on the ranking of the final SVs to find the top important data features and samples, the impact of privacy protection techniques on SV effectiveness is measured by the variance in SV ranking results before and after protection.

Figure~\ref{fig:privacy_protection} shows the results of preventing the two attacks.
In most cases, the stronger the strength of privacy protection techniques, the more effective those techniques are. 
However, stronger privacy protection renders a larger impact on SV ranking, affecting more on the effectiveness of SV for pricing, selection, weighting, or attribution.
All these results consolidate the necessity of a trade-off between the prevention of SV-driven privacy leakage and the effectiveness of SV for decision-making in DA.

For preventing FIA, DP is the most effective, while QT and DR need tuning the privacy protection strength to a high level for an apparent MAE enlargement. 
Despite this, DP renders a much larger impact on SV ranking (thus affects more on SV's effectiveness) under the setting of low or median privacy protection strength. 
These phenomena are mainly because DP adds noise to SVs of all features no matter how the privacy protection strength is varied, while QT and DR alter SVs of only 10\% (or 50\%) data features when the strength is set to a low (or median) level.

For preventing MIA, QT and DR are effective in most cases, while DP leads to AUROC larger than the results achieved without privacy protection in some cases, e.g., DV-Bank, DV-Ttt. 
Moreover, different from the results of preventing FIA, enhancing the privacy protection strength may not result in better effectiveness (lower AUROC) of the three techniques in preventing MIA. For example, in DV-Wind, QT performs the worst when setting a high privacy protection strength. 
These outcomes stem from the inability of QT, DR, and DP to rigorously enforce indistinguishability between two Shapley value distributions: the IN distribution, computed when the target sample is included in the dataset, and the OUT distribution, computed when the target sample is excluded \cite{wang2023threshold}. 
Specifically, due to randomness, DP may inject markedly dissimilar noise into SVs used for generating IN and OUT distributions and thus enlarge the difference between the two distributions, making the success of MIA much easier. 
QT and DR map the original SVs into a new value space, in which the difference between SVs used for generating IN and OUT distributions might be more distinguishable, negatively affecting the efficacy in preventing SV-driven MIA.

\textbf{Research Direction 4: Exploration on innovative techniques for tackling SV-driven privacy issues.} For practical usage of existing privacy protection measures, we suggest \textit{adjusting the strength parameter of the chosen measure}, such as the standard deviation for noise generation in DP, \textit{to its median value} to achieve the privacy-effectiveness balance. Meanwhile, we strongly advocate for \textit{innovative privacy protection measures} that prevent SV-driven attacks without compromising the computation efficiency and effectiveness of SV. 

\textbf{Research Direction 5: Exploration on new attack patterns.} Plenty of SV-driven privacy issues remain unexplored. Except for FIA and MIA, malicious adversaries may utilize SV to launch other types of attacks, such as the \textit{model extraction attack} \cite{10.1145/3624010}, reconstructing a model learned in the DA workflow by creating a substitute model that behaves very similarly in SV-based evaluations to the model under attack.
Another example is the \textit{data property inference attack} \cite{10.1145/3624010}, extracting the properties implicitly encoded as features of the dataset.

\begin{figure*}[t]
    \centering
    \begin{minipage}{0.7\columnwidth}
        \centering \includegraphics[width=\columnwidth]{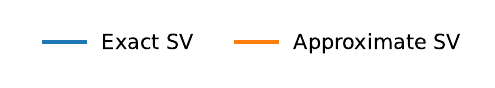}
    \end{minipage}
    \vspace{-10pt}
    \\
    \mbox{
        \includegraphics[width=0.215\columnwidth, height=0.2\columnwidth]{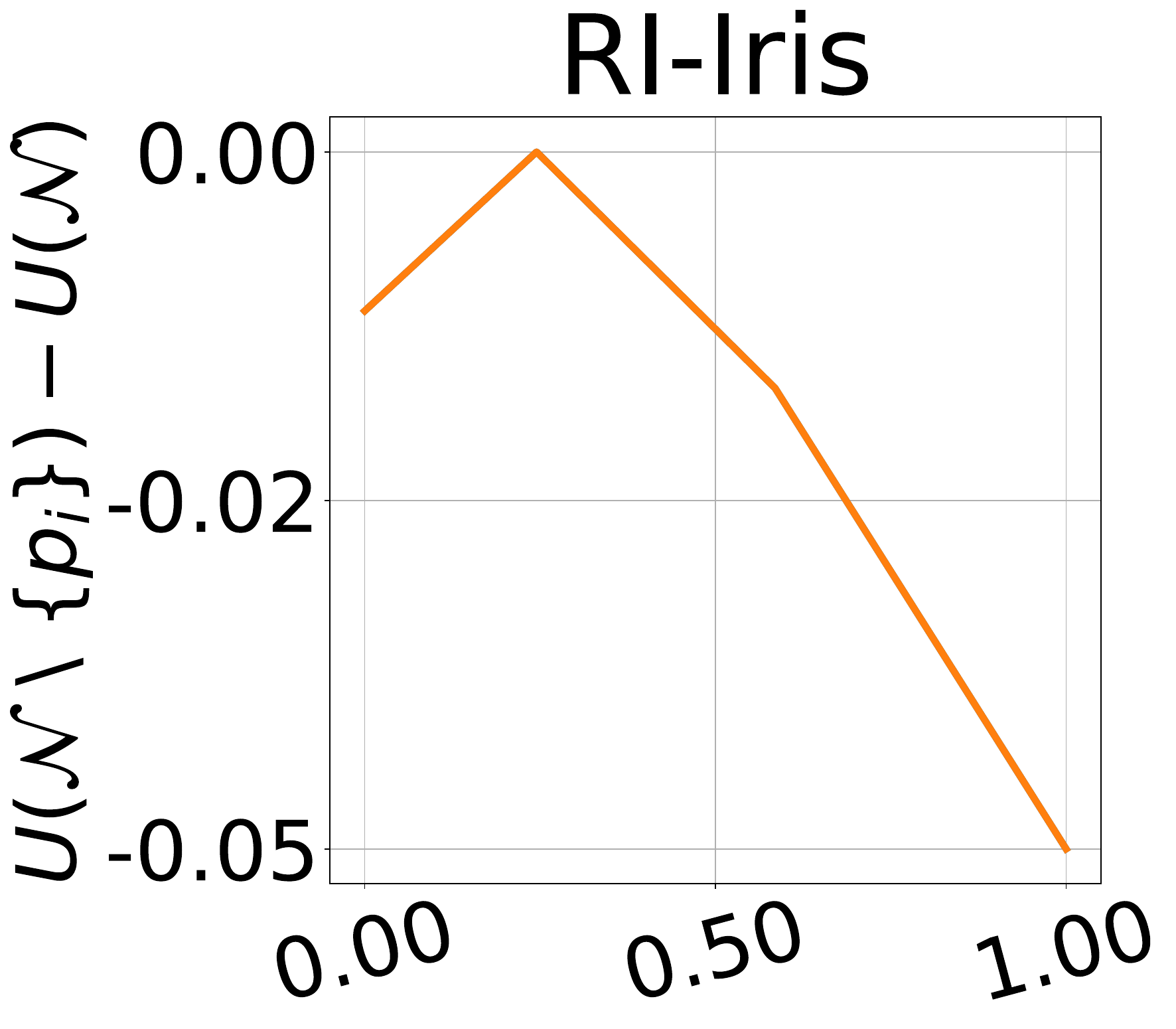}
        \includegraphics[width=0.2\columnwidth, height=0.2\columnwidth]{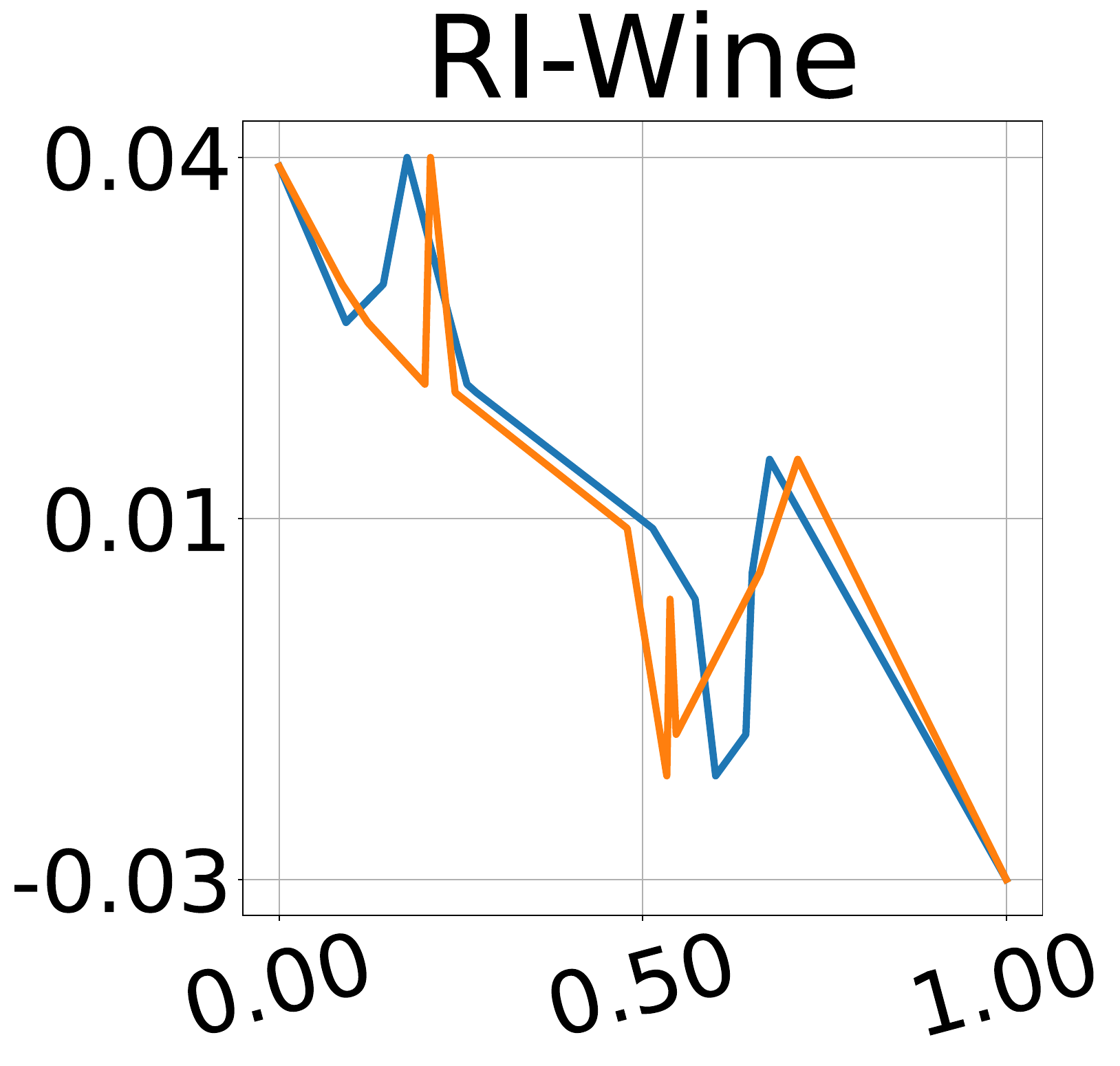}
        \includegraphics[width=0.2\columnwidth, height=0.2\columnwidth]{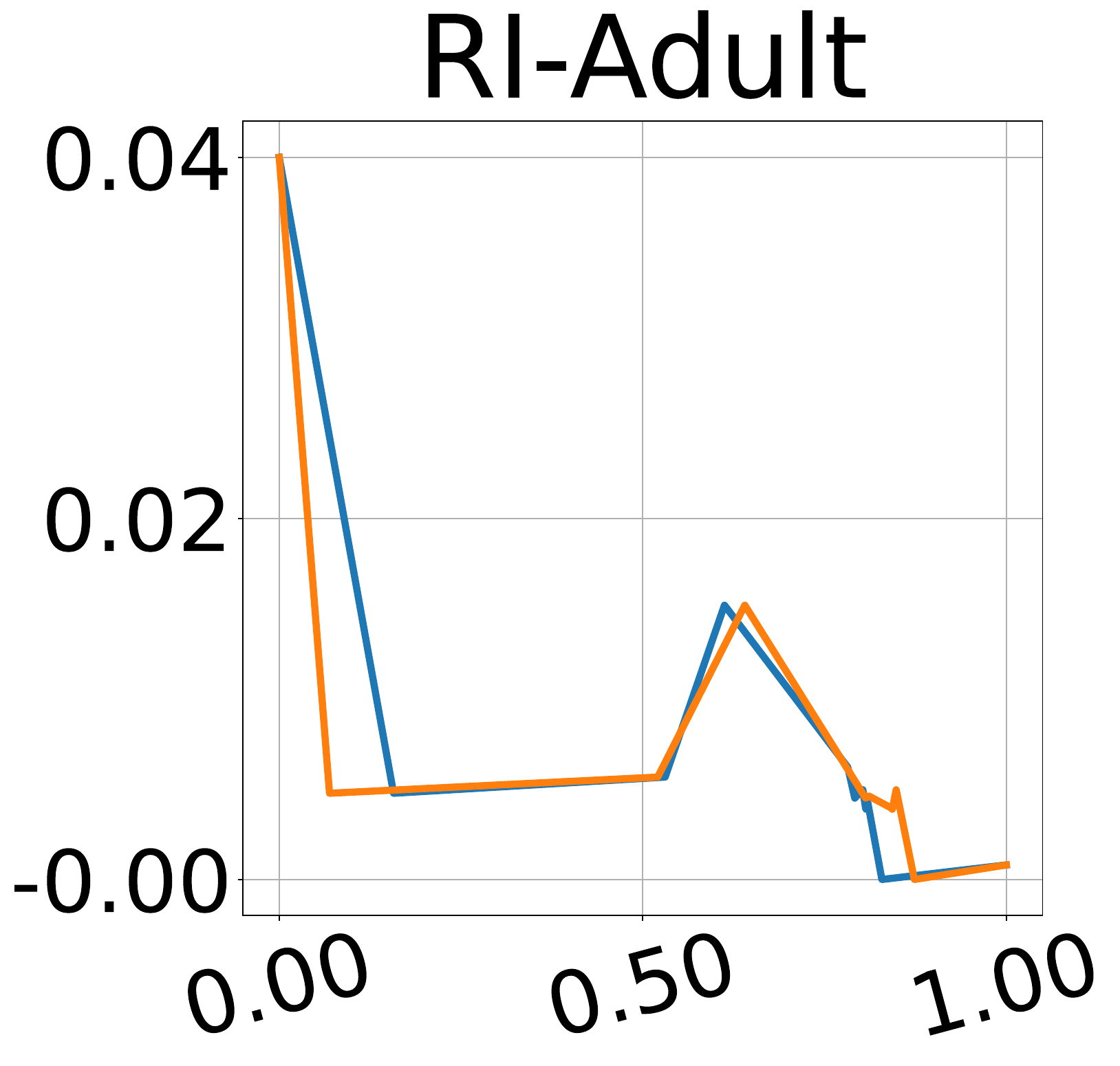}
        \includegraphics[width=0.2\columnwidth, height=0.2\columnwidth]{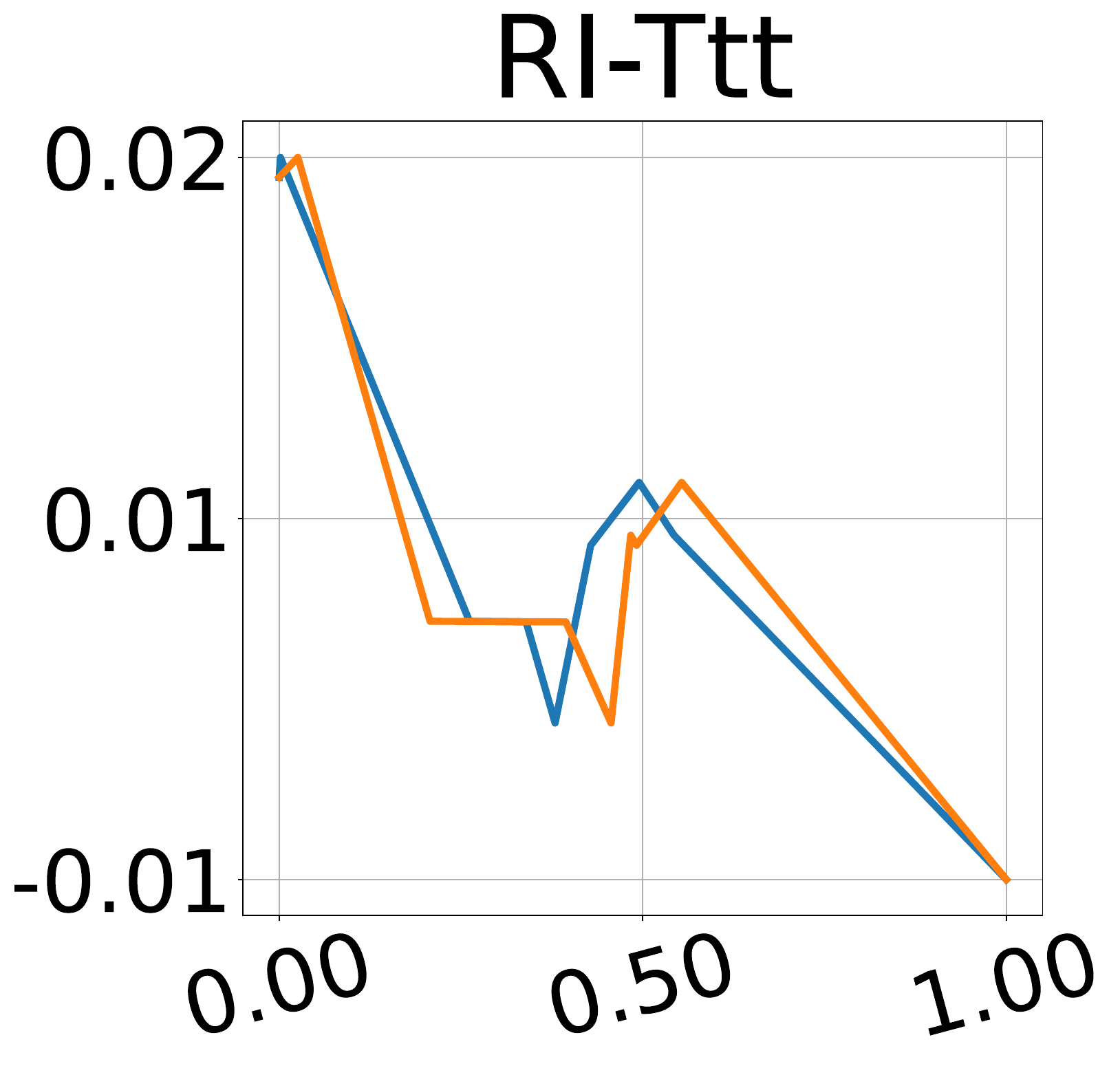}
        \includegraphics[width=0.2\columnwidth, height=0.2\columnwidth]{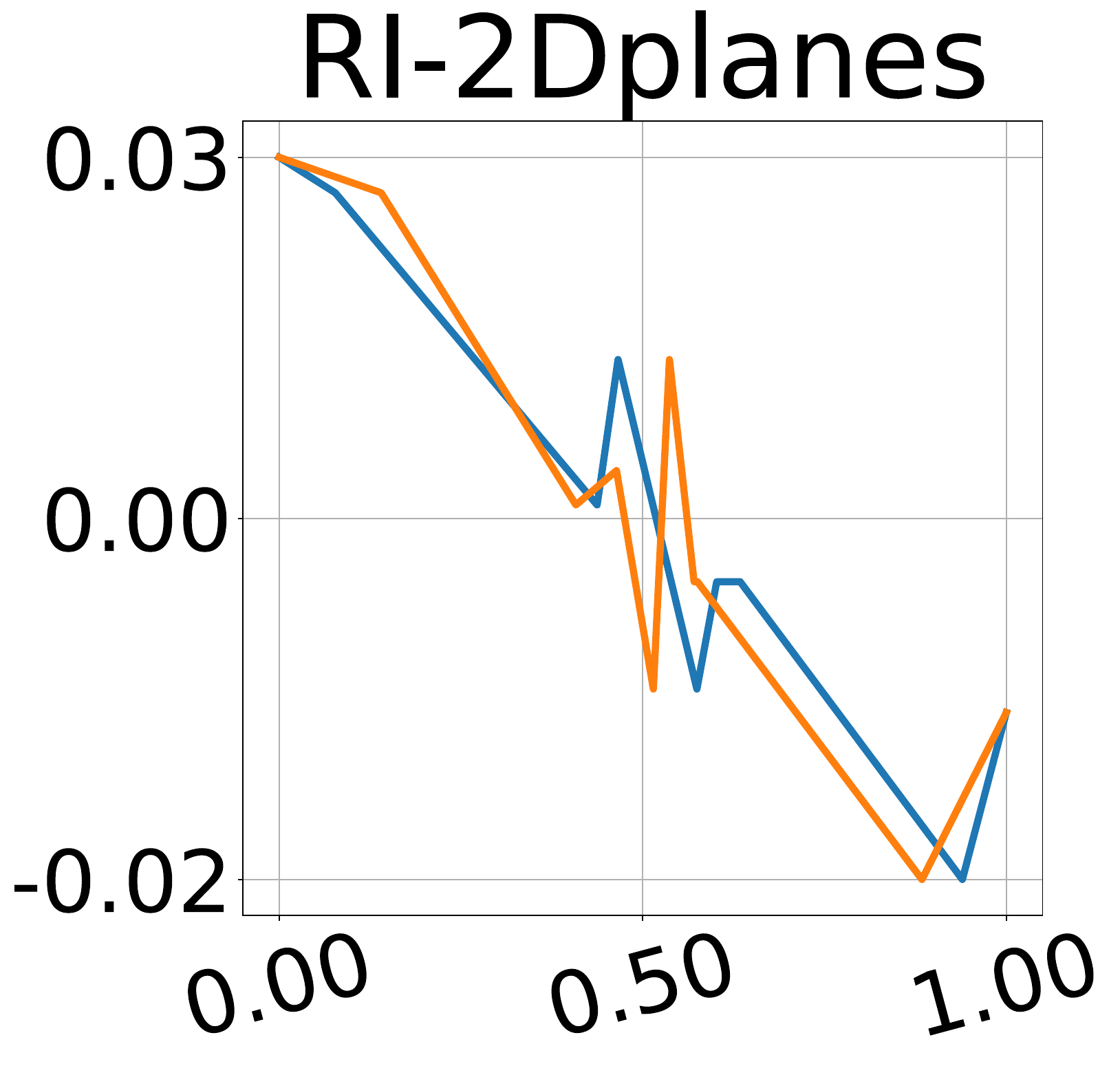}

        \includegraphics[width=0.2\columnwidth, height=0.2\columnwidth]{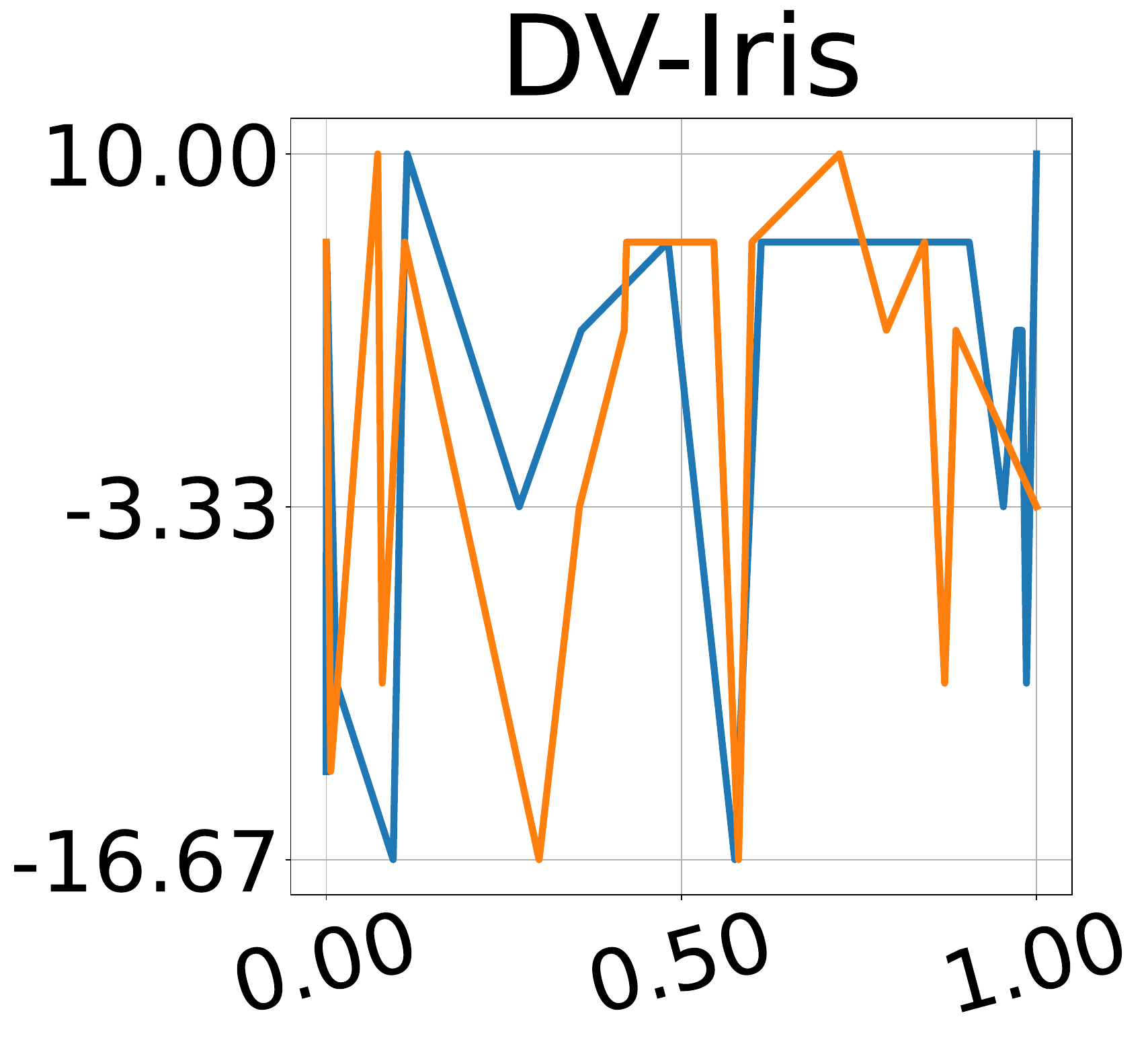}
        \includegraphics[width=0.2\columnwidth, height=0.2\columnwidth]{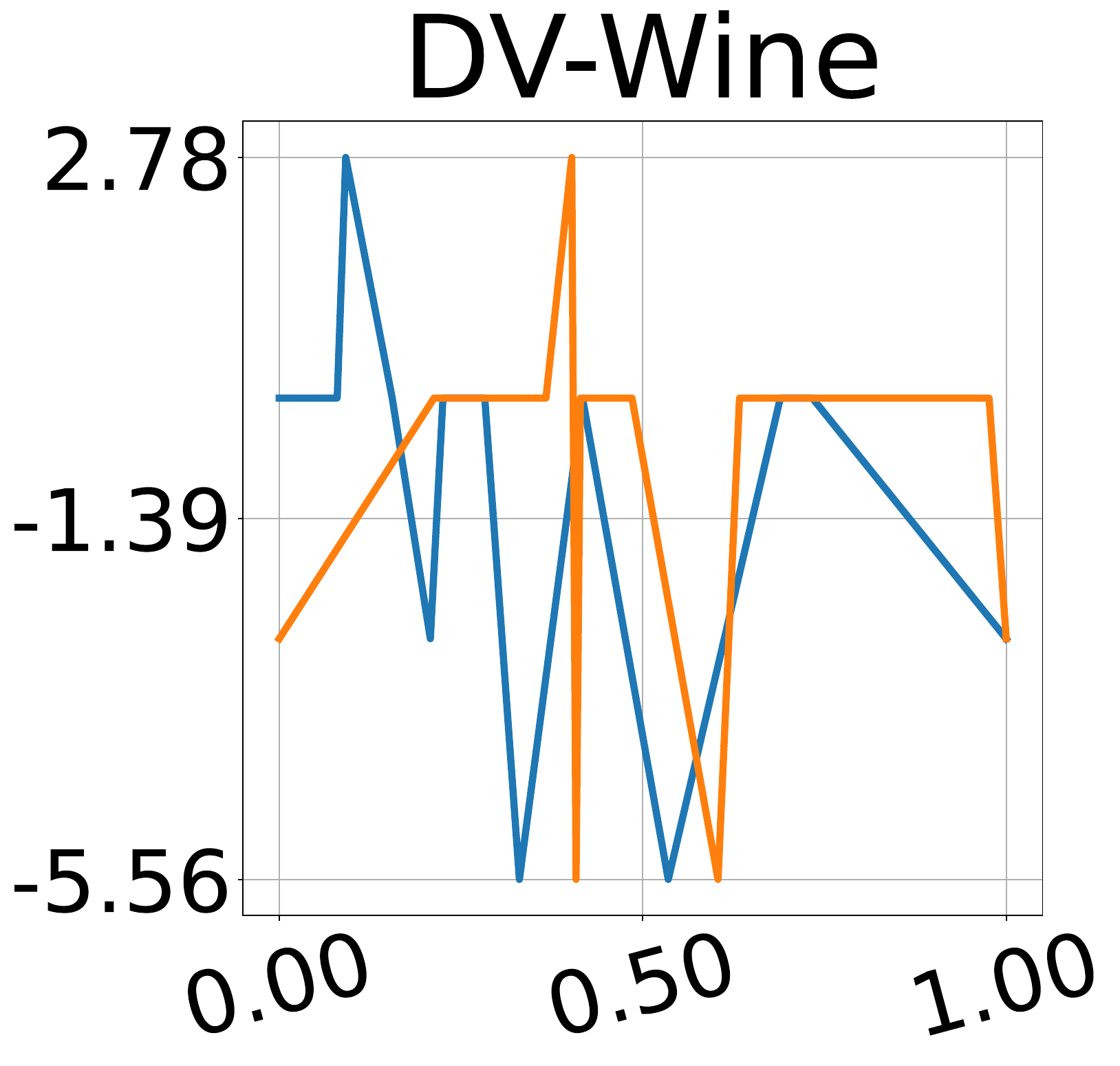}
        \includegraphics[width=0.2\columnwidth, height=0.2\columnwidth]{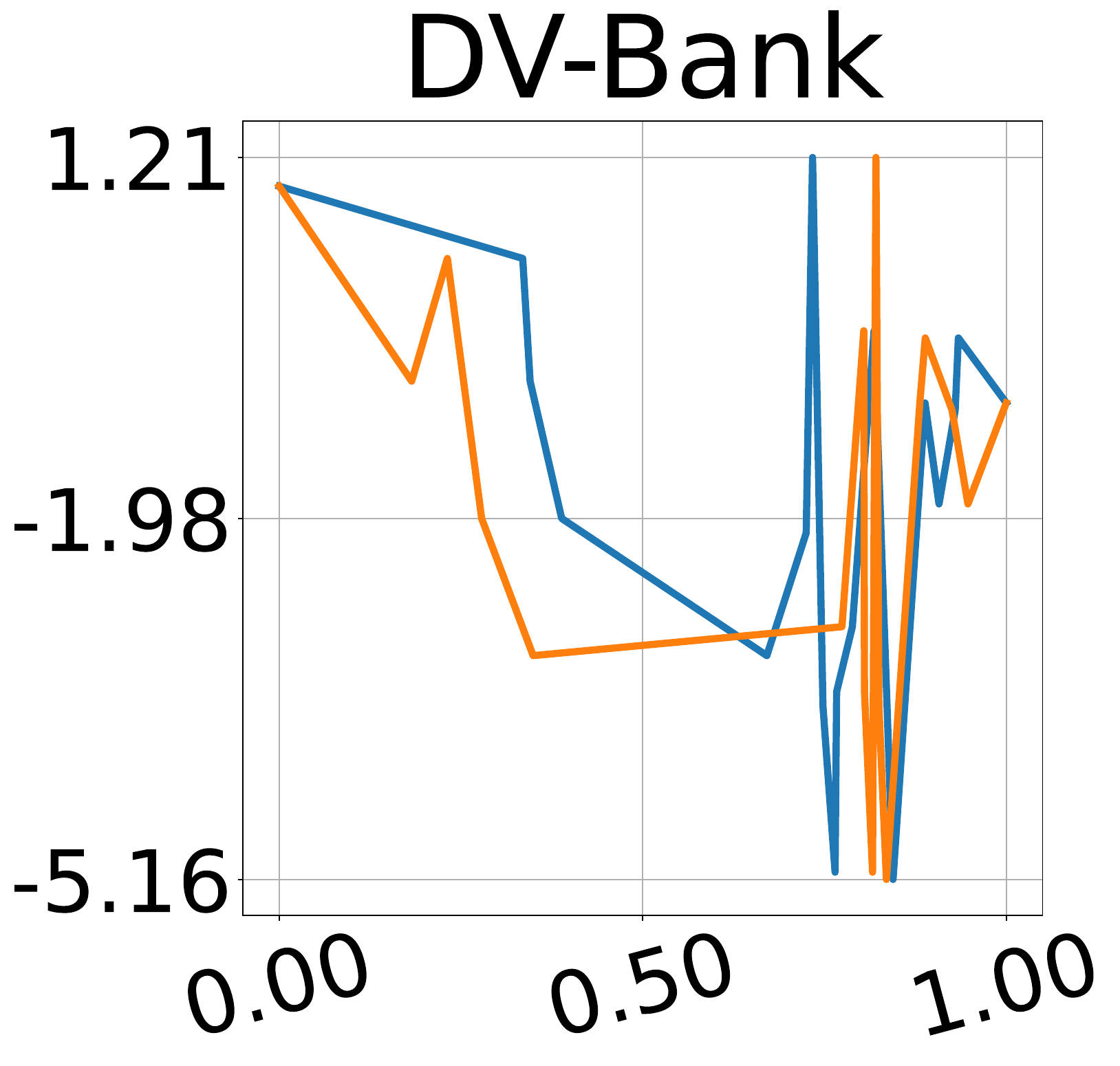}
        \includegraphics[width=0.2\columnwidth, height=0.2\columnwidth]{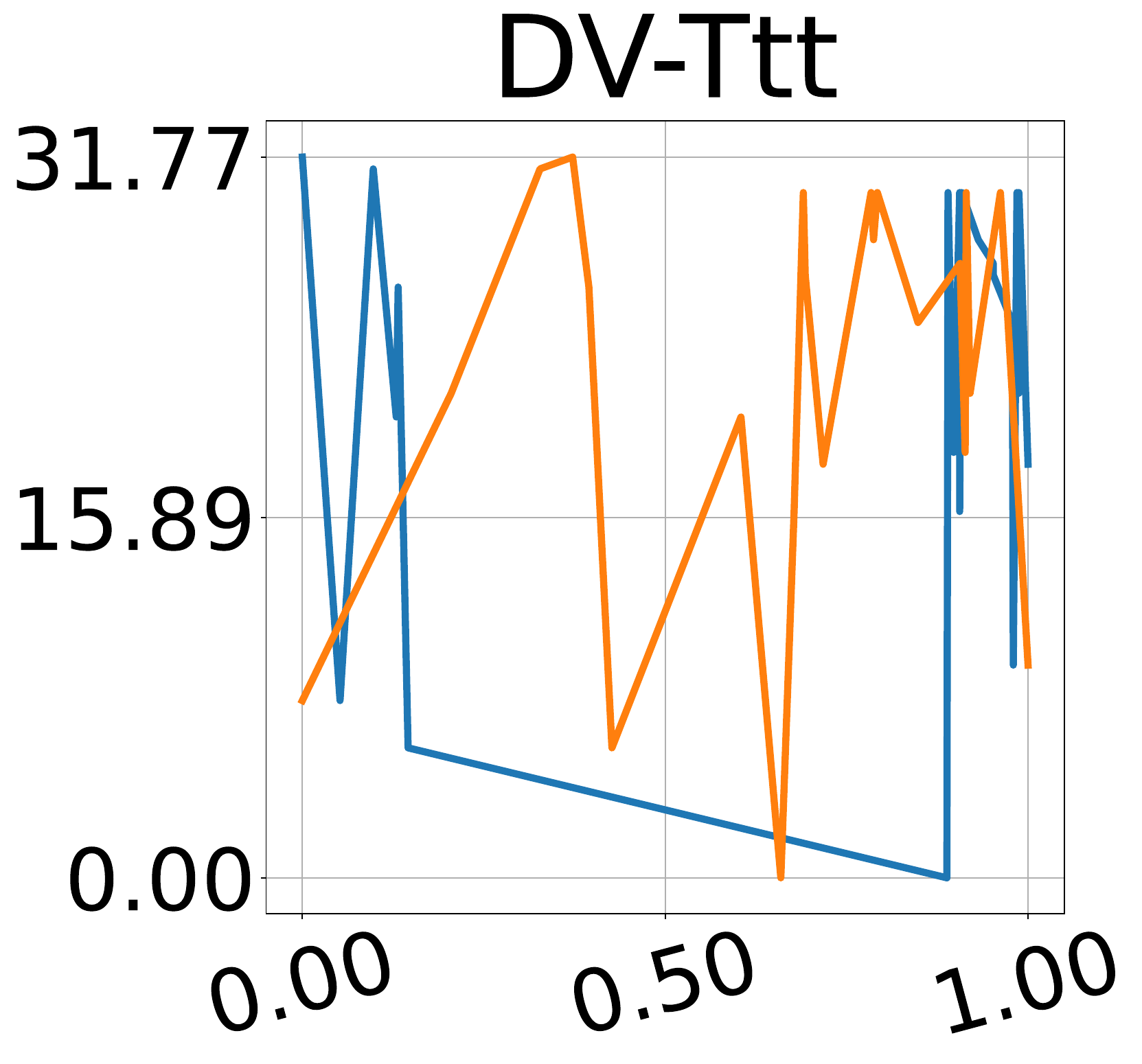}
        \includegraphics[width=0.2\columnwidth, height=0.2\columnwidth]{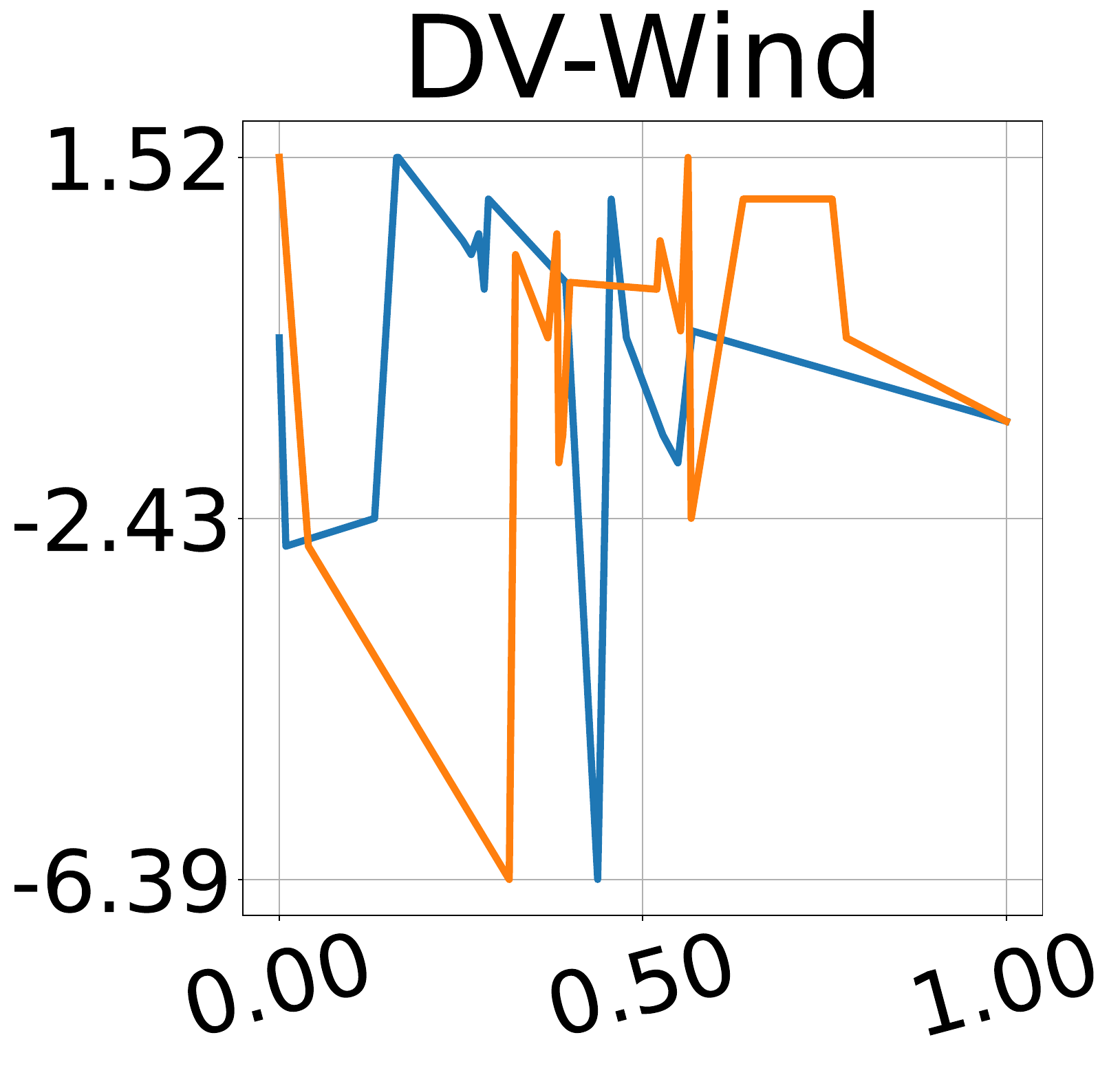}
    }
    \\
    
    \subfigure[The variation of the overall utility in different DA tasks caused by \textit{removing} players.] {
        \includegraphics[width=0.215\columnwidth, height=0.2\columnwidth]{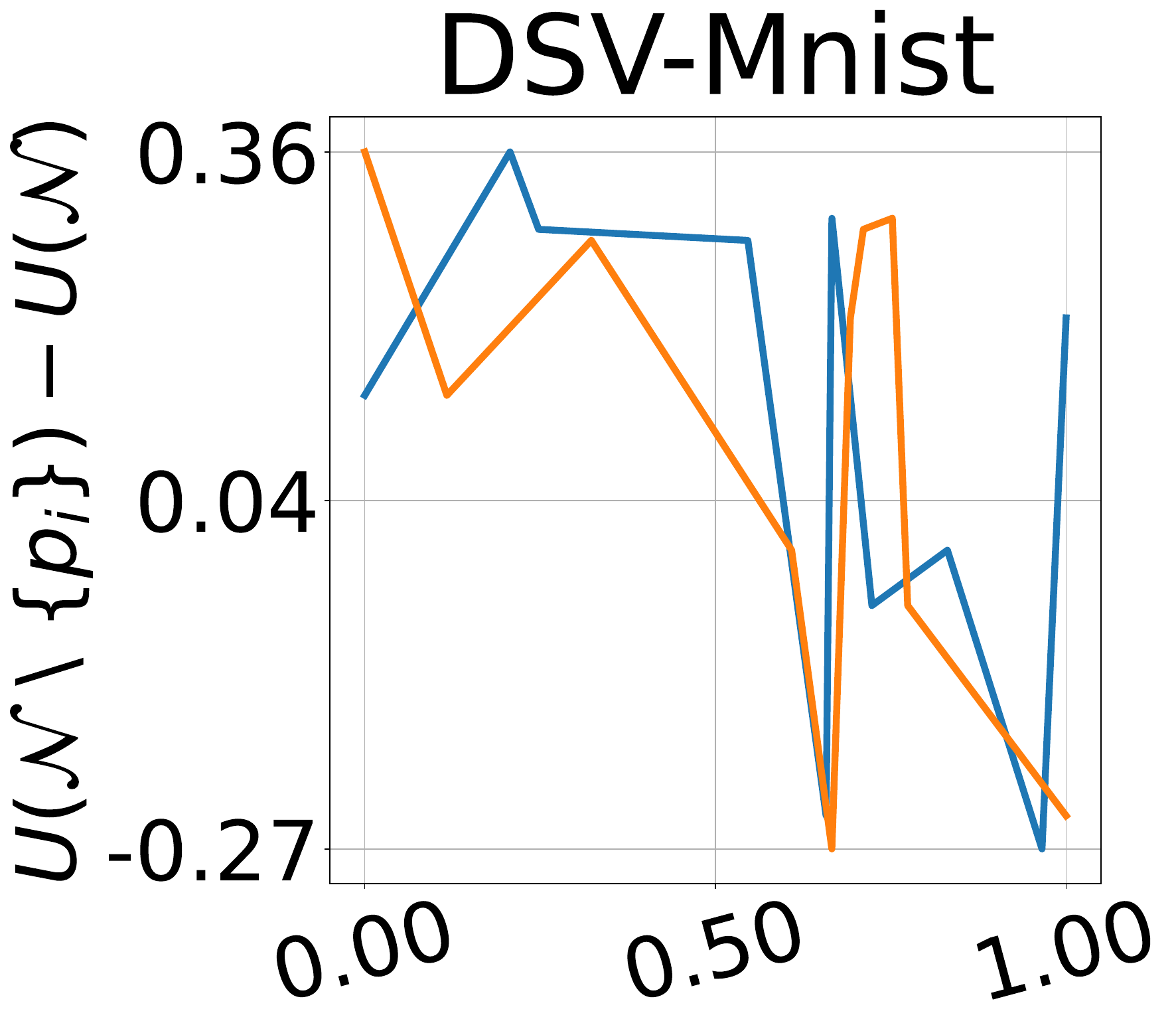}
        \includegraphics[width=0.2\columnwidth, height=0.2\columnwidth]{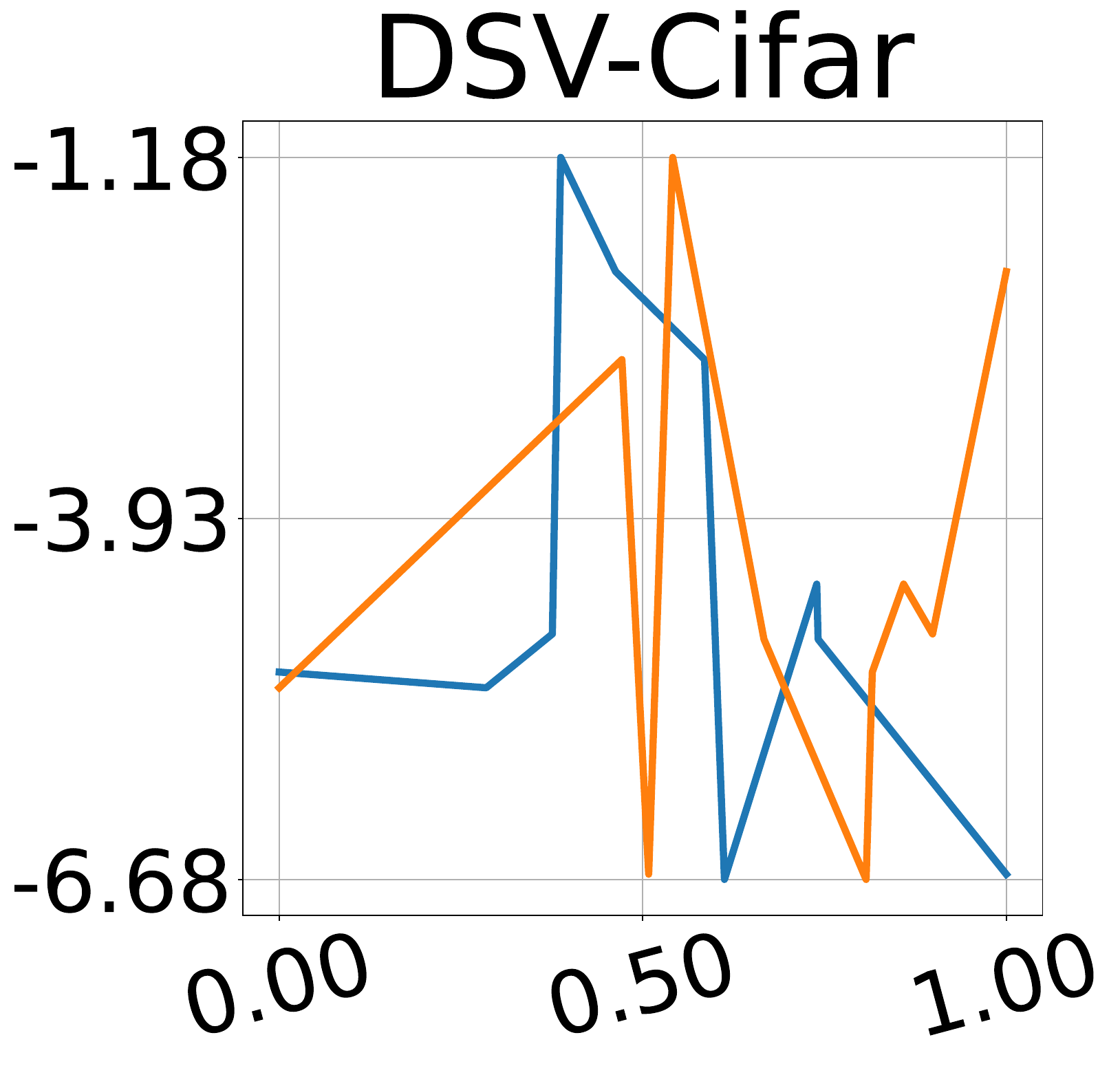}
        \includegraphics[width=0.2\columnwidth, height=0.2\columnwidth]{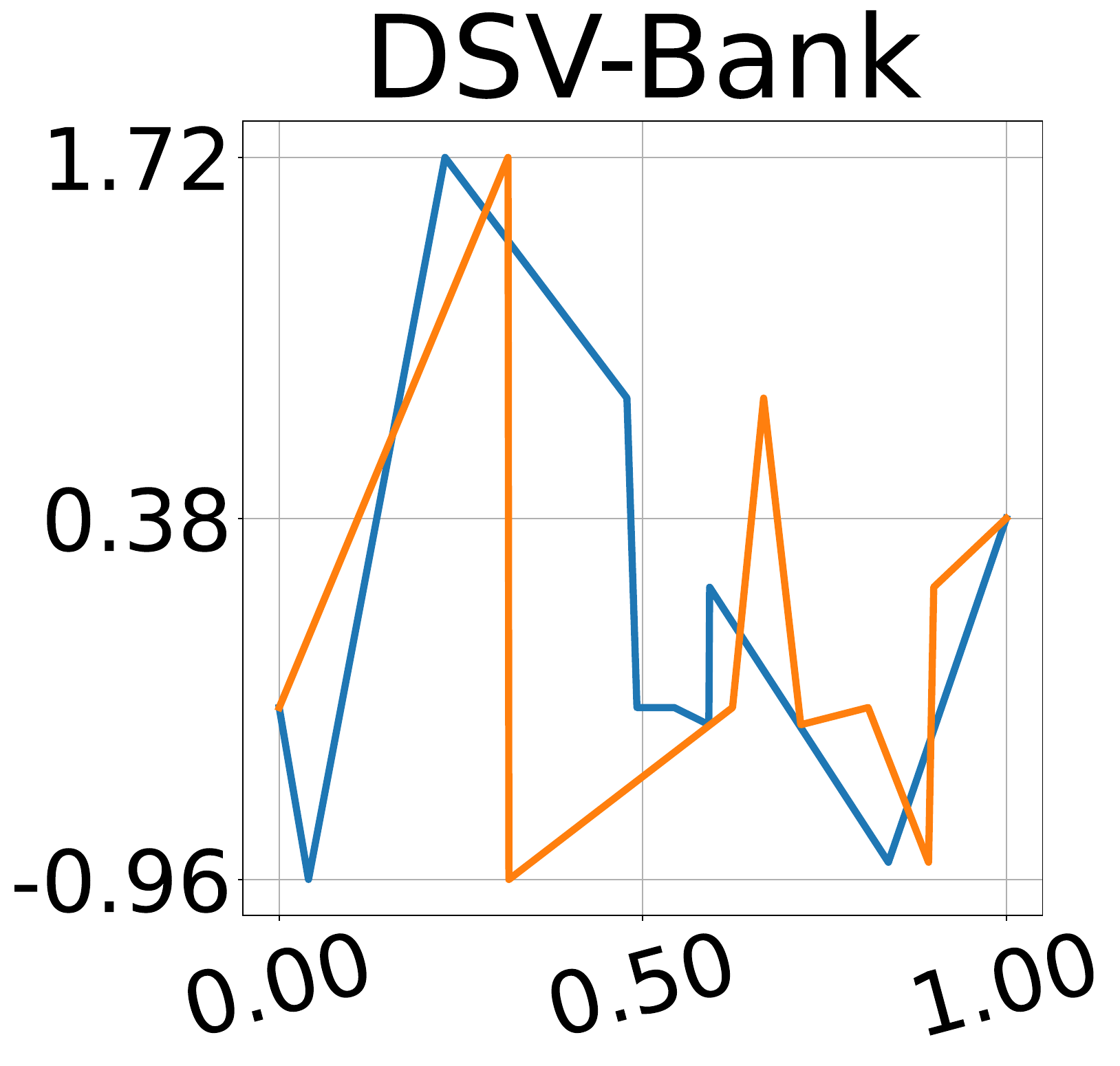}
        \includegraphics[width=0.2\columnwidth, height=0.2\columnwidth]{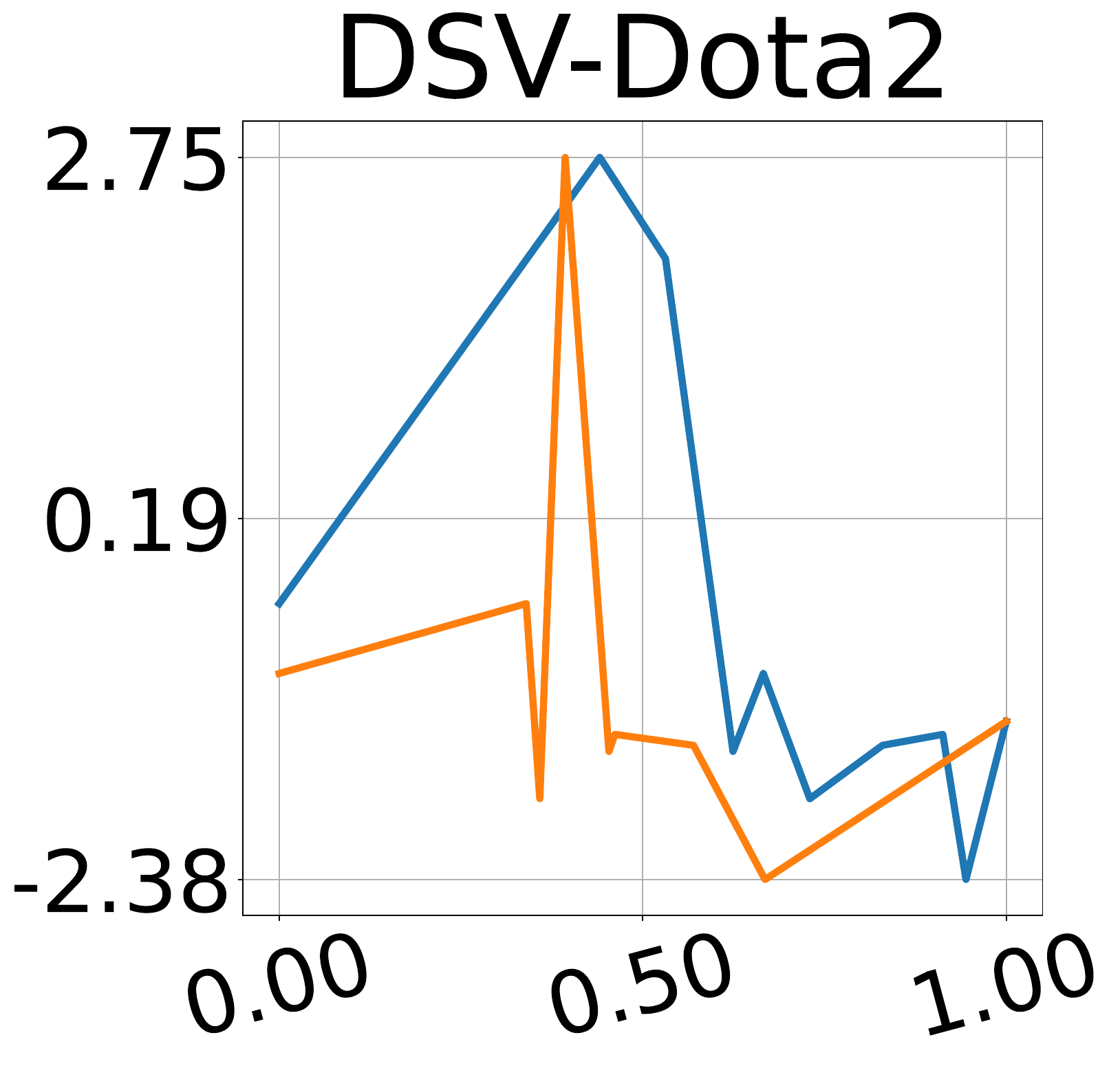}
        \includegraphics[width=0.2\columnwidth, height=0.2\columnwidth]{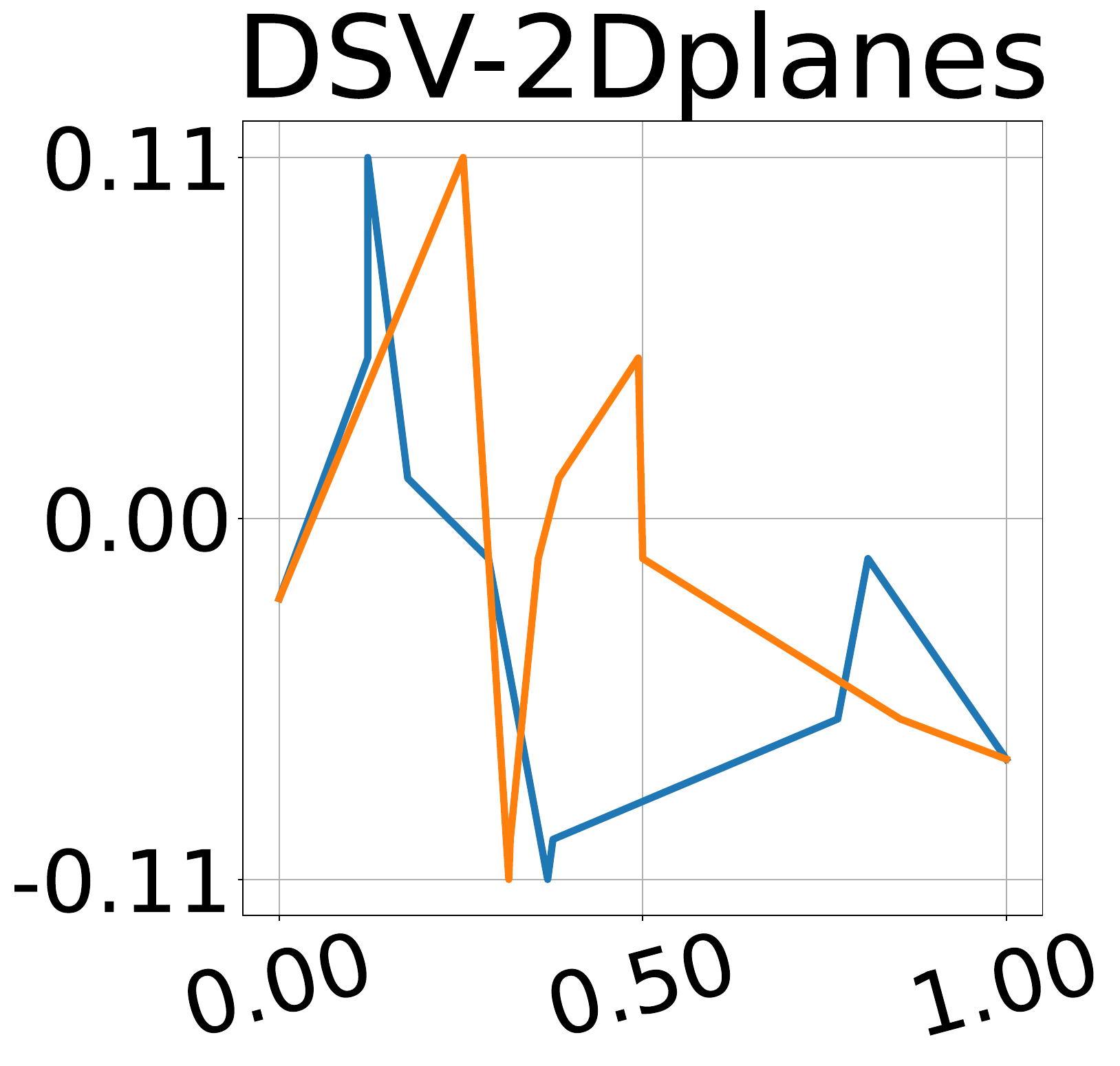}
        
        \includegraphics[width=0.2\columnwidth, height=0.2\columnwidth]{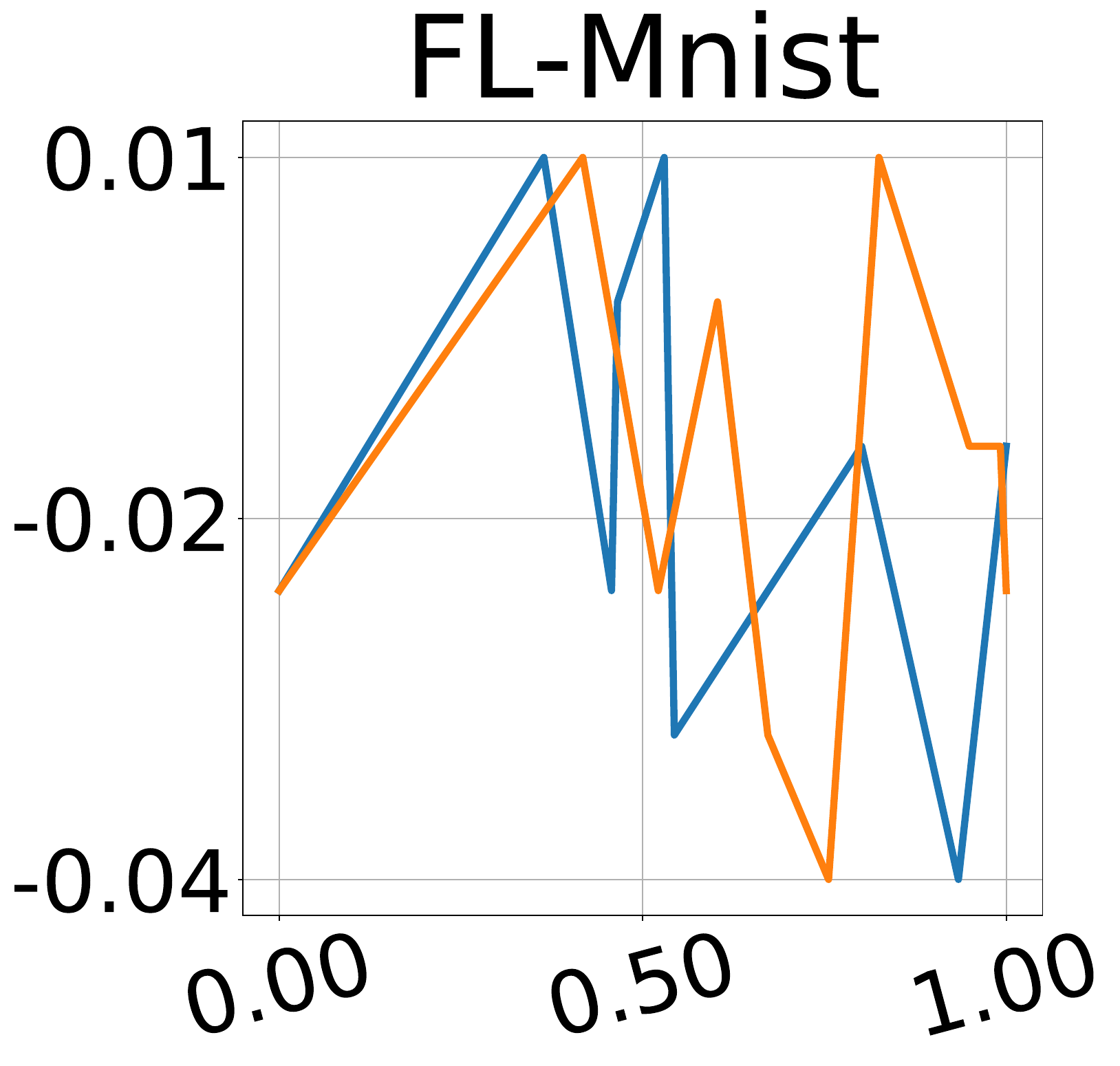}
        \includegraphics[width=0.2\columnwidth, height=0.2\columnwidth]{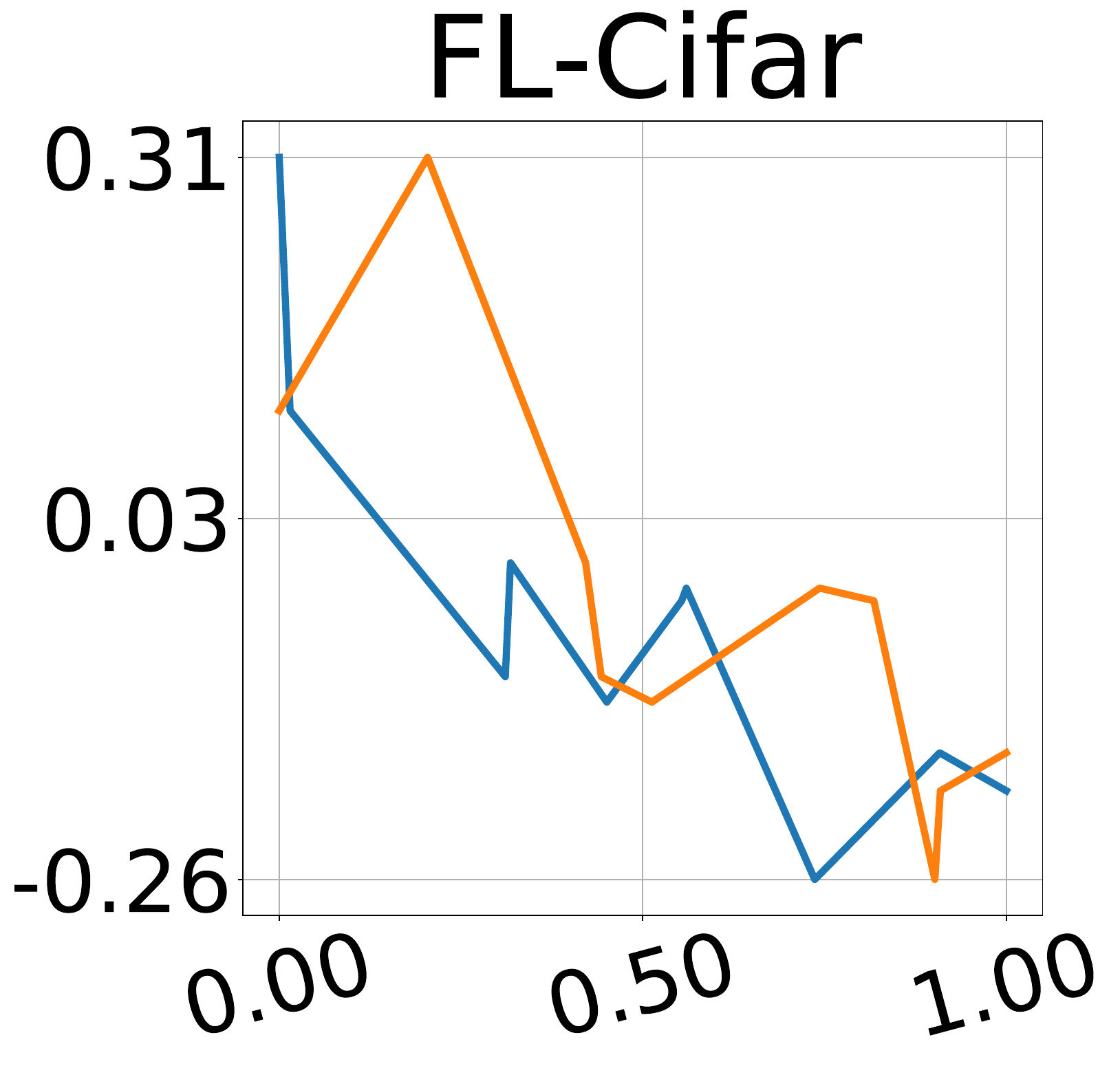}
        \includegraphics[width=0.2\columnwidth, height=0.2\columnwidth]{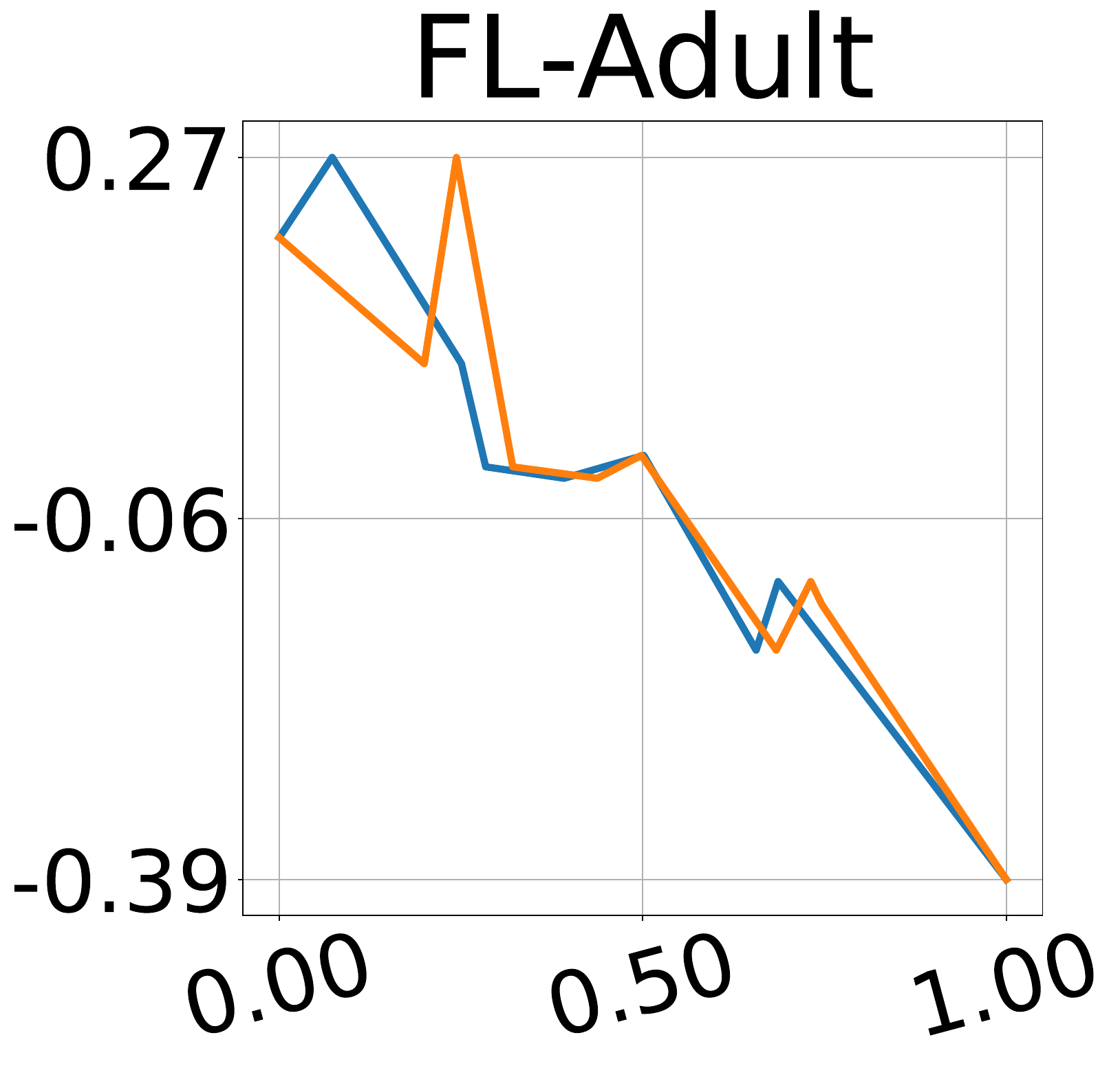}
        \includegraphics[width=0.2\columnwidth, height=0.2\columnwidth]{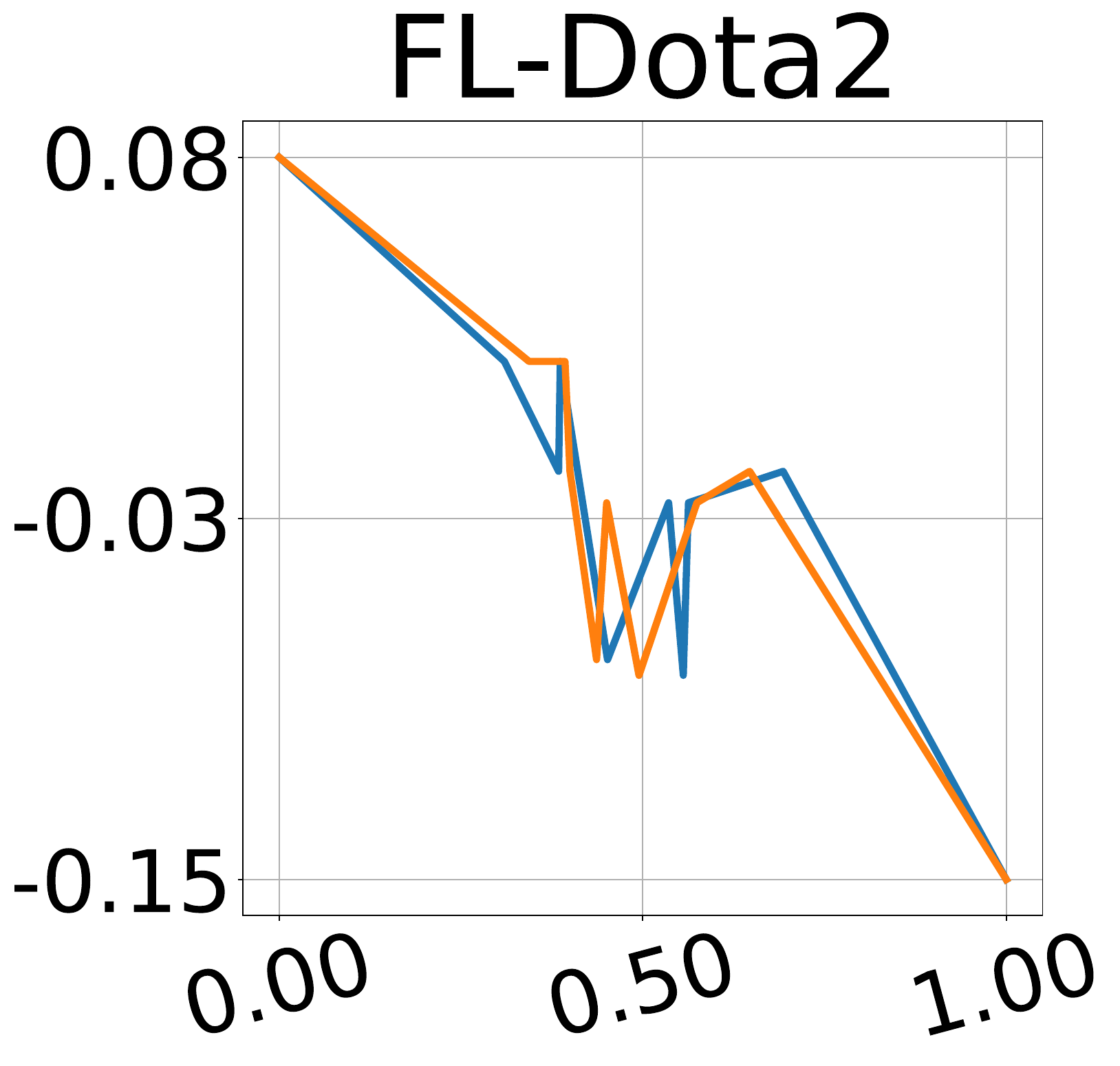}
        \includegraphics[width=0.2\columnwidth, height=0.2\columnwidth]{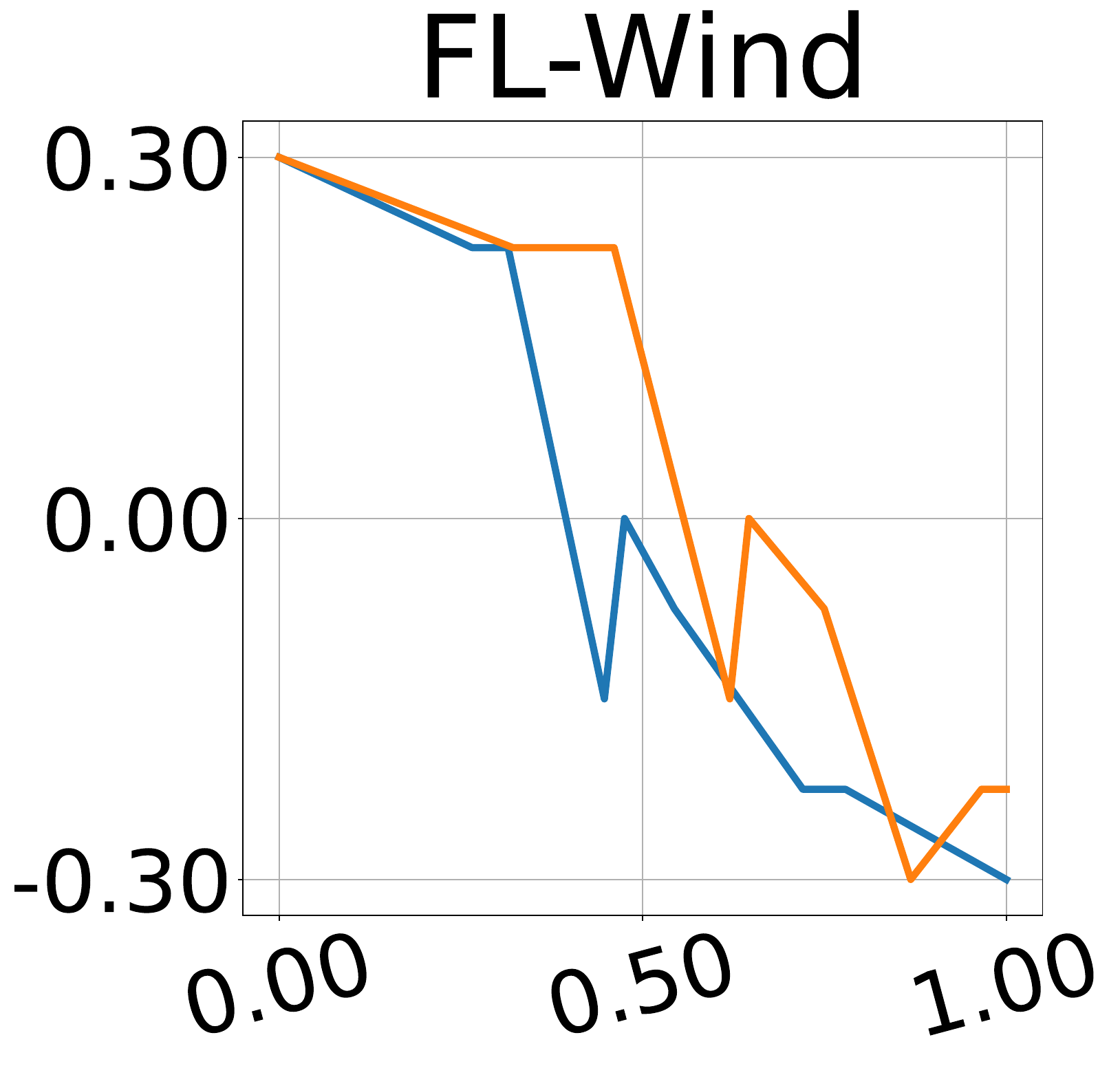}
    }  
    \\
    \mbox{
        \includegraphics[width=0.2\columnwidth, height=0.2\columnwidth]{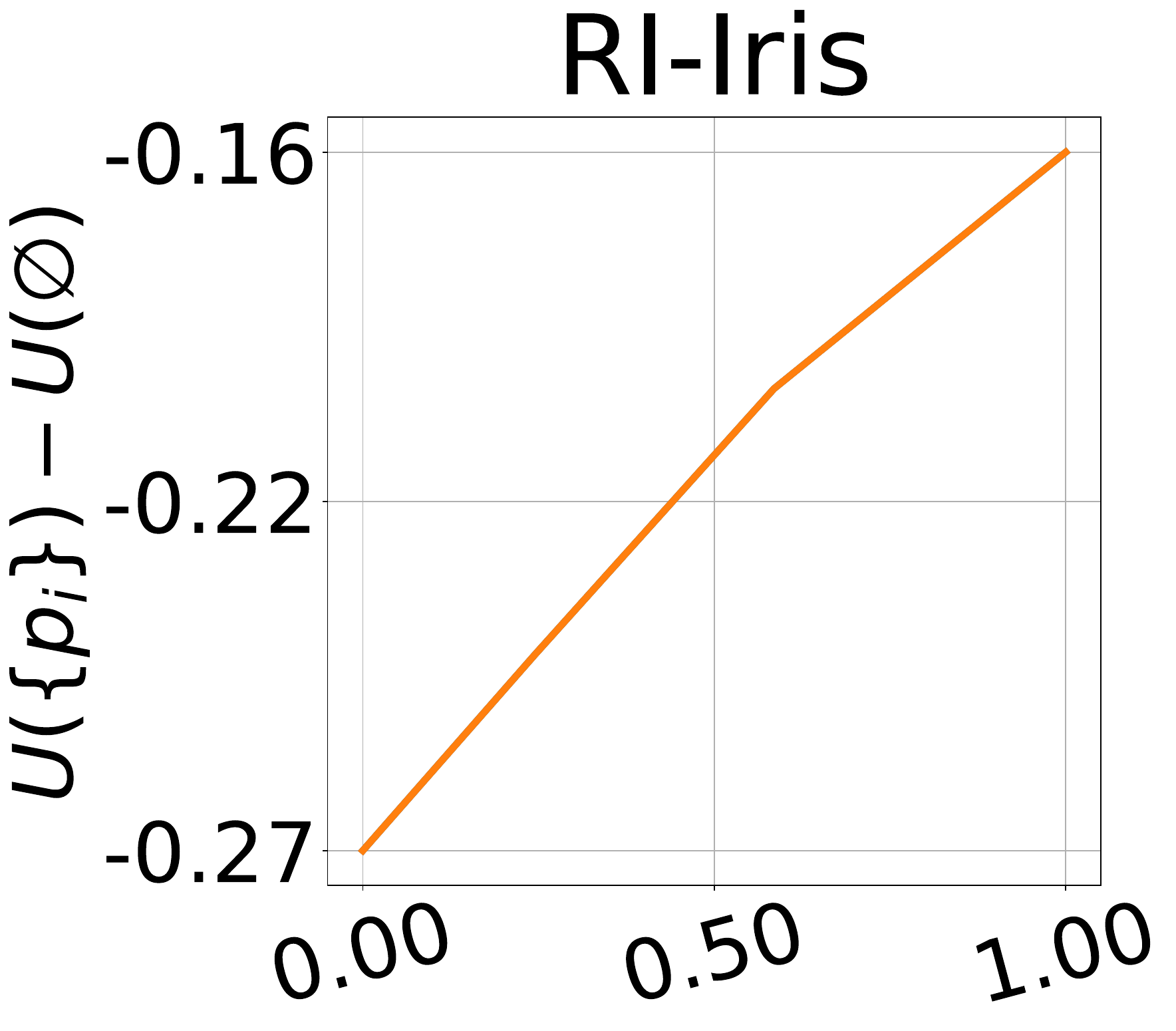}
        \includegraphics[width=0.2\columnwidth, height=0.2\columnwidth]{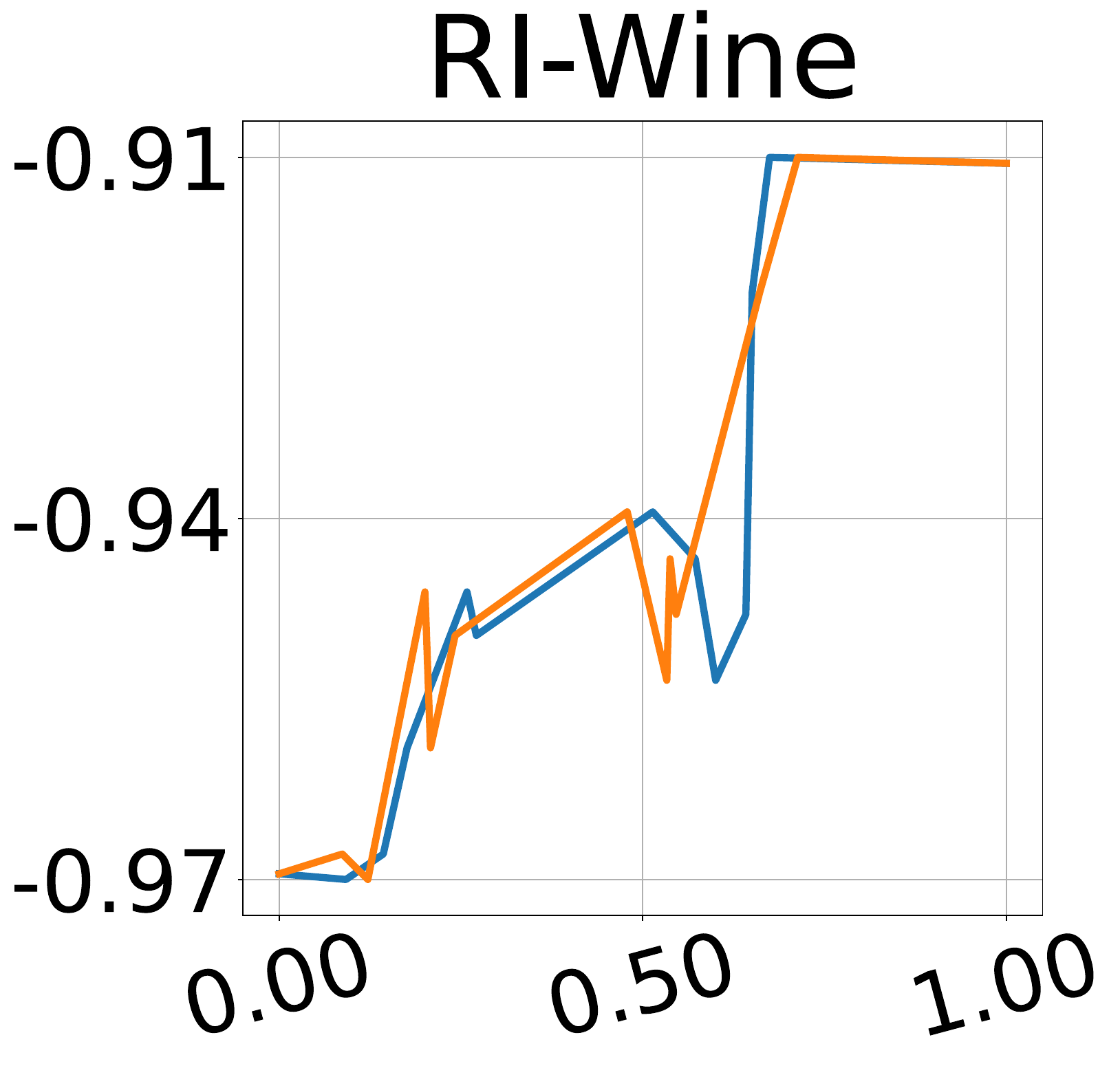}
        \includegraphics[width=0.2\columnwidth, height=0.2\columnwidth]{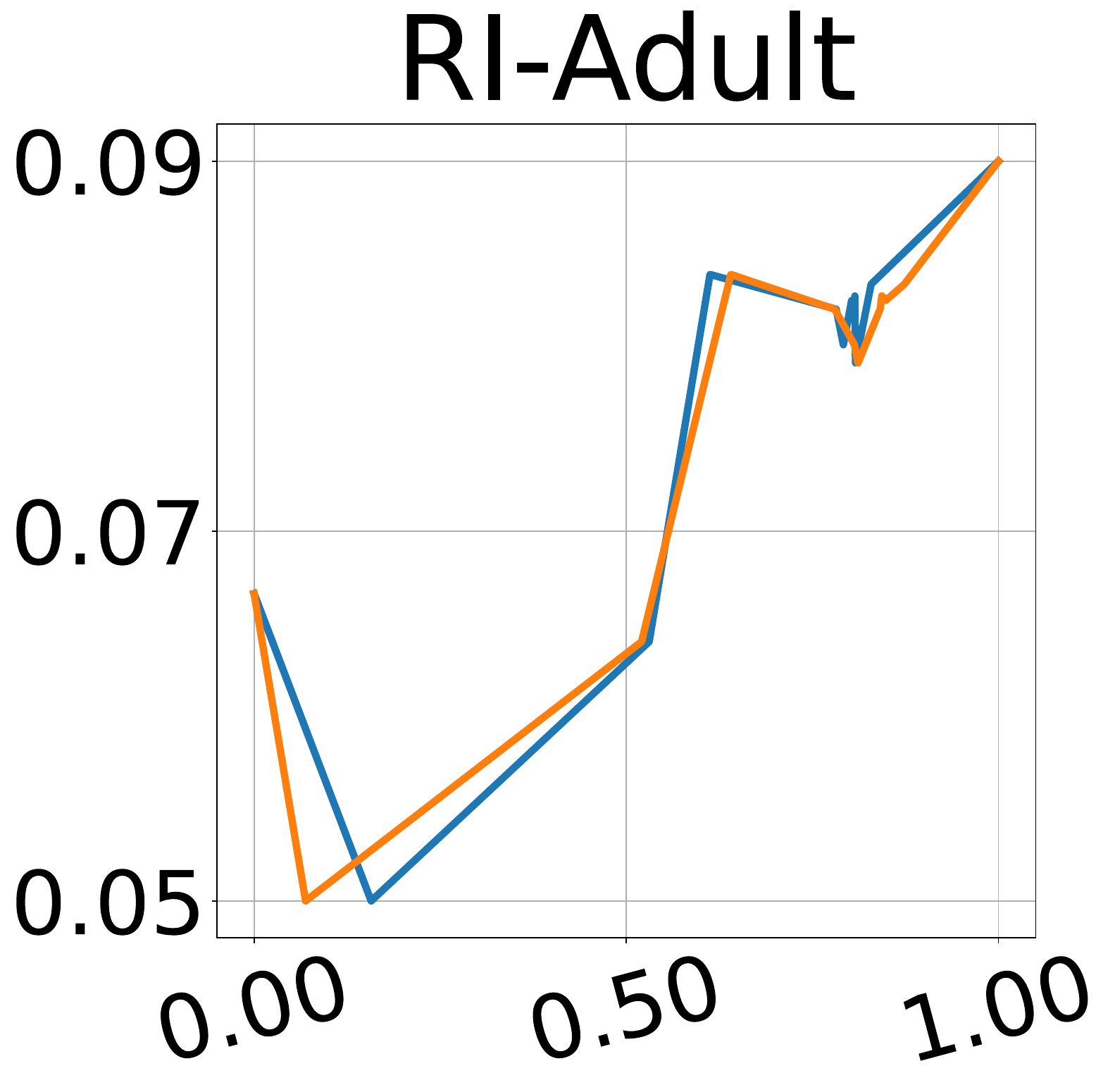}
        \includegraphics[width=0.2\columnwidth, height=0.2\columnwidth]{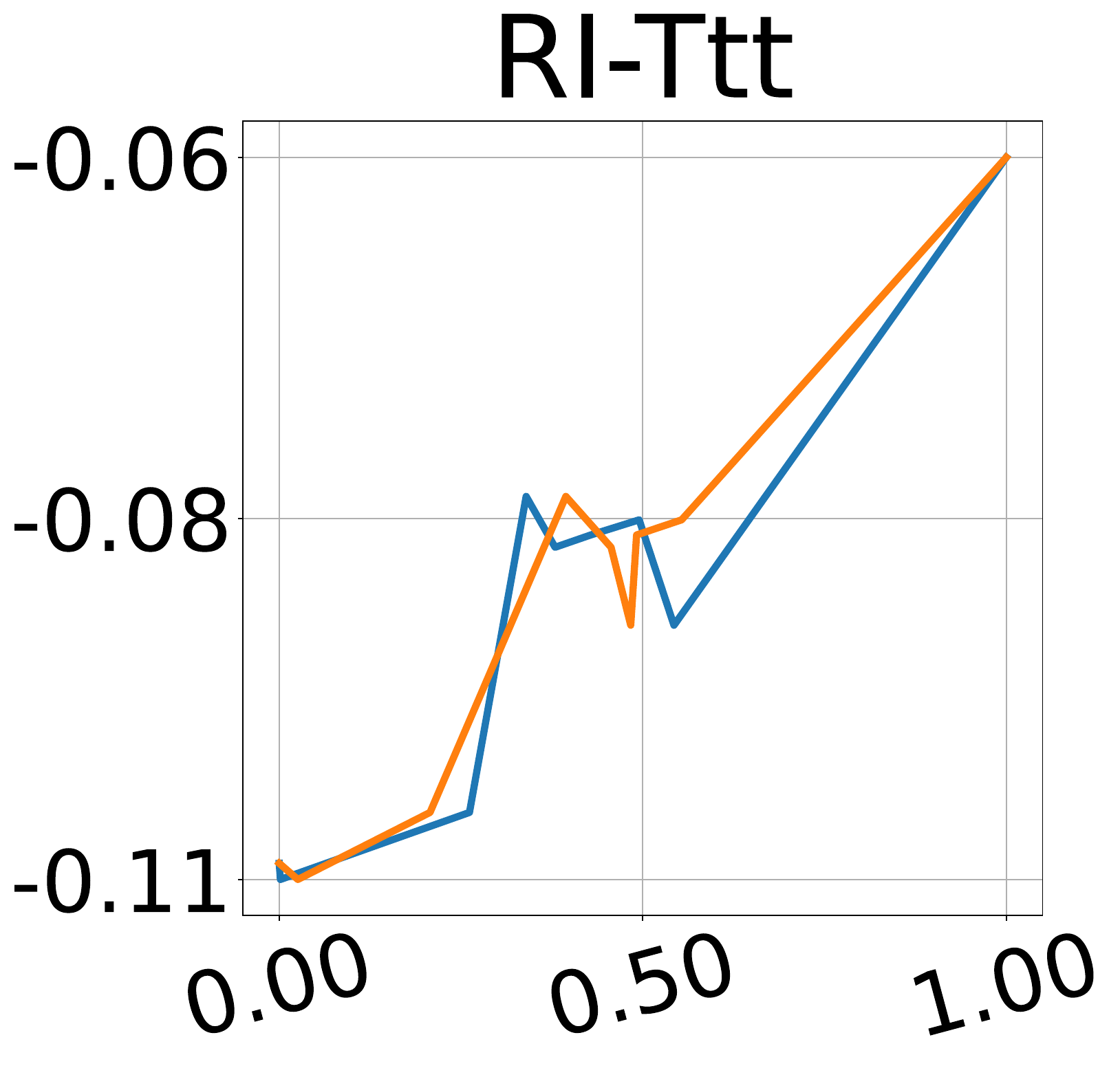}
        \includegraphics[width=0.2\columnwidth, height=0.2\columnwidth]{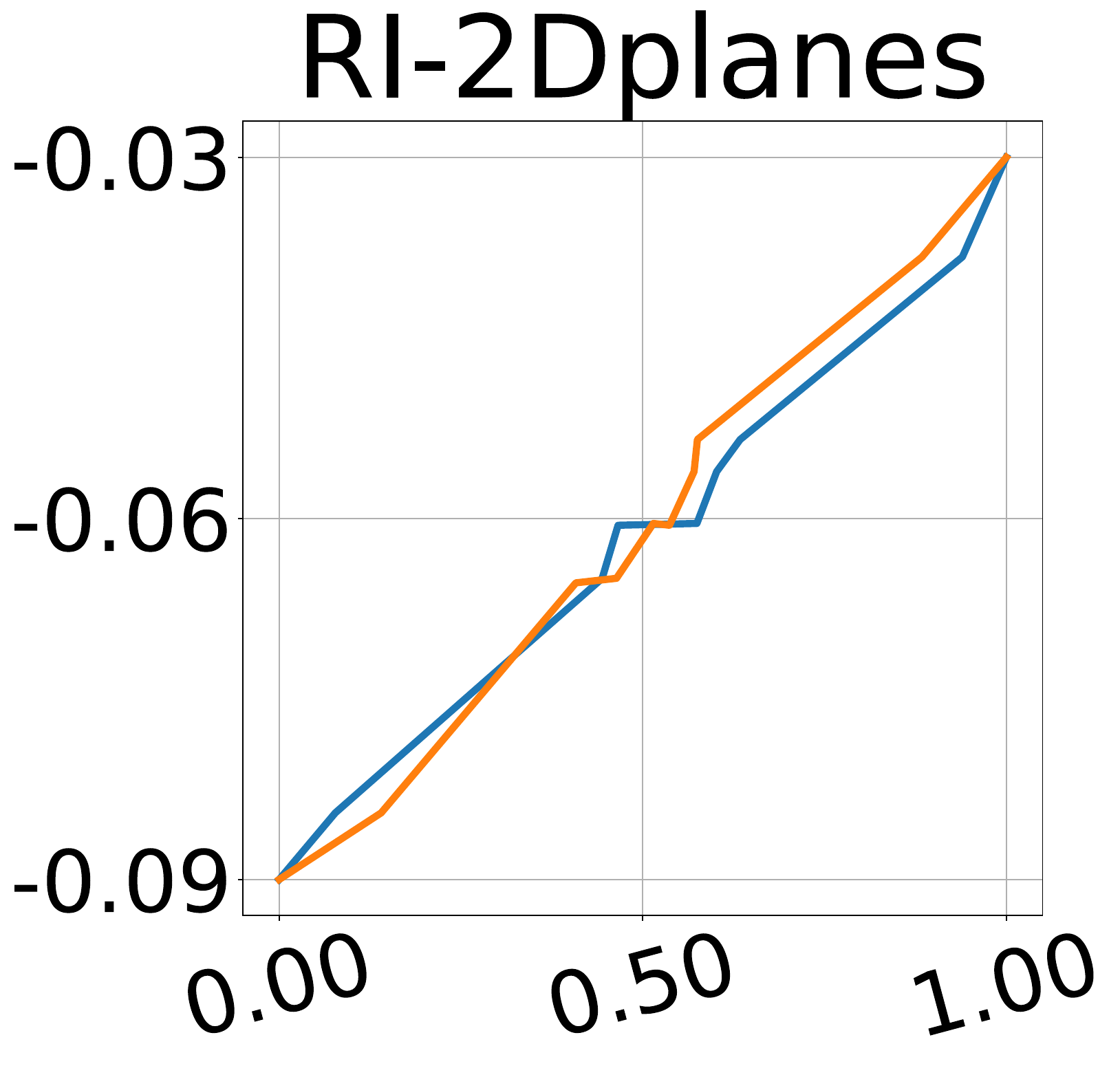}
        
        \includegraphics[width=0.2\columnwidth, height=0.2\columnwidth]{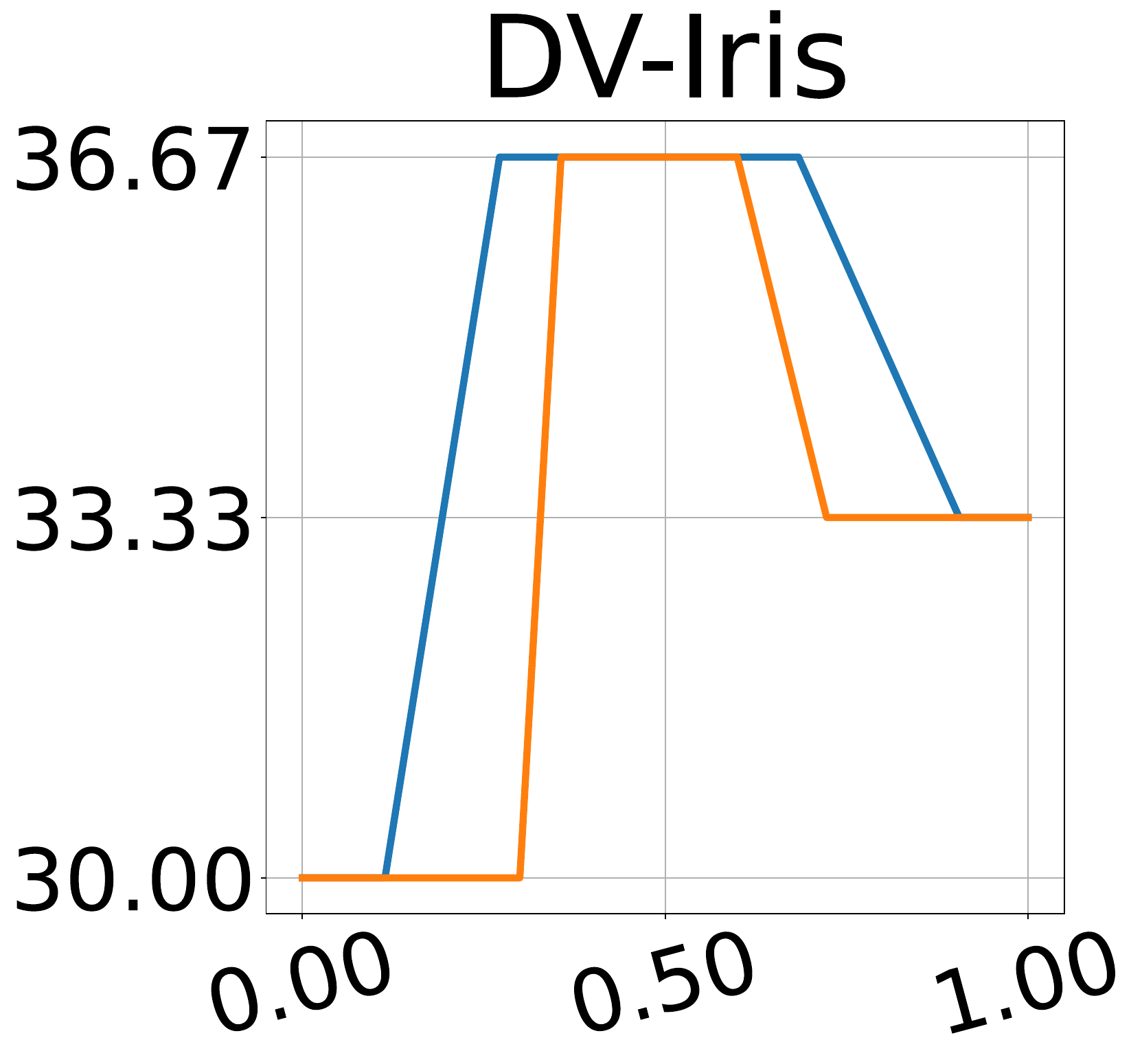}
        \includegraphics[width=0.2\columnwidth, height=0.2\columnwidth]{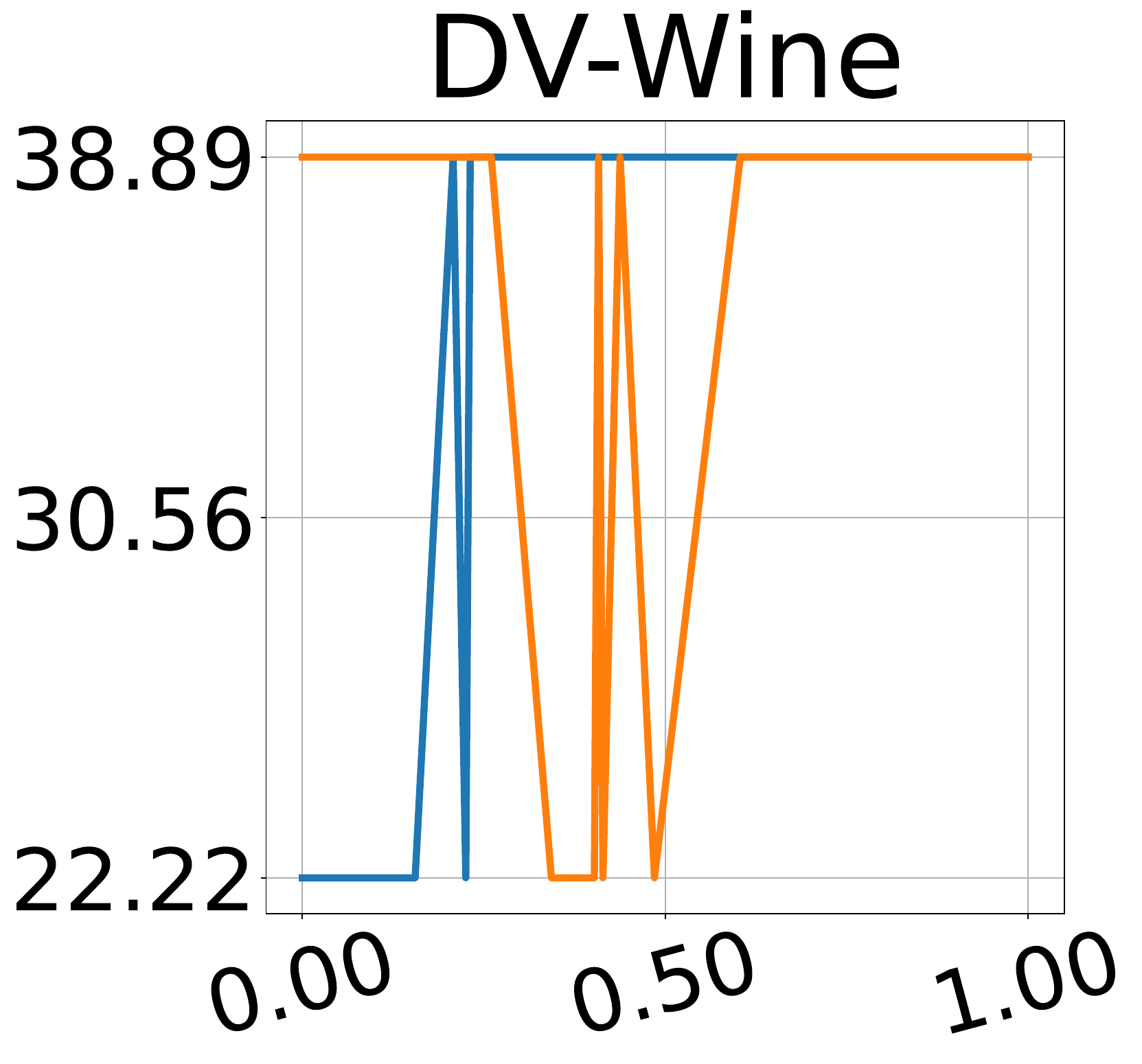}
        \includegraphics[width=0.2\columnwidth, height=0.2\columnwidth]{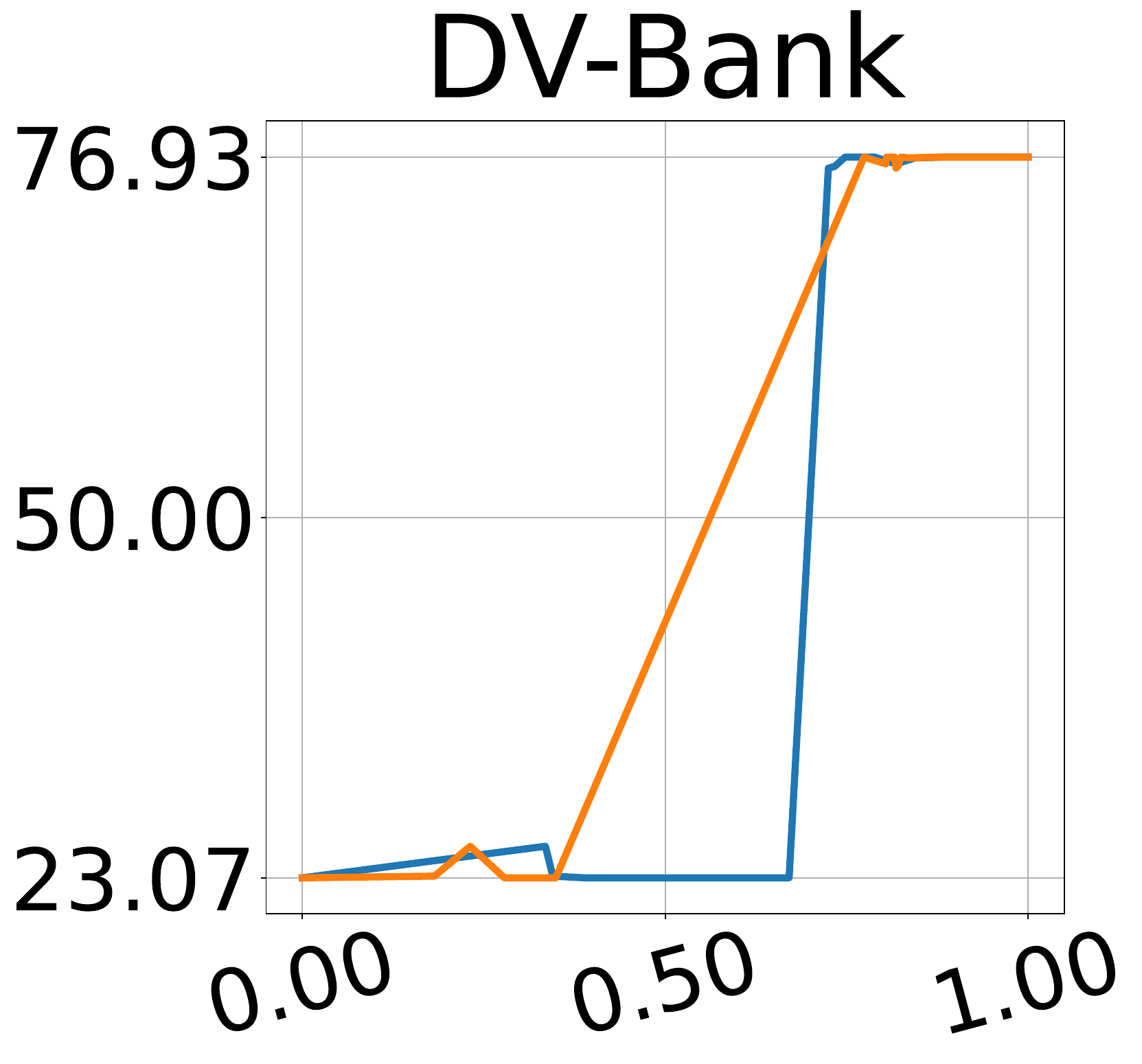}
        \includegraphics[width=0.2\columnwidth, height=0.2\columnwidth]{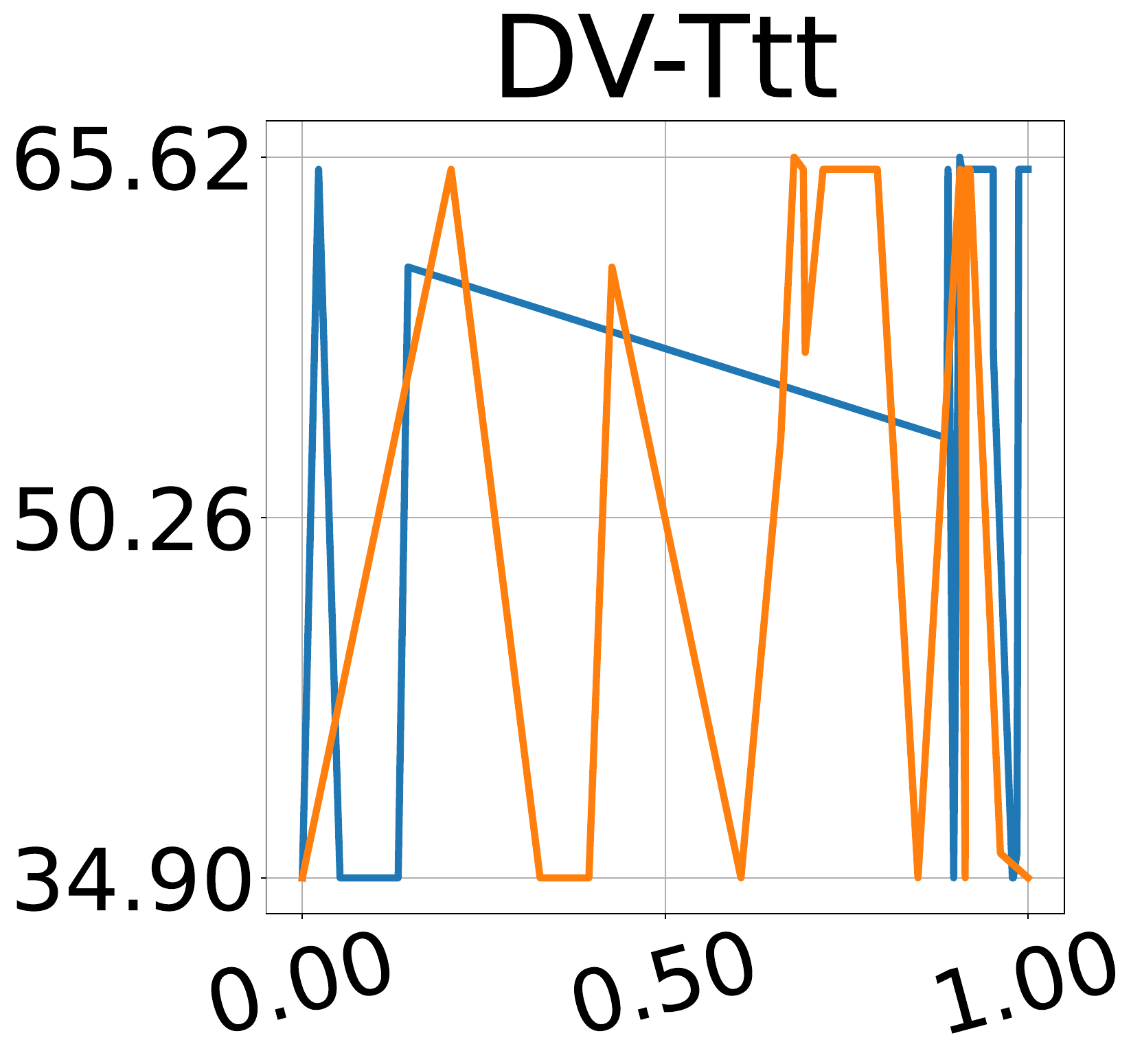}
        \includegraphics[width=0.2\columnwidth, height=0.2\columnwidth]{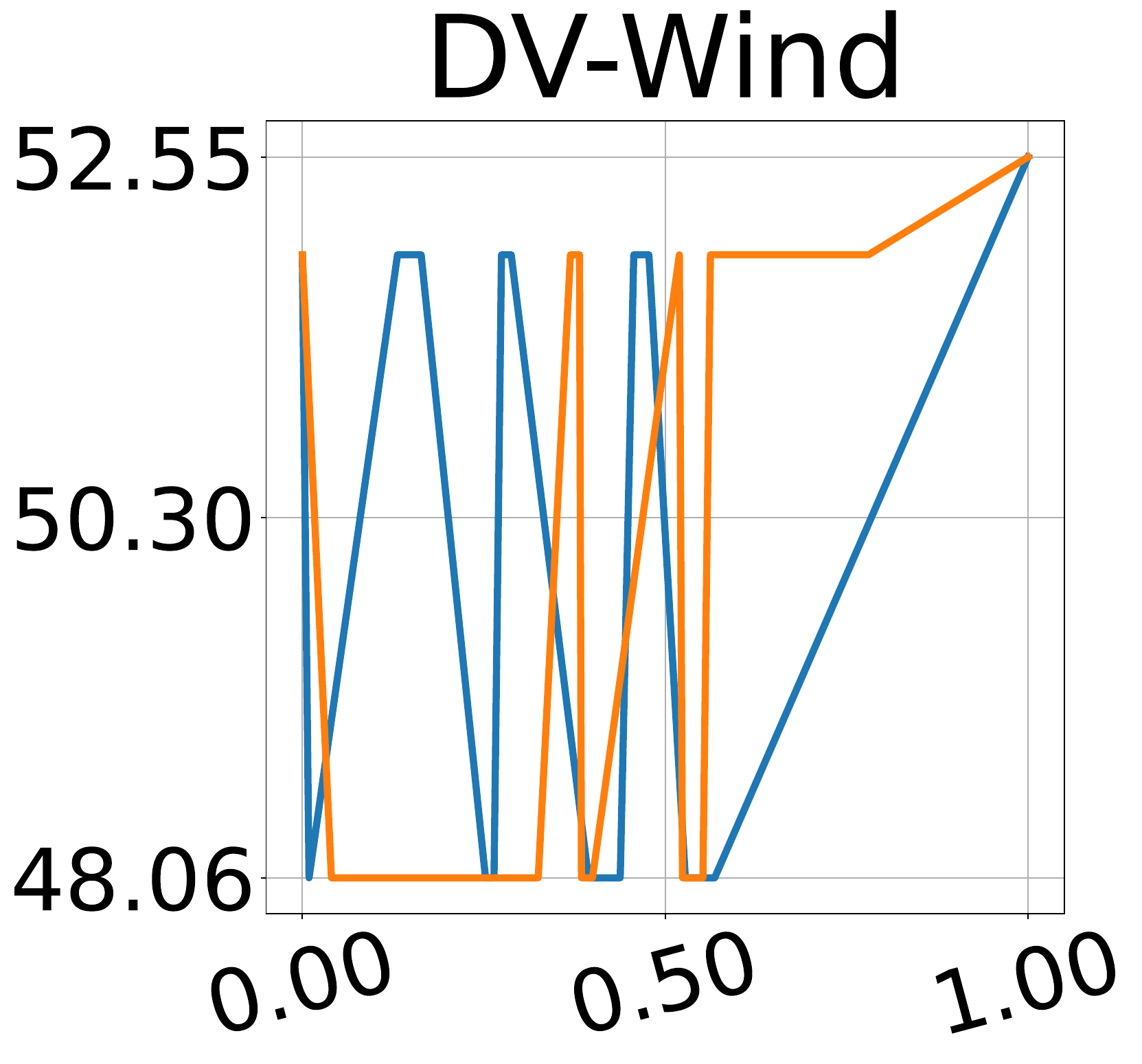}
        
    }
    \\
    \subfigure[The variation of the overall utility in different DA tasks  caused by \textit{adding} players.]  {
        \includegraphics[width=0.215\columnwidth, height=0.2\columnwidth]{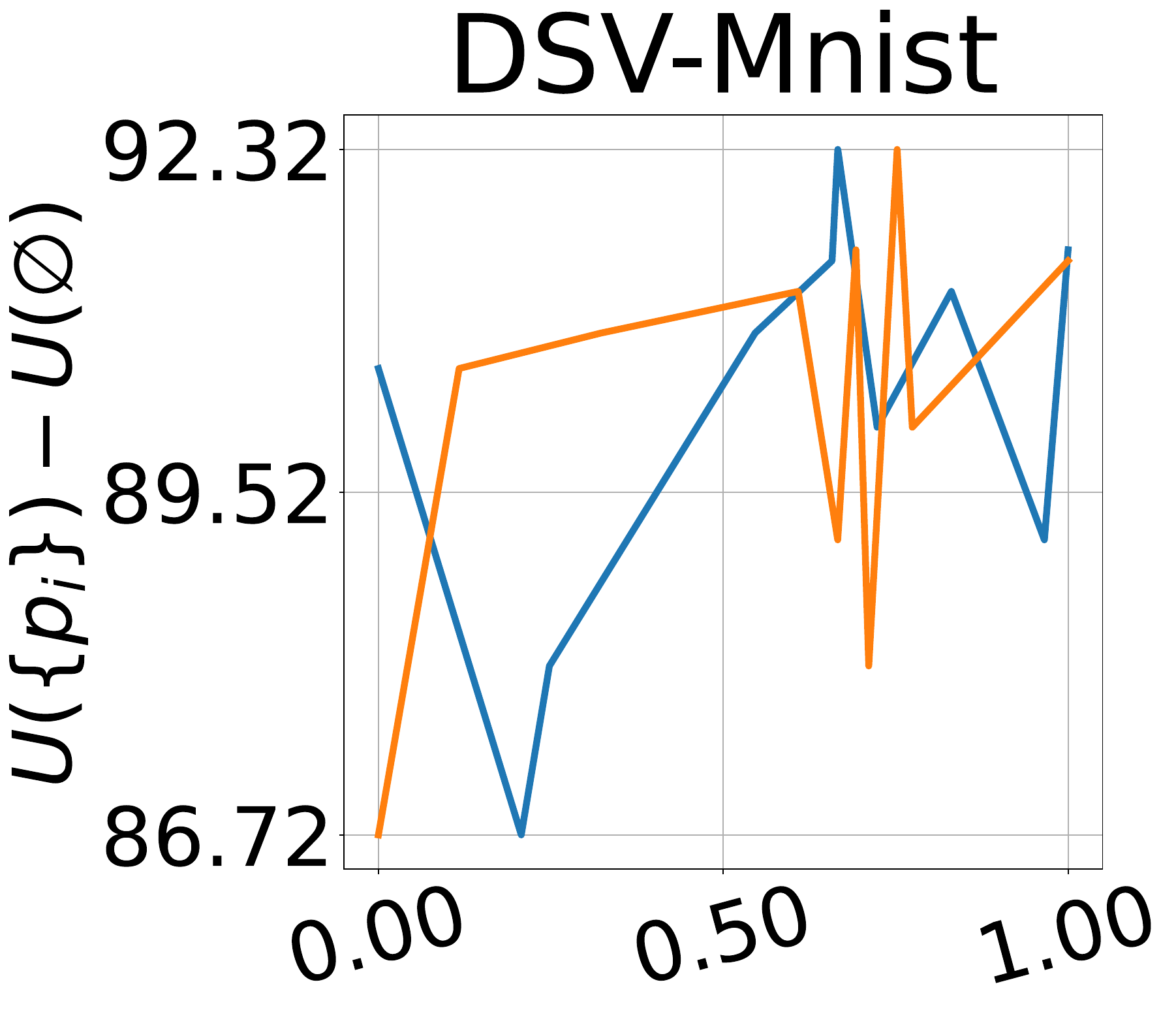}
        \includegraphics[width=0.2\columnwidth, height=0.2\columnwidth]{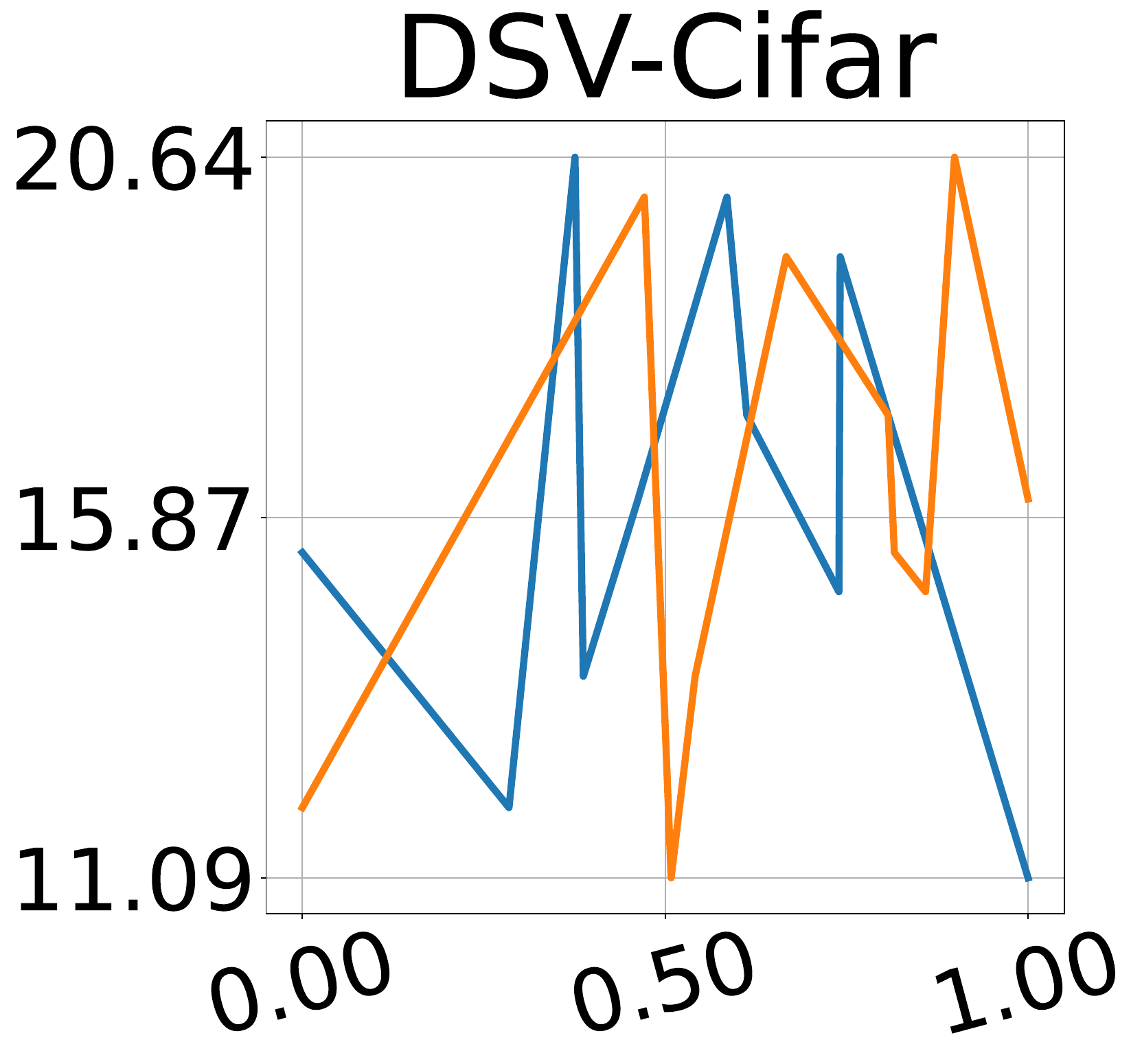}
        \includegraphics[width=0.2\columnwidth, height=0.2\columnwidth]{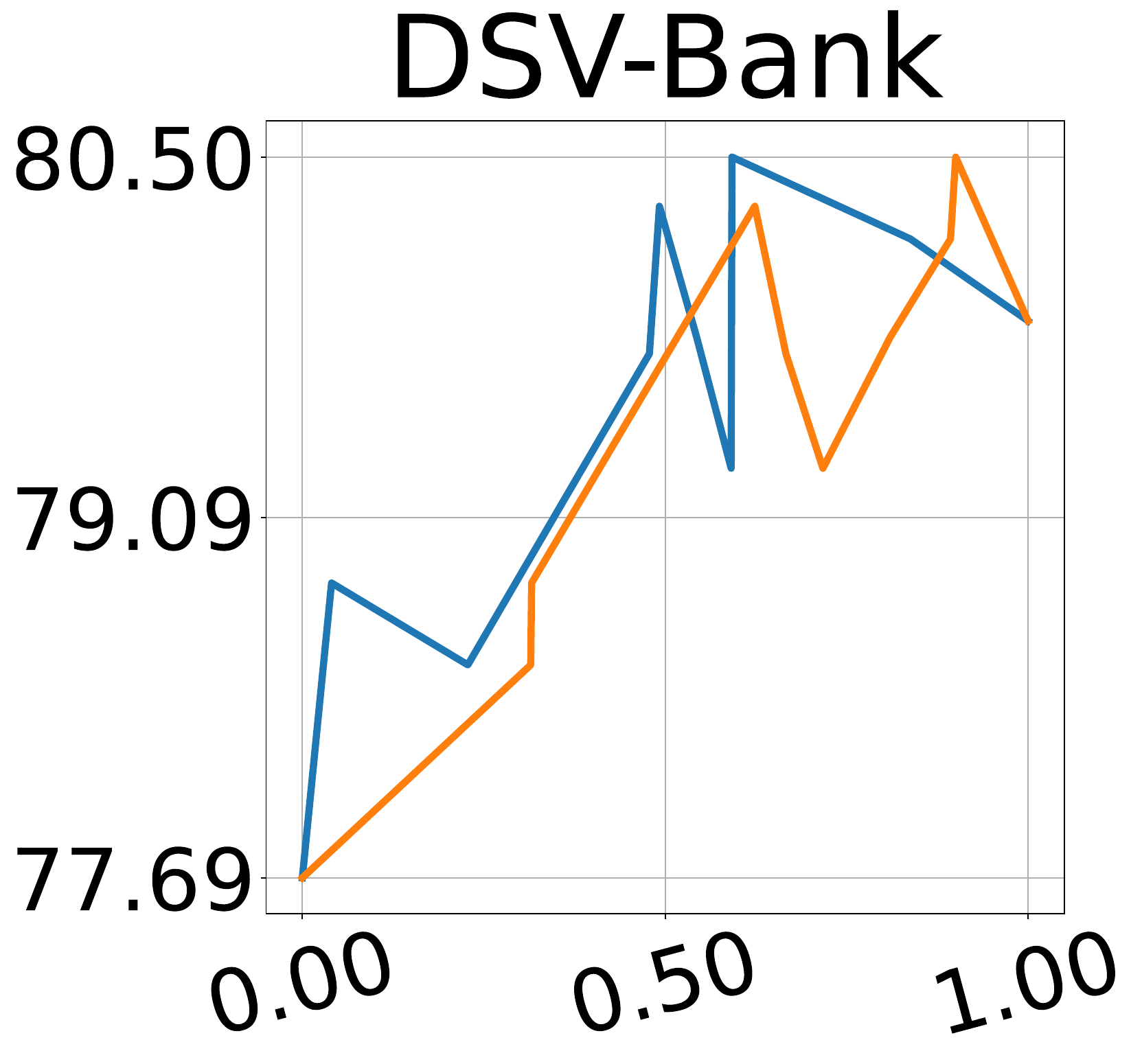}
        \includegraphics[width=0.2\columnwidth, height=0.2\columnwidth]{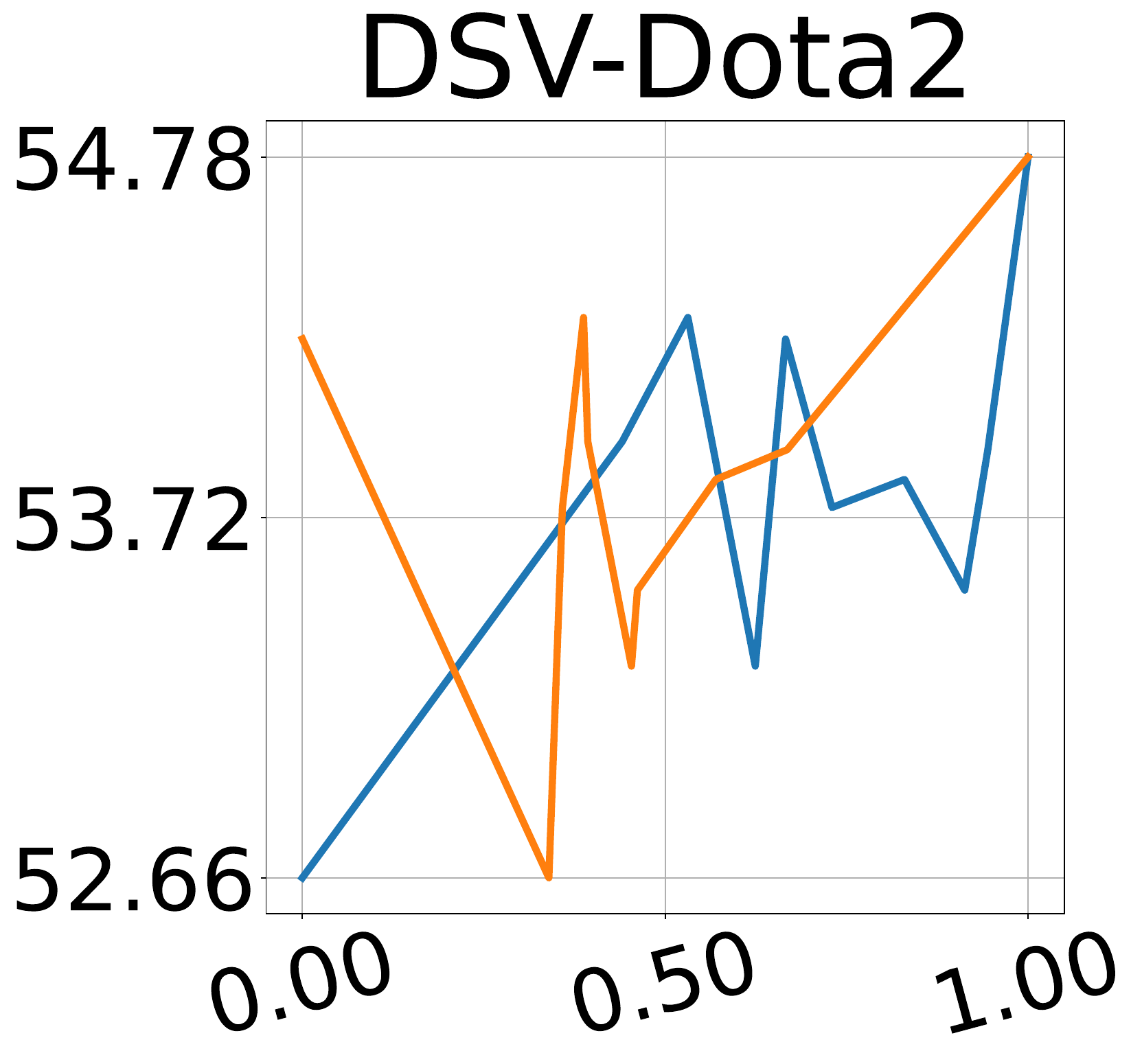}
        \includegraphics[width=0.2\columnwidth, height=0.2\columnwidth]{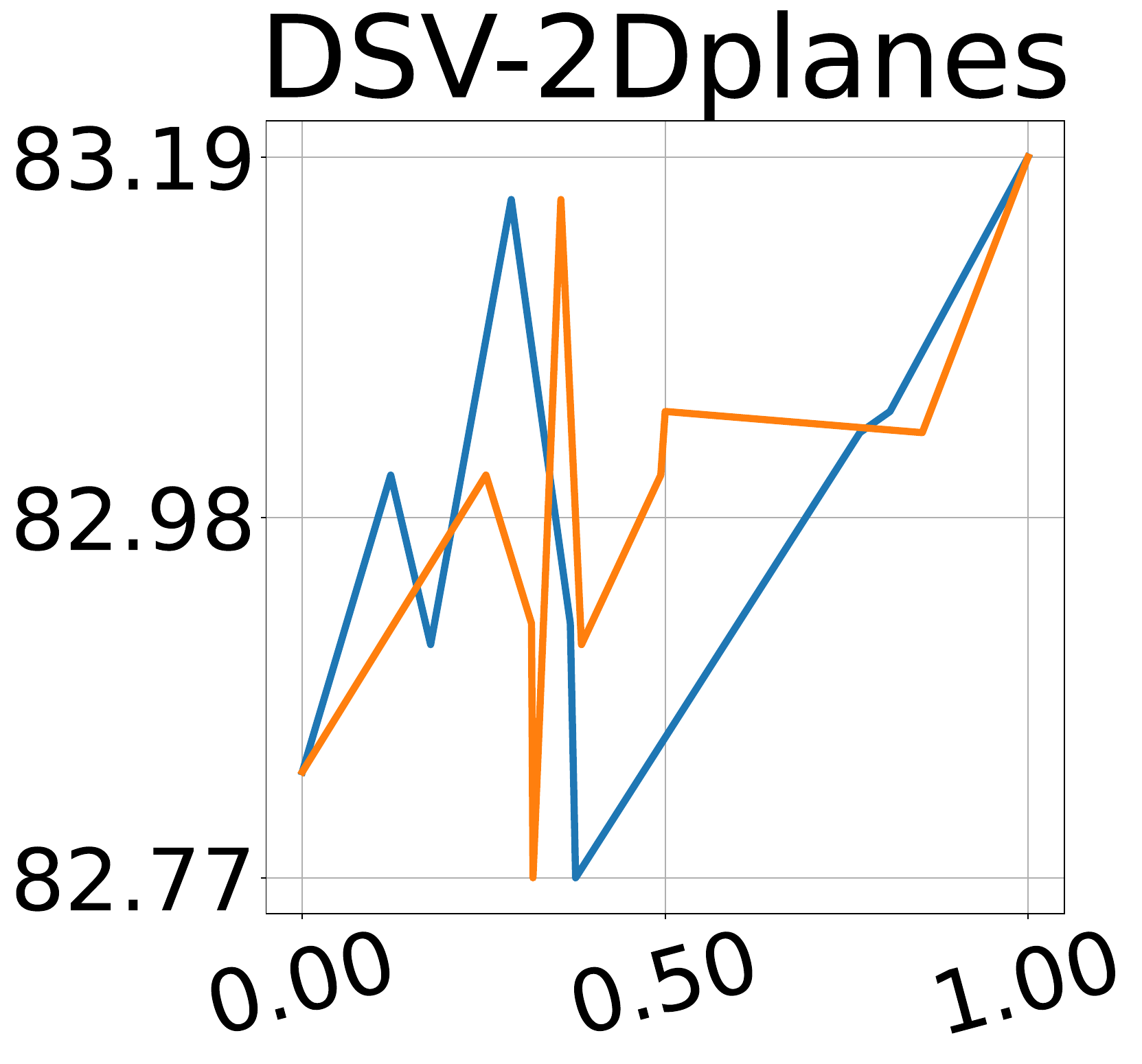}
        
        \includegraphics[width=0.2\columnwidth, height=0.2\columnwidth]{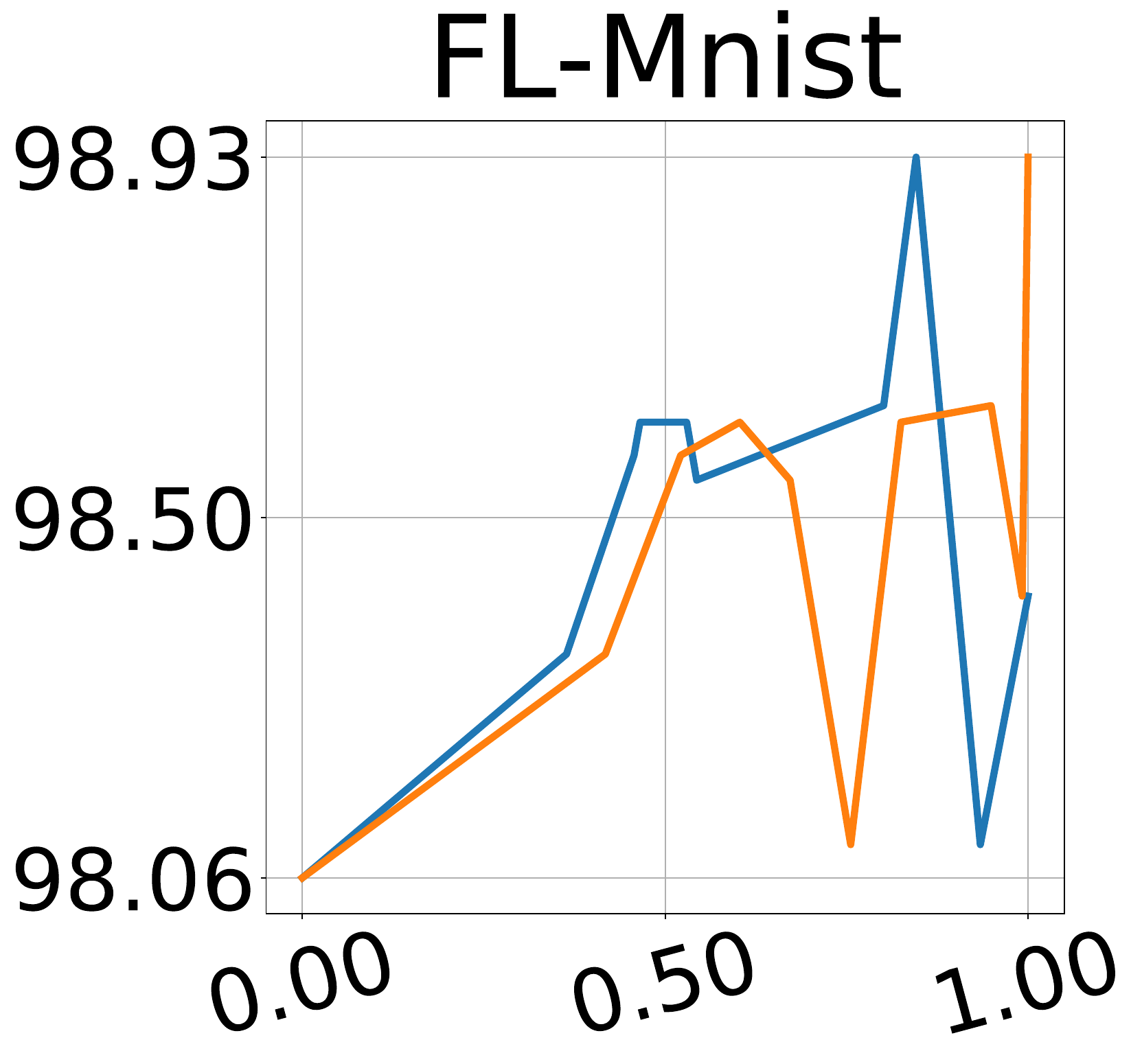}
        \includegraphics[width=0.2\columnwidth, height=0.2\columnwidth]{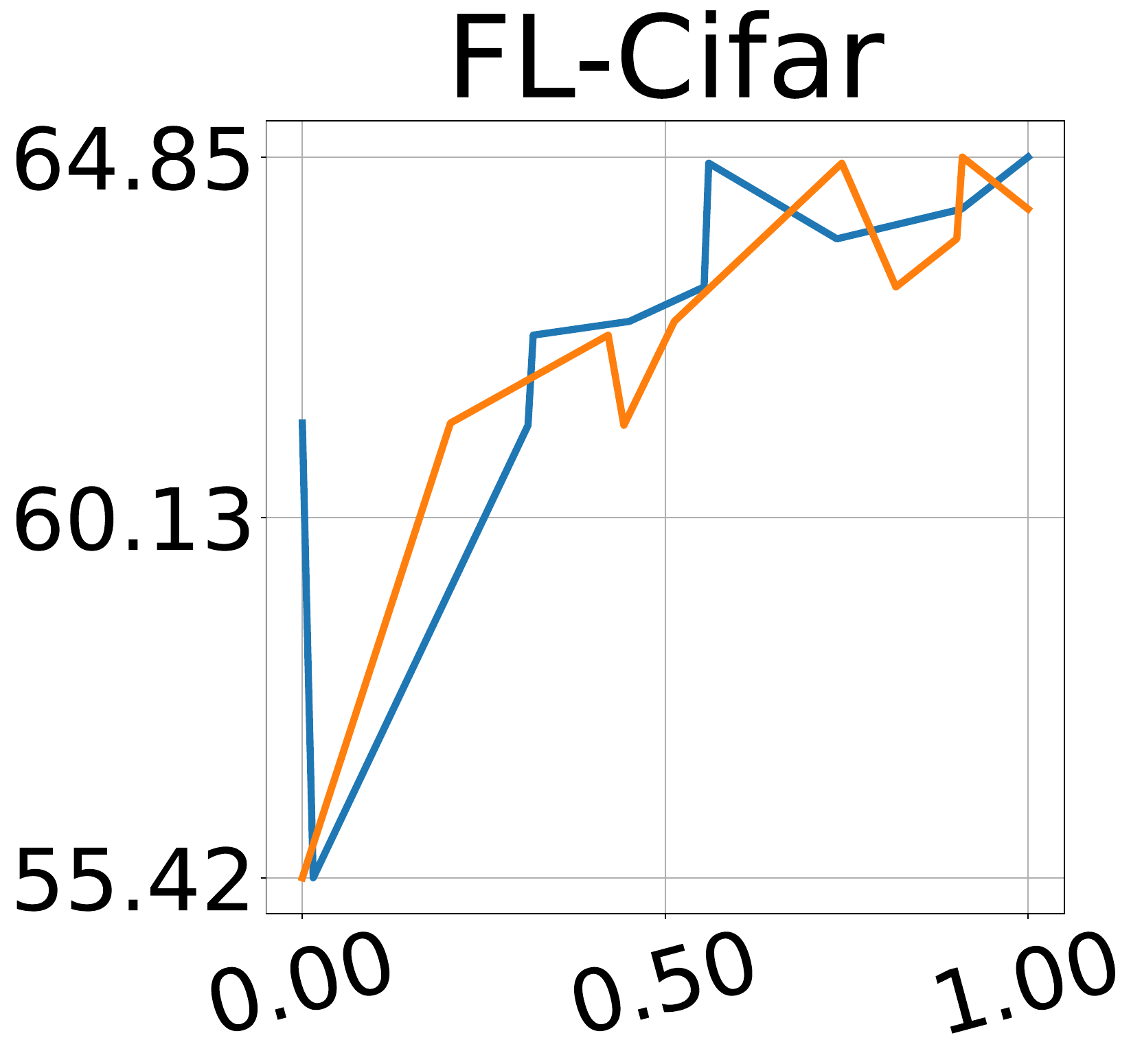}
        \includegraphics[width=0.2\columnwidth, height=0.2\columnwidth]{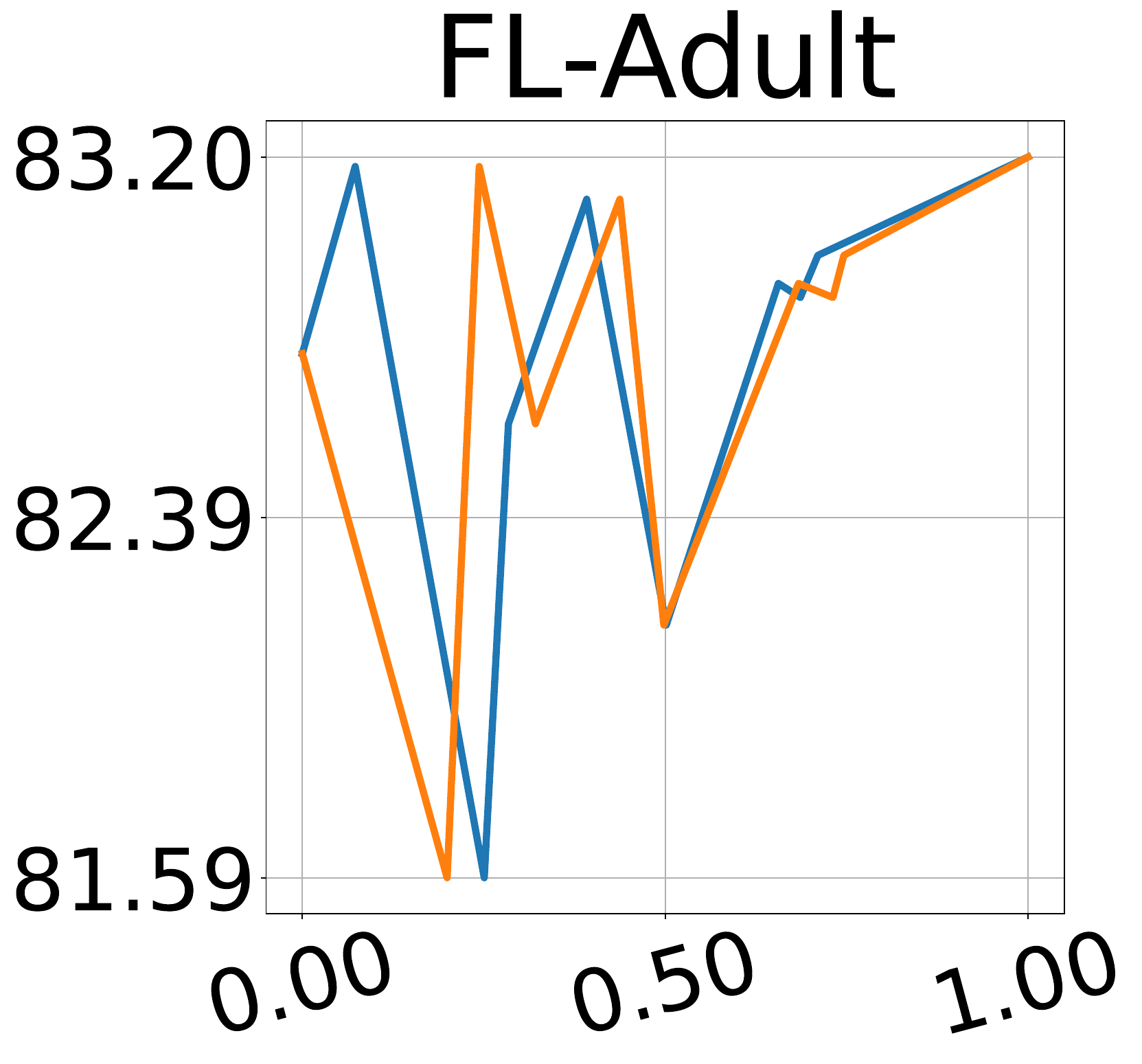}
        \includegraphics[width=0.2\columnwidth, height=0.2\columnwidth]{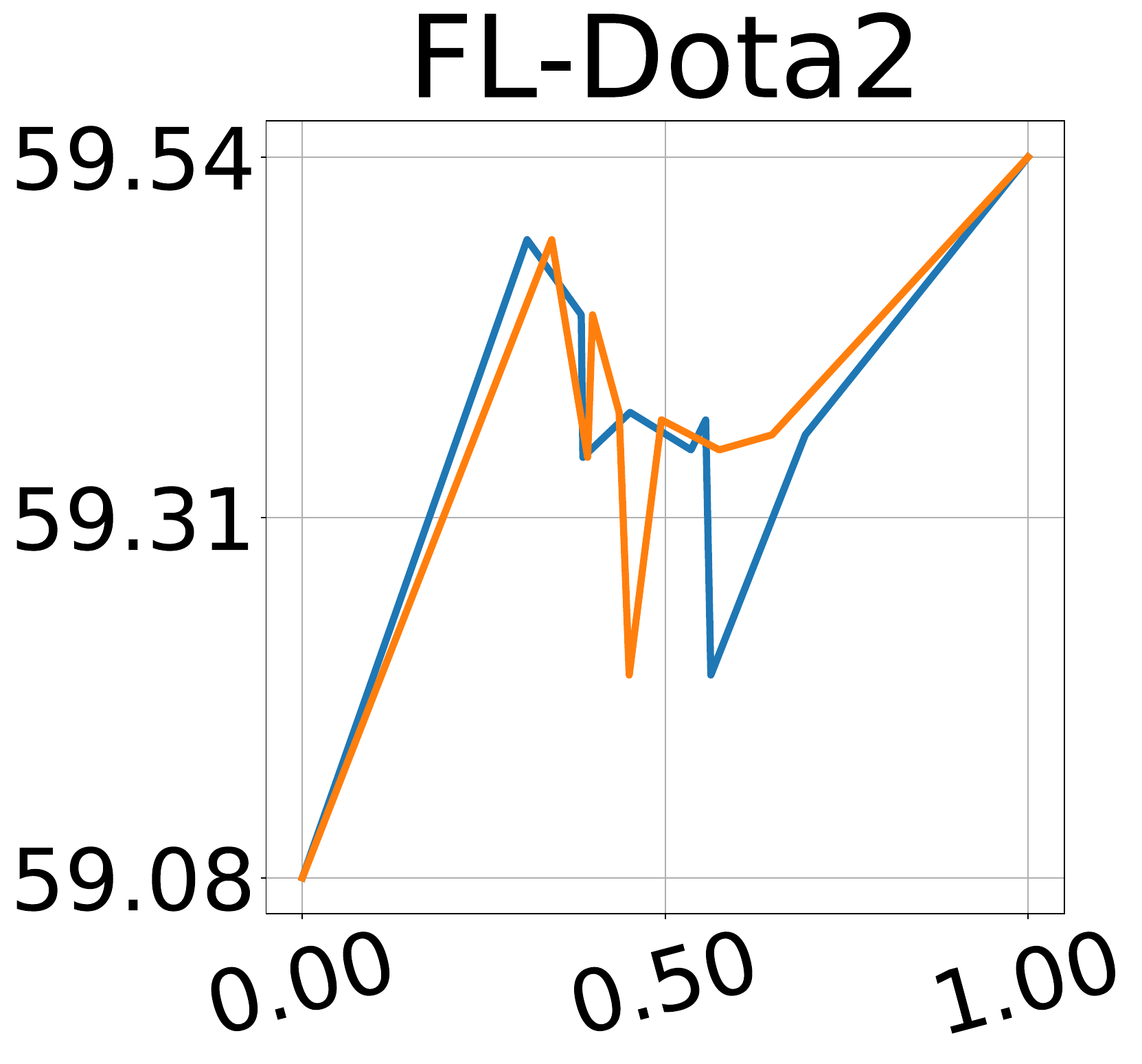}
        \includegraphics[width=0.2\columnwidth, height=0.2\columnwidth]{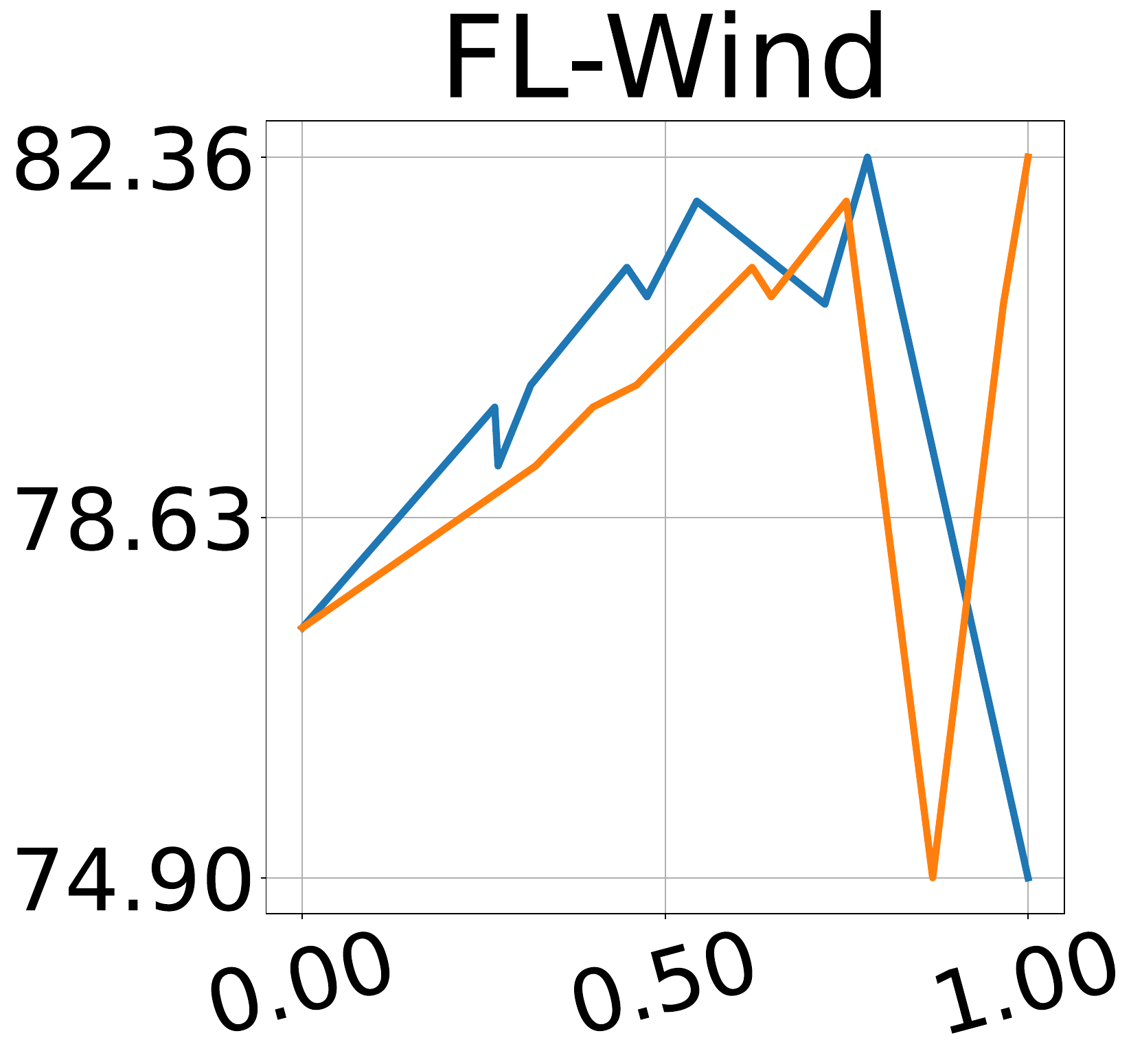}
        }  
    \caption{
    Removing or adding players (x-axis: SV of the removed/added player after min-max normalization). 
    }
    \label{fig:exp_remove_player}
\end{figure*}

\subsection{SV Interpretations} \label{subsec: exp_interpretability}
This section examines the correctness of the mainstream SV interpretation paradigm, i.e., utility-based, in explaining SVs of different types of players defined in the current DA domain. 
This paradigm points out that the larger the SV of a player, the more the impacts that the player could pose on the DA task's overall utility. Therefore, we observe the change in the overall utility of DA tasks caused by removing or adding players.

\Cref{fig:exp_remove_player} presents our observed results. 
These results show that, in all four types of DA tasks, the impacts of removing or adding a player on the task overall utilities (i.e., $U(\mathcal{N}\ \backslash \ \{p_i\})-U(\mathcal{N})$ or $U(\{p_i \})-U(\emptyset)$) fluctuate as the SV of the removed or added player, no matter whether the SV is exact or approximate. 
The fluctuating results contradict the mainstream SV interpretations, probably because the latter rely mainly on the average marginal contribution while neglecting the variance of the marginal contributions of each player.
Given a player in practical DA tasks, the variance of its marginal contributions might be much larger than its average contribution. 
For example, in DV-Wind, the variance of marginal contributions of a player ranges approximately from 11.83 to 18.00, which is much larger than the player's average marginal contribution, ranging from 0.06 to 1.43. 
Similarly, in DSV-2Dplanes, the variance ranges from 13.37 to 13.50, also larger than the average value in a range of $[0.13, 0.22]$. 
In both cases, we do not observe a strong correlation between SV and the player's contribution to a specific coalition (e.g., $\mathcal{N}$ in \Cref{fig:exp_remove_player}(a) or $\emptyset$ in \Cref{fig:exp_remove_player}(b)).
All these results indicate the necessity to consider the variance of marginal contributions when using the mainstream utility-based SV interpretations to make decisions on pricing, selection, weighting, and attribution of data. 

\textbf{Research Direction 6: In-depth investigation on key factors affecting SV interpretations.} 
SV may not be correctly interpreted or fail to serve its desired application purposes due to two points. 
One is \textit{the difference between the exact SV and the approximate SV}.
The other one is \textit{the mismatch between the DA task and the implicit assumption of SV}. For example, DV-Wind and DSV-2Dplanes tasks, as discussed above, mismatching with the assumption that the average contribution is appropriate to fairly distribute the overall utility of the task among players. 
Another example is the RI task in \Cref{fig:SV_in_DA}, where pixels in an image interact mutually to influence the model predictions on the label of this image, mismatching with the assumption that the contribution of any individual player to the cooperative game is independent of its interactions with other players. 
Future works mining the deep understanding of these factors under more types of DA tasks to enhance the effectiveness of SV are anticipated.

\textbf{Research Direction 7: Exploration on supports for complicated DA tasks.} 
The real-world DA tasks are complicated, containing \textit{mutually dependent players}  \cite{9117109, JMLR:v23:21-1413, AasNaglerJullumLøland+2021+62+81, AAS2021103502, frye2021shapley} and \textit{real-time dynamic updates} on the player set \cite{10623283,xia2025computing}. 
For example, the RI task in \Cref{fig:SV_in_DA}, where the influence of one pixel on the prediction from the image classification model tends to interrelate with nearby pixels. 
Besides, the dynamic real-time updates of the player set \cite{10623283,xia2025computing} are common in analytical tasks built upon the databases \cite{zhang2023dynamic,xia2025computing,padala2025tabshapleyidentifyingtopktabular}. However, most existing SV applications are ill-suited for these tasks due to the assumptions that the players are independent and the player set is static and immutable. 
Although conditional SV \cite{sv_related_survey} and dynamic SV \cite{zhang2023dynamic} have been devised for complicated tasks, these new types of SV do not treat the mutual dependency among players and their real-time dynamic updates \textit{as a single interrelated problem}, thus are limited in practical applications. It is expected that future work can devise more SV applications under the complicated yet realistic DA task settings.

\section{Conclusion} \label{sec: conclusion}
This paper comprehensively studied the Shapley value applied throughout the data analytics workflow. We summarized the critical variables (i.e., the player and the utility function) in designing SV applications for DA and clarified the essential functionalities of SV for data scientists. We condensed the technical challenges of applying SV in DA and discussed the related arts, qualitatively and quantitatively. The conclusions of experimental evaluations based on our development framework, \textit{SVBench}, support our findings in a synthetic review. At last, we identified the limitations of current efforts and offered insights into the directions of future work.

\begin{acks}
  This work was funded by National Key Research and Development Program of China (Grant No: 2022YFB2703100), the Pioneer R\&D Program of Zhejiang (No. 2024C01021), and Zhejiang Province ``Leading Talent of Technological Innovation Program'' (No. 2023R5214). 
  Meihui Zhang was funded by National Natural Science Foundation of China (U2441237) and the Open Research Fund of The State Key Laboratory of Blockchain and Data Security, Zhejiang University. 
  We also appreciate anonymous reviewers for their valuable feedback and thank Jiaqi Chai for proofreading this paper. 
\end{acks}


\bibliographystyle{ACM-Reference-Format}
\bibliography{sample}

\end{document}